\documentclass[final,1p,times,number]{elsarticle}

\usepackage{hyperref}
\usepackage{physics}
\usepackage{amsmath}
\usepackage{comment}
\usepackage{float}
\usepackage{units}
\usepackage{amssymb}
\usepackage[table]{xcolor}

\usepackage{caption}
\usepackage[caption=false]{subfig}
\usepackage{ytableau}

\usepackage{bbm}
\usepackage{mathtools}

\usepackage{graphicx, placeins, caption}
\usepackage{amsfonts, dsfont}
\usepackage[utf8]{inputenc}
\usepackage{color}

\newsavebox\MBox

\usepackage{amsfonts, dsfont} 

\newcommand{\be}{ \begin{equation} }
\newcommand{\ee}{\end{equation}}
\newcommand{\bea}{ \begin{eqnarray} }
\newcommand{\eea}{\end{eqnarray}}

\newcommand\calphare{\cellcolor{green!10}}
\newcommand\calphaim{\cellcolor{green!30}}
\newcommand\cbetare{\cellcolor{red!10}}
\newcommand\cbetaim{\cellcolor{red!30}}
\newcommand\cgamma{\cellcolor{blue!20}}
\newcommand\caplus{\cellcolor{yellow!40}}
\newcommand\caminus{\cellcolor{yellow!20}}

\journal{Annals of Physics}

\newcommand{\bg}{ \begin{gather} }
\newcommand{\eg}{\end{gather}}

\renewcommand{\Re}{\mathop{\rm Re}}
\renewcommand{\Im}{\mathop{\rm Im}}

\newcommand{\mfg}{\mathfrak g}
\newcommand{\mfh}{\mathfrak h}
\newcommand{\mfn}{\mathfrak n}
\newcommand{\mfp}{\mathfrak p}
\newcommand{\mfk}{\mathfrak k}
\newcommand{\mfa}{\mathfrak a}

\begin{document}

\begin{frontmatter}

\title{Generalized multifractality at spin quantum Hall transition}

\author{Jonas F.~Karcher}
\address{{Institute for Quantum Materials and Technologies, Karlsruhe Institute of Technology, 76021 Karlsruhe, Germany}}
\address{{Institut f\"ur Theorie der Kondensierten Materie, Karlsruhe Institute of Technology, 76128 Karlsruhe, Germany}}

\author{Noah Charles}
\address{Reed College, Department of Physics, 3203 Southeast Woodstock Boulevard, Portland, OR, 97202, USA}
\address{Ohio State University, Department of Physics, 191 West Woodruff Ave, Columbus OH, 43210, USA}

\author{Ilya A.~Gruzberg}
\address{Ohio State University, Department of Physics, 191 West Woodruff Ave, Columbus OH, 43210, USA}

\author{Alexander D.~Mirlin}
\address{Institute for Quantum Materials and Technologies, Karlsruhe Institute of Technology, 76021 Karlsruhe, Germany}
\address{Institute for Condensed Matter Theory, Karlsruhe Institute of Technology, 76128 Karlsruhe, Germany}
\address{L.\,D.~Landau Institute for Theoretical Physics RAS, 119334 Moscow, Russia}
\address{Petersburg Nuclear Physics Institute,188300 St.\,Petersburg, Russia.}

\begin{abstract}
Generalized multifractality characterizes scaling of eigenstate observables at Anderson-localization critical points. We explore generalized multifractality in 2D systems, with the main focus on the spin quantum Hall (SQH) transition in superconductors of symmetry class C. Relations and differences with the conventional integer quantum Hall (IQH) transition are also studied. Using the field-theoretical formalism of non-linear sigma-model, we derive the pure-scaling operators representing generalizing multifractality and then ``translate'' them to the language of eigenstate observables. Performing numerical simulations on network models for SQH and IQH transitions, we confirm the analytical predictions for scaling observables and determine the corresponding exponents. Remarkably, the generalized-multifractality exponents at the SQH critical point strongly violate the generalized parabolicity of the spectrum, which implies violation of the local conformal invariance at this critical point.

\end{abstract}


\end{frontmatter}

\tableofcontents

\section{Introduction}
\label{sec:introduction}
Anderson localization in disordered systems belongs to the most fundamental phenomena in condensed matter physics  \cite{anderson58,50_years_of_localization}.
In particular, Anderson transitions between localized and delocalized phases (or between topologically distinct localized phases) \cite{evers08} attract much attention. This interest was additionally enhanced by development of the symmetry classification of disordered systems \cite{altland1997nonstandard,zirnbauer1996riemannian,heinzner2005symmetry}, which has further extended the scope of Anderson transitions (in particular, to the area of disordered superconductors); see \cite{evers08} for review.

A remarkable hallmark of critical points of Anderson transitions is the multifractality characterizing the statistics of eigenfunctions and of the local density of states (LDOS). Multifractality of critical eigenstates at various Anderson-transition critical points has been extensivlely explored analytically and numerically \cite{evers08}.  It was demonstrated that the multifractality can be efficiently employed to determine the position of the critical point of the Anderson transition and the critical index of the localization length \cite{rodriguez2010critical, rodriguez2011multifractal}.  It was also shown that the multifractality survives  in the presence of Coulomb interaction \cite{burmistrov2013multifractality}. Recently, the multifractality was used as an efficient tool to detect and explore the emergent criticality of 2D surface states of topological superconductors \cite{ghorashi2018critical,ghorashi2020criticality,sbierski2020spectrum-wide,karcher2021how}.

Direct experimental measurement of multifractality is a highly non-trivial task. Experimental observations of multifractality near metal-insulator transition were reported for sound waves \cite{faez2009observation} and electrons in disordered semiconductors   \cite{richardella2010visualizing}. Strong fluctuations of the local density of states qualitatively analogous to multifractality were observed in experiments on superconductor-insulator transition in disordered films \cite{sacepe2008disorder-induced,noat2013unconventional}.

Recent works on interplay of multifractality and interactions have additionally emphasized the importance of multifractality. In particular, it was predicted that the multifractality may enhance the critical temperature $T_c$ of superconductiors in three-dimensional \cite{feigel2007eigenfunction,feigel2010fractal,garcia-garcia2020superconductivity} and two-dimensional (2D)
\cite{burmistrov2012enhancement,burmistrov2015superconductor,mayoh2015global,fan2020enhanced,burmistrov2021multifractallyenhanced} systems.
The multifractal enhancement of $T_c$ in 2D superconductors was indeed observed experimentally  \cite{zhao2019disorder-induced,rubio-verdu2020visualization}.
Further, it was shown that the multifractality drives instabilities of surface states of topological superconductors \cite{foster2012interaction,foster2014topological} and leads to broad distribution of Kondo temperatures in disordered metals with magnetic impurities  \cite{kettemann2012kondo,kettemann2009critical,slevin2019multifractality}.

It was shown in Refs.~\cite{mirlin94a,mirlin94b,fyodorov2004statistics,savin2005universal,fyodorov2005scattering} that, within the non-linear $\sigma$ model (which constitutes the field theory of Anderson localization),  the distribution of LDOS in Wigner-Dyson classes possesses a symmetry relating the probabilities of values $\nu$ and $\nu^{-1}$ (with the average LDOS normalized to unity).
This was used in Ref.~\cite{mirlin2006exact} to derive exact symmetry relations between multifractal exponents at Anderson transitions in the Wigner-Dyson classes.
It was later shown in Ref.~\cite{gruzberg2011symmetries} that these relations can be understood as a manifestation of Weyl-group symmetry associated with the $\sigma$-model manifold and can
be extended to unconventional symmetry classes. In Ref.~\cite{gruzberg2013classification}, a general classification of composite operators without gradients was developed, which extends that  of Refs.~\cite{hoef1986calculation,wegner1987anomalous1,wegner1987anomalous2}. Further, in Ref.~\cite{gruzberg2013classification} multiple symmetry relations between scaling exponents of composite operators were derived, which all follow from the Weyl-group symmetry. Finally,  Ref.~\cite{gruzberg2013classification}  determined a ``translation'' of these composite operators
to eigenfunction correlation functions for the symmetry class A (unitary Wigner-Dyson class).

The scaling of the whole set of composite operators characterizing critical eigenstates extends the conventional notion of multifractality (which refers only to leading exponents $\tau_q$ characterizing moments $|\psi({\bf r})|^{2q}$ of eigenstate amplitude). We will thus term it {\it ``generalized multifractality''}.  The subleading multifractal exponents manifest themselves
 in dephasing and broadening of localization transition in an interacting system with a short-range interaction that is irrelevant in the renormalization-group sense \cite{lee1996effects,wang2000short-range,burmistrov2011wave}.

The spatial dimensionality $d=2$ plays a special role in the context of Anderson localization.  In the most conventional symmetry class AI (orthogonal Wigner-Dyson class), $d=2$ is the lower critical dimension, so that a 2D system is always in the localized phase (although with an exponentially large localization length for weak disorder). The situation is, however, different in other symmetry classes, with a wealth of 2D Anderson-localization critical points \cite{evers08}. Emergence of these critical points is related either to peculiarities of the perturbative expansion (e.g., antilocalization in Wigner-Dyson symplectic class AII) or to topology. A paradigmatic example of the localization-transition critical point associated with topology is the plateau transition in the quantum Hall effect (which belongs to the Wigner-Dyson unitary class A).

Multifractality at the quantum-Hall plateau transition and, more generally, physics of the corresponding fixed point, has been attracting much interest. On the numerical side,  the multifractal spectrum is very close to parabolicity \cite{evers2001multifractality}; however, a high-precision numerics indicates small deviations \cite{Obuse-Boundary-2008, evers2008multifractality}. Many papers have attacked the problem of the character of the critical theory
\cite{zirnbauer1994towards, zirnbauer1997toward, janssen1999point-contact, zirnbauer1999conformal, kettemann1999information, bhaseen2000towards, tsvelik2001wave, tsvelik2007evidence, zirnbauer2019integer}; in particular, various versions of the Wess-Zumino-Novikov-Witten (WZNW) theory were conjectured. Quite generally, such WZNW-type theories lead to a parabolic multifractality spectrum as for free Gaussian fields.
Another important argument in favor of parabolic multifractality was provided by Ref.~\cite{bondesan2017gaussian}. In was shown in that work that, for 2D Anderson transitions, the assumption of local conformal invariance, in combination with an assumption of Abelian fusion of composite operators, leads to an exactly parabolic multifractal spectrum of moments of the local density of states: the exponent $\tau_q$ is a quadratic function of $q$.

The present paper deals with a superconducting counterpart of the quantum Hall transition---the spin quantum Hall (SQH) transition   \cite{kagalovsky1999quantum, senthil1999spin}. It was found that mapping to percolation allows one to obtain some critical exponents for the SQH transition exactly \cite{gruzberg1999exact, beamond2002quantum, mirlin2003wavefunction}. The multifractality spectrum at the SQH critical point was studied numerically in Refs.~\cite{evers2003multifractality, mirlin2003wavefunction} and very recently in Ref.~\cite{puschmann2021quartic}. It was found \cite{mirlin2003wavefunction, puschmann2021quartic} that the spectrum exhibits clear (although relatively weak) deviations from parabolicity. These deviations are very interesting physically since, in combination with results of Ref.~\cite{bondesan2017gaussian}, they hint to a possible violation of local conformal invariance.

In this work, we present a detailed analysis of the generalized multifractality at the SQH critical point, supporting and complementing analytical study by numerical simulations. Our key results are as follows:

\begin{enumerate}[(i)]

\item We explore implications of local conformal invariance in 2D systems for generalized multifractality. Generalizing the result of Ref.~\cite{bondesan2017gaussian}, we show that, under the assumptions of local conformal invariance and Abelian fusion, the generalized-multifractality spectrum exhibits ``generalized parabolicity'' and is parametrized (for a given symmetry class) by a single constant.

\item We develop explicitly two constructions of pure-scaling composite operators for $\sigma$-model of class C, invoking heighest-weight vectors and the Iwasawa decomposition, respectively. We demonstrate that these composite operators obey Abelian fusion rules.

\item We also determine pure-scaling operators in an alternative form (invariant with respect to the action of the symmetry group of the $\sigma$-model), by using one-loop renormalization group.

\item We perform a ``translation'' of the scaling operators to the language of eigenstates of the Hamiltonian and determine explicit expressions for particularly interesting eigenstate correlators exhibiting generalized multifractality. Using the network model of class C, we numerically verify that these are indeed the proper scaling operators and determine the corresponding exponents. Very remarkably, we find a {\it strong violation} of the generalized parabolicity of the generalized-multifractality spectrum. In combination with the results proven in (i) and (ii), this points out to a violation of the local conformal invariance at the SQH transition. While this is a rather surprising conclusion, we do not see any alternative plausible explanation of the numerical findings.

\end{enumerate}

The results in (ii) and a part of analytical results in (iv) represent an extension of the class-A construction of Ref.~\cite{gruzberg2013classification} to class C. It should be emphasized, however, that this extension is far from trivial, and the situation in class C turns out to be much richer than in class A. This complexity can be traced back to the additional particle-hole degree of freedom (referred as  ``spin'' in the term ``spin quantum Hall effect'') .

The structure of the paper is as follows. In Sec.~\ref{sec:general} we introduce the framework  of generalized multifractality at Anderson-localization transitions and discuss its connections with conformal field theory. Section \ref{sec:CFT-2D}  focusses on the relation between the generalized multiftactality and conformal  invariance in 2D systems. Its central result is the exact generalized parabolicity of the generalized multifractal spectra under the assumptions of local conformal invariance and Abelian fusion. In Sec.~\ref{sec:sigma} we present a derivation of the non-linear sigma model of class C that is used as the field-theoretical framework in the rest of the paper. In Sections \ref{sec:hwv}  and \ref{sec:iwasawa} we develop two constructions of pure-scaling composite operators of class-C sigma-model (based on highest-weight vectors and on the Iwasawa decomposition, respectively) that realize Abelian fusion. In Sec.~\ref{sec:rg_c} we use the one-loop renormalization group (RG) to derive another class of pure-scaling composite operators of the class-C sigma-model. These operators turn out to be particularly convenient for mapping onto wave-function observables and for establishing relations between class-C and class-A pure-scaling observables. In Sec.~\ref{sec:wave_ops} we carry out the translation of generalized-multifractality scaling observables from the sigma-model language to that of eigenstates. Performing numerical simulations on appropriate network models, we confirm the analytical predictions for the structure of eigenstate observables. Further, we determine numerical values of various exponents characterizing the generalized multifractality, for both the SQH and the IQH transitions. 
The numerical results on SQH generalized multifractality are corroborated and extended in Sec.~\ref{sec:wave_iw} where a complementary computational approach is employed (using total-density observables at variance with those for a single spin projection in Sec.~\ref{sec:wave_ops}). An important conclusion of our numerics is that the generalized parabolicity of the generalized-multifractality spectrum is strongly violated at the SQH transition point. In combination with results of Sections \ref{sec:CFT-2D}  and \ref{sec:iwasawa}, this implies violation of the local conformal invariance at the SQH critical point.
Section \ref{sec:summary} contains a summary of our findings and a short discussion of further research directions that are opened by this work.

\section{General framework}
\label{sec:general}
In this section, we present a general discussion of multifractality at Anderson transitions (ATs), including definitions of multifractal (MF) observables, their correlation functions,  the MF spectra of scaling dimensions, and relation to the field-theoretical renormalization group (RG). One central issue for us is a description of MF correlation functions in terms of conformal field theory (CFT). As discussed below, such a description in general does not hold, in view of the system-size dependence of the MF correlation functions. At the same time, a certain subclass of MF correlators satisfying a ``neutrality'' condition \eqref{neutrality} is consistent with conformal invariance.

\subsection{Multifractal exponents, their symmetries, and field theory}
\label{subsec:MF-symmetry}

A remarkable property of ATs is that critical wave
functions exhibit strong fluctuations and are characterized by an infinite set of {\it multifractal} (MF) scaling exponents. Specifically, the wave-function moments show anomalous MF scaling with respect to the system size $L$,
\begin{align}
\label{e1.1}
L^d \overline{|\psi(r)|^{2q}} &\propto L^{-\tau_q},
&
\tau_q &= d(q-1) +  \Delta_q,
\end{align}
where $d$ is the spatial dimension, the overline denotes the operation of disorder averaging, and $\Delta_q$ are anomalous MF exponents that distinguish the critical point from a simple metallic phase (where $\Delta_q \equiv 0$). General properties of the function $\tau_q$ follow from its definition: $\tau_q$ is a non-decreasing ($\tau_q' \geq 0$), convex function ($\tau_q'' \leq 0$) with $\tau_0 = -d$ and $\tau_1 = 0$ (equivalently, $\Delta_0 = \Delta_1 = 0$).

Closely related MF exponents describe the scaling of moments of the local density of states (LDOS) $\nu(r)$:
\begin{align}
\label{e1.2}
\overline{\nu^q(r)} &\propto L^{-x_q} ,
&
x_q &= \Delta_q + qx_1,
\end{align}
where $x_1$ describes the scaling of the average LDOS: $ \overline{\nu(r)} \propto L^{-x_1}$, and $x_0 = 0$. An exact symmetry relation for the MF exponents
\begin{equation}
 x_q = x_{q_*-q}
\label{symmetry-xq}
\end{equation}
was obtained (with $q_* = 1$ for Wigner-Dyson classes) in Ref.~\cite{mirlin2006exact}.

Multifractality implies the presence of infinitely many relevant (in the RG sense) operators at the fixed point of an AT. The negative dimensions $x_q$ of these operators for sufficiently large $|q|$ require us to consistently work in a finite system of size $L$ and use this size as the scaling variable. The physical meaning of negative dimensions is that the intensity $|\psi^2|$ of a critical wave function is not self-averaging, and its distribution $P(|\psi^2|)$ becomes broader and broader when the systems size $L$ grows, so that sufficiently high  moments of this distribution diverge in the limit $L \to \infty$. Let us emphasize that when we speak about a $q$-th moment, we neither require that $q$ is an integer nor that it is positive. Throughout the paper, the term ``moment'' is understood in this broad sense.

Relations between general MF spectra and field theory were considered by Duplantier and Ludwig in Ref. \cite{Duplantier-Multifractals-1991}, where the authors stressed a profound difference between convexity of the MF spectra and concavity of spectra of scaling dimensions in a unitary field theory. For single-particle disordered systems, moments of the LDOS and wave functions are related to certain operators in a $\sigma$ model \cite{mirlin00}:
\begin{align}
\label{DOS field theory operator relation}
\nu^q(r) \sim L^{dq} \nu^q \, |\psi(r)|^{2q} \sim \mathcal{O}_q(r) \,,
\end{align}
where $\nu$ (without spatial argument) is the global density of states, $ \nu = \overline{\nu(r)}$. This relation is understood as follows: correlation functions of operators $\mathcal{O}_q(r)$ in  the $\sigma$ model give the disorder averages of products of powers of the local DOS (wave function moments) at different points:
\begin{align}
 \overline{\nu^{q_1}(r_1) \ldots \nu^{q_n}(r_n)}
= \big\langle \mathcal{O}_{q_1}(r_1) \ldots \mathcal{O}_{q_n}(r_n) \big\rangle.
\end{align}
Here the angular brackets denote a field-theory expectation value.

Postponing details to later sections, let us mention here important properties of the operators $\mathcal{O}_q$ that result from their explicit construction and the symmetry relation \eqref{symmetry-xq}:
\begin{enumerate}
  \item $\mathcal{O}_q$ are exact scaling operators in the $\sigma$ model at its critical fixed point, with scaling dimensions $x_q$:
\begin{align}
\big\langle\mathcal{O}_q(r) \big\rangle \sim L^{-x_q}.
\label{1-point}
\end{align}

    \item $\mathcal{O}_{q_*}$ has vanishing scaling dimension
\begin{align}
x_{q_*} &= 0, & \big\langle\mathcal{O}_{q_*}(r) \big\rangle = 1,
\label{xq-star-0}
\end{align}
even though it describes the scaling of a nontrivial moment of the LDOS, and, thus, is different from the identity operator $\mathcal{O}_0$.

    \item $\mathcal{O}_q$ satisfy the {\it Abelian fusion} $\mathcal{O}_{q_1} \mathcal{O}_{q_2} \sim \mathcal{O}_{q_1+q_2}$.

\end{enumerate}
Making a natural assumption that there is an operator product expansion (OPE) of the MF scaling operators, we get the OPE consistent with the Abelian fusion:
\begin{align}
\label{Abelian-OPE}
\mathcal{O}_{q_1}(r_1) \mathcal{O}_{q_2}(r_2) & \, \propto \,  r_{12}^{x_{q_1 + q_2} - x_{q_1} - x_{q_2}} \mathcal{O}_{q_1+q_2} \Big(\dfrac{r_1+r_2}{2}\Big) + \ldots,
& r_{12} &= |r_1 - r_2|.
\end{align}
Here the symbol $+ \ldots$ denotes the contribution of descendant operators, i.e., those involving spatial derivatives. The manifestation of the Abelian fusion is that the only operator out of the set of  $\mathcal{O}_q$  (primary fields in the CFT language) that enters the r.h.s. of Eq.~\eqref{Abelian-OPE} is $\mathcal{O}_{q_1+q_2}$.

Using the $\sigma$ model mapping, two of us and co-authors showed that the symmetry relation \eqref{symmetry-xq} follows from the symmetry of the $\sigma$-model manifold under a Weyl group \cite{gruzberg2011symmetries}. Furthermore, the group-theoretic approach allowed us to generalize the relation \eqref{symmetry-xq} to the unconventional (Bogoliubov-de Gennes) classes CI and C, with $q_*=2$ and $q_*=3$, respectively \cite{gruzberg2011symmetries}.

The operators $\mathcal{O}_q$ form only a subset of all possible gradientless scaling operators in a $\sigma$ model. They are distinguished in that they represent the simplest observables---moments of LDOS  (or of wave-function intensities). Furthermore, they are the dominant (or most relevant) operators for each $q $ (the total power of wave function intensities) if one considers only expressions containing positive powers of amplitudes.  At the same time, $\mathcal{O}_q$  only represent ``the tip of the iceberg'' of a much larger family of gradientless composite operators. In Ref.~\cite{gruzberg2013classification}, two of us with M. Zirnbauer have developed a complete classification of exact gradientless scaling operators. We have also explicitly constructed a complete set of these operators and of the corresponding wave-function observables for symmetry class A. These generalized MF operators $\mathcal{O}_\lambda$ an their scaling dimensions $x_\lambda$ are labeled by multiple indices $q_i$ that together form a highest weight $\lambda = (q_1, q_2, \ldots, q_n)$ under the action of the Lie algebra of $G$. [In the special case when all $q_i$ are positive integers satisfying $q_1 \ge q_2 \ge \ldots \ge q_n \ge 0$, the multi-index $\lambda$ is an integer partition or a Young tableau.]

The operators $\mathcal{O}_\lambda$ satisfy an extension of the Abelian OPE \eqref{Abelian-OPE}:
\begin{align}
\label{Abelian-OPE-2}
\mathcal{O}_\lambda(r_1) \mathcal{O}_{\lambda'}(r_2) & \, \propto \,  r_{12}^{x_{\lambda + \lambda'} - x_\lambda - x_{\lambda'}} \mathcal{O}_{\lambda + \lambda'} \Big(\dfrac{r_1+r_2}{2}\Big) + \ldots
\end{align}

To avoid confusion, it is worth pointing out that $\lambda = (q_1, q_2, \ldots, q_n)$ labels irreducible representations of $G$, and there are infinitely many operators belonging to each such representation. The scaling exponent $x_\lambda$ is a property of the representation and therefore is the same for all operators in a given representation. At the same time, the OPE \eqref{Abelian-OPE-2} will in general contain in the r.h.s. an admixture of other gradientless operators if one chooses arbitrary operators from each representation. The fact that one can find $\mathcal{O}_\lambda$ operators such that they satisfy Abelian OPE (with $+ \ldots$ staying only for descendant operators, i.e., those with gradients) is very non-trivial. We will discuss it in more detail below for class C.

The Weyl group $W$ acts in the weight space, and the generalized MF dimensions $x_\lambda$ are invariant under this action:
\begin{align}
x_{w \lambda} &= x_\lambda, & \forall w \in W.
\end{align}
The important generators of the Weyl group are reflections in certain hyperplanes in the weight space. Their action on the weight $\lambda = (q_1, q_2, \ldots, q_n)$ in terms of its components $q_j$ is as follows:
\begin{enumerate}
\item[(i)] sign inversion of ${\tilde q}_j \equiv q_j + c_j/2$
for any $j\in \{1, 2, \ldots, n \}$:
\begin{equation}
 q_j \to - c_j - q_j.
 \label{Weyl-sign}
\end{equation}
A consequence of this action is the existence, for each $n$, of operators   $\mathcal{O}_{-\rho_b} = \mathcal{O}_{(-c_1, -c_2,\ldots, - c_n)}$ with vanishing scaling dimension
\begin{align}
x_{-\rho_b} = x_{(-c_1, -c_2,\ldots, - c_n)} = 0.
\label{x-rho-b}
\end{align}

\item[(ii)] interchange of ${\tilde q}_i = q_i + c_i/2$ and ${\tilde q}_j = q_j + c_j/2$ for some pair $i, j \in \{1, 2, \ldots, n \}$:
\begin{align}
q_i &\to q_j + \frac{c_j - c_i}{2}; & q_j &\to q_i + \frac{c_i - c_j}{2}.
\label{Weyl-interchange}
\end{align}
\end{enumerate}
The parameters $c_i$ are the coefficients of the expansion of the bosonic part
$\rho_b = \sum_{j=1}^{n} c_j e_j$
of the half-sum of the positive roots $\rho$ in a standard basis $e_j$. These coefficients are known for all families of symmetric superspaces. In particular,  $c_j = 1-2j$ for class A, and $c_j = 1-4j$ for class C. The usual MF dimensions $x_q$ corresponds to $\lambda = (q, 0, 0, \ldots, 0)$ [or, in a shorter equivalent notation, to $\lambda = (q)$], and $c_1 = - q_*$.

\subsection{Multifractal multipoint functions and RG}
\label{subsec:MF-RG}

Equation (\ref{1-point}) can be understood from an RG point of view. We need to run the RG from the microscopic scale $a$ up to the scale $L$, and a one-point function at this scale becomes a number of order one. The result of the RG is the appearance of the scale factor $L/a$ raised to the power $-x_q$, which is exactly Eq.~(\ref{1-point}). Similar arguments determine the behavior of multi-point functions. Let us consider the two-point function $\big\langle \mathcal{O}_{q_1}(r_1) \mathcal{O}_{q_2}(r_2) \big\rangle$. Now we run the RG up to the scale $r_{12} = |r_1 - r_2|$, which results in the renormalization factor $r_{12}^{-x_{q_1} - x_{q_2}}$. At this scale the two operators fuse to $\mathcal{O}_{q_1 + q_2}$. Then we renormalize further up to scale $L$, which results in the additional factor $(L/r_{12})^{-x_{q_1 + q_2}}$. Finally, at this scale the correlator is of order unity, and we get
\begin{align}
\label{2-point-1}
\big\langle \mathcal{O}_{q_1}(r_1) \mathcal{O}_{q_2}(r_2) \big\rangle \sim
r_{12}^{-x_{q_1} - x_{q_2}} \bigg(\frac{L}{r_{12}}\bigg)^{-x_{q_1 + q_2}}
= r_{12}^{x_{q_1 + q_2} - x_{q_1} - x_{q_2}} L^{-x_{q_1 + q_2}}.
\end{align}
The same result is obtained from the Abelian OPE \eqref{Abelian-OPE} combined with Eq. (\ref{1-point}) applied to the expectation value of $\mathcal{O}_{q_1+q_2}$.

Next consider the three-point function $\big\langle \mathcal{O}_{q_1}(r_1) \mathcal{O}_{q_2}(r_2) \mathcal{O}_{q_3}(r_3) \big\rangle$. To apply the RG argument, we need to know the hierarchy of distances between the three points. For example, let us assume that $r_{12}$ is the smallest distance:
$r_{12} \ll r_{13} \simeq r_{23} \ll L$.
Then there will be three stages of RG and two fusions resulting in
\begin{align}
\label{3-point-1}
&\big\langle \mathcal{O}_{q_1}(r_1) \mathcal{O}_{q_2}(r_2) \mathcal{O}_{q_3}(r_3) \big\rangle \sim
r_{12}^{-x_{q_1} - x_{q_2} - x_{q_3}} \big\langle \mathcal{O}_{q_1 + q_2}(r_1/r_{12}) \mathcal{O}_{q_3}(r_3/r_{12}) \big\rangle
\nonumber \\
&\sim r_{12}^{-x_{q_1} - x_{q_2} - x_{q_3}} \bigg(\frac{r_{13}}{r_{12}}\bigg)^{-x_{q_1 + q_2} - x_{q_3}} \bigg(\frac{L}{r_{13}}\bigg)^{-x_{q_1 + q_2 + q_3}}
= r_{12}^{x_{q_1 + q_2} - x_{q_1} - x_{q_2}}
r_{13}^{x_{q_1 + q_2 + q_3} - x_{q_1 + q_2} - x_{q_3}} L^{-x_{q_1 + q_2 + q_3}}.
\end{align}
Similar expressions are obtained in other cases when the distances between the points satisfy different inequalities. It is also easy to generalize these expressions to higher multipoint functions.
In particular, it should be clear from the discussion above that a generic $n$-point MF function $\big\langle \prod_{i=1}^{n}\mathcal{O}_{q_i}(r_i) \big\rangle$,
when considered as a function of system size $L$ at fixed pairwise distances $r_{ij}$  (all much smaller than $L$)  will scale with $L$ as
\begin{align}
\label{n-point}
\big\langle \mathcal{O}_{q_1}(r_1) \ldots \mathcal{O}_{q_n}(r_n) \big\rangle \propto L^{-x_{q_1 + \cdots + q_n}}.
\end{align}

\subsection{CFT correlators}

We want to see if MF multi-point correlation function can be related to correlation functions in a conformal field theory (CFT) \cite{Yellow-book}. To this end, let us summarize some CFT results.

A basic notion of CFT is that of a quasiprimary operator $\phi(r)$. Such operators transform in a specific way under global conformal transformations, and the results summarized in this section are derived assuming the quasiprimary nature of all operators involved. In a (unitary) CFT the expectation value of any nontrivial quasiprimary field with scaling dimension $x > 0$ in the infinite space is zero:
\begin{align}
\label{1-point-CFT}
\big\langle \phi(r) \big\rangle_\text{CFT} = \delta_{x,0}.
\end{align}
Only the identity operator $\mathbbm{1}$ has vanishing dimension and $\big\langle \mathbbm{1} \big\rangle_\text{CFT} = 1$.

In the presence of translation, rotation, and scale invariance, two-point functions of fields with dimensions $x_1$ and $x_2$ are
\begin{align}
\label{2-point-scale}
\big\langle \phi_1(r_1) \phi_2(r_2) \big\rangle &\sim r_{12}^{-x_1 - x_2}.
\end{align}
If one additionally assumes invariance under special conformal transformations, the two-point function is restricted further, it vanishes unless the dimensions of the two operators are equal:
\begin{align}
\label{2-point-CFT}
\big\langle \phi_1(r_1) \phi_2(r_2) \big\rangle_\text{CFT}
\sim r_{12}^{-2x_1} \delta_{x_1, x_2}.
\end{align}
Similarly, global conformal symmetry restricts three-point functions to be
\begin{align}
\label{3-point-CFT}
\big\langle \phi_1(r_1) \phi_2(r_2) \phi_3(r_3) \big\rangle_\text{CFT}
&\sim r_{12}^{x_3 - x_1 - x_2}
r_{13}^{x_2 - x_1 - x_3} r_{23}^{x_1 - x_2 - x_3}.
\end{align}

Correlation functions of four or more operators are not uniquely constrained by global conformal symmetry. For example, four-point functions can be written as
\begin{align}
\label{4-point-CFT}
\big\langle \phi_1(r_1) \ldots \phi_4(r_4) \big\rangle_\text{CFT} &=
\prod_{i<j}^{4} r_{ij}^{x/3 - x_i - x_j} F(X_1, X_2),
\qquad x = \sum_{i=1}^4 x_i \,,
\end{align}
where the function $F$ of the two cross-ratios $X_1 = r_{12} r_{34}/r_{13} r_{24}$ and $X_2 = r_{12} r_{34}/r_{14} r_{23}$
cannot be determined from conformal invariance alone.

\subsection{Relation of MF multipoint functions and CFT correlators}
\label{sec:relation-MF-CFT}

A quick comparison of MF correlators with those in a CFT indicates that not all MF multipoint functions can be represented as CFT correlators. For example, even the one-point function (\ref{1-point}) is consistent with Eq. (\ref{1-point-CFT}) only if $x_q > 0$. But as we mentioned, for an extended range of $q$ the dimensions $x_q <0$. (In fact, $x_q <0$ holds for any $q$ satisfying $q <0$ or $q > q_*$.)  In addition, as we have seen, there is a non-trivial operator $\mathcal{O}_{q_*}$, distinct from the identity operator, whose dimension is zero, and whose expectation value in the infinite system is one, see Eq. \eqref{xq-star-0}. This indicates that we cannot expect all aspects of MF multipoint functions to be describable by a CFT.

Let us look at the two-point function (\ref{2-point-1}).  In general, it is $L$-dependent, at variance with Eqs.~\eqref{2-point-scale}  and (\ref{2-point-CFT}).
We notice, however, the following: we can choose $q_2 = -q_1$, and then the dimension $x_{q_1 + q_2} = x_0 = 0$. This implies that with this choice the two-point function stops depending on the system size:
\begin{align}
\label{2-point-2}
\big\langle \mathcal{O}_{q_1}(r_1) \mathcal{O}_{- q_1}(r_2) \big\rangle \sim
r_{12}^{- x_{q_1} - x_{-q_1}} \bigg(\frac{L}{r_{12}}\bigg)^{-x_0}
= r_{12}^{- x_{q_1} - x_{-q_1}}.
\end{align}
Then the system size can be taken to infinity, and the two-point function can be compared with a two-point function in a critical field theory. We see that the above form is consistent with requirements of rotational, translational and scale invariance, Eq. (\ref{2-point-scale}), but not with conformal invariance, Eq. (\ref{2-point-CFT}), since the dimensions $x_{q_1} \neq x_{-q_1}$.

However, the presence of the special operator $\mathcal{O}_{q_*}$ allows us to make another choice. Indeed, if we choose the special value $q_2 = q_* - q_1$, the dimension $x_{q_1 + q_2} = x_{q_*} = 0$. In this case we have
\begin{align}
\label{2-point-3}
\big\langle \mathcal{O}_{q_1}(r_1) \mathcal{O}_{q_* - q_1}(r_2) \big\rangle \sim
r_{12}^{- x_{q_1} - x_{q_* - q_1}} \bigg(\frac{L}{r_{12}}\bigg)^{-x_{q_*}}
= r_{12}^{ - 2x_{q_1}}.
\end{align}
This two-point function is independent of the system size and is consistent with the CFT form
\begin{align}
\label{2-point-CFT-1}
\big\langle \phi_{q_1}(r_1) \phi_{q_* - q_1}(r_2) \big\rangle_\text{CFT} \sim r_{12}^{-2x_{q_1}},
\end{align}
since the two operators involved have the same dimensions.

This discussion and the scaling (\ref{n-point}) suggest that CFT may apply only to such multipoint MF functions $\big\langle \mathcal{O}_{q_1} \ldots \mathcal{O}_{q_n} \big\rangle$ where the sum of the indices $q_i$ is equal to $q_*$ (a sort of ``neutrality'' condition):
\begin{align}
\label{neutrality}
\sum_{i=1}^{n} q_i = q_*.
\end{align}
An alternative choice $\sum_i q_i = 0$ leads to MF multipoint functions that are well defined in the infinite system, but are inconsistent with conformal invariance.

To illustrate this point, let us make the choice (\ref{neutrality}) in the three-point function (\ref{3-point-1}):
\begin{align}
\label{3-point-2}
\big\langle \mathcal{O}_{q_1}(r_1) \mathcal{O}_{q_2}(r_2) \mathcal{O}_{q_* - q_1 - q_2}(r_3) \big\rangle \sim r_{12}^{x_{q_1 + q_2} - x_{q_1} - x_{q_2}} r_{13}^{- 2x_{q_1 + q_2}}.
\end{align}
If we specify the general CFT expression (\ref{3-point-CFT}) to our case and use the symmetry relation (\ref{symmetry-xq}) we get
\begin{align}
\label{3-point-CFT-1}
\big\langle \phi_{q_1}(r_1) \phi_{q_2}(r_2) \phi_{q_* - q_1 - q_2}(r_3) \big\rangle_\text{CFT} \, \sim \, r_{12}^{x_{q_1 + q_2} - x_{q_1} - x_{q_2}}
r_{13}^{x_{q_2} - x_{q_1} - x_{q_1 + q_2}}
r_{23}^{x_{q_1} - x_{q_2} - x_{q_1 + q_2}}.
\end{align}
If we now assume that the arrangement of points satisfies the inequality $r_{12} \ll r_{13} \simeq r_{23}$, we can replace $r_{23} \approx r_{13}$, and the three-point function simplifies to
\begin{align}
\big\langle \phi_{q_1}(r_1) \phi_{q_2}(r_2) \phi_{q_* - q_1 - q_2}(r_3) \big\rangle_\text{CFT} \sim r_{12}^{x_{q_1 + q_2} - x_{q_1} - x_{q_2}} r_{13}^{- 2x_{q_1 + q_2}},
\end{align}
which is the same as Eq. (\ref{3-point-2}). On the other hand, the choice $\sum_i q_i = 0$ leads to inconsistent expressions. The MF three-point function (\ref{3-point-1}) in this case becomes
\begin{align}
\big\langle \mathcal{O}_{q_1}(r_1) \mathcal{O}_{q_2}(r_2) \mathcal{O}_{-q_1 - q_2}(r_3) \big\rangle \sim r_{12}^{x_{q_1 + q_2} - x_{q_1} - x_{q_2}}
r_{13}^{- x_{q_1 + q_2} - x_{-q_1 - q_2}},
\end{align}
while the CFT expression (\ref{3-point-CFT}) reduces to a different form
\begin{align}
\big\langle \phi_{q_1}(r_1) \phi_{q_2}(r_2) \phi_{-q_1 - q_2}(r_3) \big\rangle_\text{CFT} \sim
r_{12}^{x_{-q_1 - q_2} - x_{q_1} - x_{q_2}} r_{13}^{- 2 x_{-q_1 - q_2}}.
\end{align}

We have thus shown that $m$-point correlation functions of CFT can be related to $m$-point MF correlation functions in an infinite system under the condition $\sum_{i=1}^m q_i = q_*$.  Let us return to the MF multipoint functions with indices $q_i$ that do not satisfy this condition. As discussed above, they in general scale with $L$ and thus should be considered in a finite system. We now point out that such generic $m$-point MF functions in a finite system can be related to $m+1$-point CFT functions in an infinite system by adding another field to ensure the condition $\sum_{i=1}^{m+1} q_i = q_*$.

As the simplest example, let us consider the two-point function (\ref{2-point-CFT-1}) (defined in an infinite system) and place the operator $\phi_{q_* - q_1}$ at a large distance $L$ (which is not the system size) from the point $r_1 = 0$. Then we can rewrite the two-point function as
\begin{align}
L^{x_{q_1}} \big\langle \phi_{q_1}(0) \phi_{q_* - q_1}(L) \big\rangle_\text{CFT} \sim L^{-x_{q_1}}.
\end{align}
This has the same form as the scaling of one-point MF correlation function in a system of size $L$, Eq. (\ref{1-point}). Similarly, if we place the operator $\phi_{q_* - q_1 - q_2}$ in the three-point function (\ref{3-point-CFT-1}) at a distance $L \gg r_{12}$ from the points $r_1$ and $r_2$, we can replace $r_{13} \approx r_{23} \approx L$, which yields
\begin{align}
\big\langle \phi_{q_1}(r_1) \phi_{q_2}(r_2) \phi_{q_* - q_1 - q_2}(L) \big\rangle_\text{CFT} \sim r_{12}^{x_{q_1 + q_2} - x_{q_1} - x_{q_2}} L^{- 2x_{q_1 + q_2}}.
\end{align}
This can be rewritten as
\begin{align}
L^{x_{q_1 + q_2}} \big\langle \phi_{q_1}(r_1) \phi_{q_2}(r_2) \phi_{q_* - q_1 - q_2}(L) \big\rangle_\text{CFT} \sim r_{12}^{x_{q_1 + q_2} - x_{q_1} - x_{q_2}} L^{-x_{q_1 + q_2}},
\end{align}
which is exactly the same as Eq. (\ref{2-point-1}). Thus, a three-point CFT function, which describes a three-point MF function with $q_1 + q_2 + q_3 = q_*$ in an infinite system, can also be related to a two-point MF function (with arbitrary $q_1$ and $q_2$) in a finite system.  We expect that such relations continue to hold for higher multi-point correlators. For example, the four-point function
$ L^{x_{q_1 + q_2 + q_3}}  \big\langle \phi_{q_1}(r_1) \phi_{q_2}(r_2) \phi_{q_3}(r_3) \phi_{q_* - q_1 - q_2 - q_3}(L) \big\rangle_\text{CFT}$ should reduce to the three-point function (\ref{3-point-1}) in the appropriate limiting case of separation of scales $r_{12} \ll r_{13} \sim r_{23} \ll L$.

Let us briefly summarize our results so far. First of all, there are multi-point MF correlation functions \eqref{n-point} with $\sum_i q_i \neq 0, q_*$ that explicitly contain the system size $L$, exemplifying strong infrared fluctuations that may grow with $L$. Secondly, there are MF correlators with $\sum_i q_i = 0$ that can be considered in an infinite system, consistent with scale invariance, but inconsistent with conformal invariance. Finally, there are ``good'' MF correlators satisfying the neutrality condition $\sum_i q_i = q_*$ that are consistent with conformal invariance.

This analysis can be extended to the generalized MF correlation functions $\langle \mathcal{O}_{\lambda_1}(r_1) \mathcal{O}_{\lambda_2}(r_2) \ldots \rangle$ involving the much broader class of gradientless operators $\mathcal{O}_\lambda$, with $\lambda = (q_1,q_2, \ldots, q_n)$,
see Sec.~\ref{subsec:MF-symmetry}. Again, almost all such correlation functions are infrared-singular with respect to the system size $L$, i.e., depend on $L$ in a power-law fashion. There is a subset of them that exhibits  $L$-independence. And within this subset, there is a still smaller subset of correlators whose scaling is consistent with conformal invariance. Specifically, the corresponding condition is $\sum_i \lambda_i = (-c_1, -c_2,\ldots, - c_n) \equiv - \rho_b$.

We stress, however, that this consistency is not a guarantee of conformal invariance at ATs. While there is a wide-spread folklore that scale invariance and locality imply conformal invariance, this is not the case in general, see Ref.~\cite{Nakayama-Scale-2015} for a comprehensive discussion of the relation between the two. All currently existing or even envisioned proofs of conformal invariance following from scale invariance involve certain technical assumptions such as unitarity (reflection positivity) of the field theory and discreteness of the spectrum of scaling dimensions. These assumptions are certainly violated in any viable theory of ATs, so conformal invariance is not guaranteed to be a feature of ATs.

Understanding the difference of scale and conformal invariance is especially important in two dimensions, where conformal invariance is especially powerful due to the existence of an infinite-dimensional Virasoro symmetry. In the next section we will see that assuming local conformal invariance in two dimensions together with Abelian fusion leads to very stringent restrictions on the spectra of MF exponents.

\section{CFT in two dimensions and exact parabolicity of MF spectra}
\label{sec:CFT-2D}
In this section we restrict our attention to ATs in two dimensions. The reason is that in two dimensions, conformal symmetry is especially powerful due to local conformal invariance and the underlying infinite-dimensional Virasoro algebra. It is commonly expected that if a 2D theory satisfies global conformal invariance, it also satisfies the---much stronger---local conformal invariance.
As we will see, this infinite symmetry implies (together with the Abelian fusion) {\it exact} parabolicity of MF spectra \cite{bondesan2017gaussian}. After reviewing the arguments of Ref.~\cite{bondesan2017gaussian} that dealt with the the leading MF spectra at the integer quantum Hall (IQH) transition, we extend them to other symmetry classes [Eq.~\eqref{exact-parabola-x-q}] and also to generalized MF exponents [Eq.~\eqref{generalized-parabolicity}].

\subsection{Exact parabolicity of $\Delta_q$ for the IQH transition under the assumption of local conformal invariance}
\label{sec:IQH-parabolicity}

Bondesan et al. \cite{bondesan2017gaussian} considered the scaling operators $\mathcal{O}_q(r)$ in the context of the IQH transition in 2D and its description in terms of the Chalker-Coddington network model. This critical point belongs to class A, so $q_* = 1$ and $x_q = \Delta_q$. The main results of \cite{bondesan2017gaussian} is that, under the assumption of local conformal invariance,  exact parabolic form of the dimensions $\Delta_q$ holds:
\begin{align}
\label{exact-parabola-Delta-q}
\Delta_q = b q(1-q),
\end{align}
with constant $b$ left undetermined.

In 2D CFT, where the points are specified by complex coordinates $z$, the two- and three-point functions factorize into holomorphc and antiholomorphic factors (with possibly different holomorphic and antiholomorphic dimensions $h$ and $\bar{h}$):
\begin{align}
\label{2-point-CFT-2D}
\big\langle \phi_1(z_1, \bar{z}_1) \phi_2(z_2, \bar{z}_2) \big\rangle_\text{CFT} &\sim z_{12}^{-2h} \bar{z}_{12}^{- 2\bar{h}},
\qquad \text{if } h_1 = h_2 = h \text{ and } \bar{h}_1 = \bar{h}_2 = \bar{h}, \\
\label{3-point-CFT-2D}
\big\langle \phi_1(z_1, \bar{z}_1) \phi_2(z_2, \bar{z}_2) \phi_3(z_3, \bar{z}_3) \big\rangle_\text{CFT}
&\sim z_{12}^{h_3 - h_1 - h_2} z_{13}^{h_2 - h_1 - h_3} z_{23}^{h_1 - h_2 - h_3} \bar{z}_{12}^{\bar{h}_3 - \bar{h}_1 - \bar{h}_2} \bar{z}_{13}^{\bar{h}_2 - \bar{h}_1 - \bar{h}_3}
\bar{z}_{23}^{\bar{h}_1 - \bar{h}_2 - \bar{h}_3}.
\end{align}
The four-point CFT correlation functions in two dimensions can be written in terms of $z_{ij} = z_i - z_j$ and the cross-ratio $\eta = z_{12} z_{34}/z_{13} z_{24}$
as follows:
\begin{align}
\label{4-point-CFT-2D}
\big\langle \phi_1(z_1, \bar{z}_1) \ldots \phi_4(z_4, \bar{z}_4) \big\rangle_\text{CFT}
&= \prod_{i<j}^{4} z_{ij}^{h/3 - h_i - h_j}
\bar{z}_{ij}^{\bar{h}/3 - \bar{h}_i - \bar{h}_j} F(\eta, \bar{\eta}), &
h &= \sum_{i=1}^{4} h_i, & \bar{h} &= \sum_{i=1}^{4} \bar{h}_i.
\end{align}
As in higher dimensions, the function $F(\eta, \bar{\eta})$ cannot be determined from conformal symmetry alone.

To prove the exact parabolicity \eqref{exact-parabola-Delta-q}, Bondesan et al. considered a certain 4-point correlation function that satisfies the neutrality condition \eqref{neutrality}. In that work the neutrality condition (\ref{neutrality}) was achieved by considering $n-1$ insertions of the operators $\mathcal{O}_{q_i}$ plus one operator $\pi_c$ describing a point contact in the network model \cite{janssen1999point-contact, Bondesan-Pure-2014}. This operator can be written as the integral over a continuum of $q$ values, and it always contributes the definite scaling dimension (or the fusion channel) necessary to have the correct total ``charge'' $q_* = 1$.

In our notations the relevant correlator is $\big\langle \mathcal{O}_{q_1}(z_1, \bar{z}_1) \mathcal{O}_{q_2}(z_2, \bar{z}_2) \mathcal{O}_{q_3}(z_3, \bar{z}_3) \mathcal{O}_{1 - q_1 - q_2 - q_3}(z_4, \bar{z}_4) \big\rangle$. The operators $\mathcal{O}_q$ are expected to have $h_q = \bar{h}_q = \Delta_q/2$, and the corresponding 4-point CFT correlator is
\begin{align}
\label{4-point-CFT-2D-1}
& \big\langle \phi_{q_1}(z_1, \bar{z}_1) \phi_{q_2}(z_2, \bar{z}_2) \phi_{q_3}(z_3, \bar{z}_3) \phi_{1 - q_1 - q_2 - q_3}(z_4, \bar{z}_4) \big\rangle_\text{CFT}
= \prod_{i<j}^{4} |z_{ij}|^{\Delta/3 - \Delta_{q_i} - \Delta_{q_j}}
F(\eta, \bar{\eta}), \\
&  \Delta = \Delta_{q_1} + \Delta_{q_2} + \Delta_{q_3} + \Delta_{q_1 + q_2 + q_3}.
\end{align}

Let us briefly summarize the argument of Ref. \cite{bondesan2017gaussian}. The authors make two essential assumptions: 1) correlation functions satisfying the neutrality condition \eqref{neutrality} exhibit at  the IQH transition point the local conformal invariance, with the corresponding CFT operators $\phi_q$ being Virasoro primaries; 2) the operators $\phi_q$ satisfy (the 2D variant of) the Abelian OPE (\ref{Abelian-OPE}):
\begin{align}
\phi_{q_1}(z_1, \bar{z}_1) \phi_{q_2}(z_2, \bar{z}_2) & \, \propto \,  |z_{12}|^{\Delta_{q_1 + q_2} - \Delta_{q_1} - \Delta_{q_2}}
\phi_{q_1+q_2}(z_2, \bar{z}_2) + \ldots
\end{align}
It is important that no other primaries enter the Abelian OPE and the ellipses stand for contributions from Virasoro descendants.

The Abelian fusion immediately implies that there is only one conformal block in the correlator (\ref{4-point-CFT-2D-1}), that is $F(\eta, \bar{\eta}) = |f(\eta)|^2$. The holomorphic function $f(\eta)$ is present in the holomorphic factor of the correlator (\ref{4-point-CFT-2D-1}):
\begin{align}
\label{4-point-CFT-holo}
G(z_1, \ldots, z_4) & = \big\langle \phi_{q_1}(z_1) \phi_{q_2}(z_2)  \phi_{q_3}(z_3)  \phi_{1 - q_1 - q_2 - q_3}(z_4) \big\rangle_\text{CFT}
= \prod_{i<j}^{4} z_{ij}^{h/3 - h_{q_i} - h_{q_j}}
f(\eta), \\
& h = h_{q_1} + h_{q_2} + h_{q_3} + h_{q_1 + q_2 + q_3}.
\end{align}
There are three possible fusion channels in this correlator, where each of the charges $q_1$, $q_2$, and $q_3$ fuses with the neutralizing charge $1 - q_1 - q_2 - q_3$. These three channels lead to simple power-law  singularities (branch points) of the function $f(\eta)$ at $\eta = 1, 0$ and $\infty$. For example, fusing $q_1$ with $q_2$, and $q_3$ with $1 - q_1 - q_2 - q_3$  (so that $|\eta| \ll 1$) gives
\begin{align}
\label{G-1}
G(z_1, \ldots, z_4) &\sim z_{12}^{h_{q_1 + q_2} - h_{q_1} - h_{q_2}}
z_{34}^{h_{q_1 + q_2} - h_{q_3} - h_{q_1 + q_2 + q_3}}
z_{13}^{-2h_{q_1 + q_2}}.
\end{align}
On the other hand, the right-hand side of Eq. (\ref{4-point-CFT-holo}) where the distances between points are chosen appropriately ($|z_{12}|, |z_{34}| \ll |z_{13}| \approx |z_{14}| \approx |z_{23}| \approx |z_{24}|$) becomes
\begin{align}
\label{G-2}
G(z_1, \ldots, z_4) &\sim f(\eta) \, z_{12}^{h/3 - h_{q_1} - h_{q_2}}
z_{34}^{h/3 - h_{q_3} - h_{q_1 + q_2 + q_3}} z_{13}^{-2h/3}.
\end{align}
Comparison of the two expressions gives that near $\eta = 0$
\begin{align}
f(\eta \sim 0) = \eta^{h_{q_1 + q_2} - h/3}[a_0 + O(\eta)].
\end{align}
The other two fusion channels give the forms of the singularities near the other two branch points:
\begin{align}
f(\eta \sim 1) &= (1 - \eta)^{h_{q_2 + q_3} - h/3}[b_0 + O(1-\eta)],
&
f(\eta \sim \infty) &= (1/\eta)^{h_{q_1 + q_3} - h/3}[c_0 + O(1/\eta)].
\end{align}

The exponents characterizing the three branch points must be related. The argument is a simple case of a more general one in Ref.~\cite{Lewellen-Constraints-1989}. Indeed, let us compactify the complex plane and go around all three branch points along a contour $\mathcal{C}$. Then, on the one hand, the total phase change (monodromy) of the function $f(\eta)$ acquired along the contour $\mathcal{C}$ is the product of the phases characterizing each of the branch points. On the other hand, the contour $\mathcal{C}$ can be deformed to a point by extending it to infinity, which means that the total monodromy is trivial (one). This gives the relation
\begin{align}
h - h_{q_1 + q_2} - h_{q_1 + q_3} - h_{q_2 + q_3} &= M,
\end{align}
where $M \geq 0$ is a non-negative integer. Then the holomorphic function $f(\eta)$ becomes
\begin{align}
f(\eta) = \eta^{h_{q_1 + q_2} - h/3} (1 - \eta)^{h_{q_2 + q_3} - h/3} [a_0 + a_1 \eta + \cdots + a_M \eta^M].
\end{align}
Choosing $q_1 = q_2 = q_3 = 0$, we see that $M=0$. This fixes the conformal block
\begin{align}
f(\eta) = a_0 \eta^{h_{q_1 + q_2} - h/3} (1 - \eta)^{h_{q_2 + q_3} - h/3}.
\end{align}

Moreover, the condition $M=0$ gives a functional equation for the dimensions $h_q$:
\begin{align}
\label{Suslov-h}
h_{q_1 + q_2 + q_3} - h_{q_1 + q_2} - h_{q_1 + q_3} - h_{q_2 + q_3} + h_{q_1} + h_{q_2} + h_{q_3} = 0,
\end{align}
or, for the full dimensions $\Delta_q$,
\begin{align}
\label{Suslov-Delta}
\Delta_{q_1 + q_2 + q_3} - \Delta_{q_1 + q_2} - \Delta_{q_1 + q_3} - \Delta_{q_2 + q_3} + \Delta_{q_1} + \Delta_{q_2} + \Delta_{q_3} = 0.
\end{align}
This equation easily leads to the result \eqref{exact-parabola-Delta-q}. First, it immediately implies $\Delta_0 = 0$, as it should be. Next, setting $q_1 = q$ and $q_2 = q_3 = \epsilon$, we get
\begin{align}
\Delta_{q + 2\epsilon} - 2\Delta_{q + \epsilon} - \Delta_{2\epsilon} + \Delta_{q} +
2\Delta_{\epsilon} = 0.
\end{align}
Now we expand up to second order in $\epsilon$ to get $(\Delta''_q - \Delta''_0)\epsilon^2 = 0$.
Thus $\Delta''_q = \Delta''_0 = \text{const}$, which implies that the function $\Delta_q$ is a quadratic polynomial. A general quadratic polynomial that vanishes at $q = 0$ and satisfies the symmetry property \eqref{symmetry-xq} is exactly of the form given in Eq. (\ref{exact-parabola-Delta-q}).

\subsection{Generalizations}

Now we can generalize the arguments of Ref.~\cite{bondesan2017gaussian} summarized in Sec.~\ref{sec:IQH-parabolicity} in two directions.

\subsubsection{Other symmetry classes}

First, the arguments are not specific to the IQH critical point but rather apply also to 2D critical points of ATs in other symmetry classes. The assumptions are the same: (1) local conformal invariance of correlation functions satisfying the neutrality condition, and (2) Abelian fusion. Under these assumptions, the arguments can be directly extended to all five symmetry classes (A, AI, AII, CI, and C)
identified in Refs. \cite{gruzberg2011symmetries, gruzberg2013classification} that exhibit the symmetry relation (\ref{symmetry-xq}), with $q_*=1$ for Wigner-Dyson classes, $q_*=2$ for class CI, and $q_*=3$ for class C.  The result is the exact parabolicity,
\begin{align}
\label{exact-parabola-x-q}
x_q = b q(q_* - q)
\end{align}
for the simple MF spectra. (Here the term ``simple'' is used in the sense of opposite to ``generalized''.)

\subsubsection{Generalized mulifractality}
\label{subsec:generalized-parabolicity}

Secondly, we can obtain a generalization of this parabolic form for all gradientless scaling operators $\mathcal{O}_\lambda$ considered in \cite{gruzberg2013classification}, i.e., to generalized MF spectrum, see Sec.~\ref{subsec:MF-symmetry}.  As was pointed out in Sec.\ref{subsec:MF-symmetry}, there is a general operator $\mathcal{O}_{-\rho_b} = \mathcal{O}_{(-c_1, -c_2,\ldots)}$ with vanishing scaling dimension \eqref{x-rho-b}.  Further, as we discussed in Sec.~\ref{sec:relation-MF-CFT}, the correlation functions satisfying $\sum_i \lambda_i = -\rho_b$ have scaling consistent with conformal invariance.
 Let us now assume, as before, that the local conformal invariance holds for these correlation functions and that
 $\mathcal{O}_\lambda$ are Virasoro primaries that satisfy the 2D version of the Abelian OPE \eqref{Abelian-OPE-2}. Then the arguments of Ref.~\cite{bondesan2017gaussian}
 presented in Sec.~\ref{sec:IQH-parabolicity}
 can be straightforwardly generalized to give the following equation for the dimensions $x_\lambda$:
\begin{align}
x_{\lambda + \lambda' + \lambda''} - x_{\lambda + \lambda'} - x_{\lambda + \lambda''} - x_{\lambda' + \lambda''} + x_\lambda + x_{\lambda'} + x_{\lambda''} = 0.
\end{align}
[This equation is a counterpart of Eq.~\eqref{Suslov-Delta}.]
Let us denote by $e_i = (0,\ldots, 1, \ldots, 0)$ (unit in the $i$-th place) the standard basis in the weight space, and choose $\lambda' = \epsilon' e_i$ and $\lambda'' = \epsilon'' e_j$:
\begin{align}
x_{\lambda + \epsilon' e_i + \epsilon'' e_j} - x_{\lambda + \epsilon' e_i} - x_{\lambda + \epsilon'' e_j} - x_{\epsilon' e_i + \epsilon'' e_j} + x_\lambda + x_{\epsilon' e_i} + x_{\epsilon'' e_j} = 0.
\label{Suslov-x-lambda}
\end{align}
Then we can expand this to second order in $\epsilon'$ and $\epsilon''$, denoting $\partial_i = \partial/\partial q_i$:
\begin{align}
x_{\lambda + \epsilon' e_i} &= x_\lambda + \partial_i x_\lambda \epsilon' + \frac{1}{2} \partial_i^2 x_\lambda \epsilon'{}^2, \qquad
x_{\lambda + \epsilon'' e_j} = x_\lambda + \partial_j x_\lambda \epsilon'' + \frac{1}{2} \partial_j^2 x_\lambda \epsilon''{}^2, \nonumber \\
x_{\lambda + \epsilon' e_i + \epsilon'' e_j} &= x_\lambda + \partial_i x_\lambda \epsilon' + \partial_j x_\lambda \epsilon'' + \frac{1}{2} \partial_i^2 x_\lambda \epsilon'{}^2 + \frac{1}{2} \partial_j^2 x_\lambda \epsilon''{}^2 + \partial_i \partial_j x_\lambda \epsilon' \epsilon''.
\end{align}
Substituting these expansions into Eq.~\eqref{Suslov-x-lambda} gives
$\partial_i \partial_j x_\lambda = \partial_i \partial_j x_0 = \text{const}$,
which implies that $x_\lambda$ is a quadratic function of $q_i$. Thus, it is also a quadratic function of the shifted ${\tilde q}_i = q_i + c_i/2$. But in these variables $x_\lambda$ must be even due to the first type of the Weyl group actions \eqref{Weyl-sign}, which restricts it to
\begin{align}
x_\lambda = \sum_i b_i {\tilde q}_i^2 + B.
\end{align}
The other type of Weyl group actions \eqref{Weyl-interchange}, which requires invariance of $x_\lambda$ with respect to the interchange $ {\tilde q}_i  \leftrightarrow {\tilde q}_j $ for any pair $i,j$, forces all coefficients $b_i$ to be equal: $b_i = -b$. Finally, the constant $B$ is found from the requirement $x_0 = 0$, which gives $B = b \sum_i c_i^2/4$, and
\begin{align}
x_\lambda = - b \sum_i q_i (q_i + c_i) = -b (\lambda, \lambda + \rho_b).
\label{generalized-parabolicity}
\end{align}
Thus, the generalized dimensions are proportional to the value of the quadratic Casimir operator in the representation labeled by the highest weight $\lambda$. Now choosing $\lambda = q e_1$, we see that the constant $b$ is the same as in the simple MF spectrum (\ref{exact-parabola-x-q}).  The generlized MF spectrum \eqref{generalized-parabolicity} is thus parametrized by a single constant $b$. We will use for the result \eqref{generalized-parabolicity} the term ``generalized parabolicity''.

\subsubsection{Generalized parabolicity of generalized MF spectra as a hallmark of local conformal invariance at 2D ATs}
\label{sec:gen-parab-hallmark-conf-inv}

Let us now discuss two conditions under which the generalized parabolicity of generalized MF spectra of 2D systems has been derived. We recall again that these are (1) local conformal invariance and (2) Abelian fusion.

We first discuss  the second condition: the Abelian fusion of operators $\mathcal{O}_\lambda$.  In Ref.~\cite{gruzberg2013classification}, the operators $\mathcal{O}_\lambda$ were explicitly constructed for class A by two approaches (Iwasawa decomposition and highest-weight vectors), and the results do satisfy the Abelian fusion requirement. Furthermore, the methods of Ref.~\cite{gruzberg2013classification} can be extended to other symmetry classes. This will be done in the present paper for class C. Specifically, the Abelian fusion of the scaling operators $\mathcal{O}_\lambda$ follows from their derivation as highest-weight vectors under the action of a Lie algebra, see Sec.~\ref{sec:hwv}. Furthermore, it is a consequence of the form of these operators as ``plane waves'' on the target space of the sigma model constructed by means of Iwasawa decomposition in Sec.~\ref{sec:iwasawa}. These results thus provide an explicit verification of the Abelian-fusion condition.

The situation with the assumption of the conformal invariance is much more delicate, as was already mentioned at the end of Sec.~\ref{sec:relation-MF-CFT}. We do not know {\it a priori} whether it holds (in the narrow sense explained above) at 2D critical points such as IQH or spin quantum Hall transitions. Therefore, the generalized parabolicity of spectra of generalized MF dimensions in fact serves as a hallmark for the local conformal invariance. By constructing explicitly eigenfunction correlations that correspond to $\mathcal{O}_\lambda$ operators and by determining the corresponding scaling dimensions numerically, one can thus test conformal invariance. This will be done below for the spin quantum Hall transition.

\subsection{Beyond 2D:  Non-parabolic MF spectra}
\label{sec:beyond2D-nonparabolic}

While this paper focusses on 2D systems, the following comment on MF spectra beyond 2D is instructive at this point. The assumption of local conformal invariance (involving the infinite-dimensional Virasoro algebra) was crucial for the above derivation of exact parabolicity of MF spectra (both simple and generalized). This assumption may or may not hold at a 2D AT critical point. However, it definitely does not hold beyond 2D. Therefore, the derivation is certainly not applicable to systems of other spatial dimensionality. Indeed, it is well known, both analytically and numerically, that the MF spectra are in general {\it not parabolic}.

On the analytical side, the MF dimensions have been calculated by perturbative $\varepsilon$-expansion of the $\sigma$ models in $2 + \varepsilon$ dimensions~\cite{wegner1987anomalous1, wegner1987anomalous2}. All these results give parabolic spectra in the leading term but exhibit deviations from exact parabolicity in higher orders in $\varepsilon$. In full consistency with the analytical predictions, numerical evaluation of MF spectra at ATs in systems of spatial dimensionality $d=3$, 4, and 5  \cite{mildenberger2002dimensionality,rodriguez2008multifractal,rodriguez2011multifractal,tarquini2017critical} demonstrate significant deviations from parabolicity that grow with increasing $d$. Furthermore, the limit  $d \gg 1$ can be addressed by using the results for the ATs on the Bethe lattice that show the MF acquires its strongest form in this limit~\cite{evers08}, with the spectrum
\begin{align}
\Delta_q \simeq d (1/2 - |q - 1/2|).
\end{align}
This spectrum is piecewise linear rather than parabolic. Numerical results indeed approach this behavior with increasing $d$  \cite{mildenberger2002dimensionality,tarquini2017critical}. A similar evolution of the MF spectrum, from a nearly-parabolic spectrum (but still with some deviations) towards a piecewise-linear spectrum, is also found (both analytically and numerically) for the model of power-law random banded matrices \cite{evers08}.

In Ref.~\cite{suslov2016strict}, it was claimed that exact parabolicity \eqref{exact-parabola-Delta-q} holds for {\it all} ATs in the standard Wigner-Dyson classes in {\it any} dimension. Clearly, this statement is incorrect since it is in contradiction with the whole body of analytical and numerical results that we have just reviewed. For the benefit of the reader, we point out the crucial flaw in the argumentation of Ref.~\cite{suslov2016strict}. The author of Ref.~\cite{suslov2016strict} obtained the relation \eqref{Suslov-Delta} from an ad-hoc assumption that {\it all} three-point  MF correlation functions are just products of simple power-law factors, see Eq. (9) in Ref.~\cite{suslov2016strict}. This assumption is in general incorrect. In reality, the three-point functions reduce to {\it different} products of power-law factors in limiting cases that correspond to different hierarchical relations among distances between the points, as follows from general RG arguments presented in Section \ref{subsec:MF-RG}; see, in particular Eq.~\eqref{3-point-1}. A similar assumption of Ref.~\cite{suslov2016strict} for {\it all} multi-point MF correlation function [Eq.~(19) there] is flawed for the same reason. Only in the presence of local conformal invariance in 2D systems, much stronger constraints emerge that allow one to prove the parabolicity.

\section{Non-linear $\sigma$ model of class C}
\label{sec:sigma}
In this section, we sketch the derivation of the class-C nonlinear sigma model. While doing this, we introduce notations and conventions employed throughout the paper. Starting from the non-interacting
Bogoliubov-de Gennes (BdG) Hamiltonian, we formulate a bosonic (2+0)D action $S_0$. We employ the replica trick \cite{wegner1979the, jungling1979effects} to average over disorder. Following the conventional procedure of Hubbard-Stratonovich decoupling and gradient expansion \cite{mirlin00, efetov1983supersymmetry, efetov1997supersymmetry, verbaarschot1985grassmann}, we arrive at the sigma model action, Eq.~\eqref{SQ-sigma-model}. The sigma-model field $Q$ lives on the symmetric space $G/K=\mathrm{SO}^*(4n)/\mathrm{U}(2n)$. The derivation largely follows similar lines as Refs. \cite{altland2000field, taras_semchuk2001quantum, liao2017response}.  For the connection to the wave function statistics in Sections  \ref{sec:wave_ops}, \ref{sec:wave_iw}, the coupling \eqref{eq:coup} between the $Q$ field of the sigma model and the bosonic field variables $\phi$ of the action $S_0$ is of crucial importance.


Systems in class C possess particle-hole and spin rotation symmetry:
\begin{align}
\pi_x H^T \pi_x &= -H, & \sigma_i H \sigma_i&= H \label{eq:sym}.
\end{align}
Here $\pi_x$ is the first Pauli matrix in the BdG space, and $\sigma_i$ are Pauli matrices in the spin space.
When written in the basis $\psi = (\psi_{\uparrow}, \psi_{\downarrow}, \psi_{\uparrow}^\dagger, \psi_{\downarrow}^\dagger)$, the Hamiltonian has the following $4 \times 4$ matrix form:
\begin{align}
H_{BdG}&=\begin{pmatrix}
\frac{1}{2m}(p-e A)^2 - \mu  & \hat{\Delta} \\
-\hat{\Delta}^* & -\frac{1}{2m}(p+e A)^2 + \mu
\end{pmatrix}_\pi \,.
\label{H-BDG}
\end{align}
Here the matrix structure in the BdG space is shown explicitly (as indicated by the subscript $\pi$); each entry is a $2 \times 2$ matrix in the spin space.
The matrix $\hat{\Delta}$ is antisymmetric $\hat{\Delta}^T=-\hat{\Delta}$ and $\sigma_y\hat{\Delta}\sigma_y = \hat{ \Delta}$ due to spin symmetry. This implies that $\hat{\Delta} = \Delta \sigma_y$, with a complex number $\Delta$. The kinetic term $h = -\frac{1}{2m}(p-e A)^2 + \mu $ is the same for each spin species. Note that $p$ changes sign upon transposition.

As usual for BdG Hamiltonians with preserved spin, Eq.~\eqref{H-BDG} decouples into two blocks. We thus get a $2\times 2$ matrix Hamiltonian for ${\bf \psi} = (\psi_\uparrow, \psi_\downarrow^\dagger)_\sigma$:
\begin{align}
H &= \begin{pmatrix}\frac{1}{2m}(p-e A)^2 - \mu  & \Delta\\
\Delta^* & -\frac{1}{2m}(p-e A)^2 + \mu
\end{pmatrix}_\sigma \label{eq:hamC}\,.
\end{align}
Here and from now on, $\sigma$ is the combined particle-hole and spin space. In general, one may have an additional matrix structure due to orbital degrees of freedom, in which case $\Delta$ is a symmetric complex matrix $\Delta^T=\Delta$. Here, we do not have additional orbital indices, so that $\Delta$ is a complex number.

In this basis, spin $\mathrm{SU}(2)$ and particle-hole symmetry combine into
\begin{align}
H &= - \sigma_y H^T \sigma_y
\label{eq:pha}.
\end{align}
This particle-hole symmetry squares to minus one, which is the defining property of class C. The network model that we study numerically below is described by a scattering matrix corresponding to a Hamiltonian of this kind.

We employ a time independent $(2+0)$D replicated action \cite{wegner1979the, jungling1979effects} to study the statistics of wavefunctions of $H$.
The action $S$ is defined in such a way that the functional integral over the corresponding fields goes with the weight $e^{-S}$.
In order to represent products of retarded and advanced Green's functions, we double the space by introducing the additional $2 \times 2$ advanced-retarded structure labelled by $\tau$. We further introduce $n$ replicas, which allows us to compute products of $m$ Greens functions (with $m \le n$) at energies $\hat{\omega} = \mathrm{diag}(\omega_1,\ldots,\omega_m,\ldots,\omega_n)$. To conveniently include finite $\omega_i$ into the action, we double the space once more by including the additional $2\times 2$ space $\Sigma$. This takes care of assigning each $\omega_i$ a particle-hole partner $-\omega_i$. The (complex) bosonic field $\phi$ has a replica index ($1, \ldots, n$), a particle-hole index $\Sigma$ and a spin index $\sigma$. We write the action in terms of a bivector
\be
{\boldsymbol \phi}=(\phi,i\sigma_y \Sigma_1\phi^*)
\label{eq:phi-bivector}
\ee
 that additionally lives in advanced-retarded $\tau$ space:
\begin{align}
S_{0}&= -i \int dr\begin{pmatrix}
\phi^\dagger & \phi^T i\sigma_y \Sigma_1
\end{pmatrix}_\tau
\begin{pmatrix}
H+\hat{\omega}\Sigma_3 + i0 & 0\\
0 & H+\hat{\omega}\Sigma_3-i0
\end{pmatrix}_\tau
\begin{pmatrix}
\phi \\
+ i\sigma_y \Sigma_1\phi^*
\end{pmatrix}_\tau \nonumber \\
&= -i \int dr \left[\phi^\dagger H \phi
+ \phi^T i\sigma_y \Sigma_1 i\sigma_y H^T i\sigma_y \Sigma_1 i\sigma_y \phi^*
+\phi^\dagger \hat{\omega}\Sigma_3 \phi
+ \phi^T i\sigma_y \Sigma_1\hat{\omega}\Sigma_3\Sigma_1 i\sigma_y \phi^*  \right. \nonumber \\
& \left. \qquad  +i0 \phi^\dagger \phi
- i0 \phi^T i\sigma_y\Sigma_1\Sigma_1 i\sigma_y \phi^*  \right] \nonumber\\
&= -i \int dr \left[2\phi^\dagger H \phi  + 2\phi^\dagger \hat{\omega}\Sigma_3 \phi +  2i0 \phi^\dagger\phi \right]\,.
\label{eq:S0}
\end{align}
The doubling represented by the additional Pauli space $\Sigma$ will be useful below to represent correlations of wavefunctions of multiple eigenstates.
The full bosonic bispinor ${\boldsymbol \phi}$ satisfies the symmetry relation:
\begin{align}
i\tau_{1}\Sigma_1\sigma_y  {\boldsymbol \phi}^* &= {\boldsymbol \phi} \,.
\label{eq:symop}
\end{align}

Up to now, we have considered a clean system, $H = H_0$. Now, we include disorder $V$, so that the full Hamiltonian reads $H = H_0 + V$.  We assume $V$ to be a white-noise distributed matrix that contains all types of randomness (not involving the momentum $p$) respecting the class-C symmetry \eqref{eq:pha}: spatial fluctuations of complex order parameter $\Delta$ (which couple to matrices $\sigma_x$ and $\sigma_y$) and of chemical potential $\mu$ (that couples to $\sigma_z$), i.e., $V=\sum_{i=1}^3 v_i\sigma_i$.
For simplicity, the disorder in all components is assumed to have the same amplitude $\sqrt{\lambda}$, so that $\langle v_i v_j\rangle=2\lambda\delta_{i,j}$.
This assumption is immaterial for the conclusions and results of this paper, since they are only based on symmetries.

Disorder averages over this ensemble have  the form of Gaussian integrals over the matrices $V$. These integrals are fully determined by the second moment \cite{zirnbauer1996riemannian}:
\begin{align}
\int d\mu(V) {\rm tr} (AV) {\rm tr} (BV) &= \lambda \: {\rm tr} (AB - Ai\sigma_y B^T i \sigma_y).
\label{eq:Vavg}
\end{align}
To convince ourselves that Eq.~\eqref{eq:Vavg} holds, it is convenient to expand the matrices $A$ and $B$ in terms of Pauli matrices in the $\sigma$ space:  $A=\sum_{i=0}^3 a_i\sigma_i$ and $B=\sum_{i=0}^3 b_i\sigma_i$.  Using ${\rm tr} (AB) = \sum_{i=0}^3 a_i b_i$,  it is straightforward to see that both sides are equal to $2\lambda \sum_{i=1}^3 a_i b_i$.

We choose $A,B$ to be the combinations $\sum_{a,\Sigma,\tau} {\boldsymbol\phi}_{\sigma, a,\Sigma,\tau}(r) {\boldsymbol\phi}^\dagger_{\sigma', \Sigma,a,\tau}(r)$, where $a$ is the replica index.
Then we can use Eq. \eqref{eq:Vavg} to average over the disorder. As usual, the averaging results in the emergence of a quartic term in the action:
\begin{align}
S_{\rm int} &= \lambda \int dr \sum_{ab,\tau\tau',\sigma\sigma',\Sigma\Sigma'} {\boldsymbol\phi}_{\sigma, a,\Sigma,\tau}(r) {\boldsymbol\phi}^\dagger_{\sigma',a,\Sigma,\tau}(r){\boldsymbol\phi}_{\sigma', b,\Sigma',\tau'}(r) {\boldsymbol\phi}^\dagger_{\sigma, b,\Sigma',\tau'}(r)\,.
\end{align}
Following the conventional route, this ``interaction'' can be decoupled by means of a Hubbard-Stratonovich transformation involving integration over a matrix field $Q$. This field couples to the fields
${\boldsymbol\phi}$ in the following way:
\begin{align}
\int d r \sum_{ab,\sigma,\Sigma\Sigma',\tau\tau'}  {\boldsymbol\phi}^\dagger_{\sigma a,\Sigma\tau } Q^{\tau,\tau'}_{\Sigma a, \Sigma'b} {\boldsymbol\phi}_{\sigma b,\Sigma'\tau'} \,.
\label{eq:coup}
\end{align}
With the matrix structure in the retarded-advanced space (for which we use the indices R and A) written explicitly, this has the form
\begin{align}
\int d r \sum_{\sigma,ab,\Sigma\Sigma'}\mathrm{tr}_\tau\left[\begin{pmatrix}
{\phi}^\dagger_{ a,\Sigma,\sigma}\phi_{b,\Sigma',\sigma} & \phi^\dagger_{a,\Sigma,\sigma}({\phi}^\dagger i\sigma_y)_{b,-\Sigma',\sigma}\\
(i\sigma_y\phi)_{a,-\Sigma,\sigma}\phi_{b,\Sigma',\sigma}& (i\sigma_y\phi )_{a,-\Sigma,\sigma}({\phi}^\dagger i\sigma_y)_{b,-\Sigma',\sigma}
\end{pmatrix}
\begin{pmatrix}
Q^{RR}_{\Sigma a,\Sigma'b} & Q^{RA}_{\Sigma a,\Sigma'b}\\
Q^{AR}_{\Sigma a,\Sigma'b} & Q^{AA}_{\Sigma a,\Sigma'b}
\end{pmatrix}\right].
\end{align}
Here we used Eq.~\eqref{eq:phi-bivector} and took into account that the matrix $\Sigma_1$ acts on the index $\Sigma$ via $\Sigma \mapsto - \Sigma$.
Note that the matrix $Q$ does not carry $\sigma$ indices and that this coupling explicitly respects the spin ($\sigma$) conservation.

Upon the Hubbard-Stratonovich decoupling, we get the action:
\begin{align}
S[{\boldsymbol \phi}, Q] &= \int d r \left[ \sum_{ab,\sigma,\sigma',\tau} {\boldsymbol \phi}^\dagger_{a\Sigma\tau\sigma}
\left(H_0\delta_{ab}\delta_{\tau\tau'} + Q^{\tau\tau'}_{\Sigma a,\Sigma'b}\delta_{\sigma\sigma'} + \Sigma_3\hat{\omega} + i0 \tau_3
\right) {\boldsymbol \phi}_{b\Sigma'\tau'\sigma'} + \lambda^{-1}{\rm tr} \: Q^2  \right]\,.
\label{eq:S-phi-Q}
\end{align}
The next step is to integrate over the $\phi$ fields. Since the action is quadratic in $ {\boldsymbol \phi}$, this integration yields an inverse determinant, and we finally get the action that only depends on $Q$:
\begin{align}
S[Q] &= {\rm tr} \log (H_0 + Q +\Sigma_3\hat{\omega} + i0 \tau_3) + \lambda^{-1} \int dr \ {\rm tr} \: Q^2.
\end{align}
This action has a manifold of saddle points
\be
\label{eq:Q-manifold}
Q=g\Lambda g^{-1} \,, \qquad \Lambda = \tau_3 \,,
\ee
which form the target space of the nonlinear sigma model, $Q \in G/K$. For the present case of class C, this target space is  $G /K=\mathrm{SO}^*(4n) / \mathrm{U}(2n)$.  This form of the target space can be understood already from the symmetry of the action \eqref{eq:S0}: transformation of ${\boldsymbol \phi}$ that
respect the symmetry constraint of $\boldsymbol\phi$ and leave the main term (involving $H$) invariant, determine the group $G$, while the term proportional to $i0$ breaks this symmetry down to $K$.

Allowing for slow spatial fluctuations of $Q=g\Lambda g^{-1}$ around the saddle point $\Lambda = \tau_3$, one obtains the usual nonlinear sigma-model action \cite{mirlin00, efetov1983supersymmetry, efetov1997supersymmetry, verbaarschot1985grassmann,altland2000field, taras_semchuk2001quantum, liao2017response}
\begin{align}
S[Q] &= - \frac{\pi \nu_0 }{8}  \int dr \ {\rm tr} \left[ D_0  (\partial Q)^2 + 4i \left( \Sigma_3\hat{\omega} + i0\tau_3  \right) Q \right].
\label{SQ-sigma-model}
\end{align}
Here $\nu_0$ and $D_0$ are the bare (ultraviolet) values of the density of states and diffusion constant, respectively, so that $\sigma_0 = \nu_0 D_0$ is the bare spin conductivity.

The sigma-model field $Q$ is a $4n\times 4n$ matrix in the advanced-retarded $\times$ particle-hole (equivalently, spin) $\times$ replica space [see also Eq.~\eqref{tensor-product-space} below]. It satisfies two constraints. The first one is
Eq.~\eqref{eq:Q-manifold}. In addition, $g$ inherits the symmetry \eqref{eq:symop}, yielding
\begin{align}
\tau_{1}\Sigma_1 g^T \Sigma_1 \tau_{1} &= g^{-1}.
\label{eq:symopT}
\end{align}
For the $Q$ field, this implies
\begin{align}
\tau_{1}\Sigma_1 Q^T \Sigma_1 \tau_{1} &= -Q.
\label{eq:symopQ}
\end{align}
Constraints \eqref{eq:Q-manifold} and \eqref{eq:symopT} [or, equivalently, \eqref{eq:Q-manifold} and \eqref{eq:symopQ}] determine the sigma-model target space $G /K=\mathrm{SO}^*(4n) / \mathrm{U}(2n)$.

In the following sections dealing with the highest weight vector (HWV) construction in Sec. \ref{sec:hwv}, the Iwasawa decomposition on $G/K$ in Sec. \ref{sec:iwasawa}, and renormalization group (RG) of generic operators composed of $Q$ in Sec. \ref{sec:rg_c}, we will also use a slightly different parametrization of $G/K$ related by a unitary basis rotation.
Specifically, the two parametrizations of $G/K$ are related by a rotation in $\Sigma$ space:
\begin{align}
Q &= U^{-1}_\Sigma \tilde{Q}U_\Sigma,
\label{Q-tilde}
\end{align}
where
\be
U_\Sigma = \mathrm{diag}(I_{2n} , \Sigma_1\otimes I_n)_\tau \,.
\label{eq:U-Sigma}
\ee
Since $U_\Sigma$ commutes with $\tau_3$, the condition  \eqref{eq:Q-manifold} holds also for $\tilde{Q}$ without any modifications.  At the same time, the condition \eqref{eq:symopQ} becomes
\begin{align}
\tau_{1}\Sigma_1 U_\Sigma^T \tilde{Q}^T U_\Sigma^{*} \Sigma_1 \tau_{1}&=-U^{-1}_\Sigma \tilde{Q}U_\Sigma,
\end{align}
which, by using $U_\Sigma^{*} \tau_1\Sigma_1 U^{-1}_\Sigma = \tau_1\otimes I_\Sigma$, simplifies to
\begin{align}
\tau_{1} \tilde{Q}^T\tau_{1}&=-\tilde{Q} \label{eq:symopQiw}.
\end{align}
With all indices written explicitly, this implies
\begin{align}
Q_{a,b}^{AA} &=  \tilde{Q}_{a,b}^{AA},
&Q_{a,b}^{AR} &=  \tilde{Q}_{a,-b}^{AR},
&Q_{a,b}^{RA} &=  \tilde{Q}_{-a,b}^{RA},
&Q_{a,b}^{RR} &=  \tilde{Q}_{-a,-b}^{RR}.
\end{align}
Here we have introduced a  convention (to be used below) that the structure in the $\Sigma$ space is represented by the sign $\pm$ of the
 the replicas $a,b$, i.e., positive replica indices $a=1, \ldots, n$ correspond to the upper component in the $\Sigma$ space, and the negative indices $-a$ to the lower component.

In the sequel, we will sometimes omit tilde and write $Q$ for $\tilde{Q}$ to simplify notations. We will explicitly point out in the corresponding sections when this will be done.

Before closing this section, we make two interrelated comments.

\begin{enumerate}[(i)]

\item Up to now, we considered the manifold $G /K=\mathrm{SO}^*(4n) / \mathrm{U}(2n)$, with $n$ being a positive integer. As is well known, the averaging over a quenched disorder requires either introducing supersymmetry or the replica trick ($n \to 0$).  The supersymmetric formalism was used in Ref.~\cite{gruzberg2013classification} where the classification of composite operators was developed. At the same time, as is clear from Ref.~\cite{gruzberg2013classification}, the most convenient way to build the composite operators is to use the boson-boson sector of the supersymmetric sigma-model. In this sense, the supersymmetry approach to this problem becomes very similar to using the replica trick for the bosonic fields. In other words,
considering the bosonic theory only is fully sufficient for our purposes. Most of the analysis can be performed for the theory with a positive integer $n$. We will point out below where the replica limit $n \to 0$ is essential (which could be equivalently replaced by making the theory supersymmetric).

\item  In general, the sigma-model \eqref{SQ-sigma-model} (when written in the supersymmetric form, i.e., also with a compact sector) may contain also a topological term. In particular, this term plays a crucial role for the emergence of the critical point of the SQH transition. However, in this paper, we will use the sigma-model only for determining the composite operators and for translating them to the language of wave-function correlators. This analysis is based entirely on the symmetry of the sigma model, for which the presence or absence of the topological term is fully irrelevant. The scaling of the corresponding correlation functions at the critical point of the SQH transition will be after this determined numerically.

\end{enumerate}

\section{Complex pure-scaling composite operators: the highest-weight vectors approach}
\label{sec:hwv}
In this section and the next one we will construct the exact scaling operators as functions on the target space $G/K = \text{SO}^*(4n)/\text{U}(2n)$ of the sigma model. As we explained in Ref. \cite{gruzberg2013classification}, the scaling operators are joint eigenfunctions of the $G$-invariant differential operators on $G/K$, also known as the Laplace-Casimir operators. There are two ways to construct these eigenfunctions. One way uses Cartan and root space decompositions of the complexification $\mfg^{\mathbb{C}}= \text{so}(4n, \mathbb{C})$ of the Lie algebra of $G$. In this way the scaling operators (the eigenfunctions) are constructed as highest-weight vectors (HWVs) for the action of $G$ on functions on $G/K$, see details below in this section. This approach produces complex-valued scaling operators, which are not easily related to wave-function correlators that we study later.

To obtain real-valued and non-negative scaling operators, that can be raised to any real powers and are related to the critical wave functions correlators, we will utilize the Iwasawa decomposition of the group $G = NAK$ and the related {\it restricted} root decomposition of the real Lie algebra $\mfg = \text{so}^*(4n)$ together with a Cartan decomposition. In this way we obtain the desired eigenfunctions as the $N$-radial spherical functions on $G/K$. This will be done in detail in Section \ref{sec:iwasawa}.

Both the complex and the real constructions are purely algebraic, and produce explicit expressions for the scaling operators in terms of matrix elements of the $Q$ matrix of the sigma model. Let us begin by collecting some definitions and notations for the symmetric space $G/K$.

\subsection{Definitions and notations for the symmetric space $G/K = {\rm SO}^*(4n)/{\rm U}(2n)$}
\label{subsec:definitions-G/K}

The basic vector space that we work with is
\begin{align}
\mathbb{C}^{4n} &= \mathbb{C}^2 \otimes \mathbb{C}^2 \otimes \mathbb{C}^n,
\label{tensor-product-space}
\end{align}
where the factors in the tensor product are the retarded-advanced (RA), spin, and replica  spaces, in this order. We will use a standard notation for the matrix units: $E_{ij}$ is the matrix with $1$ in the $i$-th row and $j$-th column, all other entries being zero. We can write this in terms of the matrix elements:
$(E_{ij})_{kl} = \delta_{ik} \delta_{jl}$.
The range of indices in this definition is the dimension of the space in which $E_{ij}$ act, and in various sections below it can be 2 for the RA space, $n$ for the replica space, or $2n$ for the product of the spin and replica spaces. The products and commutators of matrix units are easy to find:
\begin{align}
E_{ij} E_{kl} &= \delta_{jk} E_{il},
&
[E_{ij}, E_{kl}] &= \delta_{jk} E_{il} - \delta_{il} E_{kj}.
\label{matrix-units-product}
\end{align}

Let us introduce an additional notation for special matrices related to the tensor product structure \eqref{tensor-product-space}:
\begin{align}
\Sigma_{ij} &\equiv
\sigma_i \otimes \sigma_j \otimes I_n.
\label{Sigma-ij-definition}
\end{align}
Here $\sigma_i$ are the usual $2 \times 2$ Pauli matrices, and we allow the indices $i$ and $j$ to take the value $0$, which corresponds to the  identity matrix $\sigma_0 = I_2$. For example $\Sigma_{00} = I_{4n}$, and $\Sigma_{30} = \Lambda$, the usual $\Lambda$ matrix from the sigma model. The notations \eqref{Sigma-ij-definition}  are different from the ones used in Sections \ref{sec:sigma} and \ref{sec:rg_c}. The two sets are related by
\begin{align}
\tau_i &= \Sigma_{i0},
&
\Sigma_i &= \Sigma_{0i},
&
\tau_i \Sigma_j &= \Sigma_{ij}.
\end{align}

Next, we define the following group involutions:
\begin{align}
\Theta_i(g) &\equiv \Sigma_{i0} (g^{-1})^T \Sigma_{i0},
\label{group-involutions}
\end{align}
These preserve multiplication of matrices, and can be applied not only to group elements $g$, but to any matrices, including the $Q$ fields of the sigma model. Note that the involution $\Theta_1$ is closely related (but not identical) with the ``bar'' operation defined in Eq. \eqref{eq:bar-operation} in Sec. \ref{sec:rg_c}.


We use the definition \cite{Knapp-Lie-2002} of the group $\text{SO}^*(4n)$ as the subgroup of the pseudo-unitary group $\text{SU}(2n,2n)$ that preserves the symmetric bilinear from with the matrix $\Sigma_{10}$. In this definition the elements $g \in \text{SO}^*(4n)$ satisfy
\begin{align}
g^\dagger \Sigma_{30} g &= \Sigma_{30},
&
g^T \Sigma_{10} g &= \Sigma_{10}.
\label{SO-star-definition-Knapp}
\end{align}
These conditions can be rewriten using the involutions \eqref{group-involutions}:
\begin{align}
\Theta_3(g) &= g^*,
&
\Theta_1(g) &= g.
\label{SO-star-definition-Knapp-involutions}
\end{align}

The sigma model field \eqref{eq:Q-manifold} $Q = g \Lambda g^{-1}$, where $g \in \text{SO}^*(4n)$, inherits certain symmetry properties from the constraints \eqref{SO-star-definition-Knapp-involutions}. Applying the involutions $\Theta_3$ and $\Theta_1$ to $Q$ and using
$\Theta_3(\Lambda) = \Lambda$, and $\Theta_1(\Lambda) = -\Lambda$,
we get
\begin{align}
\Theta_3(Q) &= Q^*,
&
\Theta_1(Q) &= -Q.
\end{align}
Since $Q^2 = I_{4n}$, these constraints can be also written in the form
\begin{align}
\Sigma_{30} Q^T \Sigma_{30} &= Q^*,
&
\Sigma_{10} Q^T \Sigma_{10} &= -Q.
\label{Q-constraints}
\end{align}
Notice that the second constraint here is different from the one obtained in Sec. \ref{sec:sigma} in Eq. \eqref{eq:symopQ}. Rather, it is the one that appears for the transformed matrix $\tilde{Q}$ in Eq. \eqref{eq:symopQiw}, see the relevant discussion and the transformation relating the two choices for $Q$ in Sec. \ref{sec:sigma}.

Writing $Q$ as a block matrix in the RA space
\begin{align}
Q = \begin{pmatrix} Q_{RR} & Q_{RA} \\ Q_{AR} & Q_{AA} \end{pmatrix},
\end{align}
we obtain constraints on the blocks:
\begin{align}
Q_{RR}^\dagger &= Q_{RR},
&
Q_{AA}^\dagger &= Q_{AA},
&
Q_{RA}^\dagger &= - Q_{AR},
\\
Q_{RR}^T &= - Q_{AA},
&
Q_{RA}^T &= - Q_{RA},
&
Q_{AR}^T &= - Q_{AR}.
\end{align}
The anti-symmetry of the off-diagonal blocks $Q_{RA}$ and $Q_{AR}$ will be important in what follows.

\subsection{Highest-weight vector construction: a summary}
\label{subsec:HWV-summary}

In the rest of this section we focus on the HWV construction, which we now summarize. Let $\mfg^{\mathbb{C}}= \text{so}(4n, \mathbb{C})$ denote the {\it complexified} Lie algebra of the Lie group $G$. The elements $X \in \mfg^{\mathbb{C}}$ act on functions $f(Q)$ on $G/K$ as first-order differential operators $\widehat{X}$:
\begin{align}
(\widehat{X} f)(Q) = \frac{d}{dt}\Big|_{t=0} f \big(e^{-tX} Q \, e^{tX}).
\label{X-hat-action}
\end{align}
By definition, this action preserves the commutation relations: $[\widehat{X}, \widehat{Y}] = \widehat{[X,Y]}$.

Fixing a Cartan subalgebra $\mfh \subset \mfg^{\mathbb{C}}$ we get a root-space decomposition
\begin{align}
\mfg^{\mathbb{C}}= \mfn_+ \oplus \mfh \oplus \mfn_-,
\label{root-space-decpomposition}
\end{align}
where the nilpotent Lie algebras $\mfn_\pm$ are generated by positive and negative root vectors (eigenvectors of the adjoint action of $\mfh$ on $\mfg$).
Comparing with the Iwasawa decomposition of Sec.~\ref{sec:iwasawa}, we observe that $\mfn_+ = \mfn^{\mathbb{C}}$ is the same as the complexification of $\mfn$, and $\mfa$ is a subalgebra of $\mfh$, with the additional generators of $\mfh$ lying in $\mfk^{\mathbb{C}}$, the complexified Lie algebra of $K$.

Another decomposition that we need is the Cartan decomposition
\begin{align}
\mfg^{\mathbb{C}}= \mfp_- \oplus \mfk^{\mathbb{C}} \oplus \mfp_+,
\label{Cartan-decomposition-complex}
\end{align}
with commutation relations among its terms
\begin{align}
[\mfk^{\mathbb{C}}, \mfp_+] &\subseteq \mfp_+,
&
[\mfk^{\mathbb{C}}, \mfp_-] &\subseteq \mfp_-
&
[\mfp_+, \mfp_-] &\subseteq \mfk^{\mathbb{C}}
&
[\mfp_+, \mfp_+] &= [\mfp_-, \mfp_-] = 0.
&
\label{comm-rel-Cartan-complex}
\end{align}
The space $\mfn^+$ of positive root vectors decomposes as
\begin{align}
\mfn_+ = \mfp_+ \oplus (\mfn_+ \cap \mfk^{\mathbb{C}}).
\label{n+decomposition}
\end{align}

Now suppose that $\varphi_\lambda$ is a function on $G/K$ with the properties
\begin{align}
\widehat{X} \varphi_\lambda &= 0,
&& \text{for all } X \in \mfn_+,
\label{HWV-definition-1}
\\
\widehat{H} \varphi_\lambda &= \lambda(H) \varphi_\lambda,
&& \text{for all } H \in \mfh.
\label{HWV-definition-2}
\end{align}
Thus $\varphi_\lambda$ is annihilated by the raising operators from $\mfn_+$ and is an eigenfunction of the Cartan generators from $\mfh$. Such an object $\varphi_\lambda$ is called a highest-weight vector (HWV), and the eigenvalue $\lambda$ is called a highest weight. Since the Lie algebra acts on functions on $G/K$ by first-order differential operators, it immediately follows that the products of two HWVs, as well as powers of a HWV,
\begin{align}
\varphi_{\lambda_1} \varphi_{\lambda_2} &=\varphi_{ \lambda_1 + \lambda_2},
&
\varphi_\lambda^q &= \varphi_{q\lambda},
\label{Abelian-fusion-HWVs}
\end{align}
are again HWVs with highest weights $\lambda_1 + \lambda_2$ and $q\lambda$, respectively. In general the HWVs $\varphi_\lambda$ are complex functions on $G/K$, and
the power $q$ has to be quantized (a non-negative integer) so that $\varphi_\lambda^q$ is defined globally on the space $G/K$. On the other hand,
whenever we can find a {\it positive} ($\varphi_\lambda > 0$) HWV, it can be raised to an arbitrary complex power $q$.

Now recall that a Casimir invariant $C$ is a polynomial in the generators of $\mfg^{\mathbb{C}}$ with the property that $[C,X] = 0$ for all $X \in \mfg^{\mathbb{C}}$. The Laplace-Casimir operator $\widehat{C}$ is the invariant differential operator which corresponds to the Casimir invariant $C$ by the action \eqref{X-hat-action}. If a function $\varphi_\lambda$ has the highest-weight properties \eqref{HWV-definition-1} and \eqref{HWV-definition-2}, then this function is an eigenfunction of all Laplace-Casimir operators of $G$. To see this, one observes that every Casimir invariant $C$ can be expressed as
\begin{align}\label{e8.4}
C = C_\mfh + \sum_{\alpha > 0} D_\alpha X_\alpha,
\end{align}
where every summand in the second term on the right-hand side contains some $X_\alpha \in \mfn_+$ as a right factor. Thus the second term annihilates the HWV $\varphi_\lambda$. The first term, $C_\mfh$, is a polynomial in the generators of the Abelian algebra $\mfh$ and thus has $\varphi_\lambda$ as an eigenfunction by Eq. (\ref{HWV-definition-2}).

In summary, gradientless scaling operators can be constructed as functions that have the properties of HWVs. To generate bigger sets of such operators, we use the fact that powers and products of HWVs are again HWVs. Importantly, the ability to form powers and products of HWVs as in Eq. \eqref{Abelian-fusion-HWVs}, implies that the scaling operators constructed as HWVs obey the Abelian fusion rules.

To implement the HWV construction explicitly, we consider the following linear functions of the matrix elements of $Q$
\begin{align}
\mu_Y(Q) = \text{Tr}\, (YQ).
\label{mu-function}
\end{align}
These functions have a simple behavior under conjugations of $Q$ by group elements $g \in G$:
\begin{align}
\mu_Y(gQg^{-1}) = \mu_{g^{-1}Y g}(Q).
\label{mu-conjugation}
\end{align}
Then it is easy to see that the definition \eqref{X-hat-action} implies
\begin{align}\label{e8.8}
(\widehat{X} \mu_Y )(Q) = \frac{d}{dt}\Big|_{t=0} \text{Tr}\, \big(e^{tX} Y e^{-tX} Q) = \mu_{[X,Y]}(Q).
\end{align}
If we now choose $Y \in \mfp_+$, then due to the commutation relations \eqref{comm-rel-Cartan-complex}, the function $\mu_Y(Q)$ is annihilated by $\widehat{X}$ for any $X \in \mfp_+$. To construct HWVs which are annihilated by $\widehat{X}$ for all $X \in \mfn_+$, we then build certain polynomials from these linear functions $\mu_Y(Q)$, and form products of their powers.

\subsection{Construction of highest weight vectors for ${\rm so}(4n,\mathbb{C})$}
\label{sec:HWV-so-4n-C}

The complexification of $\text{so}^*(4n)$ is $\mfg^{\mathbb{C}}= \text{so}(4n,\mathbb{C})$, the complex orthogonal algebra that preserves a symmetric bilinear form in the space \eqref{tensor-product-space}. We choose the matrix for this form to be $\Sigma_{10}$, as in Section \ref{subsec:definitions-G/K}, see the second relation in Eq. \eqref{SO-star-definition-Knapp}. Then elements $Z \in \text{so}(4n,\mathbb{C})$ satisfy
\begin{align}
Z^T \Sigma_{10} + \Sigma_{10} Z &= 0.
\label{so-C-definition}
\end{align}
If we write $Z$ as block-matrices in the RA space, the constraint \eqref{so-C-definition} implies
\begin{align}
Z &= \begin{pmatrix} A & B \\ C & -A^T \end{pmatrix},
&
B^T &= - B,
&
C^T &= - C.
\label{so-C-blocks}
\end{align}
The Cartan decomposition \eqref{Cartan-decomposition-complex} contains $\mfk^\mathbb{C} = \text{gl}(2n,\mathbb{C})$, the complexified Lie algebra of $K = \text{U}(2n)$, which is embedded in $\text{so}(4n,\mathbb{C})$ as block-diagonal matrices with $B = C = 0$ in Eq. \eqref{so-C-blocks}. In the same block-matrix notation the space $\mfp_+$ consists of matrices with $A = C = 0$, and the space $\mfp_-$ contains matrices with $A = B = 0$.

We choose the following set of commuting matrices as a basis for the Cartan subalgebra $\mfh$:
\begin{align}
H_k = E_{kk} - E_{2n+k,2n+k}\quad k = 1,\ldots, 2n.
\end{align}
A generic element $H \in \mfh$ is
$H = \sum_{k=1}^{2n} h_k H_k$.
We define the basis in the dual space $\mfh^*$ as elements $x_i$ such that $x_i(H) = h_i$. [Notice that similar basis vectors were called $e_i$ in Sections \ref{sec:general} and \ref{sec:CFT-2D}.] With this choice we find the roots $x_i - x_j$, $x_i + x_j$, and $- x_i - x_j$ and the corresponding root vectors:
\begin{align}
X_{ij} &= E_{ij} - E_{2n+j,2n+i},
&
Y^+_{ij} &= E_{i,2n+j} - E_{j,2n+i},
&
Y^-_{ij} &= E_{2n+i,j} - E_{2n+j,i},
\end{align}
Then we choose $Y^+_{ij}$ with all possible $i,j$, and $X_{ij}$ with $i < j$ as positive root vectors. They generate the nilpotent subalgebra $\mfn_+$ in the root-space decomposition \eqref{root-space-decpomposition}. 
We see that $Y^+_{ij} \in \mfp_+$, while $X_{ij} \in \mfn_+ \cap \mfk^\mathbb{C}$ for $i < j$.

As we described in Section \ref{subsec:HWV-summary}, we define the functions (see Eq. \eqref{mu-function})
\begin{align}
\nu_{ij} = \mu_{Y^+_{ij}} = \text{tr}[(E_{i,2n+j} - E_{j,2n+i})Q]
= Q_{2n+j,i} - Q_{2n+i,j} = 2Q_{AR, ji},
\end{align}
where we used the anti-symmetry of the block $Q_{AR}$. Next we form an $2n \times 2n$ matrix $\nu$ whose matrix elements are given by $\nu_{ij}$. We also define the $m \times m$ submatrices
\begin{align}
\nu^{(m)} &= \begin{pmatrix}
\nu_{11} & \ldots & \nu_{1m} \\ \vdots & \ddots & \vdots \\ \nu_{m1} & \ldots & \nu_{mm}
\end{pmatrix},
&
1 \leqslant m \leqslant n,
\label{sub-matrix-nu}
\end{align}
that occupy upper-left corners of the block $Q_{AR}$, and their determinants (the principal minors of the matrix $\nu$)
\begin{align}
f_m &= \det \nu^{(m)}.
\label{fm-expansion}
\end{align}
Due to the anti-symmetry of $\nu_{ij}$ the functions $f_m$ vanish for odd $m$, while $f_{2m}$ is the square of the Pfaffian of $\nu^{(2m)}$:
\begin{align}
f_{2m} &= p_m^2,
&
p_m &= \text{Pf} \, (\nu^{(2m)})
= \frac{1}{2^m m!} \sum_{\sigma \in S_{2m}} \text{sgn}(\sigma)
\prod_{i=1}^m \nu_{\sigma(2i-1),\sigma(2i)}.
\label{p-m-expansion}
\end{align}

We will now demonstrate that $f_{2m}$ and $p_m$ are HWVs for the decomposition \eqref{root-space-decpomposition}, and find their weights. Since the root vectors $Y^+_{ij}$ commute among themselves, we only need to consider the action of $\widehat{X}_{ij}$ with $i < j$ on $f_{2m}$. To this end we use $[X_{ij}, Y^+_{kl}] = \delta_{jk} Y^+_{il} + \delta_{jl} Y^+_{ki}$ to get
\begin{align}
\widehat{X}_{ij} f_{2m} &= \frac{d}{dt} \det \big[\text{tr}(Y^+_{kl}Q + t[X_{ij},Y^+_{kl}]Q) \big]_{k,l = 1}^{2m} \Big|_{t=0}
= \frac{d}{dt} \det \big[\nu_{kl} + t (\delta_{jk} \nu_{il} + \delta_{jl} \nu_{ki}) \big]_{k,l = 1}^{2m} \Big|_{t=0}.
\end{align}
If $j > 2m$, then $\delta_{jk} = \delta_{jl} = 0$, and the determinant above does not depend on $t$, so $\widehat{X}_{ij} f_{2m} = 0$. If $i < j \leqslant 2m$, then we can rewrite
\begin{align}
\delta_{jk} \nu_{il} &= (E^{(2m)}_{ji} \nu^{(2m)})_{kl},
&
\delta_{jl} \nu_{ki} &= (\nu^{(2m)} E^{(2m)}_{ij})_{kl},
\end{align}
and use the Jacobi's formula for the derivative of a determinant
\begin{align}
\frac{d}{dt} \det [A(t)] = \det [A(t)] \, \text{tr} \Big(A^{-1}(t) \frac{dA(t)}{dt}\Big),
\end{align}
which gives
\begin{align}
\widehat{X}_{ij} f_{2m} &= \frac{d}{dt} \det \big[\nu^{(2m)} + t E^{(2m)}_{ji} \nu^{(2m)} + t \nu^{(2m)} E^{(2m)}_{ij} \big] \Big|_{t=0}
= f_{2m} \, \text{tr} \big[(\nu^{(2m)})^{-1} E^{(2m)}_{ji} \nu^{(2m)} + E^{(2m)}_{ij} \big] = 0.
\end{align}
Furthermore, since $\widehat{X}$ is a derivative, we have
\begin{align}
0 &= \widehat{X}_{ij} f_{2m} = \widehat{X}_{ij} p_m^2 = 2 p_m \widehat{X}_{ij} p_m
& \Rightarrow &&
\widehat{X}_{ij} p_m = 0.
\end{align}

Now we find the weight of $p_m$. First we have
\begin{align}
\widehat{H} \nu_{ij} = (h_i + h_j) \nu_{ij}.
\end{align}
Now we notice that every monomial in the expansion \eqref{p-m-expansion} of the Pfaffian $p_m$ contains each index from the set $1,\ldots,2m$ exactly once. Then
\begin{align}
\widehat{H} p_m &= \lambda_m(H) p_m,
&
\lambda_m(H) = \sum_{i=1}^{2m} h_i,
\end{align}
so the weight of $p_m$ is $\sum_{i=1}^{2m} x_i$. It is clear that the weight of $f_{2m}$ is twice that of $p_m$.

We have demonstrated that for all $1 \leqslant m \leqslant n$ the Pfaffians $p_m$ are HWVs, and so is the product
\begin{align}
\varphi_{(q_1,\ldots,q_n)} = p_1^{q_1 - q_2} p_2^{q_2 - q_3} \ldots p_{n-1}^{q_{n-1} - q_n} p_n^{q_n}
\end{align}
for a weakly decreasing sequence of $n$ integers $q_1 \geq q_2 \geq \ldots \geq q_n \geq 0$. The powers in this expression are restricted to be nonnegative integers, since the functions $p_m$ are complex-valued. The functions $\varphi_{(q_1,\ldots,q_n)}$ are the most general HWVs in the present situation, and their weights are
\begin{align}
\lambda_{(q_1,\ldots,q_n)}(H) = \sum_{i = 1}^{n-1}(q_i - q_{i+1})\lambda_i + q_n \lambda_n
= q_1 \lambda_1 + \sum_{i=2}^{n}q_i (\lambda_i - \lambda_{i-1})
= \sum_{i=1}^{n} q_i(h_{2i-1} + h_{2i}).
\end{align}

As we mentioned above, the scaling operators $\varphi_{(q_1,\ldots,q_n)}$ constructed in this section obey Abelian fusion. However, they are complex-valued and cannot be raised to arbitrary powers. Moreover, they are not immediately related to the wave function correlation functions of interest. Therefore, we will now turn to an alternative construction (based on the Iwasawa decomposition) that will produce non-negative scaling operators that can be raised to any powers and satisfy Abelian fusion rules. In Sec.~\ref{sec:wave_iw} we will relate these operators to wave function correlators.

\section{Pure-scaling composite operators: the Iwasawa decomposition approach}
\label{sec:iwasawa}
In this section we focus on the construction based on the Iwasawa decomposition. Let us briefly  outline basic steps in this approach. We begin with the Cartan decomposition
\begin{align}
\mfg= \mfk \oplus \mfp
\label{Cartan-decomposition-real}
\end{align}
of the real Lie algebra $\mfg= \text{so}^*(4n)$ into a maximal compact subalgebra $\mfk= \text{u}(2n)$ and the complementary subspace $\mfp$. The parts of the real Cartan decomposition satisfy the  commutation relations similar to those in the complex case \eqref{comm-rel-Cartan-complex}
\begin{align}
[\mfk, \mfk] &\subseteq \mfk,
&
[\mfk, \mfp] &\subseteq \mfp,
&
[\mfp, \mfp] &\subseteq \mfk.
\label{comm-rel-Cartan-real}
\end{align}

Then we choose a maximal Abelian subspace $\mfa \subset \mfp$
and consider the adjoint action of elements of $\mfa$ on $\text{so}^*(4n)$. The eigenvectors $E_\alpha$ of this action satisfy
\begin{align}
[H, E_\alpha] = \alpha(H) E_\alpha
\end{align}
and are called {\it restricted} root vectors, and the eigenvalues $\alpha$ are called restricted roots. The dimension $m_\alpha$ of the restricted root space
\begin{align}
\mfg_\alpha = \text{span}\, \{E_\alpha\}
\end{align}
is called the multiplicity of the restricted root $\alpha$, and can be bigger that 1. Restricted roots are linear functions on $\mfa$, and lie in the space $\mfa^*$ dual to $\mfa$. In the present context both $\mfa$ and $\mfa^*$ have dimension $n$. Basis elements in the space $\mfa^*$, specified later, will be denoted in this section by $x_i$ ($i = 1,\ldots, n$). Notice that these basis vectors were called $e_i$ in Sections \ref{sec:general} and \ref{sec:CFT-2D}.

A system of positive restricted roots is defined by choosing some hyperplane through the origin of $\mfa^*$ which divides $\mfa^*$ in two halves, and then defining one of these halves as positive. The Weyl vector $\rho$  is defined as the half-sum of positive restricted roots accounting for their multiplicities:
\begin{align}
\rho = \sum_{\alpha > 0} m_\alpha \alpha = \sum_{i=1}^{n} c_i x_i.
\end{align}
The components $c_i$ of the Weyl vector will be found later, see Eq. \eqref{Weyl-vector-components}. Positive restricted roots generate the nilpotent Lie algebra
\begin{align}
\mfn = \sum_{\alpha > 0} \mfg_\alpha.
\end{align}
The Iwasawa decomposition
\begin{align}
\mfg= \mfk\oplus \mfa \oplus \mfn
\label{Iwasawa-decomposition-algebra}
\end{align}
replaces the root-space decomposition \eqref{root-space-decpomposition} in the real setting. Similar to Eq. \eqref{n+decomposition} we have
\begin{align}
\mfn = \mfp\oplus (\mfn \cap \mfk).
\end{align}

Exponentiation of Eq. \eqref{Iwasawa-decomposition-algebra} gives the global form of the Iwasawa decomposition
\begin{align}
G = NAK,
\label{Iwasawa-decomposition-group}
\end{align}
which allows us to represent any element $g \in G$ in the form $g = nak$, with $n \in N = e^\mfn$, $a \in A = e^\mfa$, and $k \in K = e^{\mfk}$. This factorization is unique once the system of positive restricted roots is fixed, and provides a very useful parametrization of the target space $G/K$. An element $a \in A$ is fully specified by $n$ real numbers $x_i(\ln a)$, which play the role of radial coordinates on $G/K$. For simplicity, we will denote these radial coordinates simply by $x_i$. Thus $x_i$ may now have two different meanings: either its original meaning as a basis element in $\mfa^*$, or the new one as an $N$-radial function $x_i(\ln a)$ on $G/K$. It should be clear from the context which of the two meanings is being used.

Using the radial coordinates, the joint $N$-radial eigenfunctions of the Laplace-Casimir operators on $G/K$ take a very simple exponential form
\begin{align}
\varphi_\mu(Q) = e^{(\rho + \mu)(\ln a)},
\label{plane-wave-1}
\end{align}
where $a$ is the $a$-factor in the Iwasawa decomposition of $g$ in $Q = g \Lambda g^{-1}$, and
\begin{align}
\mu = \sum_{i=1}^{n} \mu_i x_i
\end{align}
is a weight vector in $\mfa^*$. We will also use the notation
\begin{align}
q_i = -(\mu_i + c_i)/2,
\end{align}
in which the exponential functions \eqref{plane-wave-1} become
\begin{align}
\varphi_\mu \equiv \varphi_{(q_1,q_2,\ldots,q_n)} = \exp \Big(-2 \sum_{i=1}^{n} q_i x_i \Big).
\label{plane-wave-2}
\end{align}

To construct the exponential $N$-radial eigenfunctions explicitly as combinations of matrix elements of $Q$, we use the key fact that there exists a choice of basis in which elements of $\mfa$ and $a \in A$ are diagonal matrices, while elements of $\mfn$ are strictly upper triangular, and elements $n \in N$ are upper triangular with units on the diagonal. This has immediate consequences for the matrix $Q \Lambda$: since elements of $K$ commute with $\Lambda$, the Iwasawa decomposition $g = nak$ leads to $Q \Lambda = n a^2 \Lambda n^{-1} \Lambda$, which is a product of an upper triangular, a diagonal, and a lower triangular matrices. In this form the principal minors of the $AA$ block of $Q \Lambda$ are simply products of diagonal elements of $a^2$, which are exponentials of the radial coordinates $x_i$ on $G/K$. These minors are basic $N$-radial spherical functions on $G/K$ which can be raised to arbitrary powers and multiplied to produce the most general exponential functions \eqref{plane-wave-1}. A great advantage of this construction is that is directly gives the general positive scaling operators that can be raised to arbitrary powers and satisfy the Abelian fusion rules.

Let us now present details of the Iwasawa construction.

\subsection{Cartan decomposition and generators}

The group $\text{SO}^*(4n)$ is connected, so all its elements can be written as $g = e^Z$, where $Z$ are elements of the Lie algebra $\text{so}^*(4n)$. These satisfy the infinitesimal versions of Eqs. \eqref{SO-star-definition-Knapp}:
\begin{align}
Z^\dagger \Sigma_{30} + \Sigma_{30} Z &= 0,
&
Z^T \Sigma_{10} + \Sigma_{10} Z &= 0,
\label{so-star-definition-Knapp}
\end{align}
In terms of the Lie algebra involutions $\theta_i$ related to the group involutions \eqref{group-involutions}
\begin{align}
\theta_i(Z) &\equiv - \Sigma_{i0} Z^T \Sigma_{i0},
\label{Lie-algebra-involutions}
\end{align}
constraints on $Z$ can be written as
\begin{align}
\theta_3(Z) &= Z^*,
&
\theta_1(Z) &= Z.
\label{so-star-definition-Knapp-involutions}
\end{align}
If we write the Lie algebra elements $Z$ as block-matrices in the RA space, the constraints \eqref{so-star-definition-Knapp} imply
\begin{align}
Z &= \begin{pmatrix} A & B \\ - B^* & A^* \end{pmatrix},
&
A^\dagger&= - A,
&
B^T & - B.
\label{so-star-Knapp}
\end{align}

The Cartan involution that determines the Cartan decomposition $\text{so}^*(4n) = \mfk \oplus \mfp$ into even and odd subspaces is $\theta_2$. It acts on generators $Z$ as
\begin{align}
\theta_2(Z) &= - \Sigma_{20} Z^T \Sigma_{20}
= \begin{pmatrix} A & -B \\ B^* & A^* \end{pmatrix}.
\end{align}
This gives us a very explicit description of the Cartan decomposition:
\begin{align}
\mfk&= \left\{ \begin{pmatrix} A & 0 \\ 0 & A^* \end{pmatrix},
\quad
A^\dagger = - A \right\},
&
\mfp&= \left\{ \begin{pmatrix} 0 & B \\ -B^* & 0 \end{pmatrix},
\quad
B^T = - B \right\}.
\label{Cartan-spaces}
\end{align}
As expected, the subalgebra $\mfk= \text{u}(2n)$, the Lie algebra of $K = \text{U}(2n)$, the maximal compact subgroup in $G = \text{SO}^*(4n)$.

Now we separate the real and imaginary parts in Eq. \eqref{Cartan-spaces}:
\begin{align}
\mfk&= \left\{ \begin{pmatrix} A_1 + iA_2 & 0 \\ 0 & A_1 - iA_2 \end{pmatrix},
\begin{array}{l} A_1^T = - A_1 \\ A_2^T = A_2 \end{array}
\right\},
&
\mfp&= \left\{ \begin{pmatrix} 0 & B_1 + i B_2 \\ -B_1 + i B_2 & 0 \end{pmatrix},
\begin{array}{l} B_1^T = - B_1 \\ B_2^T = - B_2 \end{array}
\right\}.
\end{align}
These forms allow us to write the generators of the two subsets in terms of tensors in the space \eqref{tensor-product-space}. For elements in $\mfk$, the first factor should be diagonal, so it is either $I_2$ or $i\sigma_3$. For the first choice ($I_2$), the other two tensor factors should produce a real anti-symmetric matrix, which can be done choosing one of the factors to be a symmetric matrix, and the other an anti-symmetric matrix. Likewise, for the second choice ($i\sigma_3$) we need the other two factors to produce a real symmetric matrix, so the factors can be either simultaneously symmetric or anti-symmetric. Introducing a short-hand notation for the symmetric and anti-symmetric combinations of matrix units in the replica space
\begin{align}
E^+_{ij} & = E_{ij} + E_{ji},
& i &\leqslant j,
& \frac{n(n+1)}{2} \quad \text{total},
\\
E^-_{ij} & = E_{ij} - E_{ji},
& i &< j,
& \frac{n(n-1)}{2} \quad \text{total},
\end{align}
we get the following eight groups of generators:
\begin{align}
X^{00}_{ij} &\equiv \sigma_0 \otimes \sigma_0 \otimes E^-_{ij},
&
X^{01}_{ij} &\equiv \sigma_0 \otimes \sigma_1 \otimes E^-_{ij},
&
X^{02}_{ij} &\equiv \sigma_0 \otimes i\sigma_2 \otimes E^+_{ij},
&
X^{03}_{ij} &\equiv \sigma_0 \otimes \sigma_3 \otimes E^-_{ij},
\nonumber \\
X^{30}_{ij} &\equiv i\sigma_3 \otimes \sigma_0 \otimes E^+_{ij},
&
X^{31}_{ij} &\equiv i\sigma_3 \otimes \sigma_1 \otimes E^+_{ij},
&
X^{32}_{ij} &\equiv i\sigma_3 \otimes i\sigma_2 \otimes E^-_{ij},
&
X^{33}_{ij} &\equiv i\sigma_3 \otimes \sigma_3 \otimes E^+_{ij}.
\label{k-generators}
\end{align}
The total number of generators here is $4n^2 = (2n)^2$, the dimension of $\mfk= \text{u}(2n)$, as it should be.

Similarly, for the elements in $\mfp$, the first factor should be either
$i \sigma_1$ or $i \sigma_2$, and for either of these choices the remaining two factors should give a real anti-symmetric matrix. This gives the following eight groups of generators:
\begin{align}
Y^{10}_{ij} &= i\sigma_1 \otimes \sigma_0 \otimes E^-_{ij},
&
Y^{11}_{ij} &= i\sigma_1 \otimes \sigma_1 \otimes E^-_{ij},
&
Y^{12}_{ij} &= i\sigma_1 \otimes i\sigma_2 \otimes E^+_{ij},
&
Y^{13}_{ij} &= i\sigma_1 \otimes \sigma_3 \otimes E^-_{ij},
\nonumber \\
Y^{20}_{ij} &= i\sigma_2 \otimes \sigma_0 \otimes E^-_{ij},
&
Y^{21}_{ij} &= i\sigma_2 \otimes \sigma_1 \otimes E^-_{ij},
&
Y^{22}_{ij} &= i\sigma_2 \otimes i\sigma_2 \otimes E^+_{ij},
&
Y^{23}_{ij} &= i\sigma_2 \otimes \sigma_3 \otimes E^-_{ij}.
\label{p-generators}
\end{align}
The total number of these generators is $4n^2 - 2n = 2n(2n-1)$, which is the dimension of the space $\mfp$.

\subsection{The Iwasawa decomposition}

We choose the following set of Hermitian matrices as the maximal Abelian subspace $\mfa \subset \mfp$:
\begin{align}
\mfa &= \text{span}\left\{ H_k = i\sigma_2 \otimes i\sigma_2 \otimes E_{kk}
= Y_{22,kk}/2, \quad k = 1,\ldots, n \right\}.
\end{align}
A generic element $H \in \mfa$ is
$H = \sum_{k=1}^n h_k H_k$.
We define the basis in the dual space $\mfa^*$ as elements $x_i$ such that $x_i(H) = h_i$. These elements will later play the role of the radial coordinates on the coset $\text{SO}^*(4n)/\text{U}(2n)$.

To find restricted roots and restricted root vectors, we need to find the commutators of $H$ with the generators \eqref{k-generators} and \eqref{p-generators}. We need both sets since the commutation relations \eqref{comm-rel-Cartan-real} imply that
\begin{align}
[\mfa, \mfk] &\subset \mfp,
&
[\mfa, \mfp] &\subset \mfk,
\end{align}
so the eigenvectors of the adjoint action of $\mfa$ can only be linear combinations of generators from both sets.

Straightforward computations show that the system of restricted roots consists of roots of normal length $\pm x_i \pm x_j$ ($i \neq j$) with multiplicity 4, and long roots $\pm 2x_j$ with multiplicity 1. This is the root system $C_n$ in the usual Cartan notation. The set of positive restricted roots can be chosen as (multiplicities shown in brackets)
\begin{align}
\alpha_{ij} &\equiv x_i - x_j, \quad i < j \quad (4),
&
\beta_{ij} &\equiv x_i + x_j, \quad i < j \quad (4),
&
\gamma_i &\equiv 2x_i \quad (1).
\end{align}
The Weyl vector (the half-sum of positive restricted roots accounting for their multiplicities) is
\begin{align}
\rho &= (4n - 3)x_1 + (4n-7)x_2 +\ldots+ 5x_{n-1} + x_n
= \sum_{j=1}^{n} c_j x_j,
&
c_j &= 4n + 1 - 4j.
\end{align}
In the replica limit $n \to 0$ the components of the Weyl vector become
\begin{align}
c_j = 1 - 4j.
\label{Weyl-vector-components}
\end{align}
The positive restricted root vectors corresponding to the $\alpha$,
$\beta$, and $\gamma$ roots are
\begin{align}
E_{\alpha_{ij}}^{(1)} &= X^{00}_{ij} + Y^{22}_{ij},
&
E_{\alpha_{ij}}^{(2)} &= X^{02}_{ij} - Y^{20}_{ij},
&
E_{\alpha_{ij}}^{(3)} &= X^{31}_{ij} - Y^{13}_{ij},
&
E_{\alpha_{ij}}^{(4)} &= X^{33}_{ij} + Y^{11}_{ij},
\\
E_{\beta_{ij}}^{(1)} &= X^{01}_{ij} + Y^{23}_{ij},
&
E_{\beta_{ij}}^{(2)} &= X^{03}_{ij} - Y^{21}_{ij},
&
E_{\beta_{ij}}^{(3)} &= X^{30}_{ij} - Y^{12}_{ij},
&
E_{\beta_{ij}}^{(4)} &= X^{32}_{ij} + Y^{10}_{ij},
\\
E_{\gamma_i} &= X^{30}_{ii} - Y^{12}_{ii}.
\end{align}
As we mentioned above, these vectors generate the nilpotent subalgebra $\mfn$, and we get the resulting Iwasawa decomposition \eqref{Iwasawa-decomposition-algebra}.

\subsection{Real scaling operators as $N$-radial functions}
\label{sec:N-radial-functions}

In this section we perform a unitary transformation of basis in the space \eqref{tensor-product-space} designed to bring the generators of $\mfa$ into a diagonal form. There are infinitely many such transformations, and the one we choose has an additional useful property that the resulting positive restricted root vectors are also upper-triangular in the RA space. This is achieved with the help of the unitary matrix
\begin{align}
U &=
\frac{1}{\sqrt{2}} (\sigma_0 \otimes E_{11} + \sigma_1 \otimes E_{12} - \sigma_2 \otimes E_{21} - i \sigma_3 \otimes E_{22}) \otimes I_n,
\label{U-Iwasawa}
\end{align}
where we used the $2 \times 2$ matrix units $E_{ij}$ in the spin space. While this is not a single tensor product in the RA and spin spaces, conjugation with this matrix,
\begin{align}
T(A) \equiv U A U^{-1},
\label{T-rotation}
\end{align}
keeps such single products in the basis of $\mfa$:
\begin{align}
T(H_k) = \sigma_3 \otimes \sigma_0 \otimes E_{kk},
\end{align}
in the the saddle point
\begin{align}
T(\Lambda) = - \Sigma_{12},
\label{T-Lambda}
\end{align}
as well as in the restricted root vectors:
\begin{align}
T(E_{\alpha_{ij}}^{(1)}) &= \sigma_0 \otimes \sigma_0 \otimes E^-_{ij}
+ \sigma_3 \otimes \sigma_0 \otimes E^+_{ij}
= 2 E_{11} \otimes \sigma_0 \otimes E_{ij}
- 2 E_{22} \otimes \sigma_0 \otimes E_{ji},
\nonumber \\
T(E_{\alpha_{ij}}^{(2)}) &= i\sigma_3 \otimes \sigma_1 \otimes E^+_{ij}
+ i\sigma_0 \otimes \sigma_1 \otimes E^-_{ij}
= 2 i E_{11} \otimes \sigma_1 \otimes E_{ij}
- 2 i E_{22} \otimes \sigma_1 \otimes E_{ji},
\nonumber \\
T(E_{\alpha_{ij}}^{(3)}) &= -i \sigma_0 \otimes \sigma_2 \otimes E^+_{ij} - i\sigma_3 \otimes \sigma_2 \otimes E^-_{ij}
= -2 i E_{11}\otimes \sigma_2 \otimes E_{ij}
+ 2 i E_{22} \otimes \sigma_2 \otimes E_{ji},
\nonumber \\
T(E_{\alpha_{ij}}^{(4)}) &= i\sigma_3 \otimes \sigma_3 \otimes E^+_{ij}
+ i\sigma_0 \otimes \sigma_3 \otimes E^-_{ij}
= 2 i E_{11} \otimes \sigma_3 \otimes E_{ij}
+ 2 i E_{22} \otimes \sigma_3 \otimes E_{ji},
\nonumber \\
T(E_{\beta_{ij}}^{(1)}) &= \sigma_1 \otimes \sigma_0 \otimes E^-_{ij}
+ i\sigma_2 \otimes \sigma_0 \otimes E^-_{ij}
=  2 E_{12} \otimes \sigma_0 \otimes E^-_{ij},
\nonumber \\
T(E_{\beta_{ij}}^{(2)}) &= -\sigma_2 \otimes \sigma_1 \otimes E^-_{ij}
+ i\sigma_1 \otimes \sigma_1 \otimes E^-_{ij}
= 2i E_{12} \otimes \sigma_1 \otimes E^-_{ij},
\nonumber \\
T(E_{\beta_{ij}}^{(3)}) &= -i\sigma_1 \otimes \sigma_2 \otimes E^+_{ij}
+ \sigma_2 \otimes i\sigma_2 \otimes E^+_{ij}
= - 2 i E_{12} \otimes \sigma_2 \otimes E^+_{ij},
\nonumber \\
T(E_{\beta_{ij}}^{(4)}) &= -\sigma_2 \otimes \sigma_3 \otimes E^-_{ij} + i\sigma_1 \otimes \sigma_3 \otimes E^-_{ij}
= 2 i E_{12} \otimes \sigma_3 \otimes E^-_{ij}
\nonumber \\
T(E_{\gamma_i}) &= -2i\sigma_1 \otimes \sigma_2 \otimes E_{ii}
+ 2 \sigma_2 \otimes i\sigma_2 \otimes E_{ii}
= - 4 i E_{12} \otimes \sigma_2 \otimes E_{ii}.
\end{align}
As claimed, the structure in the RA space is upper triangular: in the first factor we have either $E_{11}$ and $E_{22}$ or $E_{12}$.

We can visualize the restricted root vectors for $n=3$ as a schematic matrix diagram by indicating the matrix positions where various generators have non-zero entries. For brevity we write $\gamma_i\equiv E_{\gamma_i}$, $\alpha^{(i)}\equiv E^{(i)}_{\alpha_{kl}}$, $\beta^{(i)}\equiv E^{(i)}_{\beta_{kl}}$ with indices suppressed since they can be inferred from the matrix grid (uncolored cells have zero entries):
\begin{align}
T(E) = \left(\begin{array}{ccc|ccc|ccc|ccc}
\caplus x_1 & \calphare& \calphare \alpha^{(14)}  & & \calphaim & \calphaim \alpha^{(23)}  & & \cbetare & \cbetare \beta^{(14)} & \cgamma \gamma_1 & \cbetaim & \cbetaim \beta^{(23)}\\
 & \caplus x_2 & \calphare &  & &\calphaim  &\cbetare &   & \cbetare  & \cbetaim & \cgamma \gamma_2 & \cbetaim \\
 & & \caplus x_3  & & &  &\cbetare \beta^{(14)} & \cbetare & & \cbetaim \beta^{(23)} & \cbetaim & \cgamma \gamma_3\\
\hline
& \calphaim & \calphaim \alpha^{(23)}  & \caplus x_1 & \calphare & \calphare \alpha^{(14)}  & \cgamma  \gamma_1 & \cbetaim & \cbetaim \beta^{(23)} & & \cbetare & \cbetare \beta^{(14)}\\
& & \calphaim & & \caplus x_2 & \calphare & \cbetaim & \cgamma \gamma_2 & \cbetaim & \cbetare & & \cbetare \\
& &   & & & \caplus x_3  & \cbetaim \beta^{(23)} & \cbetaim & \cgamma \gamma_3  & \cbetare \beta^{(14)} & \cbetare &  \\
\hline
& &  & & &  & \caminus -x_1 & &  & & &\\
& & &  & & & \calphare & \caminus -x_2 & &\calphaim & & \\
& &  & & &   &\calphare \alpha^{(14)} & \calphare & \caminus -x_3  &\calphaim\alpha^{(23)} & \calphaim & \\
\hline
 & &  & & &  & & &  & \caminus -x_1 & & \\
 & &  & & &  &\calphaim & &  & \calphare & \caminus -x_2 & \\
 & &  & & &  &\calphaim \alpha^{(23)} & \calphaim &   & \calphare \alpha^{(14)} & \calphare & \caminus -x_3 \\
\end{array}\right).
\label{matrix-rrvs}
\end{align}
This matrix can be made upper triangular by an additional permutation $\pi$ of the basis vectors that can be described explicitly as follows: for $i \in 1, \ldots, n$, we have
\begin{align}
\pi(i) &= 2i-1,
&
\pi(n+i) &= 2i,
&
\pi(2n+i) &= 4n + 2 - 2i,
&
\pi(3n+i) &= 4n + 1 - 2i.
\label{permutation-pi}
\end{align}
This permutation re-sorts the diagonal entries of the matrix \eqref{matrix-rrvs} as follows:
\begin{align}
x_1, \ldots, x_n, x_1, \ldots, x_n, -x_1, \ldots, -x_n, -x_1, \ldots, -x_n
&& \to &&
x_1, x_1, \ldots, x_n, x_n, -x_n, -x_n, \ldots, -x_1, -x_1.
\end{align}
The permutation matrix $P_\pi$ that corresponds to $\pi$ has the matrix elements $(P_\pi)_{ij} = \delta_{\pi(i), j}$, and can be used to permute rows and columns of a matrix $M$ as follows:
\begin{align}
(P_\pi M)_{ij} &= M_{\pi(i), j},
&
(M P_\pi^{-1})_{ij} &= M_{i,\pi(j)}.
\end{align}
Then we can see that the conjugation of the matrix \eqref{matrix-rrvs} by $P_\pi$ makes it fully upper-triangular. Using the notation
\begin{align}
\tilde{M} \equiv P_\pi T(M) P_\pi^{-1},
\label{tilde-M}
\end{align}
for transformed matrices, we have
\begin{align}
\tilde{E} = \left(\begin{array}{cccccc|cccccc}
\caplus x_1 & & \calphare & \calphaim & \calphare & \calphaim &   \cbetare & \cbetaim & \cbetare & \cbetaim &   & \cgamma \gamma_1\\
& \caplus x_1 & \calphaim & \calphare & \calphaim & \calphare &   \cbetaim & \cbetare & \cbetaim &  \cbetare & \cgamma \gamma_1 &  \\
& & \caplus x_2 & & \calphare & \calphaim &   \cbetare & \cbetaim & & \cgamma \gamma_2 &  \cbetare & \cbetaim \\
& & & \caplus x_2 & \calphaim & \calphare &    \cbetaim  & \cbetare & \cgamma \gamma_2 &  & \cbetaim & \cbetare\\
& & & & \caplus x_3 & &    & \cgamma \gamma_3  & \cbetare & \cbetaim & \cbetare & \cbetaim \\
& & & & & \caplus x_3 &   \cgamma \gamma_3 &  & \cbetaim & \cbetare & \cbetaim &\cbetare\\
\hline
& & &  & & &  \caminus -x_3 & & \calphare & \calphaim & \calphare & \calphaim\\
& & &  & & &  & \caminus -x_3 & \calphaim & \calphare & \calphaim & \calphare\\
& & &  & & &  & & \caminus -x_2 & & \calphare & \calphaim\\
& & &  & & &  & & &\caminus -x_2 & \calphaim& \calphare\\
& & &  & & &  & &  &  & \caminus -x_1 & \\
& & &  & & &  & & &  & & \caminus -x_1 \\
\end{array}\right).
\label{matrix-rrvs-triangular}
\end{align}
In the final basis the elements of $\mfa$ are diagonal matrices, while elements of $\mfn$ are strictly upper triangular.

Let us exploit consequences of the Iwasawa decomposition of $G$ and the transformation $T$ for the sigma model field $Q$. First of all, in the original basis where $\Lambda = \Sigma_{30}$ we write $g = nak$ with $n \in N$, $a \in A$, and $k \in K$, and then
\begin{align}
\mathcal{Q} &\equiv Q \Lambda = nak \Lambda k^{-1} a^{-1} n^{-1} \Lambda
= n a^2 \Lambda n^{-1} \Lambda.
\end{align}
Here we used $k \Lambda k^{-1} = \Lambda$ and $a \Lambda a^{-1} = a^2
\Lambda$, which is a special case of the first condition in Eq. \eqref{SO-star-definition-Knapp} for a Hermitian matrix $a \in G$. Now we perform the transformation $T$, Eq. \eqref{T-rotation}, as well as the permutation $P_\pi$. Using the notation \eqref{tilde-M}, we get
\begin{align}
\tilde{\mathcal{Q}} = \tilde{n} \tilde{a}^2 \tilde{\Lambda} \tilde{n}^{-1} \tilde{\Lambda}.
\end{align}
Using Eqs. \eqref{T-Lambda} and \eqref{permutation-pi}, we can compute the matrix
\begin{align}
\tilde{\Lambda} = - P_\pi \Sigma_{12} P_\pi^{-1}
= - \sigma_1 \otimes \sigma_2 \otimes \mathcal{I}_n,
\end{align}
where $\mathcal{I}_n$ is the $n \times n$ matrix with units on the ``anti-diagonal'', that is, $(\mathcal{I}_n)_{ij} = \delta_{i, n + 1 - j}$.

As should be clear from the previous discussion, the matrices $\tilde{n}$ and $\tilde{n}^{-1}$ are upper-triangular with units on the diagonals, while $\tilde{a}$ is diagonal:
\begin{align}
\tilde{a} = \text{diag} (e^{x_1}, e^{x_1}, \ldots, e^{x_n}, e^{x_n}, e^{-x_n}, e^{-x_n},\ldots, e^{-x_1}, e^{-x_1}).
\end{align}
Conjugation by $\tilde{\Lambda}$ converts $\tilde{n}^{-1}$ into $\tilde{\Lambda} \tilde{n}^{-1} \tilde{\Lambda}$ which is {\it lower-triangular} with units on the diagonal. This results in the following structure of the $AA$ block of the matrix $\tilde{\mathcal{Q}}$:
\begin{align}
\tilde{\mathcal{Q}}_{AA} =
\begin{pmatrix} 1 & * & \ldots & * & * \\
0 & 1 & \ldots & * & * \\
\vdots & \vdots & \ddots & \vdots & \vdots\\
0 & 0 & \ldots & 1 & * \\
0 & 0 & \ldots & 0 & 1 \end{pmatrix}
\begin{pmatrix} e^{-2x_n} & 0 & \ldots & 0 & 0 \\
0 & e^{-2x_n} & \ldots & 0 & 0 \\
\vdots & \vdots & \ddots & \vdots &\vdots \\
0 & 0 & \ldots & e^{-2x_1} & 0 \\
0 & 0 & \ldots & 0 & e^{-2x_1} \end{pmatrix}
\begin{pmatrix} 1 & 0 &\ldots &0 &0 \\
* & 1 & \ldots & 0 & 0 \\
\vdots & \vdots & \ddots & \vdots & \vdots \\
* & * & \ldots & 1 & 0 \\ * & * & \ldots & * & 1 \end{pmatrix}.
\label{Q-tilde-AA}
\end{align}
Determinants of the lower-right $2m \times 2m$ submatrices of $\tilde{\mathcal{Q}}_{AA}$ give the basic positive $N$-radial eigenfunctions
\begin{align}
d_{2m} = \prod_{i = 1}^{m} e^{-4x_i}
= \exp \Big(- 4 \sum_{i=1}^{m} x_i \Big).
\label{basic-d-2m}
\end{align}
We can form the most general $N$-radial eigenfunctions as products
\begin{align}
\varphi_{(q_1,\ldots,q_n)} = d_2^{(q_1 - q_2)/2} d_4^{(q_2 - q_3)/2}
\ldots d_{2(n-1)}^{(q_{n-1} - q_n)/2} d_{2n}^{q_n/2},
\label{d-product}
\end{align}
where we may take the $q_i$ to be arbitrary complex numbers. It is easy to see that the product \eqref{d-product} is the same as the exponential eigenfunction \eqref{plane-wave-2}, while the basic function $d_{2m}$ is $\varphi_{(2,2,\ldots)}$ with $m$ twos in the subscript.

Notice that the doubling of the diagonal entries $e^{-2x_i}$ for each $i$ in Eq. \eqref{Q-tilde-AA} compelled us to take determinants of sub-matrices of even size and raise the resulting functions to powers written as $(q_i - q_{i+1})/2$. This is very similar to the appearance of minors of even size in Section \ref{sec:HWV-so-4n-C}. Those minors were of an anti-symmetric matrix, and thus were squares of Pfaffians, see Eq. \eqref{p-m-expansion}. In the Iwasawa formalism it is also possible to obtain directly the ``Pfaffian'' solutions $\varphi_{(1,1,\ldots)}$. Pfaffians can be defined only for anti-symmetric matrices, so we need to look at transformation properties of various matrices under transposition. For group elements $g$ this property is given in the second equation in \eqref{SO-star-definition-Knapp}. We can use this equation to derive the following consequence:
\begin{align}
T(g)^T &= (U \Sigma_{10} U^T)^{-1} T(g)^{-1} (U \Sigma_{10} U^T).
\end{align}
A simple computation shows that $U \Sigma_{10} U^T = \Sigma_{10}$. Thus the rotated matrix $T(g)$ satisfies the same second equation in \eqref{SO-star-definition-Knapp}:
\begin{align}
T(g)^{-1} = \Sigma_{10} T(g)^T \Sigma_{10}.
\end{align}

Let us now look at
\begin{align}
T(Q \Lambda) = T(g) T(\Lambda) T(g)^{-1} T(\Lambda)
= T(g) \Sigma_{12} \Sigma_{10} T(g)^T \Sigma_{10} \Sigma_{12}
= T(g) \Sigma_{02} T(g)^T \Sigma_{02},
\end{align}
which means that $T(Q \Lambda) \Sigma_{02}$ is anti-symmetric. Now we can employ the Iwasawa decomposition $T(g) = T(n) T(a) T(k)$ and drop the last factor, since in $T(Q\Lambda)$ the group element $g$ enters only in the product $g \Lambda g^{-1}$, and $k$ commutes with $\Lambda$:
\begin{align}
T(Q \Lambda) \Sigma_{02}
= T(n) T(a) \Sigma_{02} [T(n) T(a)]^T.
\end{align}
Next, we restrict everything to the $AA$ block using the upper triangular structure of $T(n)$ and the diagonal nature of $T(a)$ in the RA space, see Eq. \eqref{matrix-rrvs}. In this block $\Sigma_{02,AA} = \sigma_2 \otimes I_n$, and we have
\begin{align}
T(Q \Lambda)_{AA} (\sigma_2 \otimes I_n)
= T(n)_{AA} T(a)_{AA} (\sigma_2 \otimes I_n) [T(n)_{AA} T(a)_{AA}]^T.
\end{align}
We can use the property of the Pfaffians $\text{Pf} \, (M A M^T) = \det(M) \, \text{Pf} \, A$, where $A$ is anti-symmetric and $M$ arbitrary, to obtain
\begin{align}
\text{Pf} \, [T(Q \Lambda)_{AA} (\sigma_2 \otimes I_n)]
= \det[T(n)_{AA}] \det[T(a)_{AA}] \, \text{Pf} \, (\sigma_2 \otimes I_n).
\end{align}
Since the determinant is basis-independent, we have
\begin{align}
\det[T(n)_{AA}] &= \det[\tilde{n}_{AA}] = 1,
&
\det[T(a)_{AA}] &= \det[\tilde{a}_{AA}] = \prod_{i=1}^{n} e^{-2x_i}.
\end{align}
Finally, we get the Pfaffian $N$-radial functions
\begin{align}
p_n \equiv \varphi_{(1,1,\ldots)} = \exp \Big(-2 \sum_{i=1}^{n} x_i \Big)
= \frac{\text{Pf} \, [T(Q \Lambda)_{AA} (\sigma_2 \otimes I_n)]}{\text{Pf} \, (\sigma_2 \otimes I_n)} \label{eq:nradial}.
\end{align}
This can be done for arbitrary number of replicas $n$, so we get all basic Pfaffians this way, and the most general $N$-radial functions
\begin{align}
\varphi_{(q_1,\ldots,q_n)} = p_1^{q_1 - q_2} p_2^{q_2 - q_3}
\ldots p_{n-1}^{q_{n-1} - q_n} p_n^{q_n}.
\label{p-product}
\end{align}

We should stress again the achievement of this section: the general $N$-radial eigenfunctions in Eqs. \eqref{d-product} and \eqref{p-product} are parametrized by {\it arbitrary} complex numbers $q_1,\ldots, q_n$ and satisfy the Abelian fusion. It was the combination of these two properties together with the local 2D conformal invariance that allowed us to derive the generalized parabolicity \eqref{generalized-parabolicity} in Section \ref{subsec:generalized-parabolicity}.

\section{Renormalization group and invariant scaling operators}
\label{sec:rg_c}
In this section, we employ one more approach---the field-theoretic RG---in order to determine the pure-scaling operators in class C. The approach is based on the invariance property of the RG transformation that works on functions (composite operators) $O[Q]$ on the symmetric space $G/K$. Importantly, this invariance holds to any order of RG. In view of this, eigenfunctions of one-loop RG are the pure-scaling operators, cf. previous works for class A \cite{burmistrov2011wave} and class AI \cite{burmistrov2016mesoscopic}. 

We use the one-loop RG to determine the class-C eigenoperators $\mathcal{P}^C_\lambda[Q]$ that are polynomials in $Q$. These operators correspond to Young diagrams $ \lambda = (q_1, q_2, \ldots , q_n)$  with integers $q_1, \ldots, q_n$ satisfying  $q_1 \ge q_2 \ge \ldots q_n > 0$. Clearly, in each order $|\lambda| \equiv q_1 + q_2 + \ldots + q_n$  of the polynomial, there is a finite number of such Young diagrams. For example, for $|\lambda|=2$ there are two Young diagrams, (2) and (1,1);  for $|\lambda|=3$ there are three of them, (3), (2,1), and (1,1,1), and so on. Further, RG preserves the order $|\lambda|$ of the polynomial. Thus, for each $|\lambda|$, the action of RG reduces to a matrix in the corresponding subspace. Diagonalization of this matrix yields pure-scaling operators $\mathcal{P}^C_\lambda[Q]$. These operators are particularly convenient for mapping onto wave-function observables performed  in Sec.~\ref{sec:wave_ops}.  (To avoid confusion, let us point out that the operators $\mathcal{P}^C_\lambda[Q]$ do not satisfy the Abelian fusion, at variance with the $N$-radial operators \eqref{p-product} constructed in Sec.~\ref{sec:iwasawa} by means of the Iwasawa decomposition.)

While our main focus is on the class C, for which $G/K=\mathrm{SO}^*(4n)/\mathrm{U}(2n)$, we also briefly  review the RG analysis for class A, for which $G/K=\mathrm{U}(n,n)/\mathrm{U}(n)\times \mathrm{U}(n)$.  This allows us to point out important differences in eigenoperators between classes C and A. Furthermore, we establish relations between the eigenoperators of both classes, which will be useful in Sec.~\ref{sec:wave_ops} for the numerical evaluation of scaling exponents of the SQH transition. 

In Sec.~\ref{sec:rg_deriv} we introduce the RG procedure and use it to determine eigenoperators with the  lowest non-trivial degree of polynomial, $|\lambda|=2$. In Sec.~\ref{sec:rg_gen} this analysis is extended to eigenoperators of higher degree, $|\lambda| >2$; we explicitly determine them for $|\lambda| = 3$ and  $|\lambda| = 4$. 

Whereas our goal in this section is to determine pure-scaling operators by RG means, we also calculate eigenvalues of the one-loop RG. Specifically, after having inspected examples of not too large $|\lambda| $ in Secs. \ref{sec:rg_2} and \ref{sec:rg_gen}, we prove in Sec. \ref{sec:onelaplacian} a general statement that the one-loop RG operator is proportional to Laplace-Beltrami operator on $G/K$. This implies that eigenvalues of one-loop RG satisfy generalized parabolicity. 

\subsection{Renormalization procedure}
\label{sec:rg_deriv}

The RG procedure  is implemented in the standard way by splitting the matrix $g$ that defines the sigma-model field $Q  = g\Lambda g^{-1}$ in the slow (subscript ``$s$'') and fast (subscript ``$f$'') components, 
$g = g_s g_f$.  (This approach is also known as background field method; e.g., recent papers \cite{burmistrov2016mesoscopic,Burmistrov-Magnetic-2018}
and references therein.) The matrix $g$ is antiunitary, $g^{-1} = \tau_3 g^\dagger\tau_3$, and the same condition holds for $g_s$ and $g_f$.
The fast field 
\be
g_f =e^{-\mathcal{X}}= 1 - \mathcal{X}+\frac12\mathcal{X}^2 +\ldots
\ee
is expanded in $\mathcal{X}$.  The matrix $\mathcal{X}$ does not contain the ``gauge'' (group $K$) degrees of freedom, i.e., it anticommutes with $\Lambda \equiv \tau_3$:
\begin{align}
\mathcal{X} &= \begin{pmatrix}
0 & X\\
X^\dagger& 0
\end{pmatrix}_{\tau} \label{eq:xdecomp}
\end{align}
We thus have
\begin{align}
\label{eq:RG-Q-expansion}
Q &= g\Lambda g^{-1} \nonumber\\
&= g_s g_f \Lambda g_f^{-1} g_s^{-1}\nonumber\\
&= Q_s + 2g_s\Lambda\mathcal{X} g_s^{-1} + 2g_s\Lambda\mathcal{X}^2 g_s^{-1} + \ldots,
\end{align}
where $Q_s = g_s \Lambda g_s^{-1}$.  For the one-loop analysis, it is sufficient to keep terms up to the order $\mathcal{X}^2$. 

To obtain the Gaussian fast-mode action $S_f$, we expand the sigma-model action \eqref{SQ-sigma-model} to lowest (quadratic) order in the fast fields $\mathcal{X}$:
\begin{align}
S_f[\mathcal{X}] &= \dfrac{\pi\sigma_0}{2}\int d^dx\left[ \mathrm{tr} \: (\nabla \mathcal{X})^2  + h^2  \mathrm{tr} \:   \mathcal{X}^2\right]  \nonumber \\
& = \pi\sigma_0 \int d^dx\left[ \mathrm{tr} \: (\nabla X^\dagger)(\nabla X)   + h^2  \mathrm{tr} \:  X^\dagger X \right].
\label{eq:fastaction}
\end{align}
Here we have included the infrared cutoff $h^2$ that regularizes the divergence in spatial dimensionality $d\leq 2$. In $d=2$ dimensions, the one-loop integrals are logarithmic with respect to the ratio of the ultraviolet and infrared cutoffs; in the bare theory, the ultraviolet regularization is usually provided by the lattice spacing. One can also consider the theory in $d=2+\epsilon$ dimensions, with $\epsilon>0$,  \cite{wegner1987anomalous1,wegner1987anomalous2}. The analysis that we perform to determine the eigenoperators of RG is based solely on symmetry, so that it is equally applicable in any spatial dimensionality. As we will see below, the loop integral will enter as a constant $I_f$, whose value will be immaterial for the analysis. 

We define $ \delta O $ to be the one loop renormalization contribution of an operator $O(Q)$ with respect to this fast action. It is obtained by expanding $O$ up to second order in $\mathcal{X}$, 
\be
O[\mathcal{X}] = O^{(0)} + O^{(1)}[\mathcal{X}] + \frac{1}{2}O^{(2)}[\mathcal{X},\mathcal{X}]+\ldots \,,
\ee
where $O^{(n)}$  a $n$-linear form in $\mathcal{X}$,
and then performing the Gaussian averaging $\langle \ldots \rangle$ with respect to $\mathcal{X}$ with the action \eqref{eq:fastaction}. We thus have
\begin{align}
\delta O = \frac{1}{2} \int \mathcal{D}\mathcal{X} \: O^{(2)}[\mathcal{X},\mathcal{X}] \: e^{-S_f[\mathcal{X}]}  \equiv
\frac{1}{2}\langle O^{(2)}[\mathcal{X},\mathcal{X}]\rangle .
\label{eq:ren}
\end{align}
Every term in the RG flow $\delta O$ thus originates from a single contraction of two fast fields and is proportional to the fast-field propagator. Since the composite operators that we are considering are local, 
this propagator is taken at coinciding spatial points and thus involves the loop integral
\begin{align}
I_f & =  \frac{1}{\pi\sigma_0} \int \dfrac{d^{2+\epsilon}p}{p^2+h^2}.
\label{eq:if}
\end{align}
At one loop level, every term in $\delta O$  is thus proportional to $I_f$ by construction. In order to determine the eigenoperators, we will not need the value of $I_f$, since it will simply stay as an overall prefactor in matrices that we will have to diagonalize. 

\subsection{RG for operators quadratic in $Q$.}
\label{sec:rg_2}

We consider first the operators that are quadratic with respect to $Q$. The corresponding Young diagrams $\lambda$ are characterized by $|\lambda|=2$. Obviously, there are two such Young diagrams: (2) and (1,1).   To derive the RG equations, we consider the renormalization of the operators $\mathrm{tr}(A Q B Q)$ and $\mathrm{tr}(A Q)\mathrm{tr}( B Q)$ with generic matrices $A$ and $B$. 
Using the expansion \eqref{eq:RG-Q-expansion}, we get
\begin{align}
\mathrm{tr}(A Q B Q) &= \mathrm{tr}(A g_s Q_f g_s^{-1} B g_s Q_f g_s^{-1}) \nonumber\\
&\simeq \mathrm{tr}(A Q_s B Q_s) + 4\mathrm{tr}(g_s^{-1} A g_s \Lambda\mathcal{X} g_s^{-1} B g_s \Lambda\mathcal{X})
\nonumber\\ &
\quad + 2\mathrm{tr}( A g_s \Lambda\mathcal{X}^2 g_s^{-1} B Q_s)+ 2\mathrm{tr}( A Q_s B g_s \Lambda\mathcal{X}^2 g_s^{-1}), \nonumber \\
& = \mathrm{tr}(A Q_s B Q_s) + 4\mathrm{tr} (\tilde{A } \mathcal{X} \tilde{B} \mathcal{X} ) + 2 \mathrm{tr} ( \tilde{B} \tilde{A } \mathcal{X}^2 ) + 2 \mathrm{tr} ( \tilde{A} \tilde{B } \mathcal{X}^2 ) \,;  \nonumber \\
\mathrm{tr}(A Q)\mathrm{tr}( B Q) &= \mathrm{tr}(A g_s Q_f g_s^{-1})\mathrm{tr}( B g_s Q_f g_s^{-1}) \nonumber\\
& \simeq \mathrm{tr}(A Q_s)\mathrm{tr}( B Q_s) + 4\mathrm{tr}(g_s^{-1} A g_s \Lambda\mathcal{X})\mathrm{tr}( g_s^{-1} B g_s \Lambda\mathcal{X}) \nonumber\\
& \quad + 2\mathrm{tr}( A g_s \Lambda\mathcal{X}^2 g_s^{-1} )\mathrm{tr}(B Q_s)+ 2\mathrm{tr}( A Q_s)\mathrm{tr}( B g_s \Lambda\mathcal{X}^2 g_s^{-1}) \nonumber \\
& =  \mathrm{tr}(A Q_s)\mathrm{tr}( B Q_s)  +  4\mathrm{tr} (\tilde{A } \mathcal{X} )  \mathrm{tr} (\tilde{B} \mathcal{X} ) + 
2\mathrm{tr} (\tilde{A } \mathcal{X}^2 )  \mathrm{tr} (BQ_s) + 2\mathrm{tr} (\tilde{B} \mathcal{X}^2 )  \mathrm{tr} (AQ_s) \,,
\label{qdecomp}
\end{align}
where we have retained only terms of zeroth and second order in $X$ and
\be
\tilde{A} =  g_s^{-1}Ag_s\Lambda \,, \qquad \tilde{B} = g_s^{-1}Bg_s\Lambda \,.
\label{eq:transf}
\ee

To calculate $\delta [{\rm tr}(AQ) {\rm tr}(BQ)] $ and $\delta {\rm tr}(AQBQ) $, we should average according to Eq.~\eqref{eq:ren} the terms of second order in $\mathcal{X}$ in Eq.~\eqref{qdecomp}. 
This involves the following averages (with 1,2 being the indices in the RA space):
\begin{align}
\langle {\rm tr}(\tilde{A }\mathcal{X}) {\rm tr}(\tilde{B }\mathcal{X}) \rangle &= \left\langle\left( {\rm tr}(\tilde{A }_{12}X^\dagger) + {\rm tr}(\tilde{A }_{21}X)\right) \left({\rm tr}(\tilde{B}_{12}X^\dagger) + {\rm tr}(\tilde{B}_{21}X)\right)\right\rangle \nonumber\\
&= \tilde{A }_{12}^{\beta\alpha}\tilde{B}_{12}^{\delta\gamma}\left\langle X^\dagger_{\alpha\beta}  X^\dagger_{\gamma\delta} \right\rangle
+ \tilde{A }_{12}^{\beta\alpha}\tilde{B}_{21}^{\delta\gamma}\left\langle X^\dagger_{\alpha\beta}  X_{\gamma\delta} \right\rangle
+ \tilde{A }_{21}^{\beta\alpha}\tilde{B}_{12}^{\delta\gamma}\left\langle X_{\alpha\beta}  X^\dagger_{\gamma\delta} \right\rangle
+ \tilde{A }_{21}^{\beta\alpha}\tilde{B}_{21}^{\delta\gamma}\left\langle X_{\alpha\beta}  X_{\gamma_\delta} \right\rangle, \label{eq:rg1}\\
\langle {\rm tr}(\tilde{A }\mathcal{X}\tilde{B}\mathcal{X}) \rangle &= \left\langle{\rm tr}(\tilde{A }_{12}X^\dagger \tilde{B}_{12}X^\dagger + \tilde{A }_{22}X^\dagger \tilde{B}_{11} X)+ {\rm tr}(\tilde{A }_{11}X\tilde{B}_{22}X^\dagger+\tilde{A }_{21}X\tilde{B}_{21}X)\right\rangle \nonumber\\
&= \tilde{A }_{12}^{\delta\alpha}\tilde{B}_{12}^{\beta\gamma}\left\langle X^\dagger_{\alpha\beta}  X^\dagger_{\gamma\delta} \right\rangle
+ \tilde{A }_{11}^{\delta\alpha}\tilde{B}_{22}^{\beta\gamma}\left\langle X^\dagger_{\alpha\beta}  X_{\gamma\delta} \right\rangle
+ \tilde{A }_{22}^{\delta\alpha}\tilde{B}_{11}^{\beta\gamma}\left\langle X_{\alpha\beta}  X^\dagger_{\gamma\delta} \right\rangle
+ \tilde{A }_{21}^{\delta\alpha}\tilde{B}_{21}^{\beta\gamma}\left\langle X_{\alpha\beta}  X_{\gamma\delta} \right\rangle,
\label{eq:rg2} 
\end{align}
where  $\alpha$, $\beta$, $\gamma$, and $\delta$ are replica indices and we use the convention of summation over repeated indices.
Further, replacing $\tilde{B}$ in Eq.~\eqref{eq:rg2} by the unit matrix, we obtain
\be
\langle {\rm tr}(\tilde{A }\mathcal{X}^2) \rangle = \tilde{A }_{11}^{\delta\alpha}\left\langle X^\dagger_{\alpha\beta}  X_{\beta\delta} \right\rangle
+ \tilde{A }_{22}^{\delta\alpha}\left\langle X_{\alpha\beta}  X^\dagger_{\beta\delta} \right\rangle.
\label{eq:rg3}
\ee

The analysis up to now was general and thus was applicable to both classes A and C. To proceed further, we should specify the fast-mode propagators, which are different for classes A and C. 

\subsubsection{Class A}

In class A, the manifold is $G/K = \text{U}(n,n)/\text{U}(n)\times \text{U}(n)$ and the only condition on $g$ is pseudo-unitarity, $g^{-1} = \tau_3 g^\dagger\tau_3$. The field $X$ in Eq. \eqref{eq:xdecomp} is correspondingly a generic complex matrix with the propagator
\begin{align}
\langle X^\dagger_{\alpha \beta} X_{\gamma \delta} \rangle & =I_f \delta_{\alpha\delta}\delta_{\beta\gamma} \,, \nonumber\\
\langle X_{\alpha \beta} X_{\gamma \delta} \rangle & = \langle X^\dagger_{\alpha \beta} X^\dagger_{\gamma \delta} \rangle = 0 \,.
\label{eq:X-propagators-A}
\end{align}
Using Eq.~\eqref{eq:X-propagators-A} for the propagator in Eqs.\eqref{eq:rg1}  and \eqref{eq:rg2}, we get
\bea
\langle {\rm tr}(\tilde{A }\mathcal{X}) {\rm tr}(\tilde{B }\mathcal{X}) \rangle &=& I_f \: {\rm tr} (\tilde{A }_{12} \tilde{B}_{21} + \tilde{A }_{21} \tilde{B}_{12})
= I_f \: {\rm tr} \left( \tilde{A } P_- \tilde{B} P_+  + \tilde{A } P_+ \tilde{B} P_- \right) \nonumber \\
&=& \frac{1}{2}I_f \: {\rm tr} ( \tilde{A }  \tilde{B } -  \tilde{A } \Lambda \tilde{B} \Lambda) \,; 
\label{eq:rg1a} \\
\langle {\rm tr}(\tilde{A }\mathcal{X}\tilde{B}\mathcal{X}) \rangle & = &  I_f \: ( {\rm tr} \tilde{A }_{11}  {\rm tr} \tilde{B}_{22} + {\rm tr} \tilde{A }_{22}  {\rm tr} \tilde{B}_{11} )
\nonumber \\
&=& I_f \: \left[ {\rm tr} \left( \tilde{A } P_+ \right)  {\rm tr} \left( \tilde{B} P_- \right) 
+ {\rm tr} \left( \tilde{A } P_- \right)  {\rm tr} \left( \tilde{B} P_+ \right)  \right] \nonumber \\
&=& \frac{1}{2}I_f \:  [{\rm tr} \tilde{A } \, {\rm tr}  \tilde{B } -  {\rm tr} (\tilde{A } \Lambda) \,   {\rm tr} (\tilde{B} \Lambda) ] \,,
\label{eq:rg2a} 
\eea
where we have introduced the projectors 
$$P_{\pm}=  \frac{\mathds{1}\pm \Lambda}{2}$$ 
on the advanced and retarded sectors.
Substituting here now Eq.~\eqref{eq:transf} for $\tilde{A }$ and  $\tilde{B }$, we find
\bea
\langle {\rm tr}(\tilde{A }\mathcal{X}) {\rm tr}(\tilde{B }\mathcal{X}) \rangle & = & \frac{1}{2}I_f {\rm tr} (Q_sA Q_s B - AB) \,; 
\label{eq:rg1b} \\
\langle {\rm tr}(\tilde{A }\mathcal{X}\tilde{B}\mathcal{X}) \rangle & = & \frac{1}{2}I_f  \left[ {\rm tr} (Q_sA) \,  {\rm tr} (Q_s B) - {\rm tr} A \,  {\rm tr} B \right].
\label{eq:rg2b}
\eea
Below we will focus on the choice of matrices $A$ and $B$ satisfying (in the replica limit $ n \to 0$)  ${\rm tr} A = {\rm tr} B = {\rm tr} (AB) =0$, so that we discard the $Q_s$-independent terms in the r.h.s. of Eqs. \eqref{eq:rg1b}, \eqref{eq:rg2b}. We note, however, that these terms play only a minor role also for more generic matrices $A$ and $B$: they only lead to an admixture of constant ($Q$-independent) terms to eigenoperators, without affecting either the quadratic-in-$Q$ structure of eigenoperators or the corresponding eigenvalues.

Equation \eqref{eq:rg3} yields, with the class-A propagator \eqref{eq:X-propagators-A},
\be 
\langle {\rm tr}(\tilde{A }\mathcal{X}^2) \rangle = 
I_f \, n \, ( {\rm tr} \tilde{A }_{11}  + {\rm tr} \tilde{A }_{22}  ) = I_f \, n \,  {\rm tr} \tilde{A }\,.
\label{eq:rg3a}
\ee
Since we are interested in the replica limit $n \to 0$, the terms of the type  $\langle {\rm tr}( \ldots \mathcal{X}^2) \rangle$ thus do not give any contribution. 
Using Eqs.~\eqref{eq:rg1b} and \eqref{eq:rg2b} for the remaining averages in Eq.~\eqref{qdecomp}, we obtain the following RG flow (that is conveniently presented in the matrix form) for the operators ${\rm tr}(AQ) {\rm tr}(BQ)$ and $ {\rm tr}(AQ BQ)$:
\begin{align}
\delta\begin{pmatrix}
{\rm tr}(AQ) {\rm tr}(BQ)\\
{\rm tr}(AQ BQ)
\end{pmatrix} &=(-2I_f)\cdot\underbrace{\begin{pmatrix}
	0 & -1 \\
	-1 & 0
	\end{pmatrix}}_{=:M_2^A}
\cdot\begin{pmatrix}
{\rm tr}(AQ) {\rm tr}(BQ)\\
{\rm tr}(AQ BQ)
\end{pmatrix} .
\label{eq:rgresA}
\end{align}
Consequently the eigenoperators are (the superscript ``A'' refers to class A)
\begin{align}
\mathcal{P}^A_{(2)AB} &={\rm tr}(A Q){\rm tr}(B Q)+{\rm tr}(A Q B Q),\nonumber \\
\mathcal{P}^A_{(1,1)AB} &={\rm tr}(A Q) {\rm tr}(B Q)- {\rm tr}(A Q B Q) \,,
\label{eq:opsbA_AB}
\end{align}
with eigenvalues $2I_f > 0$ and $-2I_f < 0$, respectively. The first of them (increasing under RG) belongs to the representation (equivalently, Young diagram) $\lambda = (2)$ and the second one (decreasing under RG) to $\lambda = (1,1)$. Let us emphasize that the eigenvalues do not depend on the choice of the matrices $A$ and $B$: the operators $\mathcal{P}^A_{(2)}$  with any $A$ and $B$ belong to the same representation (2), and 
the operators $\mathcal{P}^A_{(1,1)}$  with any $A$ and $B$ belong to the representation (1,1). 

One important choice of the matrices $A$ and $B$ is $A = B = \Lambda$, which yields the operators
\begin{align}
\mathcal{P}^A_{(2)} &={\rm tr}(\Lambda Q){\rm tr}(\Lambda Q)+{\rm tr}(\Lambda Q \Lambda Q),\nonumber \\
\mathcal{P}^A_{(1,1)} &={\rm tr}(\Lambda Q) {\rm tr}(\Lambda Q)- {\rm tr}(\Lambda Q\Lambda Q) \,.
\label{eq:opsbA}
\end{align}
The special feature of this choice is that the operators \eqref{eq:opsbA} are $K$-invariant:  they are invariant with respect to rotations $Q \to U Q U^{-1}$ with $U \in K$, since all such $U$ commute with $\Lambda$. 

\subsubsection{Class C}
\label{sec:RG-quadratic-class-C}

We extend now the above analysis to the class C.  There is then an additional symmetry operation, 
\be
\overline{O} \equiv \tau_{1}O^T\tau_{1}\,,
\label{eq:bar-operation}
\ee
that constrains $g$:
\begin{align} 
\overline{g} = g^{-1} \,. 
\label{eq:symopg}
\end{align}
This means $\overline{\mathcal{X}} = -\mathcal{X}$.  Since $\overline{\Lambda}=-\Lambda$, this corresponds to $Q$ being odd with respect to the operation \eqref{eq:bar-operation},
\be
\overline{Q}=-Q \,,
\ee
see Eq.~\eqref{eq:symopQiw}. Note that $Q$ in this section is the same as $\tilde{Q}$ introduced in the end of Sec.~\ref{sec:sigma}, see Eq.~\eqref{Q-tilde}.

Combining Eq. \eqref{eq:xdecomp} with $\overline{\mathcal{X}} = -\mathcal{X}$, we find that $X$ is an antisymmetric complex matrix,  $X= - X^T$. 
Therefore,  expectation values with the Gaussian action \eqref{eq:fastaction} yield the following propagator:
\begin{align}
\langle X^\dagger_{\alpha \beta} X_{\gamma \delta} \rangle & =I_f(\delta_{\alpha\delta}\delta_{\beta\gamma}  - \delta_{\alpha\gamma}\delta_{\beta\delta}) \,, \nonumber\\
\langle X_{\alpha \beta} X_{\gamma \delta} \rangle & = \langle X^\dagger_{\alpha \beta} X^\dagger_{\gamma \delta} \rangle = 0 \,.
\label{eq:X-propagators-C}
\end{align}
We recall, that, as discussed in the end of Sec.~\ref{sec:sigma}, the particle-hole index $\Sigma$ is included in the replica index, so that we have $2n$ replicas.

Substituting Eq.~\eqref{eq:X-propagators-C} into Eqs.~\eqref{eq:rg1}, \eqref{eq:rg2}, we find for the fast-field correlation functions that emerge in the RG analysis \eqref{qdecomp}, with two factors of $\mathcal{X}$ originating from different $Q$ fields:
\begin{align}
\langle {\rm tr}(\tilde{A}\mathcal{X}) {\rm tr}(\tilde{B}\mathcal{X}) \rangle &= I_f\left[
{\rm tr}(\tilde{A}_{12}\tilde{B}_{21})-{\rm tr}(\tilde{A}_{21}\tilde{B}_{12}^T)+
{\rm tr}(\tilde{A}_{21}\tilde{B}_{12})-{\rm tr}(\tilde{A}_{12}\tilde{B}_{21}^T)\right]\nonumber\\
&=\frac{I_f}2 \left[
{\rm tr}(\tilde{A}P_- \tilde{B}P_+)-{\rm tr}(\tilde{A}P_+ \overline{\tilde{B}}P_- )+{\rm tr}(\tilde{A}P_+\tilde{B}P_-)
-{\rm tr}(\tilde{A}P_-\overline{\tilde{B}}P_+)\right]\nonumber\\
&=\frac{I_f}2 \left[-
{\rm tr}(\tilde{A}\Lambda \tilde{B}\Lambda)+{\rm tr}(\tilde{A}\tilde{B})
-{\rm tr}(\tilde{A}\overline{\tilde{B}})+{\rm tr}(\tilde{A}\Lambda \overline{\tilde{B}}\Lambda )\right]; \nonumber\\
\langle {\rm tr}(\tilde{A}\mathcal{X}\tilde{B}\mathcal{X}) \rangle &=  \dfrac{I_f}{2}\left[{\rm tr}(\tilde{A}_{22}){\rm tr}(\tilde{B}_{11})-{\rm tr}(\tilde{A}_{22}\tilde{B}_{11}^T)+
{\rm tr}(\tilde{A}_{11}){\rm tr}(\tilde{B}_{22})-{\rm tr}(\tilde{A}_{11}\tilde{B}_{22}^T)\right]\nonumber\\
&=\dfrac{I_f}{2}\left[{\rm tr}(\tilde{A}P_-){\rm tr}(\tilde{B}P_+)-{\rm tr}(\tilde{A}P_-\overline{\tilde{B}}P_-)+{\rm tr}(\tilde{A}P_+){\rm tr}(\tilde{B}P_-)-{\rm tr}(\tilde{A}P_+\overline{\tilde{B}}P_+)\right]\nonumber\\
&=\dfrac{I_f}{2}\left[{\rm tr}(\tilde{A}){\rm tr}(\tilde{B})-{\rm tr}(\tilde{A}\Lambda){\rm tr}(\tilde{B}\Lambda)-{\rm tr}(\tilde{A}\overline{\tilde{B}})-{\rm tr}(\tilde{A}\Lambda\overline{\tilde{B}}\Lambda)\right].
\end{align}
Substituting now Eq.~\eqref{eq:transf} for $\tilde{A }$ and  $\tilde{B }$, we obtain
\begin{align}
\langle {\rm tr}(\tilde{A}\mathcal{X}) {\rm tr}(\tilde{B}\mathcal{X}) \rangle & =
\frac{I_f}{2}\left[-{\rm tr}(AB)+{\rm tr}(AQ_s BQ_s)+{\rm tr}(A\overline{B})-{\rm tr}(AQ_s\overline{ B}Q_s)\right]
,\nonumber\\
\langle {\rm tr}(\tilde{A}\mathcal{X}\tilde{B}\mathcal{X}) \rangle &= 
\dfrac{I_f}{2}\left[{\rm tr}(AQ_s){\rm tr}(BQ_s)-{\rm tr}(A){\rm tr}(B)+{\rm tr}(A\overline{B})+{\rm tr}(AQ_s\overline{B}Q_s)\right].
\label{eq:rg-C1}
\end{align}

In addition, Eq.~\eqref{qdecomp} contain terms of the type \eqref{eq:rg3}, with both factors of $\mathcal{X}$ originating from the same $Q$ field, i.e., entering in the form of $\mathcal{X}^2$. 
Substituting Eq.~\eqref{eq:X-propagators-C} into  Eq.\eqref{eq:rg3}, we get
\begin{align}
\langle {\rm tr}(\tilde{A }\mathcal{X}^2)\rangle &= I_f(- 1+ 2n)  \,{\rm tr} \, \tilde{A} = - I_f \,{\rm tr} \, \tilde{A} \,.
\label{eq:rg-C2}
\end{align}
In the last equality sign, we have taken the replica limit $n \to 0$.  At variance with the class A [see Eq.~\eqref{eq:rg3a}], such contributions remain non-zero in the replica limit in class C. In particular, such a term is responsible for renormalization of operators of first order in $Q$ [e.g., ${\rm tr}(\Lambda Q)$] in class C:
\begin{align}
\delta\, {\rm tr}(\Lambda Q)&=2 \langle{\rm tr}(g_s\Lambda g_s^{-1}\Lambda \mathcal{X}^2)\rangle = -2I_f \, {\rm tr}(\Lambda Q).
\label{eq:rg-C-LambdaQ}
\end{align}
This operator belongs to the representation (1) and determines the scaling of the average LDOS. The corresponding eigenvalue is $-2I_f < 0$, which means that the average LDOS is suppressed within one-loop RG. 

We return to the RG for operators that are quadratic in $Q$. Substituting Eqs.~\eqref{eq:rg-C1} and \eqref{eq:rg-C2} into Eq.~\eqref{qdecomp}, we obtain the RG equations
\begin{align}
\delta [{\rm tr}(AQ) {\rm tr}(BQ)] &= 2I_f\left[-{\rm tr}(AB)+{\rm tr}(AQ BQ)-\left(-{\rm tr}(A\overline{B})+{\rm tr}(AQ\overline{ B}Q)\right)- 2{\rm tr}(AQ) {\rm tr}(BQ)\right],
\nonumber\\
\delta {\rm tr}(AQBQ)  &= 2I_f\left[{\rm tr}(AQ){\rm tr}(BQ)-{\rm tr}(A){\rm tr}(B)+\left({\rm tr}(A\overline{B})+{\rm tr}(AQ\overline{B}Q)\right)- 2{\rm tr}(AQBQ)\right].
\label{eq:RG-C3}
\end{align}
In full analogy with the class A [see comment after Eq.~\eqref{eq:rg2b}], we drop constant contributions (those that do not contain $Q$). Furthermore, we assume that, like the matrix $Q$, the matrices $A$ to $B$ are odd 
\footnote{It is straightforward to check that choosing $A$ and $B$ to be even, $\overline{A}=A$ and $\overline{B}=B$, does not bring anything new. Specifically, the operator ${\rm tr}(AQ) {\rm tr}(BQ)$ is then zero, and the operator  ${\rm tr}(AQBQ)$  renormalizes to itself with the eigenvalue $-2I_f$ and thus belongs to the same representation (2) as the operator $\mathcal{P}^C_{(2)}$ from Eq.~\eqref{eq:opsb}.}
with respect to the symmetry operation \eqref{eq:bar-operation}:
\be
\overline{A}=-A \,, \qquad \overline{B}=-B \,.
\label{eq:RG-classC-A-B-odd}
\ee
Two choices of $A$ and $B$ will be particularly important for us. The first one is $A=B=\Lambda$, which yields gauge-invariant operators [i.e., operators invariant with respect to the group $K = \text{U}(2n)$].  The second choice is $A=E_{aa}\Lambda$ and  $B=E_{bb}\Lambda$, with $E_{aa}$ and $E_{bb}$ being projectors on two distinct replica indices $a$ and $b$. This choice of $A$ and $B$ will be useful for establishing connections with wavefunction correlators in Sec. \ref{sec:wave_ops}. In both cases, $A$ and $B$ are odd, i.e., satisfy Eq.~\eqref{eq:RG-classC-A-B-odd}.
The RG equations \eqref{eq:RG-C3} then reduce to
\begin{align}
\delta\begin{pmatrix}
{\rm tr}(AQ) {\rm tr}(BQ)\\
{\rm tr}(AQ BQ)
\end{pmatrix} &=(- 2I_f)\cdot\underbrace{\begin{pmatrix}
	2 & - 2 \\
	- 1 & 3
	\end{pmatrix}}_{=:M_2}
\cdot\begin{pmatrix}
{\rm tr}(AQ) {\rm tr}(BQ)\\
{\rm tr}(AQ BQ)
\end{pmatrix}.
\label{eq:rgres}
\end{align}
To determine pure-scaling operators, we write an operator $O$ as a linear combination
\be 
O = c_{(1,1)}{\rm tr}(A Q) {\rm tr}(B Q)+c_{(2)}{\rm tr}(A Q B Q).
\ee
Equation \eqref{eq:rgres} then is translated into the RG flow of the coefficients $c_{(1,1)}$ and $c_{(2)}$:
\begin{align}
\delta\begin{pmatrix}
c_{(1,1)}\\
c_{(2)}
\end{pmatrix}&=(- 2I_f)\cdot
\underbrace{\begin{pmatrix}
	2 & - 1\\
	- 2 & 3
	\end{pmatrix}}_{=M_2^T} \cdot 
\begin{pmatrix}
c_{(1,1)}\\
c_{(2)}
\end{pmatrix}
\label{eq:RG-matrixM2}
\end{align}
Note that $M_2$ from Eq.~\eqref{eq:rgres} acts transposed on the vector of coefficients. (We emphasize this since, at variance with class A, the matrix $M_2$ is not symmetric for class C.)
The eigenvectors of $M_2^T$ yield the operators 
\begin{align}
\mathcal{P}^C_{(2)AB} &={\rm tr}(A Q){\rm tr}(B Q)+{\rm tr}(A Q B Q),\nonumber \\
\mathcal{P}^C_{(1,1)AB} &={\rm tr}(A Q) {\rm tr}(B Q)- 2{\rm tr}(A Q B Q) \,,
\label{eq:opsb-AB}
\end{align}
which are eigenoperators of the RG flow with eigenvalues $-2I_f$ and $-8I_f$, respectively. The superscript ``C'' serves to distinguish them from class-A eigenoperators \eqref{eq:opsbA_AB}.
Importantly, the coefficients in Eq.~\eqref{eq:opsb-AB} do not depend on the choice of matrices $A$ and $B$.
In particular, for $A=B=\Lambda$ we find the $K$-invariant pure-scaling operators
\begin{align}
\mathcal{P}^C_{(2)} &={\rm tr}(\Lambda Q){\rm tr}(\Lambda Q)+{\rm tr}(\Lambda Q \Lambda Q),\nonumber \\
\mathcal{P}^C_{(1,1)} &={\rm tr}(\Lambda Q) {\rm tr}(\Lambda Q)- 2{\rm tr}(\Lambda Q\Lambda Q) \,.
\label{eq:opsb}
\end{align}
For another important choice mentioned above, $A=E_{aa}\Lambda$ and $B=E_{bb}\Lambda$ with two different replicas $a$ and $b$, we get the pure-scaling operators
\begin{align}
\mathcal{P}^C_{(2)ab} &={\rm tr}( \Lambda Q_{aa}){\rm tr}( \Lambda Q_{bb})+{\rm tr}(\Lambda Q_{ab} \Lambda Q_{ba}),\nonumber \\
\mathcal{P}^C_{(1,1)ab} &={\rm tr}(\Lambda Q_{aa}) {\rm tr}(\Lambda Q_{bb})- 2{\rm tr}(\Lambda Q_{ab} \Lambda Q_{ba}) \,,
\label{eq:opsb-ab}
\end{align}
where $Q_{aa}$, $Q_{bb}$, $Q_{ab}$, and $Q_{ba}$ are projections of $Q$ on the corresponding replica subspaces. 

From the Weyl-symmetry relations,  Sec.~\ref{subsec:MF-symmetry}, we know that the eigenoperators belonging to the representations (1) and (2) should scale in the same way under RG. Our results are in full agreement with this symmetry constraint:  the operators $\mathcal{P}^C_{(2)AB}$ from the representation (2) have the same eigenvalue $-2I_f$ as the operator representing the average LDOS, i.e., representation (1),  see Eq.~\eqref{eq:rg-C-LambdaQ}.  The operator $\mathcal{P}^C_{(1,1)AB}$ that has the eigenvalue $-8I_f$ (i.e., is ``more irrelevant'' in the RG sense) belongs to the representation (1,1). As we discuss below, the ratio 4 between the two eigenvalues is a manifestation of the exact generalized parabolicity that is a property of one-loop RG. 

\subsubsection{Gauge-invariant operators by $K$-averaging}
\label{sec:u_avg}

In Sec.~\ref{sec:RG-quadratic-class-C}, we have determined the quadratic-in-$Q$ pure scaling operators \eqref{eq:opsb-AB}. We have discussed two important choices of the matrices $A$ and $B$. One of them was to take $A=B=\Lambda$, which yields $K$-invariant eigenoperators \eqref{eq:opsb} that are linear combinations of the operators
\be
O_{(2)} = \mathrm{tr}(\Lambda Q \Lambda Q) \,, \qquad O_{(1,1)}=\mathrm{tr}(\Lambda  Q) \mathrm{tr}( \Lambda Q) \,.
\label{eq:RG-O2-invariant}
\ee
The other choice was $A=E_{aa}\Lambda$ and $B=E_{bb}\Lambda$, which leads to the eigenoperators \eqref{eq:opsb-ab}, which are linear combinations of
\be
O_{(2)}^{ab} = \mathrm{tr}(\Lambda Q_{ab} \Lambda Q_{ba})\,, \qquad O_{(1,1)}^{ab}=\mathrm{tr}(\Lambda Q_{aa}) \mathrm{tr}( \Lambda Q_{bb})
\label{eq:RG-Oab}
\ee
with the same coefficients. While the operators \eqref{eq:RG-Oab} are not $K$-invariant, one can make $K$-invariant operators out of them by performing the averaging over the group $K$. This yields an alternative way to $K$-invariant pure-scaling operators. The result should be the same as $K$-invariant operators \eqref{eq:opsb} determined above. In this subsection we show explicitly  that this is indeed the case, which constitutes a non-trivial consistency check of our analysis. 

Let $O[Q]$ be a composite operator. We can construct out of it an operator $\left\langle O[Q]\right\rangle_K$ by averaging over the group $K= \text{U}(2n)$:
\begin{align}
\left\langle O[Q]\right\rangle_K &= \int_K d\mu(U) O[ U^{-1} Q U].
\label{eq:OQ-U-averaged}
\end{align}
The operator \eqref{eq:OQ-U-averaged} is $K$-invariant, i.e., it is invariant under transformations $Q \mapsto V^{-1} Q V$ with $V \in K$, as immediately follows from the invariance of the Haar measure $d\mu(U)$ on $K=\text{U}(2n)$. 

The elements of $K$ commute with $\Lambda$ and thus have the block-diagonal form in the retarded-advanced ($\tau$) space, $U=\mathrm{diag}(U_1,U_2)_{\tau}$.
In order to perform the $K$-averaging of the operators \eqref{eq:RG-Oab}, we thus have to evaluate the following averages
\begin{align}
\left\langle\mathrm{tr}(\Lambda Q_{ab}\Lambda Q_{ba} )\right\rangle_K &= \left \langle (U^{-1}_1Q_{11}U_1)_{ab}(U^{-1}_1Q_{11}U_1)_{ba}+
(U^{-1}_2Q_{22}U_2)_{ab}(U^{-1}_2Q_{22}U_2)_{ba} \right. \nonumber\\
&   \left. -(U^{-1}_1Q_{12}U_2)_{ab}(U^{-1}_2Q_{21}U_1)_{ba}-
(U^{-1}_2Q_{21}U_1)_{ab}(U^{-1}_1Q_{12}U_2)_{ba} \right \rangle_K \,,
\nonumber\\
\label{eq:ab-Un-averaging1} \\
\left\langle\mathrm{tr}(\Lambda Q_{aa}) \, \mathrm{tr}(\Lambda Q_{bb} )\right\rangle_K&= \left \langle (U^{-1}_1Q_{11}U_1)^{aa}(U^{-1}_1Q_{11}U_1)_{bb}+
(U^{-1}_2Q_{22}U_2)_{aa}(U^{-1}_2Q_{22}U_2)_{bb}  \right. \nonumber\\
&  \left. -(U^{-1}_1Q_{11}U_1)_{aa}(U^{-1}_2Q_{22}U_2)_{bb}-
(U^{-1}_2Q_{22}U_2)_{aa}(U^{-1}_1Q_{11}U_1)_{bb} \right\rangle_K \,.
\label{eq:ab-Un-averaging2}
\end{align}
Here the indices 11, 12, 21, and 22 correspond to retarded-advanced space, while $a$ and $b$ are two given distinct replica indices. 
For class A, $U_1$ and $U_2$ are independent, and $K = \text{U}(n) \times \text{U}(n)$.  In class C, with the sigma-model field satisfying the additional symmetry constraint \eqref{eq:symopQiw}, we have $U_1 = U_2^*$, and  $K = \text{U}(2n)$. To evaluate the averages  in Eqs.~\eqref{eq:ab-Un-averaging1}, \eqref{eq:ab-Un-averaging2}, we need averages of products of matrix elements of matrices $U$ and $U^*$ over the unitary group $\text{U}(2n)$. 
The required averages are as follows \cite{mello1990averages} (see also Appendix of Ref. \cite{burmistrov2011wave}):
\begin{align}
\langle U^*_{a\alpha} U_{b \beta} \rangle_{\text{U}(2n)} &= V_1\delta_{ab}\delta_{\alpha\beta},\\
\langle U^*_{a\alpha} U_{b \beta} U^*_{c\gamma} U_{d \delta} \rangle_{\text{U}(2n)} &= V_{1,1}\left(\delta_{a b}\delta_{\alpha\beta} \delta_{cd}\delta_{\gamma\delta} + \delta_{a d}\delta_{\alpha\delta} \delta_{cb}\delta_{\gamma\beta}\right) + V_{2}\left(\delta_{c b}\delta_{\alpha\beta} \delta_{ad}\delta_{\gamma\delta} + \delta_{a b}\delta_{\alpha\delta} \delta_{cd}\delta_{\gamma\beta}\right),
\label{eq:Un-averages}
\end{align}
where $V_\lambda$ are the Weingarten functions \cite{weingarten1978asymptotic},
\be
V_1 = \frac{1}{2n} \,, \qquad V_{1,1} = \frac{1}{4n^2-1}\,, \qquad V_{2} = -\frac{1}{2n(4n^2-1)}\,.
\ee 

Let us first use Eq.~\eqref{eq:Un-averages} to evaluate the averages in the second line of Eq.~\eqref{eq:ab-Un-averaging2} (we denote $U_1 = U_2^* \equiv U$):
\begin{align}
&\left\langle(U^{-1}_1Q^{11}U_1)_{aa}(U^{-1}_2Q^{22}U_2)_{bb}\right\rangle_{\text{U}(2n)} \nonumber\\
=& \left\langle U^{-1}_{a\alpha} Q^{11}_{\alpha\beta}U_{\beta a}(U^*)^{-1}_{b \gamma}Q^{22}_{\gamma\mu}(U^*)_{\mu b}\right\rangle_{\text{U}(2n)} \nonumber\\
=& \left\langle U^{*}_{\alpha a} U_{\beta a} U^{*}_{\nu\lambda}U_{\sigma\rho}\right\rangle_{\text{U}(2n)} \delta_{b \rho}\delta_{\sigma \gamma}\delta_{\mu \nu}\delta_{\lambda b}Q^{11}_{\alpha\beta}Q^{22}_{\gamma\mu}\nonumber\\
=&\left( V_{1,1}\left(\delta_{\alpha\beta}\delta_{aa} \delta_{\nu\sigma}\delta_{\lambda\rho} + \delta_{\alpha\sigma}\delta_{a\rho} \delta_{\nu \beta}\delta_{\lambda a}\right) + V_{2}\left(\delta_{\nu \beta}\delta_{aa} \delta_{\alpha\sigma}\delta_{\lambda\rho} + \delta_{\alpha\beta}\delta_{a\rho} \delta_{\nu\sigma}\delta_{\lambda a}\right)\right)\delta_{b,\rho}\delta_{\sigma \gamma}\delta_{\mu \nu}\delta_{\lambda b}Q^{11}_{\alpha\beta}Q^{22}_{\gamma\mu}\nonumber\\
=& V_{1,1}\left(\delta_{b \rho}\delta_{\rho,b}Q^{11}_{\alpha\alpha}\delta_{\nu \gamma}Q^{22}_{\gamma\mu}\delta_{\mu \nu} + \delta_{b,a}\delta_{\sigma \gamma}\delta_{\mu \beta}\delta_{ab}Q^{11}_{\sigma\beta}Q^{22}_{\gamma\mu}\right) \nonumber\\
&+ V_{2}\left( \delta_{b \lambda}\delta_{\lambda b}Q^{11}_{\alpha\beta}\delta_{\alpha\gamma}Q^{22}_{\gamma\mu}\delta_{\mu\beta} + \delta_{\alpha\beta}\delta_{a\rho} \delta_{\nu\sigma}\delta_{\lambda a}\delta_{b\rho}\delta_{\sigma\gamma}\delta_{\mu\nu}\delta_{\lambda,b}Q^{11}_{\alpha\beta}Q^{22}_{\gamma\mu}\right)\nonumber \\
=&V_{1,1}\mathrm{tr}\left(Q^{11}\right)\mathrm{tr}\left(Q^{22}\right)+V_{2}\mathrm{tr}\left(Q^{11}\left(Q^{22}\right)^T\right)\nonumber\\
=& - V_{1,1}\left(\mathrm{tr}\left(Q^{11}\right)\right)^2 - V_{2}\mathrm{tr}\left(\left(Q^{11}\right)^2\right).
\label{eq:ab-Un-averaging3}
\end{align}
Further, we calculate in the same way the terms in the second line of Eq.~\eqref{eq:ab-Un-averaging1}:
\begin{align}
&\left\langle(U^{-1}_1Q^{12}U_2)_{ab}(U^{-1}_2Q^{21}U_1)_{ba}\right\rangle_{\text{U}(2n)} \nonumber\\
=& \left\langle U^{-1}_{a\alpha} Q^{12}_{\alpha\beta}(
U^*
)_{\beta b}(U^*)^{-1}_{b \gamma}Q^{21}_{\gamma\mu}U_{\mu a}\right\rangle_{\text{U}(2n)} \nonumber\\
=& \left\langle U^{*}_{\alpha a} U_{\mu a} U^{*}_{\nu\lambda}U_{\sigma\rho}\right\rangle_{\text{U}(2n)}  \delta_{b \rho}\delta_{\sigma \gamma}\delta_{\beta \nu}\delta_{\lambda b}Q^{12}_{\alpha\beta}Q^{21}_{\gamma\mu}\nonumber\\
=&\left( V_{1,1}\left(\delta_{\alpha\mu}\delta_{aa} \delta_{\nu\sigma}\delta_{\lambda\rho} + \delta_{\alpha\sigma}\delta_{a\rho} \delta_{\nu \mu}\delta_{\lambda a}\right) + V_{2}\left(\delta_{\nu \mu}\delta_{aa} \delta_{\alpha\sigma}\delta_{\lambda\rho} + \delta_{\alpha\mu}\delta_{a\rho} \delta_{\nu\sigma}\delta_{\lambda a}\right)\right)\delta_{b \rho}\delta_{\sigma \gamma}\delta_{\beta \nu}\delta_{\lambda b}Q^{12}_{\alpha\beta}Q^{21}_{\gamma\mu}\nonumber\\
=& V_{1,1}\left( \delta_{b \rho}\delta_{\rho b}Q^{12}_{\mu\beta}\delta_{\beta \nu}\delta_{\nu \gamma}Q^{21}_{\gamma\mu} +  \delta_{ba}\delta_{ab}\delta_{\alpha\gamma}Q^{12}_{\alpha\beta}Q^{21}_{\gamma\nu}\delta_{\beta,\nu}\right) \nonumber\\
&+ V_{2}\left(  \delta_{b\rho}\delta_{\rho b}\delta_{\alpha \gamma}Q^{12}_{\alpha\beta}Q^{21}_{\gamma\nu}\delta_{\beta \nu} +  \delta_{b,a}\delta_{\nu \gamma}\delta_{\beta \nu}\delta_{ab}Q^{12}_{\sigma\beta}Q^{21}_{\gamma\alpha}\right)\nonumber \\
=&V_{1,1}\mathrm{tr}\left(Q^{12}Q^{21}\right)+V_{2}\mathrm{tr}\left(Q^{12}\left(Q^{21}\right)^T\right)\nonumber\\
=&\left(V_{1,1}-V_{2}\right) \mathrm{tr}\left(Q^{12}Q^{21}\right)\nonumber\\
=&\left(V_{1,1}-V_{2}\right)\left[ \mathrm{tr}\left(\mathds{1}_{4n\times 4n}\right)- \mathrm{tr}\left(\left(Q^{11}\right)^2\right)\right].
\label{eq:ab-Un-averaging4}
\end{align}
Finally, there is a  term of the following type both in Eq.~\eqref{eq:ab-Un-averaging1} and Eq.~\eqref{eq:ab-Un-averaging2}:
\begin{align}
&\left\langle(U^{-1}_1Q^{11}U_1)_{ab}(U^{-1}_1Q^{11}U_1)_{cd}\right\rangle_{\text{U}(2n)} \nonumber\\
=& \left\langle U^{-1}_{a \alpha} Q^{11}_{\alpha\beta}U_{\beta b}U^{-1}_{c \gamma}Q^{11}_{\gamma\mu}U_{\mu,d}\right\rangle_{\text{U}(2n)}\nonumber\\
=& \left\langle U^{*}_{\alpha a} U_{\beta b} U^{*}_{\gamma c}U_{\mu d}\right\rangle_{\text{U}(2n)}  Q^{11}_{\alpha\beta}Q^{11}_{\gamma\mu}\nonumber\\
=&\left[ V_{1,1}\left(\delta_{\alpha\beta}\delta_{ab} \delta_{\gamma\mu}\delta_{c d} + \delta_{\alpha\mu}\delta_{ad} \delta_{\gamma \beta}\delta_{c b}\right) + V_{2}\left(\delta_{\gamma \beta}\delta_{ab} \delta_{\alpha\mu}\delta_{c d} + \delta_{\alpha\beta}\delta_{ad} \delta_{\gamma\mu}\delta_{c b}\right)\right] Q^{11}_{\alpha\beta}Q^{22}_{\gamma\mu}\nonumber\\
=& V_{1,1}\left[(\mathrm{tr}Q^{11})^2\delta_{ab}\delta_{c d} + \mathrm{tr}(Q^{11}Q^{11})\delta_{ad}\delta_{c b}\right] + V_{2}\left[\mathrm{tr}(Q^{11}Q^{11})\delta_{ab}\delta_{c d} + (\mathrm{tr}Q^{11})^2\delta_{ad}\delta_{c b}\right] \nonumber\\
=& \left[ V_{1,1}\left(\mathrm{tr}Q^{11}\right)^2+V_{2}\mathrm{tr}\left(\left(Q^{11}\right)^2\right)\right] \delta_{ab}\delta_{cd} + 
\left[V_{1,1}\mathrm{tr}\left(\left(Q^{11}\right)^2\right)+V_{2}\left(\mathrm{tr}Q^{11}\right)^2\right] \delta_{ad}\delta_{cb} \,.
\label{eq:ab-Un-averaging5}
\end{align}
In the above transformations, we have used the  symmetry property of $Q$,
\begin{align}
\begin{pmatrix}
Q_{11} & Q_{12}\\
Q_{21} & Q_{22}
\end{pmatrix}=Q = -\overline{Q} = \begin{pmatrix}
-Q_{22}^T & -Q_{12}^T\\
-Q_{21}^T & -Q_{11}^T
\end{pmatrix} \,,
\end{align}
as well as the constraints
\begin{align}
\mathds{1} &= Q^2 = \begin{pmatrix}
Q^{11}Q^{11}+Q^{12}Q^{21} & Q^{11}Q^{12}+Q^{12}Q^{22}\\
Q^{21}Q^{11}+Q^{22}Q^{21} & Q^{22}Q^{22}+Q^{21}Q^{12}
\end{pmatrix},&  0&=\mathrm{Tr}Q=\mathrm{Tr}Q^{11}+\mathrm{Tr}Q^{22} \,.
\label{eq:red}
\end{align}

Substituting Eqs. \eqref{eq:ab-Un-averaging3}, \eqref{eq:ab-Un-averaging4}, and \eqref{eq:ab-Un-averaging5} into  Eqs.~\eqref{eq:ab-Un-averaging1} and \eqref{eq:ab-Un-averaging2}, we obtain
\begin{align}
\left\langle\mathrm{tr}(\Lambda Q_{aa}) \, \mathrm{tr}(\Lambda Q_{bb} ) \right\rangle_{\text{U}(2n)}  &=
2V_{1,1}\left(\mathrm{tr}Q^{11}\right)^2+2V_{2}\mathrm{tr}\left(\left(Q^{11}\right)^2\right),
\label{eq:ab-Un-averaging6}
\\
\left\langle \mathrm{tr}(\Lambda Q_{ab}\Lambda Q_{ba} ) \right\rangle_{\text{U}(2n)} & =
V_{2}\left(\mathrm{tr}Q^{11}\right)^2+(2V_{1,1}-V_2)\mathrm{tr}\left(\left(Q^{11}\right)^2\right)-\left(V_{1,1}+V_{2}\right) \mathrm{tr}\left(\mathds{1}_{4n\times 4n}\right).
\label{eq:ab-Un-averaging7}
\end{align}
In the replica limit $n \to 0$, the last term in Eq.~\eqref{eq:ab-Un-averaging7} vanishes, and we obtain
\begin{align}
\left\langle\begin{pmatrix}
O_{(1,1)}^{ab}\\
O_{(2)}^{ab}
\end{pmatrix}\right\rangle_{\text{U}(2n)} &= \underbrace{\begin{pmatrix}
	2V_{1,1} & 2V_2 \\
	V_2 & 2V_{1,1}-V_2
	\end{pmatrix}}_{=:K_2}\cdot 
\begin{pmatrix}
O_{(1,1)} \\
O_{(2)}
\end{pmatrix} \,,
\label{eq:RG-matrixK2}
\end{align}
where notations for basis operators introduced in Eqs. \eqref{eq:RG-O2-invariant} and \eqref{eq:RG-Oab} were used. 

Equation \eqref{eq:RG-matrixK2} is a central result of this subsection. Let us discuss its implications. Consider a $K$-invariant operator $O$ that is quadratic in $Q$. We can expand it in the basis 
$O_{(1,1)}, \, O_{(2)}$ or, alternatively, in the basis  $\langle O_{(1,1)}^{ab} \rangle_{\text{U}(2n)}, \, \langle O_{(2)}^{ab} \rangle_{\text{U}(2n)}$:
\be
O = c_{(1,1)} O_{(1,1)} + c_{(2)} O_{(2)} = c_{(1,1)}'   \langle O_{(1,1)}^{ab} \rangle_{\text{U}(2n)}+ c_{(2)}'  \langle O_{(2)}^{ab} \rangle_{\text{U}(2n)} \,.
\label{eq:invariant-q2-1}
\ee
According to Eq.~\eqref{eq:RG-matrixK2}, the corresponding vectors of coefficients are related by the matrix $K_2^T$:
\be
\begin{pmatrix}
	c_{(1,1)} \\
	c_{(2)}
\end{pmatrix} = K_2^T
\begin{pmatrix}
	c_{(1,1)}' \\
	c_{(2)}'
\end{pmatrix} \,.
\label{eq:invariant-q2-2}
\ee
We know from Sec.~\ref{sec:RG-quadratic-class-C} that, in order for the operator $O$ to be an invariant operator, the vector $(c_{(1,1)}, c_{(2)})^T$ should be an eigenvector of the matrix $M_2^T$ defined in Eq.~\eqref{eq:RG-matrixM2}.  Furthermore, the operation of ${\text{U}(2n)}$ averaging commutes with RG, so that the RG flow of the coefficients $c_{(1,1)}'$,  $c_{(2)}'$ is described by the same equation
\eqref{eq:RG-matrixM2}. Therefore, for $O$ to be an invariant operator, $(c_{(1,1)}' , c_{(2)}')^T$ should also be an eigenvector of $M_2^T$. In view of Eq.~\eqref{eq:RG-matrixK2}, both conditions are compatible if and only if the matrix $K_2^T$ has the same eigenvectors as $M_2^T$. In other words, $K_2^T$ should commute with $M_2^T$, or, equivalently, $K_2$ should commute with $M_2$. 
To demonstrate that this is indeed the case, we represent the matrix $K_2$ in the form
\begin{align}
K_2 &= 2(V_{1,1}+V_2)\begin{pmatrix}
1 &  \\ 
& 1
\end{pmatrix} - V_2\cdot  \underbrace{\begin{pmatrix}
	2 & -2 \\ 
	-1& 3
	\end{pmatrix}}_{=M_2} \,,
\end{align}
which obviously commutes with $M_2$. 

Our analysis thus shows that the eigenoperators can be also obtained as eigenvectors of the matrix $K_2^T$.  Explicit verification of the fact that $M_2$ and $K_2$ commute serves as a non-trivial check of the calculation.

\subsection{Generalization to operators of higher degree, $q \equiv |\lambda| > 2$}
\label{sec:rg_gen}

We extend now the RG analysis to determine the polynomial pure-scaling operators of higher degree,  $q \equiv |\lambda| > 2$. 

Let us begin with considering the basis of $K$-invariant operators
\begin{align}
O_{(q_1,\ldots, q_l)} &= \prod_{i=1}^l\mathrm{tr} \left((\Lambda Q)^{q_i}\right) \label{eq:rg-kinv} \,,
\end{align}
where $(q_1, \ldots, q_l)$ are partitions of $q$, i.e.,
$q_1+q_2 + \ldots + q_l = q$ and $q_1 \ge q_2 \ge \ldots \ge q_l > 0$.  This is an extension of the basis \eqref{eq:RG-O2-invariant} to arbitrary integer $q \ge 2$. As we are going to discuss, the RG does not  mix operators of different degree $q=|\lambda|$.  Within given $q$ it acts according to 
\begin{align}
\delta \begin{pmatrix}
O_{(1,\ldots 1)}\\
\vdots\\ 
O_{(q)}
\end{pmatrix} &= M_q  \begin{pmatrix}
O_{(1,\ldots 1)}\\
\vdots\\ 
O_{(q)}
\end{pmatrix}.
\label{eq:RG-Mq-def}
\end{align} 
We will derive the rules for calculating the matrix $M$ for classes A and C for any $q$ below, and will present
it explicitly for $q=3$ and 4.  

Any invariant operator of degree $q$ can be expanded in terms of the basis operators,
\be
O=\sum_{(q_1,\ldots, q_l)}   c_{(q_1,\ldots, q_l)}  O_{(q_1,\ldots, q_l)} \,,
\ee
where the sum goes over the partitions of $q$.  The RG evolution of the coefficients is determined by the matrix $M^T$:
\begin{align}
\delta \begin{pmatrix}
c_{(1,\ldots 1)}\\
\vdots\\ 
c_{(q)}
\end{pmatrix} &= M_q^T  \begin{pmatrix}
c_{(1,\ldots 1)}\\
\vdots\\ 
c_{(q)}
\end{pmatrix}.
\label{eq:RG-MqT}
\end{align} 
Thus, determining eigenvectors of the matrix $M_q^T$, one finds invariant pure-scaling operators of degree $q$. 

In order to derive the RG rules, it is useful to inspect an example.  Consider the operator $\mathrm{tr} \left(A Q B Q\right)  \mathrm{tr} \left(C Q\right)$. Splitting $Q$ in slow and fast modes and expanding in fast fields according to Eq.~\eqref{eq:RG-Q-expansion}, and keeping only the terms of zeroth and second order in $\mathcal{X}$, we get 
\begin{align}
& \mathrm{tr} \left(A Q B Q\right) 
\mathrm{tr} \left(C Q\right)  = \mathrm{tr} \left(A Q_s B Q_s\right) 
\mathrm{tr} \left(C Q_s\right)   \nonumber \\
& +  2 \mathrm{tr} \left(A g_s\Lambda \mathcal{X}^2 g_s^{-1}B Q_s\right) 
\mathrm{tr} \left(C Q_s\right) 
+ 2 \mathrm{tr} \left(A Q_s B g_s\Lambda \mathcal{X}^2 g_s^{-1} \right) 
\mathrm{tr} \left(C Q_s\right)
+ 2 \mathrm{tr} \left(A Q_s B Q_s \right) 
\mathrm{tr} \left(C g_s\Lambda \mathcal{X}^2 g_s^{-1}\right)
\nonumber\\&+ 
4 \mathrm{tr} \left(A g_s\Lambda \mathcal{X} g_s^{-1} B g_s\Lambda \mathcal{X} g_s^{-1}\right) 
\mathrm{tr} \left(C Q_s\right) + 
4 \mathrm{tr} \left(A g_s\Lambda \mathcal{X} g_s^{-1} B Q_s\right)  \mathrm{tr} \left(C g_s\Lambda \mathcal{X} g_s^{-1}\right) \nonumber \\
& + 4 \mathrm{tr} \left(A Q_s  B g_s\Lambda \mathcal{X}g_s^{-1} \right) 
\mathrm{tr} \left(C g_s\Lambda \mathcal{X} g_s^{-1}\right).
\end{align}
Averaging now over the fast fields $\mathcal{X}$, one finds contributions of three types. In the contributions of the first type, both factors  $\mathcal{X}$ originate from the same $Q$, thus entering
as $\mathcal{X}^2$. In the contribution of the second type, two factors of $\mathcal{X}$ come from two different $Q$s under the same trace. Finally, in the terms of third type, 
two $\mathcal{X}$s originate from different $Q$s under different traces. We have already encountered all three types of contributions 
[see Eq.~\eqref{qdecomp}] when developing the RG for $q=2$ operators in Sec.~\ref{sec:rg_2}. By using this, we can generalize the results obtained there,
Eqs. \eqref{eq:rgres} and  \eqref{eq:rgresA}, to the RG rules valid for any $q$. We do it now separately for classes A and C.

\subsubsection{Class A}
\label{sec:inv-op-class-A}

We first derive general RG rules in class A. The contributions $\delta_d ( \ldots )$ originating from two $\mathcal{X}$ factors coming from two distinct traces  fuse these two traces into a single one, see the first line of Eq. \eqref{eq:rgresA} :
\be
\label{eq:RG-general-classA-1}
\delta_d \left[ \mathrm{tr} \left((\Lambda Q)^{q_i}\right) \mathrm{tr} \left((\Lambda Q)^{q_j}\right) \right] = 2I_f q_i q_j \cdot \mathrm{tr} \left((\Lambda Q)^{q_i+q_j}\right).
\ee
The prefactor $q_i q_j$ accounts for $q_i$ places where the fast field $\mathcal{X}$ can occur in $(\Lambda Q)^{q_i}$ and $q_j$ places where another $\mathcal{X}$ can occur in $(\Lambda Q)^{q_j}$. For every of these (identical) $q_iq_j$ terms, we use the first line of Eq. \eqref{eq:rgresA} with $A= (\Lambda Q_s)^{q_i-1}\Lambda$ and $B= (\Lambda Q_s)^{q_j-1}\Lambda$, which yields the r.h.s. of Eq.~\eqref{eq:RG-general-classA-1}. If the operator is a product of a larger number of traces, each pair of them will produce such a contribution.  

Further, the contributions $\delta_s (\ldots)$ originating from two $\mathcal{X}$ factors coming from the same trace cut this trace into two pieces in all possible ways, see the second line of Eq. \eqref{eq:rgresA} :
\bea
\label{eq:RG-general-classA-2}
\delta_s \mathrm{tr} \left((\Lambda Q)^{q_i}\right) &=& 2I_f \sum_{q_j=1}^{q_i-1} q_j \cdot \mathrm{tr} \left((\Lambda Q)^{q_j}\right) \mathrm{tr} \left((\Lambda Q)^{q_i-q_j}\right) \nonumber \\
& = & I_f  \, q_i  \sum_{q_j=1}^{q_i-1}  \mathrm{tr} \left((\Lambda Q)^{q_j}\right) \mathrm{tr} \left((\Lambda Q)^{q_i-q_j}\right).
\eea
Indeed, the two fast fields can come from $Q$ matrices in arbitrary positions $k$ and $k+q_j$, where $1 \le k < k+q_j \le q_i$.  For each such contribution, we use the second line of \eqref{eq:rgresA} with $A= (\Lambda Q)^{q_j-1}\Lambda $ and $B= (\Lambda Q)^{q_{i}-q_j-1}\Lambda $. Note that contributions originating from the $\mathcal{X}^2$ factors (i.e. with both $\mathcal{X}$ fields coming from the same $Q$) vanish in the class A due to the replica limit, see Eq.~\eqref{eq:rg3a} and the comment after it. 
If the operator is a product of several traces, then each of them will produce a contribution $\delta_s(\ldots)$ according to 
Eq.~\eqref{eq:RG-general-classA-2}.

Applying these rules, we can find the matrix $M_q^A$ determining the renormalization of operators of degree $q$, Eq.~\eqref{eq:RG-Mq-def}. For $q=3$ and 4 we get:
\begin{align}
(M_2^A)^T &=(2I_f)\left(
\begin{array}{cc}
0 & 1 \\
1 & 0 \\
\end{array}
\right), &
(M_3^A)^T &=(2I_f)\left(
\begin{array}{ccc}
0 & 1 & 0 \\
3 & 0 & 3 \\
0 & 2 & 0 \\
\end{array}
\right),\nonumber\\
(M_4^A)^T &=(2I_f)\left(
\begin{array}{ccccc}
0 & 1 & 0 & 0 & 0 \\
6 & 0 & 2 & 3 & 0 \\
0 & 1 & 0 & 0 & 2 \\
0 & 4 & 0 & 0 & 4 \\
0 & 0 & 4 & 3 & 0 \\
\end{array}
\right).
\label{eq:MT-234}
\end{align}
For convenience, we have also included here the $q=2$ result, Eq.~\eqref{eq:rgresA}. We have listed in Eq.~\eqref{eq:MT-234} the transposed matrices $(M_q^A)^T$ since,
according to Eq.~\eqref{eq:RG-MqT}, eigenvectors of $(M_q^A)^T$ determine the eigenoperators of RG, i.e.,  pure-scaling operators. 
The results for the $K$-invariant pure scaling operators of class A read
\begin{align}
\begin{pmatrix}
\mathcal{P}^A_{(1,1)}\\
\mathcal{P}^A_{(2)}
\end{pmatrix}
&=\underbrace{\begin{pmatrix}
1 & -1 \\
1 & 1\\
\end{pmatrix}}_{\equiv P^A_2}
\begin{pmatrix}
\mathrm{tr} (\Lambda Q)\mathrm{tr} (\Lambda Q)\\
\mathrm{tr} (\Lambda Q\Lambda Q)
\end{pmatrix}, \nonumber \\
\begin{pmatrix}
\mathcal{P}^A_{(1,1,1)}\\
\mathcal{P}^A_{(2,1)} \\
\mathcal{P}^A_{(3)}
\end{pmatrix}
&=\underbrace{\begin{pmatrix}
1 & -3 & 2\\
1 & 0 & -1\\
1 & 3 & 2
\end{pmatrix}}_{\equiv P^A_3}
\begin{pmatrix}
\mathrm{tr} (\Lambda Q)\mathrm{tr} (\Lambda Q)\mathrm{tr} (\Lambda Q) \\
\mathrm{tr} (\Lambda Q)\mathrm{tr} (\Lambda Q\Lambda Q) \\
\mathrm{tr} (\Lambda Q\Lambda Q\Lambda Q)
\end{pmatrix}, \nonumber \\
\begin{pmatrix}
\mathcal{P}^A_{(1,1,1,1)}\\
\mathcal{P}^A_{(2,1,1)}\\
\mathcal{P}^A_{(2,2)}\\
\mathcal{P}^A_{(3,1)}\\
\mathcal{P}^A_{(4)}
\end{pmatrix}
&=\underbrace{\begin{pmatrix}
1 & - 6 & 3 & 8 & -6\\
1 & - 2 & -1 & 0 & 2\\
1 & 0 & 3 & 4 & 0\\
1 & - 2 & 1 & 0 & -2\\
1 &  6 & 3 & 8 & 6
\end{pmatrix}}_{\equiv P^A_4}
\begin{pmatrix}
\mathrm{tr} (\Lambda Q)\mathrm{tr} (\Lambda Q)\mathrm{tr} (\Lambda Q) \mathrm{tr} (\Lambda Q) \\
\mathrm{tr} (\Lambda Q)\mathrm{tr} (\Lambda Q)\mathrm{tr} (\Lambda Q\Lambda Q) \\
\mathrm{tr} (\Lambda Q\Lambda Q)\mathrm{tr} (\Lambda Q\Lambda Q) \\ \mathrm{tr} (\Lambda Q\Lambda Q\Lambda Q)\mathrm{tr} (\Lambda Q)\\
\mathrm{tr} (\Lambda Q\Lambda Q\Lambda Q\Lambda Q)
\end{pmatrix}.
\label{eq:ops_a}
\end{align}
We fix the freedom in the overall prefactor in each of the eigenoperators by choosing the coefficient in front of $O_{(1,1,\ldots,1)}$ to be unity. As proven below in Sec.~\ref{sec:onelaplacian}, eigenvalues 
$m^A_{(q_1,\ldots,q_n)}$
of the one-loop RG matrix $(M_q^A)^T$ are proportional to quadratic Casimir invariants,
\be
m^A_{(q_1,\ldots,q_n)} = - I_f \sum_j q_j(-c_j^{A}-q_j),
\label{eq:mA-gen-parabol}
\ee
where $c_j^A = 1-2j$, see Sec.~\ref{subsec:MF-symmetry}. This serves as an additional control of numerical calculations and allows us to unambiguously associate Young diagrams with the eigenvectors of $M_q^T$.

\subsubsection{Class C}
\label{sec:RG-class_C-arbitrary_q}

Now we perform the analysis for class C.  For $q=2$, the RG rules are given by Eq.~\eqref{eq:rgres}. In order to generalize them to operators with higher $q$, it is useful to trace the origin of various terms in 
Eq.~\eqref{eq:rgres} by inspecting the derivation,  Eqs.~\eqref{eq:rg-C1}--\eqref{eq:RG-C3}. The upper right element $-2$ of the matrix $M_2$ in Eq.~\eqref{eq:rgres} originates from two fast fields $\mathcal{X}$ coming from different traces. It is analogous to class A but contains an additional factor of two. We thus have 
\be
\label{eq:RG-general-classC-1}
\delta_d \left[ \mathrm{tr} \left((\Lambda Q)^{q_i}\right) \mathrm{tr} \left((\Lambda Q)^{q_j}\right) \right] = 4I_f q_i q_j \cdot \mathrm{tr} \left((\Lambda Q)^{q_i+q_j}\right).
\ee
The lower left element 1 of the matrix $M_2$ in Eq.~\eqref{eq:rgres}, as well as the contribution $-1$ to the lower right element, originate from two fast fields $\mathcal{X}$ coming from the same trace but different $Q$ fields. Finally, the remaining contribution $-2$ to the lower right element originates from two fast fields $\mathcal{X}$ coming from the same $Q$ matrix. This leads to 
\begin{align}
\delta_s \langle\mathrm{tr} \left((\Lambda Q)^{q_i}\right)  \rangle &= 2I_f\sum_{q_j=1}^{q_i-1} q_j \, \mathrm{tr} \left((\Lambda Q)^{q_j}\right) \mathrm{tr} \left((\Lambda Q)^{q_i-q_j}\right)-I_f q_i(q_i-1) \mathrm{tr} \left((\Lambda Q)^{q_i}\right) - 2 I_f q_i \,\mathrm{tr} \left((\Lambda Q)^{q_i}\right) \nonumber \\
& = I_f q_i \sum_{q_j=1}^{q_i-1} \mathrm{tr} \left((\Lambda Q)^{q_j}\right) \mathrm{tr} \left((\Lambda Q)^{q_i-q_j}\right)-I_f q_i(q_i +1) \, \mathrm{tr} \left((\Lambda Q)^{q_i}\right).
\label{eq:rg4}
\end{align}
In the first line of Eq.~\eqref{eq:rg4}, the first two terms come from a contraction of two $\mathcal{X}$ fields originating from different $Q$ fields, and the last term from the contraction within the $\mathcal{X}^2$ factor originating from a single $Q$ field. 

Using these rules, we determine the matrices $M_q$ that govern the renormalization of operators of degree $q$ in class C according to Eq.~\eqref{eq:RG-Mq-def}. The results for  $q=3$ and 4 read (the $q=2$ result found above is also included for completeness):
\begin{align}
M_2^T &=(-2I_f)\begin{pmatrix}
2 &  -1\\
-2 & 3
\end{pmatrix}, &
M_3^T &=(-2I_f)\left(
\begin{array}{ccc}
3 & -1 & 0 \\
-6 & 4 & -3 \\
0 & -4 & 6 \\
\end{array}
\right),\nonumber\\
M_4^T &=(-2I_f)\left(
\begin{array}{ccccc}
4 & -1 & 0 & 0 & 0 \\
-12 & 5 & -2 & -3 & 0 \\
0 & -2 & 6 & 0 & -2 \\
0 & -8 & 0 & 7 & -4 \\
0 & 0 & -8 & -6 & 10 \\
\end{array}
\right).
\label{eq:RG-matrices-M234}
\end{align}
These are class-C counterparts  of class-A results given by Eq.~\eqref{eq:MT-234}.  We recall that the matrices  $M_q^T$ control the renormalization of coefficients $c_{(q_1,q_2,\ldots)}$, see Eq.~\eqref{eq:RG-MqT}. Calculating their eigenvectors, we find the eigenoperators:
\begin{align}
\begin{pmatrix}
\mathcal{P}^C_{(1,1)}\\
\mathcal{P}^C_{(2)}
\end{pmatrix}
&= \underbrace{\begin{pmatrix}
1& -2\\
1& 1
\end{pmatrix}}_{\equiv P^C_2}
\begin{pmatrix}
\mathrm{tr} (\Lambda Q)\mathrm{tr} (\Lambda Q) \\
\mathrm{tr} (\Lambda Q\Lambda Q)
\end{pmatrix}  , \nonumber\\
\begin{pmatrix}
\mathcal{P}^C_{(1,1,1)}\\
\mathcal{P}^C_{(2,1)} \\
\mathcal{P}^C_{(3)}
\end{pmatrix}
&= \underbrace{\begin{pmatrix}
1& -6& 8\\
1& -1& -2\\
1& 3& 2
\end{pmatrix}}_{\equiv P^C_3}
\begin{pmatrix}
\mathrm{tr} (\Lambda Q)\mathrm{tr} (\Lambda Q)\mathrm{tr} (\Lambda Q) \\
\mathrm{tr} (\Lambda Q\Lambda Q)\mathrm{tr} (\Lambda Q) \\
\mathrm{tr} (\Lambda Q\Lambda Q\Lambda Q)
\end{pmatrix} , \nonumber\\
\begin{pmatrix}
\mathcal{P}^C_{(1,1,1,1)}\\
\mathcal{P}^C_{(2,1,1)}\\
\mathcal{P}^C_{(2,2)}\\
\mathcal{P}^C_{(3,1)} \\
\mathcal{P}^C_{(4)}
\end{pmatrix}
&= \underbrace{\begin{pmatrix}
1 & -12 & 12 & -32 & 48 \\
1 & -5 & -2 & -4 & -8 \\
1 & -2 & 7 & 8 & -2 \\
1 & 1 & -2 & 2 & 4 \\
1 & 6 & 3 & -8 & -6 
\end{pmatrix}}_{\equiv P^C_4}
\begin{pmatrix}
\mathrm{tr} (\Lambda Q)\mathrm{tr} (\Lambda Q)\mathrm{tr} (\Lambda Q) \mathrm{tr} (\Lambda Q) \\
\mathrm{tr} (\Lambda Q)\mathrm{tr} (\Lambda Q)\mathrm{tr} (\Lambda Q\Lambda Q)  \\
\mathrm{tr} (\Lambda Q\Lambda Q)\mathrm{tr} (\Lambda Q)\mathrm{tr} (\Lambda Q) \\ \mathrm{tr} (\Lambda Q\Lambda Q\Lambda Q)\mathrm{tr} (\Lambda Q) \\
\mathrm{tr} (\Lambda Q\Lambda Q\Lambda Q\Lambda Q)
\end{pmatrix}.
\label{rg:higher_ord}
\end{align}
In analogy with class A, the eigenvalues 
$m_{(q_1,\ldots,q_n)}$ of the class-C one-loop RG matrix $(M_q)^T$ are proportional to quadratic Casimir invariants,
\be
m_{(q_1,\ldots,q_n)} = - I_f \sum_j q_j(-c_j^{C}-q_j),
\label{eq:mC-gen-parabol}
\ee
where $c_j^C = 1-4j$, see Sec.~\ref{subsec:MF-symmetry}. This statement (which we have verified numerically up to the order $q=15$) will be proven in full generality in Sec.~\ref{sec:onelaplacian},

\subsubsection{Invariant operators in class C by $\text{U}(2n)$ averaging for $q > 2$. }

As was demonstrated in Sec.~\ref{sec:u_avg}, and alternative way to obtain the $K$-invariant operators is to perform the $K \equiv \text{U}(2n)$ averaging of fixed-replica operators (replica indices $a_1, \ldots, a_q$). This yields an alternative basis in the space of $K$-invariant operators. Any $K$-invariant operator can be expanded in the one and the other basis, and the corresponding coefficients are related by the matrix $K_q^T$, see Eqs.~\eqref{eq:RG-matrixK2}, \eqref{eq:invariant-q2-1}, and \eqref{eq:invariant-q2-2} for $q=2$, which are directly extended to arbitrary $q$:
\be
\Bigg\langle\begin{pmatrix}
	O_{(1,1, \ldots,1)}^{a_1\ldots a_q}\\
	\vdots \\
	O_{(q)}^{a_1\ldots a_q}
\end{pmatrix}\Bigg\rangle_{\text{U}(2n)}= K_q 
\begin{pmatrix}
	O_{(1,1, \ldots,1)} \\
	\vdots\\
	O_{(q)}
\end{pmatrix},
\qquad \qquad
\begin{pmatrix}
	c_{(1,1, \ldots,1)} \\
	\vdots\\
	c_{(q)}
\end{pmatrix}  = K_q^T 
\begin{pmatrix}
	c_{(1,1, \ldots,1)}'\\
	\vdots \\
	c_{(q)}'
\end{pmatrix},
\label{eq:RG-invariant-arbitrary_q-1}
\ee
where
\be
O = c_{(1,1, \ldots, 1)} O_{(1,1, \ldots, 1)} + \ldots + c_{(q)} O_{(q)} = c_{(1,1, \ldots, 1)}'   \langle O_{(1,1, \ldots, 1)}^{a_1, \ldots, a_q} \rangle_{\text{U}(2n)} + 
c_{(q)}'  \langle O_{(q)}^{a_1, \ldots, a_q} \rangle_{\text{U}(2n)} \,.
\label{eq:RG-invariant-arbitrary_q-2}
\ee

For $q=3$, we find
\begin{align}
K_3^T=\left(
\begin{array}{ccc}
1 & 0 & 0 \\
0 & 1 & 0 \\
0 & 0 & 1 \\
\end{array}
\right)V_{(1,1,1)}
+\left(
\begin{array}{ccc}
0 & \frac{1}{2} & 0 \\
3 & -\frac{1}{2} & \frac{3}{2} \\
0 & 2 & -\frac{3}{2} \\
\end{array}
\right)V_{(2,1)}
+\left(
\begin{array}{ccc}
0 & 0 & \frac{1}{4} \\
0 & 1 & -\frac{3}{4} \\
2 & -1 & 1 \\
\end{array}
\right)V_{(3)},
\label{eq:RG-invariant-K3}
\end{align}
where, as before, $V_\lambda$ are Weingarten functions. As was explained in Sec.~\ref{sec:u_avg}, the consistency requires that $K_q^T$ commutes with the RG matrix $M_q^T$. Indeed, a straightforward check shows that all the matrices in front of the Weingarten coefficients $V_\lambda$ in Eq.~\eqref{eq:RG-invariant-K3} commute with $M_3^T$, Eq.~\eqref{eq:RG-matrices-M234}.

An analogous calculation for $q=4$ yields
\begin{align}
K_4^T&=\left(
\begin{array}{ccccc}
1 & 0 & 0 & 0 & 0 \\
0 & 1 & 0 & 0 & 0 \\
0 & 0 & 1 & 0 & 0 \\
0 & 0 & 0 & 1 & 0 \\
0 & 0 & 0 & 0 & 1 \\
\end{array}
\right)V_{(1,1,1,1)}
+\left(
\begin{array}{ccccc}
0 & \frac{1}{2} & 0 & 0 & 0 \\
6 & -\frac{1}{2} & 1 & \frac{3}{2} & 0 \\
0 & 1 & -1 & 0 & 1 \\
0 & 4 & 0 & -\frac{3}{2} & 2 \\
0 & 0 & 4 & 3 & -3 \\
\end{array}
\right)V_{(2,1,1)}\nonumber\\
&+\left(
\begin{array}{ccccc}
0 & 0 & \frac{1}{4} & 0 & 0 \\
0 & \frac{1}{2} & -\frac{1}{2} & 0 & \frac{1}{2} \\
3 & -\frac{1}{2} & \frac{5}{4} & 0 & -\frac{1}{4} \\
0 & 0 & 0 & \frac{3}{2} & -1 \\
0 & 2 & -1 & -\frac{3}{2} & \frac{5}{4} \\
\end{array}
\right)V_{(2,2)}
+\left(
\begin{array}{ccccc}
0 & 0 & 0 & \frac{1}{4} & 0 \\
0 & 2 & 0 & -\frac{3}{4} & 1 \\
0 & 0 & 0 & \frac{3}{2} & -1 \\
8 & -2 & 4 & \frac{5}{2} & -2 \\
0 & 4 & -4 & -3 & 4 \\
\end{array}
\right)V_{(3,1)}\nonumber\\
&+\left(
\begin{array}{ccccc}
0 & 0 & 0 & 0 & \frac{1}{8} \\
0 & 0 & 1 & \frac{3}{4} & -\frac{3}{4} \\
0 & 1 & -\frac{1}{2} & -\frac{3}{4} & \frac{5}{8} \\
0 & 2 & -2 & -\frac{3}{2} & 2 \\
6 & -3 & \frac{5}{2} & 3 & -\frac{5}{2} \\
\end{array}
\right)V_{(4)}.
\label{eq:RG-invariant-K4}
\end{align}
Again, we have verified that the matrices multiplying the Weingarten coefficients $V_\lambda$ in Eq.~\eqref{eq:RG-invariant-K4} commute with $M_4^T$, Eq.~\eqref{eq:RG-matrices-M234}.
This is an explicit demonstration of the fact that eigenoperators can also be obtained as eigenvectors of $K_q^T$, see Sec.~\ref{sec:u_avg}.

\subsection{Generalized parabolicity of one-loop RG eigenvalues}
\label{sec:onelaplacian}

In Ref. \cite{friedan1980nonlinear}, it was shown that the one-loop $\beta$ function characterizing the RG flow of the coupling constant in non-linear sigma models over Riemannian manifolds $\mathcal{M}$ is given by a purely geometrical property of the manifold---the Ricci curvature. Here we prove that the one-loop RG scaling dimensions describing the flow of gradientless operators $f(Q)$ in sigma models over symmetric spaces $G/K$ are given (up to an overall constant) by Casimir invariants or, equivalently, by eigenvalues of the Laplace-Beltrami operator on $G/K$ [that acts on functions $f(Q)$]. A consequence of this result is the exact generalized parabolicity of one-loop RG scaling dimensions.

Consider an arbitrary function (``composite operator'')  $f(Q)$, which is a map $G/K\rightarrow \mathbb{C}$.  We begin the proof by noticing that the one-loop RG operator $\mathcal{D}$ acting on an function $f(Q)$  can be written in the form
\be
\mathcal{D}f(Q) = \frac{1}{2} \int_{\mathfrak{p}} d\mu(\mathcal{X}) \left.\dfrac{d^2}{dt^2}f( g_s e^{-t\mathcal{X}} \Lambda e^{t\mathcal{X}}  g_s^{-1})\right|_{t=0}e^{-S_f[\mathcal{X}]},
\label{eq:one-loop-gen-parab1}
\ee
where $S_f[\mathcal{X}]$ is the action \eqref{eq:fastaction} and  the integration runs over the tangent space $\mathfrak{p}$ to the manifold $G/K$, see Eq.~\eqref{Cartan-decomposition-real}.
The second derivative in $t$ extracts two fast fields $\mathcal{X}$ in all possible ways from all $Q$ fields in $f(Q)$, and the Gaussian integral over the manifold $G/K$ then yields the one-loop contractions. 

We choose a  basis $\{\mathcal{X}_i\}$  in $\mathfrak{p}$ satisfying the symmetry constraints of the class-C sigma-model,  $\mathcal{X}_i^\dagger = \mathcal{X}_i$ and $\tau_3 \mathcal{X}_i^\dagger\tau_3 = -\mathcal{X}_i$. 
[A natural choice of this basis was given in Eq.~\eqref{p-generators}.]
Then $\mathcal{X} = \sum_i y_i \mathcal{X}_i$, with real and independent coefficients $y_i$. Further we can demand orthonormality with respect to the trace ${\rm tr} \left(\mathcal{X}_i  \mathcal{X}_j\right) = \delta_{ij}$. The fast-mode action \eqref{eq:fastaction} then is diagonal in $y_i$,
\begin{align}
S_f[\{y_i\}] &=  \frac{\pi \sigma_0}{2} \sum_i \left(\nabla y_i \nabla y_i + h^2  y_i y_i\right) .
\label{eq:one-loop-gen-parab2}
\end{align}
Using for brevity $e^{t\mathrm{ad}_{\mathcal{X}}}O \equiv e^{-t\mathcal{X}}O e^{t\mathcal{X}}$, we can rewrite Eq.~\eqref{eq:one-loop-gen-parab1} as 
\begin{align}
\mathcal{D}f(Q)  &= \left. \frac{1}{2} \dfrac{d^2}{dt^2} \int_{\mathfrak{p}} d\mu(\{y_i\})  f(g_s (e^{t\sum_i y_i\mathrm{ad}_{\mathcal{X}_i}}\Lambda)g_s^{-1})  \cdot e^{- S_f[\{y_i\}]} \right|_{t=0}.
\label{eq:one-loop-gen-parab3}
\end{align}

We rewrite the second derivative via
\begin{align}
\left.\dfrac{d^2}{dt^2} f(   g_s (e^{t\sum_i y_i\mathrm{ad}_{\mathcal{X}_i}}\Lambda)g_s^{-1} ) \right|_{t=0}
&= \sum_{ij}y_i y_j \hat{\mathcal{X}_i'} \hat{\mathcal{X}_j' } f(Q) \,,
\label{eq:one-loop-gen-parab4}
\end{align}
where $\hat{\mathcal{X}_i'}$ is the derivative in the direction $\mathcal{X}'_i = g_s \mathcal{X}_i g_s^{-1}$ on the manifold $G/K$:
\be
\hat{\mathcal{X}'_i}f(Q) = 
\partial_t f( g_s (e^{t \, \mathrm{ad}_{\mathcal{X}_i}}\Lambda) g_s^{-1}  )|_{t=0} =
\partial_t f( e^{t \, \mathrm{ad}_{\mathcal{X}'_i}} Q )|_{t=0} .
\label{eq:one-loop-gen-parab5}
\ee
It is easy to see that $\{\mathcal{X}'_i\}$ has a meaning of the basis in the tangent space to sigma-model manifold at the point $Q$. 

Substituting Eq.~\eqref{eq:one-loop-gen-parab4} into Eq.~\eqref{eq:one-loop-gen-parab3}, we obtain
\begin{align}
\mathcal{D} f(Q)&= \sum_{ij} \left(\hat{\mathcal{X}}'_i\hat{\mathcal{X}}'_jf(Q)\right) \cdot \int_{\mathfrak{p}} d\mu(\{y_l\})  \, y_i y_j \, e^{-S_f[\{y_l\}]} 
\end{align}
The gaussian integral here is exactly the one-loop integral defined by Eq.~\eqref{eq:if},
\begin{align}
\int_{\mathfrak{p}} d\mu(\{y_l\})  \, y_i y_j \, e^{-S_f[\{y_l\}]} = I_f \delta_{ij} \,,
\end{align}
so that
\begin{align}
\mathcal{D} &= I_f \sum_{i} \hat{\mathcal{X}}'_i\hat{\mathcal{X}}'_i \,.
\end{align}
Thus, the one-loop RG operator is proportional to the Laplacian $\sum_{i} \hat{\mathcal{X}}'_i\hat{\mathcal{X}}'_i$. Therefore, the RG eigenvalues are proportional to the quadratic Casimir invariants.
It follows that the spectrum of one-loop RG eigenvalues satisfies the generalized parabolicity, as has been already stated (and numerically verified) for classes A and C, Eqs.~\eqref{eq:mA-gen-parabol} and \eqref{eq:mC-gen-parabol}. As the above proof is very general, we believe that it is applicable to all ten symmetry classes. 

Let us emphasize that the generalized parabolicity of the one-loop RG is violated by higher-loop contributions. In general, the generalized multifractal spectra are not parabolic, see also the discussion in Sec.~ \ref{sec:beyond2D-nonparabolic}.

\section{From $\sigma$-model to wave function correlators}
\label{sec:wave_ops}
In this section, we derive eigenfunction correlators that correspond to pure-scaling sigma-model operators determined in Sec.~\ref{sec:rg_c}. Let us emphasize, that this ``translation'' to the wave-function language is not unique:  there are many wave-function observables that correspond to the same sigma-model composite operator, i.e., belong to the same representation and scale in the same way. In the present section, we focus on correlation functions that involve only one spin component of wave functions (say, spin up). 
 In Sec. \ref{sec:trans}, we perform the necessary steps: (i) express correlators of  wave functions $\psi_\sigma$ in terms of retarded and advanced Greens functions;  (ii) find linear combinations $S_\sigma$ of the bosonic action variables $\phi_\sigma$ that map one-to-one (in the sense of correlation functions) to $\psi_\sigma$; (iii) determine the $Q$-field combinations that remain after integrating out $S$ and connect them to operators whose RG flow has been derived above. Finally,  (iv) we have to take into account connections between the different parametrizations $Q$ and $\tilde{Q}$ of the sigma-model manifold. Overall, we find that the single-spin wave-function combinations translate into $Q$-field operators in class C in the same manner as in the class-A sigma-model. 
 
 Having derived pure-scaling eigenstate correlation functions (Sec. \ref{sec:trans}), we can determine numerically the scaling exponents by simulating the appropriate network model. While this perfectly works at the conventional quantum-Hall transition, it turns out that there is a major computational difficulty for implementing this with a high accuracy at the SQH transition. The point is that, apart from the simplest case of the representations $(q)$ corresponding to the conventional multifractality,  the eigenstate correlations for class C obtained in this way have an indefinite sign for an individual disorder realization and exhibit very strong fluctuations. As a result, one has to perform averaging over a very large number of disorder realizations in  order to reach a reasonable accuracy for the exponents. 
 We do this for the (1,1) exponent in Sec. \ref{sec:numerics}. At the same time, for exponents corresponding to Young diagrams with $q>2$ [such as (2,1), (1,1,1), (3,1), (2,1,1), (2,2), etc] this would require truly outstanding numerical efforts, which go beyond this work. In view of this, we develop two further alternative approaches  to determine numerically these exponents. 
 
 One of these approaches is presented in Sec. \ref{sec:young}. We focus there on Young-symmetrized wave function combinations $|\Psi_\lambda|^2$ of $\psi_{\uparrow}$ that are manifestly positive and show pure scaling of representation $\lambda$ in class-A systems \cite{gruzberg2013classification}. This makes them very suitable for numerical averaging. In class C,  these combinations 
 [except for LDOS moments $\lambda = (q)$] do not map to pure-scaling operators of the sigma-model. One thus could naively think that studying numerically these correlators does not bring any useful information about the class-C scaling. Luckily, the situation turns out to be much more favorable.
 Using our results for the coefficients of pure scaling operators $\mathcal{P}_\lambda^A$ \eqref{eq:ops_a} in class A and $\mathcal{P}_\lambda^C$ \eqref{rg:higher_ord} in class C as obtained in Sec. \ref{sec:rg_gen}, we derive in Sec. \ref{sec:young} mixing matrices relating the two. Remarkably, these matrices have many zero entries, implying that we are able to access numerically some subleading exponents of class C by investigating the scaling of eigenoperators of class A. 
 
 In Sec. \ref{sec:numerics} we present numerical results for the exponents accessible within this framework. Since the finite-size effects and the need for ensemble averaging proliferate with increasing $q$, we restrict ourselves here to $q\leq 4$. The key results are the scaling exponents collected in Table \ref{tab:lC}. Crucially, we find that the generalized parabolicity is violated strongly: the exponents deviate up to $50\%$ from the values that they would take for a spectrum satisfying the generalized parabolicity. For comparison, we present the results of the class-A network-model numerics in Table \ref{tab:lA}. There, deviations from parabolicity are much weaker (of the order of several percent). 

A second approach for obtaining class-C generalized MF exponents from sign-definite correlation functions is developed in Sec. \ref{sec:wave_iw} below. The idea is to explore correlation functions that involve both spin projections on the same site. As we show there, correlation functions involving the total density $|\psi|=\sqrt{|\psi|^2_\uparrow + |\psi|^2_\downarrow}$ turn out to be very useful in this connection. 

\subsection{Translation dictionary}
\label{sec:trans}
In this subsection, we are going to derive a ``dictionary'' to translate correlations of wave functions $\psi$ into those of bosonic field variables $\phi$ of the replicated action and further to the sigma-model ($Q$-field) language.
We choose linear combinations $S$ of the bosonic variables $\phi$, such that the translation of $S$ to $\psi$ is one-to-one. We then restrict the wave-function combination to a singe spin component (e.g., spin up). 
Mapping the single-spin expressions to $Q$-field correlation functions and averaging over a certain subgroup $\text{U}_d$ of the gauge group $K=\text{U}(2n)$, we obtain  sigma-model composite operators as studied in Sec.~\ref{sec:rg_c}. 

\subsubsection{From wave functions to Green's functions}

We introduce retarded and advanced Green's function:
\begin{align}
\hat{G}^{R/A}(\epsilon,r_1,r_2) &= \langle r_1|\dfrac{1}{\epsilon-\hat{H}\pm i0}|r_2\rangle.
\end{align}
The hat indicates the matrix structure representing the two-by-two combined spin and particle-hole space of the Hamiltonian \eqref{eq:hamC}.
To extract eigenstate correlations, we use a connection between the eigenstates $\psi_{\alpha,\sigma}(r)$ (where $\sigma$ is the spin index)  and the  Green's functions in a given disorder configuration 
\begin{align}
G^R_{\sigma_1\sigma_2}(\epsilon,r_1,r_2)-G^A_{\sigma_1\sigma_2}(\epsilon,r_1,r_2) = 2\pi i \sum_{\alpha} \delta(\epsilon-\omega_\alpha)\psi^*_{\alpha,\sigma_1}(r_1)\psi_{\alpha,\sigma_2}(r_2) \,.
\end{align}
Here $\omega_\alpha$ are exact eigenenergies in that disorder configuration. This implies the correspondence \cite{mirlin00,gruzberg2013classification}
\begin{align}
\dfrac{1}{2\pi i\nu(\epsilon)}\left( G^R_{\sigma_1\sigma_2}(\epsilon,r_1,r_2)-G^A_{\sigma_1\sigma_2}(\epsilon,r_1,r_2)\right)\ \longleftrightarrow \ \left. \psi^*_{\alpha,\sigma_1}(r_1)\psi_{\alpha,\sigma_2}(r_2)\right |_{\omega_\alpha \approx \epsilon} \label{eq:Gpsi}
\end{align}
in averages $\langle \ldots \rangle$ over disorder configurations $V$. Here $\nu(\epsilon)$ is the average density of states 
This correspondence is extended to eigenstate composite objects of higher order and allows one to relate the corresponding averages to averaged products of 
Green's functions \cite{mirlin00,gruzberg2013classification}. 

The components of the Green's function are not independent.  In view of the particle-hole symmetry \eqref{eq:pha}  of the Hamiltonian, the Green's functions satisfy 
\begin{align}
i\sigma_y\hat{G}^R(\epsilon,r_1,r_2)i\sigma_y &= \hat{G}^A(-\epsilon,r_2,r_1)^T \,.
\end{align}
When written explicitly in terms of the components, this reads
\begin{align}
\begin{pmatrix}
-G^R_{\downarrow\downarrow}(\epsilon,r_1,r_2) & G^R_{\downarrow\uparrow}(\epsilon,r_1,r_2)
\\
G^R_{\uparrow\downarrow}(\epsilon,r_1,r_2) & -G^R_{\uparrow\uparrow}(\epsilon,r_1,r_2)
\end{pmatrix}&=
\begin{pmatrix}
G^A_{\uparrow\uparrow}(-\epsilon,r_2,r_1) & G^A_{\downarrow\uparrow}(-\epsilon,r_2,r_1)
\\
G^A_{\uparrow\downarrow}(-\epsilon,r_2,r_1) & G^A_{\uparrow\uparrow}(-\epsilon,r_2,r_1)
\end{pmatrix} .
\end{align}
We see that the symmetry operation connects the retarded Green's function at positive energy to the advanced one at negative energy and opposite spin components.

\subsubsection{From Green's functions to averages over bosonic vector fields}

In order to express composite objects built out of Green's functions in terms of field-theoretic averages, it is useful to introduce  the following linear combinations of the bosonic integration variables $\phi$:
\begin{align}
S_{\uparrow,a} &= \phi_{\uparrow,a} +ie^{i\alpha_a}\phi_{\downarrow,-a}^* \,,\nonumber \\
S_{\downarrow,a} &= \phi_{\downarrow,a} -ie^{i\alpha_a}\phi_{\uparrow,-a}^* \,,
\end{align}
with arbitrary phases $\alpha_a$. 
These fields have the useful property of only having ``particle-conserving'' contractions with respect to the Gaussian action $S_0[\phi^*,\phi; V]$, Eq.~\eqref{eq:S0}, defined for the Hamiltonian $H=H_0+V$ in a given disorder configuration $V$:
\begin{align}
i\langle S^*_{\downarrow,a} (r_1) S_{\downarrow,a}(r_2)\rangle_{S_0[\phi^*,\phi; V]} &= G_{\downarrow\downarrow}^{R}(\omega_a;r_1,r_2)+G_{\uparrow\uparrow}^{R}(-\omega_a;r_2,r_1) 
\nonumber \\
& = G_{\downarrow\downarrow}^{R}(\omega_a;r_1,r_2)-G_{\downarrow\downarrow}^{A}(\omega_a;r_1,r_2), \nonumber\\
i\langle S_{\downarrow,a}^* (r_1)S_{\uparrow,a}(r_2)\rangle_{S_0[\phi^*,\phi; V]} &= G_{\downarrow\uparrow}^{R}(\omega_a;r_1,r_2)-G_{\downarrow\uparrow}^{R}(-\omega_a;r_2,r_1)
\nonumber \\
& = G_{\downarrow\uparrow}^{R}(\omega_a;r_1,r_2)-G_{\downarrow\uparrow}^{A}(\omega_a;r_1,r_2).
\end{align}
Other non-zero contractions can be obtained by complex conjugation and the particle-hole operation. 
Combining this with Eq. \eqref{eq:Gpsi}, we get
\begin{align}
\left\langle\left\langle S^*_{\downarrow,a} (r_1) S_{\downarrow,a}(r_2) \right\rangle_{S_0[\phi^*,\phi; V]}\right\rangle_V & =  \left.\dfrac{-1}{2\pi \nu(\epsilon)}\left\langle \psi^*_{\alpha,\downarrow}(r_1)\psi_{\alpha,\downarrow}(r_2)\right\rangle_V\right |_{\omega_\alpha\approx\epsilon} \,,  \nonumber \\
\left\langle\left\langle S_{\downarrow,a}^* (r_1)S_{\uparrow,a}(r_2) \right\rangle_{S_0[\phi^*,\phi; V]}\right\rangle_V & =   \left.\dfrac{-1}{2\pi \nu(\epsilon)}\left\langle \psi^*_{\alpha,\downarrow}(r_1)\psi_{\alpha,\uparrow}(r_2)\right\rangle_V\right |_{\omega_\alpha\approx \epsilon} 
 \,.
\label{eq:psiS}
\end{align}
This shows a one-to-one correspondence between $\psi$ and $S$ (since we are only interested in scaling, we omit the prefactor):
\be
  \psi^*_{\alpha,\sigma_1}(r_1)\psi_{\alpha,\sigma_2}(r_2)
 \ \longleftrightarrow \  S^*_{\sigma_1,a} (r_1) S_{\sigma_2,a}(r_2) \,.
 \ee
The correspondence is understood in the sense of averages that are explicitly shown in Eq.~\eqref{eq:psiS}.
This correspondence is straightforwardly extended to higher-order products by using the fact that the action $S_0[\phi^*,\phi; V]$ is Gaussian and diagonal in replicas. As an example,
\be
  \psi^*_{\alpha,\sigma_1}(r_1)\psi_{\alpha,\sigma_2}(r_2)   \psi^*_{\beta,\sigma_3}(r_3)\psi_{\beta,\sigma_4}(r_4)
 \ \longleftrightarrow \  S^*_{\sigma_1,a} (r_1) S_{\sigma_2,a}(r_2) S^*_{\sigma_3,b} (r_3) S_{\sigma_4,b}(r_4) \,.
 \ee
Note that the wave-function indices $\alpha, \beta, \gamma, \ldots$ are translated into the replica indices $a, b, c, \ldots$. For this reason, we will use below the replica indices $a, b, c, \ldots$ also as eigenfunction labels. 


\subsubsection{From bosonic vector fields  to the sigma-model field $Q$}
\label{sec:from-phi-to-Q}

The next step is to find the dictionary that translates the vector fields $S$ of the replica field theory  into the matrix field $Q$ of the sigma-model. Upon the Hubbard-Stratonovich transformation, we obtain the theory with the action $S[\phi,\phi^*,Q]$, see Eq.~\eqref{eq:S-phi-Q}. Intergrating over the fields $\phi,\phi^*$, one translates any composite object $O[\phi,\phi^*]$ into its sigma-model counterpart 
$\langle O[\phi,\phi^*]\rangle_{S[\phi,\phi^*,Q]}$ expressed in terms of the field $Q$. For the objects that are bilinear in $S$ fields (or, equivalently in $\phi$ fields), we have
\begin{align}
\langle S^*_{\downarrow,a} (r_1) S_{\downarrow,b}(r_1)\rangle_{S[\phi,\phi^*,Q]} &=  ie^{-i\alpha_a}Q^{AR}_{-b,-a}-ie^{i\alpha_b}Q^{RA}_{a,b}+Q^{RR}_{a,b}+e^{i\alpha_b-i\alpha_a}Q^{RR}_{-b,-a} \label{eq:trans2} \\
&=-ie^{-i\alpha_a}Q^{AR}_{a,b}-ie^{i\alpha_b}Q^{RA}_{a,b}+Q^{RR}_{a,b}-e^{i\alpha_b-i\alpha_a}Q^{AA}_{a,b}\equiv \mathcal{Q}_{ab}^{00},
\label{eq:trans1}\\
\langle S^*_{\uparrow,a} (r_1) S_{\uparrow,b}(r_1)\rangle_{S[\phi,\phi^*,Q]} &=  -ie^{-i\alpha_a}Q^{AR}_{a,b}+ie^{i\alpha_b}Q^{RA}_{-b,-a}+Q^{RR}_{a,b}+e^{i\alpha_b-i\alpha_a}Q^{RR}_{-b,-a} \nonumber\\
&=-ie^{-i\alpha_a}Q^{AR}_{a,b}-ie^{i\alpha_b}Q^{RA}_{a,b}+Q^{RR}_{a,b}-e^{i\alpha_b-i\alpha_a}Q^{AA}_{a,b}\equiv \mathcal{Q}_{ab}^{00},  \\
\langle S_{\downarrow,a} (r_1) S_{\downarrow,b}(r_1)\rangle_{S[\phi,\phi^*,Q]} &= 0,\\
\langle S_{\downarrow,a}^* (r_1) S_{\uparrow,b}^*(r_1)\rangle_{S[\phi,\phi^*,Q]} &= Q^{AR}_{b,-a}e^{-i\alpha_a-i\alpha_b}+Q^{RA}_{a,-b}-ie^{-i\alpha_b}Q^{RR}_{a,-b}+ie^{-i\alpha_a}Q^{RR}_{b,-a} \nonumber\\
&=-Q^{AR}_{a,-b}e^{-i\alpha_a-i\alpha_b}+Q^{RA}_{a,-b}-ie^{-i\alpha_b}Q^{RR}_{a,-b}-ie^{-i\alpha_a}Q^{AA}_{a,-b}\equiv \mathcal{Q}_{ab}^{01},  \\
\langle S_{\downarrow,a} (r_1) S_{\uparrow,b}(r_1)\rangle_{S[\phi,\phi^*,Q]} &= Q^{AR}_{-a,b}+Q^{RA}_{-b,a}e^{i\alpha_a+i\alpha_b}-iQ^{RR}_{-a,b}e^{i\alpha_a}+iQ^{RR}_{-b,a}e^{i\alpha_b} \nonumber\\
&=Q^{AR}_{-a,b}-Q^{RA}_{-a,b}e^{i\alpha_a+i\alpha_b}-iQ^{RR}_{-a,b}e^{i\alpha_a}-iQ^{AA}_{-a,b}e^{i\alpha_a}\equiv \mathcal{Q}_{ab}^{10}, 
\label{eq:trans}\\
\langle S_{\downarrow,a}^* (r_1)S_{\uparrow,b}(r_1)\rangle_{S[\phi,\phi^*,Q]} &= 0.  \label{eq:trans3}
\end{align}
The right-hand sides can be straightforwardly read off from the coupling \eqref{eq:coup} of $Q$ and $\phi$ fields in the action $S[\phi,\phi^*; Q]$. We have introduced here the short-hand notations 
$\mathcal{Q}_{ab}^{pp'}$ with $p, p' = 0,1$. 

Translating composite objects of higher order in $S, S^*$ fields amounts to using Wick's theorem, since the action $S[\phi,\phi^*,Q]$ is Gaussian with respect to $\phi, \phi^*$ fields. However, since this action is not diagonal in replicas (due to the coupling to the $Q$ field), many terms arise.  A drastic simplification occurs if one assumes (which is fully sufficient for the purposes of calculating scaling exponents) that all $r_i$ are far apart from each other in comparison to the microscopic scale $a$ controlling the decay of averaged single-particle Green's function, $|r_i-r_j|\gg a$.  Then one only has to retain contractions that are diagonal in spatial coordinates. In the special case of the network model that is used for numerical simulations in this paper, the condition $r_i\neq r_j$ is sufficient since all contractions between different spatial points are identically equal to zero. With only one spin component, further simplifications occur. First, we only need the contraction rule \eqref{eq:trans1}.
Second, by averaging over the phases $\alpha_a, \ \alpha_b, \ldots$, i.e., over the  diagonal subgroup $\text{U}_d=\text{U}(1)^{2n}$ of the gauge group $\text{U}(2n)$, 
we end up in the space of sigma-model composite operators spanned by products of traces of the type $\mathrm{tr} \, (E_{aa}\Lambda Q E_{bb}\Lambda Q \ldots)$, where $E_{aa}$ is the projector on replica $a$ as defined above. As the simplest example, we show this for composite operators that are of fourth order in $S, S^*$ fields and thus map on sigma-model composite operators that are quadratic in $Q$ fields (i.e., $q=2$ operators). There are two such independent composite operators (we skip the spin index, which is the same for all $S$ fields), $|S_a(r_1)|^2 |S_b(r_2)|^2$ and $S_a^*(r_1)S_b(r_2)S_b(r_1)S_a^*(r_2)$; we will call them the Hartree and Fock terms, respectively, for obvious reason. For the Hartree term, we get, according to Eq.~\eqref{eq:trans1},
\begin{align}
\langle\mathcal{Q}^{00}_{aa}
\mathcal{Q}^{00}_{bb}\rangle_{\text{U}_d} &= \int_0^{2\pi}\int_0^{2\pi}\dfrac{d\alpha_ad\alpha_b}{(2\pi)^2}\left(-ie^{-i\alpha_a}Q^{AR}_{a,a}-ie^{i\alpha_a}Q^{RA}_{a,a}+Q^{RR}_{a,a}-Q^{AA}_{a,a}\right)\nonumber\\&\times
\left(-ie^{-i\alpha_b}Q^{AR}_{b,b}-ie^{i\alpha_b}Q^{RA}_{b,b}+Q^{RR}_{b,b}-Q^{AA}_{b,b}\right)\nonumber\\
&=\left(Q^{RR}_{a,a}-Q^{AA}_{a,a}\right)\left(Q^{RR}_{b,b}-Q^{AA}_{b,b}\right)=\mathrm{tr}\left(E_{aa}\Lambda Q\right)\mathrm{tr}\left(E_{bb}\Lambda Q\right)\equiv O_{(1,1)}^{ab} \,.
\end{align}
Similarly, we have for the Fock term:
\begin{align}
\langle\mathcal{Q}^{00}_{ab}
\mathcal{Q}^{00}_{ba}\rangle_{\text{U}_d} &= \int_0^{2\pi}\int_0^{2\pi}\dfrac{d\alpha_ad\alpha_b}{(2\pi)^2}\left(-ie^{-i\alpha_a}Q^{AR}_{a,b}-ie^{i\alpha_b}Q^{RA}_{a,b}+Q^{RR}_{a,b}-e^{i\alpha_b-i\alpha_a}Q^{AA}_{a,b}\right)\nonumber\\&\times
\left(-ie^{-i\alpha_b}Q^{AR}_{b,a}-ie^{i\alpha_a}Q^{RA}_{b,a}+Q^{RR}_{b,a}-e^{i\alpha_a-i\alpha_b}Q^{AA}_{b,a}\right) \, \nonumber\\
&=Q^{RR}_{a,b}Q^{RR}_{b,a}+Q^{AA}_{a,b}Q^{AA}_{b,a}-Q^{AR}_{a,b}Q^{RA}_{b,a}-Q^{RA}_{a,b}Q^{AR}_{b,a}=\mathrm{tr}\left(E_{aa}\Lambda QE_{bb}\Lambda Q\right)\equiv O_{(2)}^{ab} \,.
\end{align}
Thus, after the $\text{U}_d$ averaging, we have obtained  the operators $O_{(1,1)}^{ab}$ and $O_{(2)}^{ab}$ as defined in Eq.~\eqref{eq:RG-Oab}. 

This is straightforwardly generalized to $q>2$ composite operators, and we obtain a mapping of the composite operators of $S$ fields to the sigma-model operators of the type
\be
O_{(i,j,\ldots )}^{a_1\ldots a_q}=\mathrm{tr}\left( E_{a_1a_1}\Lambda Q\cdots  E_{a_ia_i} \Lambda Q\right)\mathrm{tr}\left( E_{a_{i+1}a_{i+1}}\Lambda Q\cdots  E_{a_{i+j}a_{i+j}} \Lambda Q\right)\ldots. 
\label{eq:sigma_mod_basis_comp_ops}
\ee
Recalling the direct correspondence between the $S$ fields and eigenstates $\psi$, we thus get a correspondence between eigenstate observables and sigma-model composite operators:
\begin{align}
O_{(1,1)}^{ab}& \ \longleftrightarrow\  |\psi_a(r_1)|^2 |\psi_b(r_2)|^2 \,, \nonumber \\
O_{(2)}^{ab} &  \ \longleftrightarrow\   \psi_a^*(r_1) \psi_b(r_1)\psi_b^*(r_2) \psi_a(r_2) \,, \nonumber \\
O_{(1,1,1)}^{abc}&  \ \longleftrightarrow\   |\psi_a(r_1)|^2 |\psi_b(r_2)|^2  |\psi_c(r_3)|^2 \,, \nonumber \\
O_{(2,1)}^{abc}&  \ \longleftrightarrow\   \psi_a^*(r_1) \psi_b(r_1)\psi_b^*(r_2) \psi_a(r_2)   |\psi_c(r_3)|^2 \,, \nonumber \\
O_{(3)}^{abc}&  \ \longleftrightarrow\   \psi_a^*(r_1)\psi_b(r_1)\psi_b^*(r_2) \psi_c(r_2) \psi_c^*(r_3)\psi_a(r_3) \,, \nonumber \\
O_{(1,1,1,1)}^{abcd}&  \ \longleftrightarrow\   |\psi_a(r_1)|^2 |\psi_b(r_2)|^2  |\psi_c(r_3)|^2 |\psi_d(r_4)|^2  \,, \nonumber\\
O_{(2,1,1)}^{abc}&  \ \longleftrightarrow\   \psi_a^*(r_1) \psi_b(r_1)\psi_b^*(r_2) \psi_a(r_2)   |\psi_c(r_3)|^2 |\psi_d(r_4)|^2  \,, \nonumber\\
O_{(2,2)}^{abcd}&  \ \longleftrightarrow\   \psi_a^*(r_1) \psi_b(r_1)\psi_b^*(r_2) \psi_a(r_2) \psi_c^*(r_3) \psi_d(r_3)\psi_d^*(r_4) \psi_c(r_4) \,, \nonumber\\
O_{(3,1)}^{abcd} &  \ \longleftrightarrow\   \psi_a^*(r_1) \psi_b(r_1)\psi_b^*(r_2) \psi_c(r_2) \psi_c^*(r_3)\psi_a(r_3) |\psi_d(r_4)|^2 \,, \nonumber \\
O_{(4)}^{abcd}&  \ \longleftrightarrow\   \psi_a^*(r_1) \psi_b(r_1)\psi_b^*(r_2) \psi_c(r_2) \psi_c^*(r_3)\psi_d(r_3) \psi_d^*(r_4)\psi_a(r_4) \,.
\label{eq:O-psi-correspondence}
\end{align}
The spin index for wave functions is suppressed here (and in the rest of Sec.~\ref{sec:wave_ops})  since it is the same for all of them (e.g., all wave functions have spin up).
We denote the symmetrized version of the operators $O_{(i,j,\ldots )}^{a_1\ldots a_q}$ over the symmetric group $S_q$ by $O_{(i,j,\ldots )}^{S\{a_1\ldots a_q\}}$:
\begin{align}
O_{(i,j,\ldots )}^{S\{a_1\ldots a_q\}}=\dfrac{1}{q!}\sum_{\sigma\in S_q} \mathrm{tr}\left( E_{\sigma(a_1)\sigma(a_1)}\Lambda Q\cdots  E_{\sigma(a_i)\sigma(a_i)} \Lambda Q\right)\mathrm{tr}\left( E_{\sigma(a_{i+1})\sigma(a_{i+1})} \Lambda Q\cdots  E_{\sigma(a_{i+j})\sigma(a_{i+j})} \Lambda Q\right)\ldots. 
\label{eq:ops_rep}
\end{align}
(For $q=2$ the symmetrization is redundant: $O_{(1,1)}^{ab} \equiv O_{(1,1)}^{S\{ab\}}$ and $O_{(2)}^{ab} \equiv O_{(2)}^{S\{ab\}}$.)
The symmetrized operators $O_{(i,j,\ldots )}^{S\{a_1\ldots a_q\}}$ obey the same RG equations of Sec.~\ref{sec:RG-class_C-arbitrary_q} as the $K$-invariant operators. 
Thus, pure-scaling operators are constructed from them according to Eq.~(\ref{rg:higher_ord}). At the same time, as we have just shown, the operators $O_{(i,j,\ldots )}^{S\{a_1\ldots a_q\}}$ correspond to eigenstate observables \eqref{eq:O-psi-correspondence} (with $S_q$-symmetrization over energies corresponding to symmetrization over replica indices). Therefore, by using Eq.~\eqref{rg:higher_ord}, we also obtain the pure-scaling combinations of eigenstates. This is done in Sec.~\ref{sec:young}.

There is the following technical subtlety in the above derivation. The choice of the sigma-model manifold that was used in the derivation of the sigma-model in Sec.~\ref{sec:sigma} is slightly different from that used in Sec.~\ref{sec:RG-quadratic-class-C} for derivation of RG equations. The connection was explained in detail in the end of Sec.~\ref{sec:sigma} where the former parametrization was denoted $Q$ and the latter $\tilde{Q}$, see Eq.~\eqref{Q-tilde} for explicit relation between them. In the present section (Sec.~\ref{sec:wave_ops}) we used the $Q$ parametrization, since it is most conveneint for the mapping of eigenstate observables to the sigma-model composite operators. At the same time, at the last step, we used the fact that composite operators \eqref{eq:ops_rep} renormalize according to RG equations derived in Sec.~\ref{sec:RG-quadratic-class-C}. To demonstrate this fully rigorously, we should reexpress Eq.~\eqref{eq:ops_rep} in terms of the $\tilde{Q}$-field and verify that these operators satisfy the conditions for validity of the RG equations of Sec.~\ref{sec:RG-quadratic-class-C}. This is done in  \ref{appendix:Q-to-tilde-Q}.

\subsection{Generalized multifractality: Pure-scaling eigenstate observables}
\label{sec:young}

We are now in a position to write down the pure-scaling observables in terms of eigenfunctions. Let us first demonstrate this for the case of $q=2$.  We then have two eigenstates (labeled by $a$ and $b$ corresponding to replica indices in the field theory) and two spatial point $r_1$ and $r_2$. Two basis combinations are the Hartree ($H$) and Fock ($F$) terms, as given by Eq.~\eqref{eq:O-psi-correspondence} (with symmetrization over indices $a, b$)
\begin{align}
{O}_{(1,1)}^{ab}& \ \longleftrightarrow\  \frac{1}{2} \left(  |\psi_a(r_1)|^2 |\psi_b(r_2)|^2  +  |\psi_b(r_1)|^2 |\psi_a(r_2)|^2 \right) \equiv H \,, \nonumber \\
{O}_{(2)}^{ab} &  \ \longleftrightarrow\   \frac{1}{2} \left(   \psi_a^*(r_1) \psi_b(r_1)\psi_b^*(r_2) \psi_a(r_2) +  \psi_b^*(r_1) \psi_a(r_1)\psi_a^*(r_2) \psi_b(r_2) \right)  \equiv F\,. 
\label{O-H-F}
\end{align}
Using Eq.~\eqref{eq:opsb-ab}  [the same matrix of coefficients  appears also in Eqs.~\eqref{eq:opsb-AB} and \eqref{eq:opsb}, and is denoted $P_2^C$ in Eq.\eqref{rg:higher_ord}], we can write down the $q=2$ pure-scaling observables as combinations of these two basis combinations:
\bea
\mathcal{P}^C_{(2)} & \longleftrightarrow &  H + F \,, \nonumber \\
\mathcal{P}^C_{(1,1)} & \longleftrightarrow &  H -2 F \,.
\label{eq:H-minus-2F}
\eea
The operator $\mathcal{P}^C_{(2)}$ is a representative of the conventional multifractality  and can be realized also with a single eigenfunction, as $ |\psi_a(r)|^4$.  On the other hand, the subleading operator $\mathcal{P}^C_{(1,1)}$ goes beyond the conventional multifractality and is a representative of the generalized multifractalty. In the same way, pure-scaling operators for $q>2$ can be obtained. 

We have verified numerically, by using the class-C network model, that  $\mathcal{P}^C_{(1,1)}$  is realized by $H-2F$ and determined in this way the corresponding scaling exponent, see Sec.~\ref{sec:numerics}. There is, however, the following computational difficulty in determining the exponents of subleading operators in this way. For an individual realization of disorder, the quantity $H-2F$ has an indefinite sign, and its typical magnitude is of the same order as $H$ and $F$.  Only after the averaging one obtains $\langle H - 2F \rangle$ that is strongly suppressed with respect to $\langle H \rangle$ and $\langle F \rangle$. This should be contrasted to the case of class A: the operator $\mathcal{P}^A_{(1,1)}$ corresponds to $H-F$ which can be written as an absolute value squared, is thus strictly positive and small for any realization of disorder. In view of this, the numerical evaluation of the scaling exponent $\mathcal{P}^C_{(1,1)}$  by calculating the average $\langle H - 2F \rangle$  requires much larger computational efforts than the analogous calculation  (of the average $\langle H - F \rangle$) in class A. For this reason, we have numerically implemented this approach in the present work only for the (1,1) operator but not for subleading operators with $q>2$ for which still more extensive computational efforts are needed. We describe below (in this section and in Sec.~\ref{sec:wave_iw})  two alternative approaches that we used to determine some of subleading exponents with higher $q$. 

The first of these alternative approaches is based on expressing the sigma-model scaling operators $\mathcal{P}^C_\lambda$ in terms of a linear combination of the class-A scaling operators $\mathcal{P}^A_{\lambda'}$. 
According to the analysis of Ref.~\cite{gruzberg2013classification}, the wave-function observable $|\Psi_\lambda|^2$, with $\Psi_\lambda$ obtained by a combination of symmetrization and antisymmetrization  of the product of eigenstate amplitudes according to a Young diagram $\lambda$ (``Young symmetrization''), maps onto $\mathcal{P}^A_\lambda$ in the sigma-model language. Thus, expanding $\mathcal{P}^C_\lambda$ over $\mathcal{P}^A_{\lambda'}$, we simultaneously obtain an expansion of the pure-scaling eigenstate observable of class C corresponding to the diagram $\lambda$ over the pure-scaling eigenstate observables $|\Psi_{\lambda'}|^2$ of class A. 
The idea is to invert this expansion and to determine numerically the scaling of $|\Psi_\lambda|^2$, i.e., of $\mathcal{P}^A_\lambda$, in a system of class C. Since $|\Psi_\lambda|^2$ is strictly positive, this can be done quite efficiently. This scaling will be determined by the most relevant (in the RG sense) operator $\mathcal{P}^C_{\lambda'}$ out of those that contribute to $\mathcal{P}^A_\lambda$. This alows us to determine some (although not all) of the subleading scaling expionents in class C.

We briefly sketch the construction of $|\Psi_\lambda|^2$ corresponding to a Young diagram $\lambda $; see Appendix A of Ref.~\cite{gruzberg2013classification} for a detailed exposition. 
Consider a Young diagram $\lambda= (q_1, q_2, \ldots)$ with $|\lambda|=q$, as well as $q$ distinct eigenfunctions $\psi_1,\ldots, \psi_q$ and $q$ points $r_1,\ldots, r_q$.
We can put the indices of $\psi$ and $r$ in two Young tableaux of shape $\lambda$ (i.e., with the first row having the length $q_1$, the second row the length $q_2$, and so on). As an illustration, let us consider the following example for $\lambda = (3,2,1)$:
\begin{align}
\ytableausetup{boxsize=1.25em}
T_\psi &= \ytableaushort[\psi_]{246,15,3} && T_r =  \ytableaushort[{\bf r}_]{623,15,4} \;.
\end{align}
One associates with the pair of these two tableaus the corresponding (unsymmetrized) product $\Psi_{(3,2,1)}(T_\psi,T_r)$ of wave function amplitudes:
\begin{align*}
\Psi_{(3,2,1)}(T_\psi, T_r) = \psi_1({\bf r}_1) \psi_2({\bf r}_6) \psi_3({\bf r}_4) \psi_4({\bf r}_2) \psi_5({\bf r}_5) \psi_6({\bf r}_3).
\end{align*}
The next step is to introduce the operator $a_\lambda$ of symmetrization within the rows of a Young tableau and the operator $b_\lambda$ of antisymmetrization within the columns.  The Young symmetrizers are defined  as $c_\lambda = b_\lambda a_\lambda$ and $\tilde{c}_\lambda = a_\lambda b_\lambda$.  We consider the Young-symmetrized combinations
\be
\Psi_\lambda (c_\lambda T_\psi, T_r) =  \Psi_\lambda ( T_\psi, \tilde{c}_\lambda T_r)
\ee
and 
\be
\Psi_\lambda ( T_\psi, c_\lambda T_r) =  \Psi_\lambda ( \tilde{c}_\lambda T_\psi,  T_r) \,.
\ee
As shown in Ref.~\cite{gruzberg2013classification}, the squared absolute values of these Young-symmetrized products of eigenfunctions
map, in the case of class A, to the pure-scaling sigma-model operator $\mathcal{P}^A_\lambda$. 

We list absolute values squared of Young-symmetrized eigenstate combinations (which are pure-scaling observables for class A)  for the first few Young diagrams:
\begin{align}
|\Psi_{(2)}|^2 &= \frac12|\psi_a(r_1)\psi_b(r_2) + \psi_b(r_1)\psi_a(r_2)|^2 \,, \nonumber \\
|\Psi_{(1,1)}|^2 & = \frac12|\psi_a(r_1)\psi_b(r_2) - \psi_b(r_1)\psi_a(r_2)|^2 \,,
\nonumber\\ 
|\Psi_{(3)}|^2 &= \frac1{6}|\psi_a(r_1)\psi_b(r_2) \psi_c(r_3)+ \psi_b(r_1)\psi_a(r_2) \psi_c(r_3)+\psi_c(r_1)\psi_b(r_2) \psi_a(r_3) \,, \nonumber\\
&+\psi_b(r_1)\psi_c(r_2) \psi_a(r_3)+\psi_c(r_1)\psi_a(r_2) \psi_b(r_3)+\psi_a(r_1)\psi_c(r_2) \psi_b(r_3)|^2 \,,
\nonumber\\
|\Psi_{(2,1)}|^2 &= \frac14|\psi_a(r_1)\psi_b(r_2)\psi_c(r_3) - \psi_c(r_1)\psi_b(r_2)\psi_a(r_3)+\psi_b(r_1)\psi_a(r_2)\psi_c(r_3) - \psi_c(r_1)\psi_a(r_2)\psi_b(r_3)|^2 \,, \nonumber \\
\vdots\label{eq:waves}
\end{align}
Since we will discuss linear combinations of these expressions, it is important to comment on their normalization. Let us denote by $Y(\lambda)$ the number of terms in $\Psi_\lambda$ produced by the Young symmetrization. Up to order $q=4$, the values of $Y(\lambda)$ are given in the following table:
\begin{center}
\begin{tabular}{cc|cc}
$\lambda$ & $Y(\lambda)$	&$\lambda$ & $Y(\lambda)$\\
(1,1) & $2$ & (1,1,1,1) & $24$\\
(2) & $2$ & (2,1,1) & $12$\\
(1,1,1) & $6$ & (2,2) & $16$\\
(2,1) & $4$ & (3,1) & $12$\\
(3) & $6$ & (4) & $24$\\
\end{tabular}	
\end{center}
Writing down $|\Psi_{\lambda}|^2$ as a sum of monomials, we get exactly $Y(\lambda)$ terms of modulus-squared type. We choose the prefactor of $|\Psi_\lambda|^2$ in  Eq.~\eqref{eq:waves} to be equal to $Y(\lambda)^{-1}$, so that the total prefactor of these terms is unity. This has the consequence that $|\Psi_\lambda|^2$ maps precisely to $\mathcal{P}^A_{\lambda}$ given by Eq.~\eqref{eq:ops_a}. 
(We recall that all entries in the first column of matrices in Eq.~\eqref{eq:ops_a} have been fixed to be equal to unity.)

The next step is to establish a connection between the class-A eigenoperators $\mathcal{P}^A_{\lambda}$ and class-C eigenoperators $\mathcal{P}^C_{\lambda}$.  Using Eq.~\eqref{eq:ops_a} for $\mathcal{P}^A_{\lambda}$ and Eq. \eqref{rg:higher_ord} for $\mathcal{P}^C_{\lambda}$, we find
\begin{align}
\begin{pmatrix}
\mathcal{P}^C_{(1,1)}\\
\mathcal{P}^C_{(2)}
\end{pmatrix}
&= \begin{pmatrix}
1 & -\frac{1}{3} \\
0 & 1 
\end{pmatrix}
\begin{pmatrix}
\mathcal{P}^A_{(1,1)}\\
\mathcal{P}^A_{(2)}
\end{pmatrix}    \,, \nonumber\\ 
\begin{pmatrix}
\mathcal{P}^C_{(1,1,1)}\\
\mathcal{P}^C_{(2,1)}\\
\mathcal{P}^C_{(3)}
\end{pmatrix}
&= \begin{pmatrix}
1 & -\frac{4}{5} & \frac{1}{5} \\
0 & 1 & -\frac{1}{4} \\
0 & 0 & 1 
\end{pmatrix}
\begin{pmatrix}
\mathcal{P}^A_{(1,1,1)} \\
\mathcal{P}^A_{(2,1)} \\
\mathcal{P}^A_{(3)}
\end{pmatrix} \,, \nonumber \\
\begin{pmatrix}
\mathcal{P}^C_{(1,1,1,1)}\\
\mathcal{P}^C_{(2,1,1)}\\
\mathcal{P}^C_{(2,2)}\\
\mathcal{P}^C_{(3,1)} \\
\mathcal{P}^C_{(4)}
\end{pmatrix}
&= \begin{pmatrix}
1 & -\frac{9}{7} & -\frac{4}{35} & \frac{27}{35} & -\frac{1}{7} \\
0 & 1 & -\frac{2}{9} & -\frac{4}{9} & \frac{1}{9} \\
0 & 0 & 1 & -\frac{1}{2} & 0 \\
0 & 0 & 0 & 1 & -\frac{1}{5} \\
0 & 0 & 0 & 0 & 1 \\
\end{pmatrix}
\begin{pmatrix}
\mathcal{P}^A_{(1,1,1,1)}\\
\mathcal{P}^A_{(2,1,1)}\\
\mathcal{P}^A_{(2,2)}\\
\mathcal{P}^A_{(3,1)}\\
\mathcal{P}^A_{(4)}
\end{pmatrix} 
\label{eq:mixing}.
\end{align}
Replacing the sigma-model operators $\mathcal{P}^A_{\lambda'}$ on the right-hand side of these equations by  $|\Psi_{\lambda'}|^2$, we obtain wave-function combinations that scale like $\mathcal{P}^C_{\lambda}$. For example, for $\lambda = (1,1)$, we get $\mathcal{P}^C_{(1,1)} \: \longleftrightarrow \: |\Psi_{(2)}|^2-\frac13 |\Psi_{(1,1)}|^2$.
This is, of course, exactly the $H-2F$ expression from Eq.~\eqref{eq:H-minus-2F} (up to an overall factor). The corresponding numerical results for $\mathcal{P}^C_{(1,1)}$ are shown below in Sec.~\ref{sec:numerics}
(see Fig.~\ref{fig:AC}). As has been already explained, this type of numerics suffers from very strong fluctuations, which become even stronger with increasing $|\lambda| \equiv q$.  
We therefore choose a different route for $q=3,4$. Inverting the matrices in Eq.~\eqref{eq:mixing}, we find
\begin{align}
\begin{pmatrix}
\mathcal{P}^A_{(1,1,1)}\\
\mathcal{P}^A_{(2,1)} \\
\mathcal{P}^A_{(3)}
\end{pmatrix}
&= \begin{pmatrix}
\frac{1}{15} & -\frac{1}{10} & 0 \\
0 & \frac{1}{8} & -\frac{1}{24} \\
0 & 0 & \frac{1}{6} \\
\end{pmatrix}
\begin{pmatrix}
\mathcal{P}^C_{(1,1,1)}\\
\mathcal{P}^C_{(2,1)}\\
\mathcal{P}^C_{(3)}
\end{pmatrix} \,, \nonumber \\
\begin{pmatrix}
\mathcal{P}^A_{(1,1,1,1)}\\
\mathcal{P}^A_{(2,1,1)}\\
\mathcal{P}^A_{(2,2)}\\
\mathcal{P}^A_{(3,1)}\\
\mathcal{P}^A_{(4)}
\end{pmatrix}
&= \begin{pmatrix}
\frac{1}{35} & -\frac{1}{14} & -\frac{1}{40} & 0 & 0 \\
0 & \frac{1}{18} & \frac{1}{72} & -\frac{1}{18} & 0 \\
0 & 0 & \frac{1}{16} & -\frac{1}{20} & \frac{1}{80} \\
0 & 0 & 0 & \frac{1}{10} & -\frac{1}{40} \\
0 & 0 & 0 & 0 & \frac{1}{8} \\
\end{pmatrix}
\begin{pmatrix}
\mathcal{P}^C_{(1,1,1,1)}\\
\mathcal{P}^C_{(2,1,1)}\\
\mathcal{P}^C_{(2,2)}\\
\mathcal{P}^C_{(3,1)}\\
\mathcal{P}^C_{(4)}
\end{pmatrix} \,.
\label{eq:mixing2}
\end{align}
Naively, one could expect that, in class C, all $\mathcal{P}^A_\lambda$ will be dominated by the most relevant operator $\mathcal{P}^C_{(q)}$. This is not the case, however, due to zero entries at the upper right corner of the matrix. Equations \eqref{eq:mixing2} thus lead to non-trivial predictions amenable to numerical investigations. Specifically, when computing the $q=3$ eigenstate combinations $|\Psi_\lambda|^2$ corresponding to $\mathcal{P}^A_\lambda$ in a system of class C, we should see that $\mathcal{P}^A_{(3)}$ and $\mathcal{P}^A_{(2,1)}$ exhibit both the leading $(3)$ scaling
(with the exponent $x_\lambda$ equal to zero due to Weyl symmetry), while 
$\mathcal{P}^A_{(1,1,1)}$  has the leading contribution from the representation $(2,1)$.   By Weyl symmetry, the scaling exponent $x_\lambda$ of $(2,1)$ should be identical to the exponent of  $(1,1)$ obtained for $q=2$. For $q=4$, there should be three operators---$\mathcal{P}^A_{(4)}$, $\mathcal{P}^A_{(3,1)}$, and $\mathcal{P}^A_{(2,2)}$---showing the leading $(4)$ scaling, the operator $\mathcal{P}^A_{(2,1,1)}$
should be dominated by representation $(3,1)$,  and the operator $\mathcal{P}^A_{(1,1,1,1)}$ by the representation $(2,2)$. 

In Sec.~\ref{sec:numerics}, these predictions are verified and used for determining the scaling exponents by numerical simulations on the network model.

\subsection{Numerical computations}
\label{sec:numerics}

In this subsection, we present results of numerical evaluation of generalized multifractal exponents by using the Chalker-Coddington network models [U(1) model for class A and its SU(2) generalization for class C].  For class C, we use two approaches outlined in Sec.~\ref{sec:young}: the direct evaluation of a class-C pure scaling operator and the path going through strictly  positive Young-symmetrized wave-function combinations $|\Psi_\lambda|^2$.  Our numerics fully confirms all existing analytical predictions, including those based on Weyl symmetry as well as those resulting from relations between class-A and class-C scaling observables derived in Sec.~\ref{sec:young}. Obtained numerical results for the exponents at the SQH transition are summarized in Table \ref{tab:lC}. Remarkably, we find strong deviations from generalized parabolicity for the subleading exponents [(1,1), (2,1), (3,1), (2,2)].   In Sec. \ref{sec:wave_iw} these results will be corroborated by an alternative approach (based on obervables involving both spin projections), where also further subleading exponents will be determined numerically. 

While our main focus in this work is on class C, we have also determined numerically all exponents for Young diagrams with $q \le 4$ for the conventional (class-A) IQH transition, see Table \ref{tab:lA} below. At variance with the SQH transition, we find here only weak deviations from the generalized parabolicity. 

\subsubsection{The Chalker-Coddington network models}
An intuitive picture of quantum Hall criticality is imagining the electrons to drift along equipotential lines of the disorder potential. Near saddle points of the potential, tunneling between distinct lines is allowed.  The Chalker-Coddington network model \cite{chalker1988percolation} adds further simplification to this picture: the random geometry derived from extremal points of the potential landscape is replaced by a square network, with electrons acquiring a random U(1) phase when traveling along a link. The SU(2) version of the Chalker-Coddington network model describes spin quantum hall criticality \cite{gruzberg1999exact,kagalovsky1999quantum}.  The network models have been to be very useful for numerical investigation of localization transitions in 2D systems of various symmetry classes, see
the reviews \cite{kramer2005random,evers08}. 

We analyze multifractality of eigenstates of the network scattering matrix $\mathcal{U}$ numerically using the conventional sparse-matrix techniques \cite{klesse1995universal, mirlin2003wavefunction, evers2003multifractality}. For each realization of disorder, we determine four eigenstates with lowest positive energies. (When speaking about the energy in the context of the network model, we actually mean the quasienergy $\epsilon$ corresponding to the eigenvalue $e^{i\epsilon}$ of the network evolution operator $\mathcal{U}$.)
 The systems studied are of linear size $L=128\ldots 1024$, and we perform the ensemble  averaging over at least $10^5$ configurations. For each realization of disorder, we also average over $L^2$ positions of the ``center of mass'' of  the set of involved spatial points $r_i$. 

\subsubsection{Scaling considerations}

To investigate the scaling of an operator with $q\le 4$, we consider eigenstate combinations defined on $q$ lowest-energy eigenstates and at $q$ distinct points $r_1, \ldots, r_q$, see Sec.~\ref{sec:young}.
The exponents of interest control the power-law scaling of the corresponding observables with $L$, see general discussion of scaling in Sec.~\ref{sec:general}. To improve the statistics, we also consider the scaling with the distance $r$. We keep all the distances $|r_i-r_j|$ between the involved $q$ points to be of the same order: for $q=4$ the points form a square, and for $q=3$ a triangle with approximately equal side lengths.
When investigating a combination $\mathcal{O}[\psi]$ of order $q$ that is dominated by a pure-scaling operator $\mathcal{P}_{(q_1,q_2,\ldots)}$, we have (see Sec.~\ref{sec:general})
\begin{align}
L^{2q} \mathcal{O}[\psi] (r_1,r_2,\ldots)&\sim \left(\dfrac{|r_i-r_j|}{L}\right)^{\Delta_{(q_1,q_2, \ldots)} }  \,, \qquad \qquad
\Delta_{(q_1,q_2,\ldots)}  \equiv x_{(q_1,q_2,\ldots)} - qx_\nu \,,
\label{eq:scal}
\end{align}
where $x_\nu\equiv x_{(1)}$ is the scaling dimension of the operator $Q$ that controls the scaling of the density of states. In class A (and all other Wigner-Dyson classes) the density of states is not critical, i.e.,
$x_{(1)} = 0$.  For the SQH transition,  $x_{(1)}^C = \frac14$ is known exactly from mapping to percolation. 
The slope of the log-log plot of  $L^{2q} \mathcal{O}[\psi]$ versus $r/L$ thus yields the exponent $\Delta_{(q_1,q_2, \ldots)}$. 

As discussed in Sec.~\ref{sec:CFT-2D}, if the theory satisfies local conformal invariance, the generalized MF spectrum would obey generalized parabolicity. (The second assumption is the abelian fusion that has been shown explicitly.) For the IQH and SQH transitions, the generalized parabolicity would imply the following form of the dimensions $x_{(q_1,q_2,\ldots)}$ of scaling operators:
\begin{align}
\text{IQH:} & \qquad x^{A,\,\text{para}}_{(q_1,q_2,\ldots)} = b\left[q_1(1-q_1) + q_2(3-q_2)+ q_3 (5-q_3) + \ldots\right],
\label{eq:iqh-gen-parab} \\
\text{SQH:} & \qquad x^{C,\,\text{para}}_{(q_1,q_2,\ldots)} = \frac{1}{8}\left[q_1(3-q_1) + q_2(7-q_2)+ q_3 (11-q_3)+ \ldots\right]. 
\label{eq:xq}
\end{align}
For the SQH transition, the exact values of the exponents $x_{(1)}^C = x_{(2)}^C = \frac14$ imply that, if the generalized parabolicity \eqref{eq:xq} holds, the prefactor should be exactly $1/8$. 
For the IQH transition, none of the exponents is rigorously known analytically, which leaves a freedom in the prefactor $b$.  It was conjectured in a recent paper \cite{zirnbauer2019integer} 
that the IQH critical point is described by a certain model of WZNW type, which would give strict generalized parabolicity with $b=\frac14$.
As already pointed out in Sec.~\ref{sec:introduction}, high-precision numerical studies reveal small but clear deviations from parabolicity of the standard MF spectrum $x_{(q)}$, both for the IQH  \cite{Obuse-Boundary-2008, evers2008multifractality} and the SQH \cite{mirlin2003wavefunction, puschmann2021quartic} transitions. For the IQH transition, if one looks for the best parabolic approximation to
the spectrum $x_{(q)}$, one finds $b$ in the range 0.26 -- 0.27, depending on the range of $q$ where the fit is performed.

\subsubsection{Numerical results}
\label{sec:numerical_results}

\begin{figure}
	\centering
	\includegraphics[width = .46\textwidth]{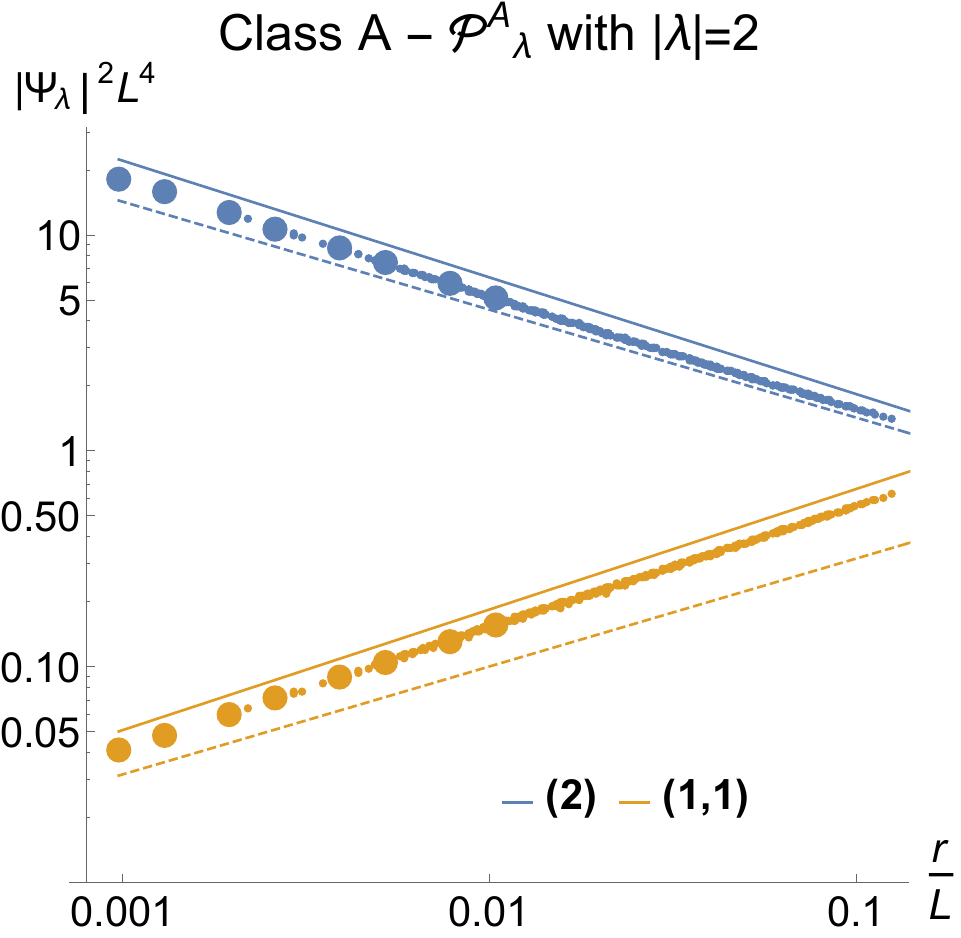}
	\includegraphics[width = .46\textwidth]{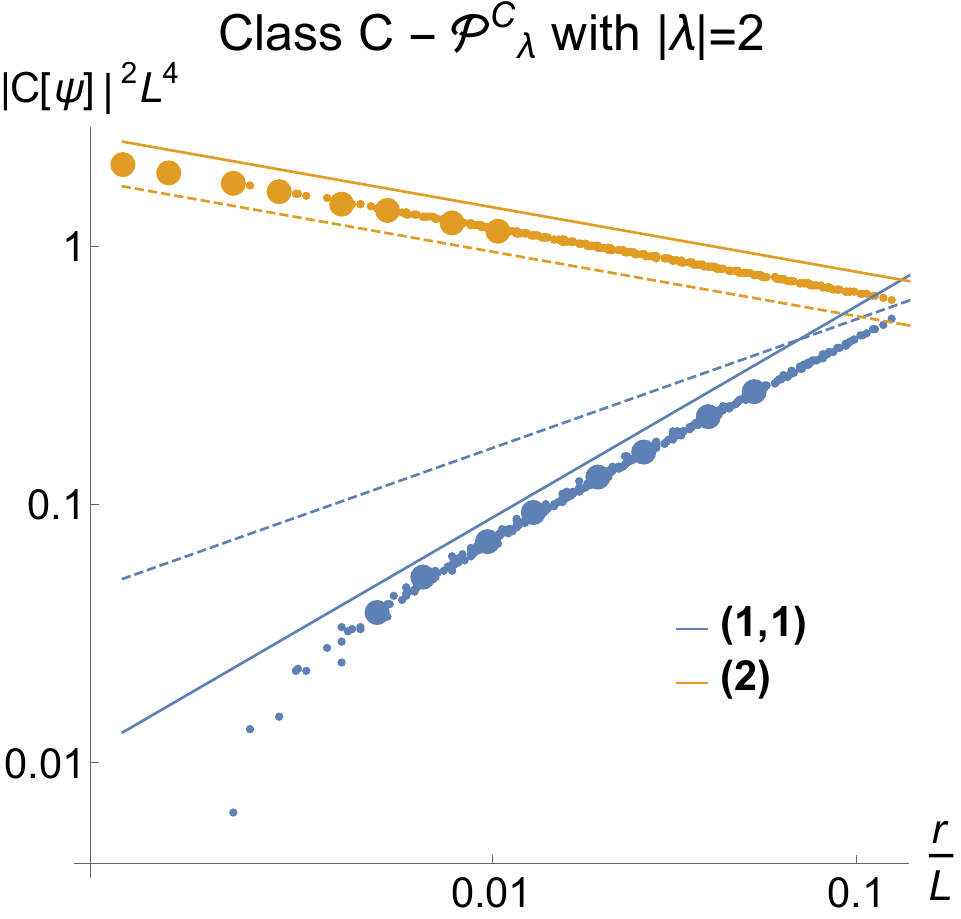}
	\includegraphics[width = .46\textwidth]{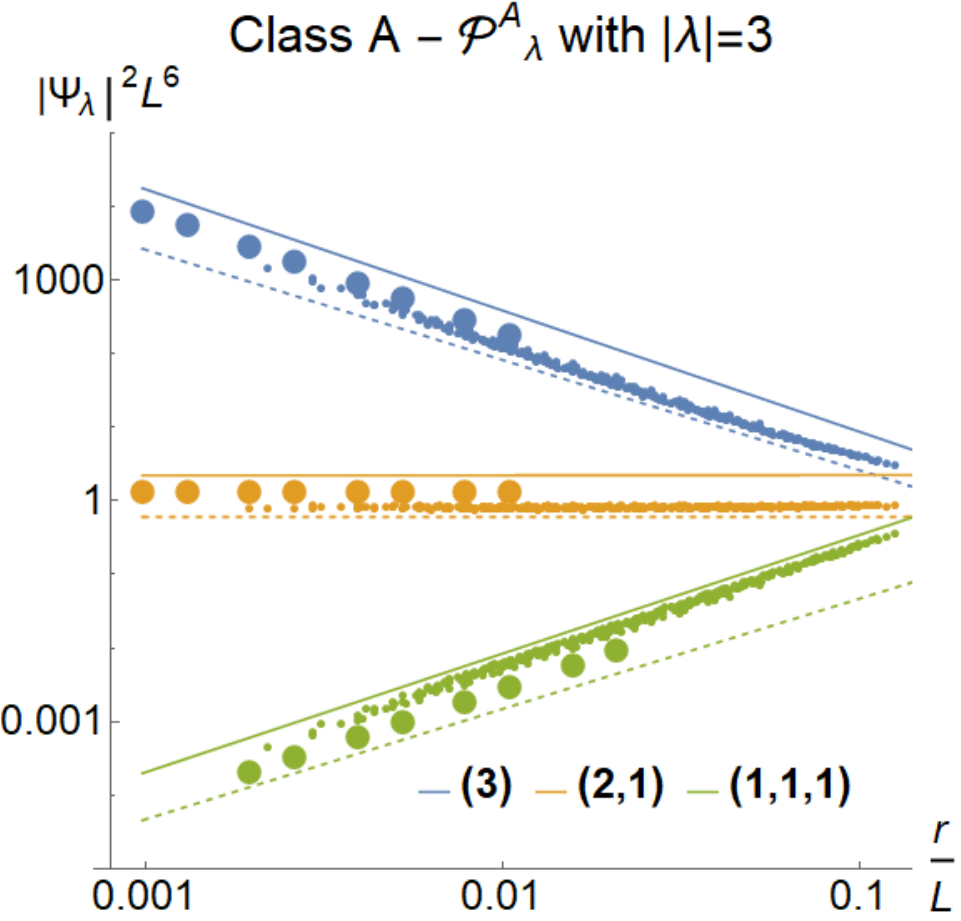}
	\includegraphics[width = .46\textwidth]{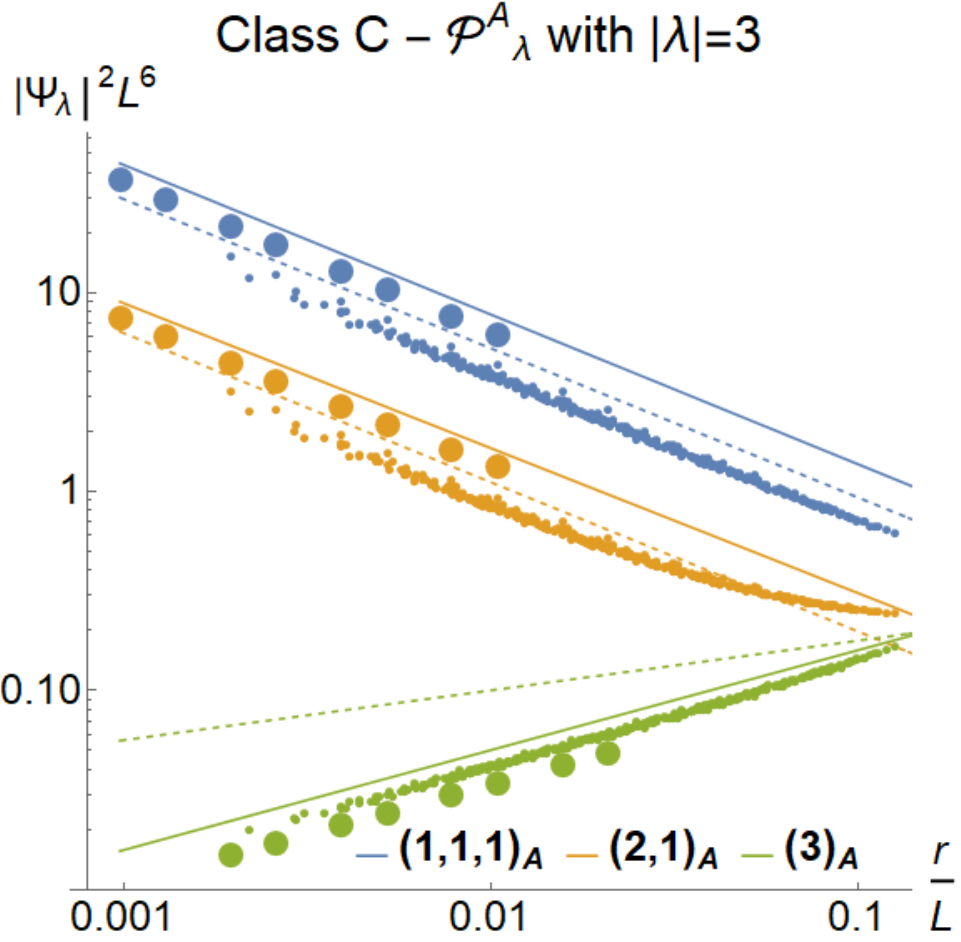}
	\includegraphics[width = .46\textwidth]{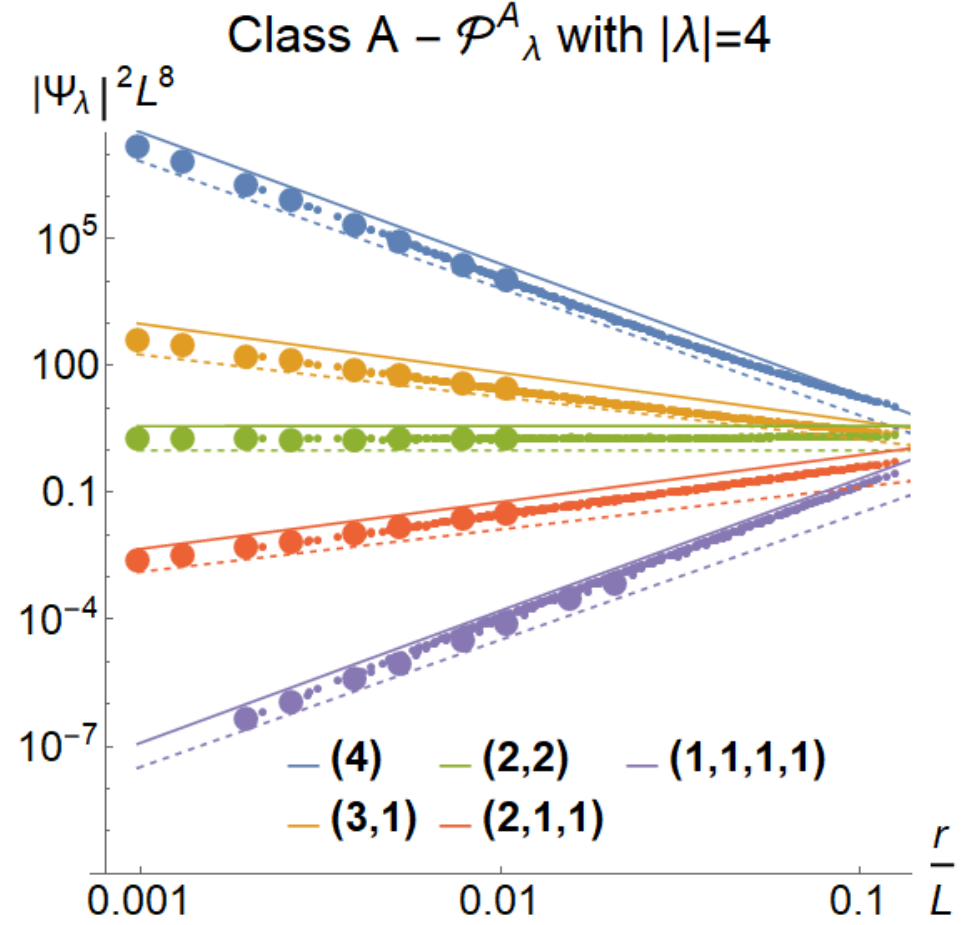}
	\includegraphics[width = .46\textwidth]{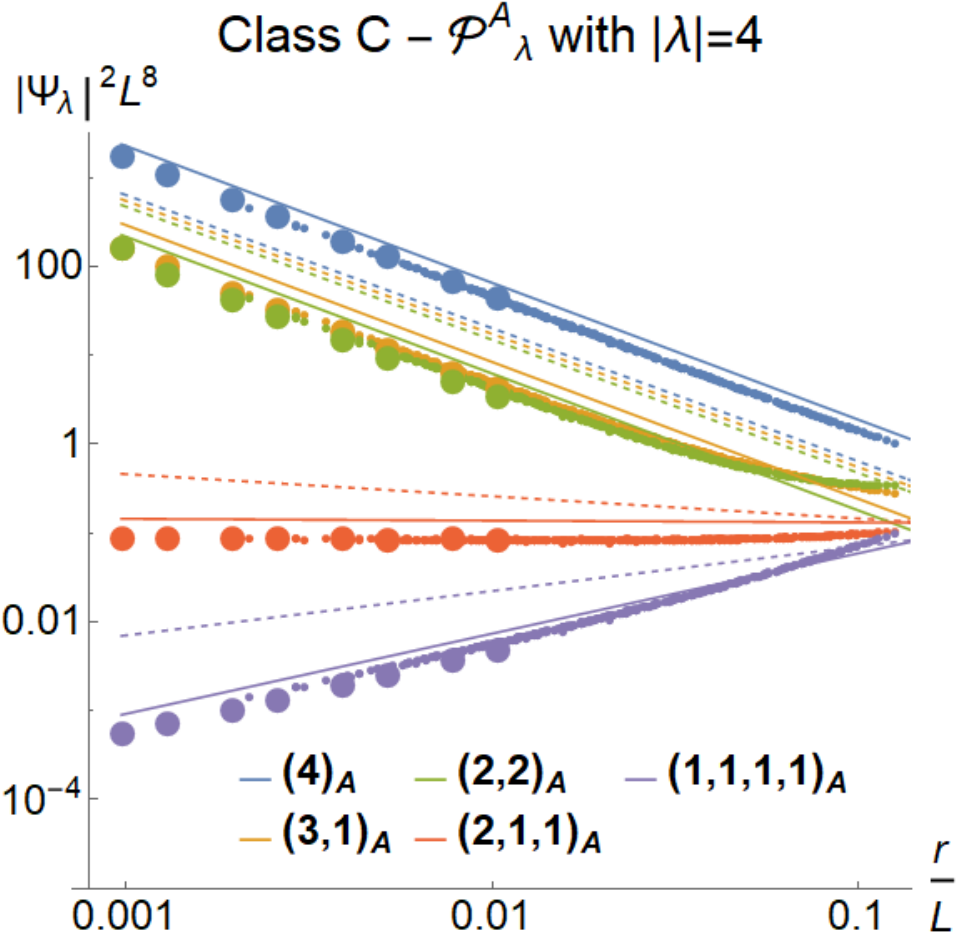}
	\caption{Numerical determination of generalized parabolicity at IQH (left) and SQH (right) transitions, for $q=2$ (top), $q=3$ (middle), and $q=4$ (bottom) eigenstate observables. Data result from simulations on class-A (left) and class-C (right) network models, with averaging over $10^5$ realizations of disorder. Left panels show Young-symmetrized eigenstate combinations $L^{2q} |\Psi_\lambda|^2$, which correspond to pure-scaling operators $\mathcal{P}^A_\lambda[Q]$ of class A, collapsed as functions of $r/L$. Full lines are fits to the data; the corresponding exponents $\Delta^{\rm num}_\lambda$ are given in Table \ref{tab:lA}. Dashed lines correspond to the generalized parabolic spectrum \eqref{eq:iqh-gen-parab} with $b=1/4$, as would follow from the WZNW model proposed in Ref.~\cite{zirnbauer2019integer}; the corresponding values are given in Table \ref{tab:lA} as $\Delta^{{\rm para}, \, b=1/4}_\lambda$.  For each $\lambda$, data points for one fixed value of $r$ of order unity are highlighted as bold dots, in order to visualize the $L$-dependence at a fixed $r$. Upper right panel shows the class-C pure-scaling observables \eqref{eq:H-minus-2F} with $q=2$. The middle and bottom right panels display the scaling of Young-symmetrized combinations $L^{2q} |\Psi_\lambda|^2$ (which correspond to $\mathcal{P}^A_\lambda[Q]$) with $q=3$ and 4 in a class-C network model. 
They allow to access the class-C exponents for the representations (3), (2,1), (4), (3,1), and (2,2). The extracted exponents $\Delta^{\rm num}_\lambda$ (full lines) are collected in Table \ref{tab:lC}. The dashed lines in the right panels correspond to generalized parabolicity \eqref{eq:xq}; the corresponding exponents $\Delta^{\rm para}_\lambda$ are also listed in Table \ref{tab:lC}. A strong violation of the generalized parabolicity at the SQH transition is evident.}
\label{fig:AC}
\end{figure}

In Fig. \ref{fig:AC} we show numerical results for the scaling of $L^{2q} |\Psi_\lambda|^2$ with $r/L$, where $\Psi_\lambda$ are Young-symmetrized eigenstate combinations as defined in Sec.~\ref{sec:young}. For class C, the spin projection is fixed, $\psi\equiv\psi_\uparrow$, for all combinations considered here. In both the class A and the class C, these combinations map to the operators $\mathcal{P}^A_\lambda[Q]$ of the sigma-model. 

In the case of IQH transition (class A),  $\mathcal{P}^A_\lambda[Q]$ are pure-scaling operators. The corresponding results for $L^{2q} |\Psi_\lambda|^2$ are shown in the left panels of Fig. \ref{fig:AC}  for the operators with $q=2$ (top), $q=3$ (middle), and $q=4$ (bottom). The numerical results provide an excellent confirmation of the prediction that $L^{2q} |\Psi_\lambda|^2$ exhibit scaling corresponding to the representation $\lambda$: we observe nice fans representing distinct scaling exponents. The fitted slopes are shown by full lines; the corresponding values of numerically determined exponents
$\Delta_{\lambda}^{A, \rm num}$  are collected in Table \ref{tab:lA}.   For those IQH exponents that have been found numerically previously, our results are in full agreement with previous data. In particular, our value $\Delta_{(2)}^{A, \rm num}\approx - 0.54$ is in excellent agreement with previous values \cite{evers2001multifractality,Obuse-Boundary-2008, evers2008multifractality}, which are in the range between $- 0.54$ and $- 0.55$ with error bars within $\pm 0.01$. (In Refs.~\cite{Obuse-Boundary-2008} and \cite{evers2008multifractality}, the values of $\Delta_{(q)}^{A, \rm num} / q(1-q)$ were shown only up to $q=1.5$ and $q=1.75$, respectively, so that a small  extrapolation to $q=2$ is needed.) For $\lambda=(1,1)$ we find $\Delta_{(1,1)}^{A, \rm num}\approx 0.57$, which is slightly below the value  $\Delta_{(1,1)}^{A, \rm num}\approx 0.62 \pm 0.05$ of Ref. \cite{burmistrov2011wave} but agrees within the error bars given there. The exponents corresponding to $\lambda=(2,1)$ and (2,2) represent a very useful test for the numerics, since they are exactly zero by virtue of the Weyl symmetry. Thus, their obtained numerical values provide an estimate for numerical errors associated with finite-size effects and finite statistics. For $\lambda=(2,1)$ we get $\Delta_{(2,1)}^{A, \rm num}\approx 0.01$, indicating an excellent accuracy of the numerics, $\pm 0.01$, for $q \le 3$.  For $\lambda=(2,2)$ we get a somewhat larger deviation from the exact value, $\Delta_{(2,2)}^{A, \rm num}\approx 0.04$, which is not surprising, since errors increase with increasing $q$.  We see, however, that the numerics remains very good also for $q=4$. 

When compared to the conjecture of Ref.~\cite{zirnbauer2019integer} (exact generalized parabolicity with $b=1/4$),
our results for the IQH transition are fully consistent with those of Refs.~\cite{Obuse-Boundary-2008, evers2008multifractality}: the exponents deviate from the spectrum proposed in Ref.~\cite{zirnbauer2019integer}  by $\sim 10\%$.  At the same time, our numerical values of the exponents are rather close to those that would follow from generalized parabolicity with $b \approx 0.27$. 
Deviations from parabolicity were observed in Refs.~\cite{Obuse-Boundary-2008, evers2008multifractality} where the MF spectrum $\Delta_{(q)}$ was studied with a high accuracy for fractional $q$ in the range from $q=-0.5$ to $q=1.75$. This is favorable for detecting relatively small deviations, since the numerical accuracy is especially high in this range of $q$. In principle, this analysis can be extended also to generalized MF exponents with continuously changing $q$.  Since our main focus in this paper is on generalized multifractality at SQH transition, we leave further studies of the IQH transition to later work. 

In the right panels of Fig. \ref{fig:AC} we show the results of numerical simulations for a network model of class C. In the top panel, the scaling of eigenfunction combinations corresponding to class-C eigenoperators,  $H+F \sim \mathcal{P}^C_{(2)}$ and $H-2F \sim \mathcal{P}^C_{(1,1)}$, is shown, see Eq.~\eqref{eq:H-minus-2F} and the first line of Eq.~\eqref{eq:mixing}. For the subleading operator $\mathcal{P}^C_{(1,1)}$, we observe stronger fluctuations and deviations from scaling at large $L$.  As explained above, they are related to the fact that the corresponding eigenstate combination $H-2F$ is sign-indefinite, and, as a result, the averaging becomes insufficient for large $L$.  Since these difficulties become more severe with increasing $q$, we use for $q=3$ (middle panel) and $q=4$ (bottom panel) the alternative approach outlined in Sec.~\ref{sec:young}. Specifically, we plot the $r/L$ dependence of strictly positive Young-symmetrized combinations $L^{2q} |\Psi_\lambda|^2$.
According to Eq.~\eqref{eq:mixing2}, this allows us to access the exponents corresponding to representations (3) and (2,1) for $q=3$ as well (4), (3,1), and (2,2) for $q=4$, see the discussion below Eq.~\eqref{eq:mixing2}. 

The data shown in Fig. \ref{fig:AC} fully confirm the analytical predictions. The combinations  $|\Psi_{(3)}|^2$  and $|\Psi_{(2,1)}|^2$   for $q=3$, as well as 
$|\Psi_{(4)}|^2$, $|\Psi_{(3,1)}|^2$,  and $|\Psi_{(2,2)}|^2$  for $q=4$ indeed show the leading scaling. At the same time, the combinations
$|\Psi_{(1,1,1)}|^2$, $|\Psi_{(2,1,1)}|^2$  and $|\Psi_{(1,1,1,1)}|^2$  exhibit the subleading scaling, yielding the exponents $\Delta^C_{(2,1)}$,    $\Delta^C_{(3,1)}$, and $\Delta^C_{(2,2)}$,  respectively. 

The values of the class-C exponents are shown in Table \ref{tab:lC}.  We recall that the eigenfunction exponents $\Delta^C_\lambda$ are related to the field-theory exponents $x^C_\lambda$ via $\Delta^C_\lambda = x^C_\lambda - qx^C_{(1)}$, with $x^C_{(1)} = 1/4$ for the SQH transition. We see that the numerical data very well respect all analytically known exact values and relations (from the mapping to percolation and from Weyl symmetry). These include $x^C_{(2)} = 1/4$, $x^C_{(3)} = 0$, and $x^C_{(1,1)} = x^C_{(2,1)}$. Some deviation between the numerically found $x^C_{(1,1)}$ and $x^C_{(2,1)}$ is due to a somewhat larger error in $x^C_{(1,1)}$ related to strong fluctuations of the $H-2F$ combination, as explained above. In Sec.~\ref{sec:wave_iw} we will present an alternative (and more accurate) numerical calculation of $x^C_{(1,1)}$, which perfectly fulfils the identity $x^C_{(1,1)} = x^C_{(2,1)}$.

The central observation from the obtained values of the exponents  is a very strong violation of generalized parabolicity by the subleading exponents  [representations (1,1), (2,1), (3,1), and (2,2)].  
This will be fully confirmed and reinforced in Sec.~\ref{sec:wave_iw} where these and further exponents will be obtained by a complementary approach. The deviations of numerical values $\Delta^{\rm num}_\lambda$  from the values $\Delta^{\rm para}_\lambda$ that would follow from generalized parabolicity (also shown in the table) are very big---of the order of $\Delta^{\rm para}_\lambda$. These deviations are more than an order of magnitude larger than numerical uncertainties (that can be estimated from the accuracy with which analytically known exponents and Weyl-symmetry relations are reproduced). This strong violation of generalized parabolicity in class C represents one of central results of this paper. As was explained in Sec.~\ref{sec:gen-parab-hallmark-conf-inv}, it has a very fundamental implication:  a violation of local conformal invariance at the SQH transition.

\begin{table}
	\centering 
	\begin{tabular}{c|cc|c}
		& rep. $\lambda$  & $\Delta_\lambda^{\rm num}$ & $\Delta_\lambda^{{\rm para},\, b=1/4}$\\[5pt]
		\hline 
		\hline 
		&&&\\[-5pt]
		$q=2$ & (2) & $-0.54$ & $-\frac12$ \\[3pt]
		& (1,1) & $0.57$ & $\frac12$
		\\[5pt]
		\hline 
		&&&\\[-5pt]
		$q=3$ & (3) & $-1.66$ & -$\frac32$ \\[3pt]
		& (2,1) & $0.01$ & {\bf 0}\\[3pt]
		& (1,1,1) & $1.61$ & $\frac32$
		\\[5pt]
		\hline 
		&&&\\[-5pt]
		$q=4$ & (4) & $-3.12$ & -3 \\[3pt]
		& (3,1) & $-1.10$ & -1\\[3pt]
		& (2,2) & $0.04$ & {\bf 0}\\[3pt]
		& (2,1,1) & $1.10$ & 1\\[3pt]
		& (1,1,1,1) & $3.12$ & 3
	\end{tabular}
	\caption{Scaling exponents of generalized multifractality at the IQH transition (class A) for eigenstate observables from representations $\lambda = (q_1,q_2, \ldots)$ with $q \equiv q_1 + q_2 + \ldots = 2$, 3, and 4. The exponents $\Delta_\lambda^{\rm num}$ are determined numerically as shown in the left column of Fig.~\ref{fig:AC}. For comparison, we also present the exponents $\Delta^{{\rm para}, \, b=1/4}_\lambda$ corresponding to the generalized parabolicity \eqref{eq:iqh-gen-parab} with $b=1/4$, as would result from the WZNW theory conjectured in Ref.~\cite{zirnbauer2019integer}. The values $\Delta_{(2,1)} = \Delta_{(2,2)} = 0$ highlighted by boldface are exact (enforced by Weyl symmetry; independent of presence or absence of generalized parabolicity). }
	\label{tab:lA}
\end{table}

\begin{table}
	\centering 
	\begin{tabular}{c|ccc|c}
		& rep. $\lambda$ & $\Delta_\lambda^{\rm num}$ & $\Delta_\lambda^{\rm num,|\psi|}$ & $\Delta^{\rm para}_\lambda $\\[5pt]
		\hline 
		\hline 
		&&&&\\[-5pt]
		$q=2$ & (2) & $-0.25$ & $-0.25$ & $-\mathbf{\frac14}$\\[3pt]
		& (1,1) &  $0.79$ & $0.74$         & $\frac12$
		\\[5pt]
		\hline 
		&&&&\\[-5pt]
		$q=3$ & (3) & $-0.75$ & $-0.75$ & $-\mathbf{\frac{3}{4}}$\\[3pt]
		& (2,1) & $0.50$ & $0.49$ & $\frac{1}{4}$\\[3pt]
		& (1,1,1) & --- & $1.17$ & $\frac{3}{2}$
		\\[5pt]
		\hline 
		&&&&\\[-5pt]
		$q=4$ & (4) & $-1.5$ & $-1.5$  & -$\frac{3}{2}$\\[3pt]
		& (3,1) & $-0.01$ & $-0.02$ & -$\frac14$\\[3pt]
		& (2,2) & $0.91$ & $0.86$ &  $\frac{1}{2}$\\[3pt]
		& (2,1,1) & --- & --- &  $\frac{5}{4}$\\[3pt]
		& (1,1,1,1) & --- & --- & $3$
	\end{tabular}
	\caption{Scaling exponents of generalized multifractality at the SQH transition (class C) for eigenstate observables with $q\equiv |\lambda|\leq 4$. 
The exponents $\Delta_\lambda$ shown in the table are related to the field-theoretical exponents $x_\lambda$ via $\Delta_\lambda = x_\lambda - qx_{(1)}$  with $x_{(1)}= 1/4$. 	
The exponents $\Delta_\lambda^{\rm num}$ are determined numerically by using eigenstate combinations with a single spin projection, see right panels of Fig.~\ref{fig:AC}.
The exponents $\Delta_\lambda^{\rm num,|\psi|}$ are obtained by a complementary numerical approach using observables involving the total density $|\psi|$ (spin up and spin down), Fig.~\ref{fig:CS}. 
The symbol ``---'' means that the exponent was not determined by the corresponding approach.
The agreement between both sets of numerical exponents is very good. Some deviation  between two numerical values of the (1,1) exponent is attributed mainly to the error in the value $\Delta_{(1,1)}^{\rm num}$ because of strong fluctuations of the corresponding sign-indefinite combination $H-2F$.
The Weyl symmetry	implies exact relations $\Delta_{(2,1)} = \Delta_{(1,1)} - 1/4$ and $\Delta_{(2,1,1)} = \Delta_{(1,1,1)} - 1/4$; the first of them is nicely fulfilled by the numerically found exponents. 
The last column displays the exponents $\Delta_\lambda^{\rm para}$ corresponding to the generalized parabolic spectrum, Eq.~\eqref{eq:xq}. 
The values $\Delta_{(2)} = - 1/4$ and $\Delta_{(3)} = - 3/4$ shown in boldface are exact  and thus independent of the status of generalized parabolicity. 	
Strong deviations from the generalized parabolicity are clearly seen in all subleading exponents that have been determined numerically.}
\label{tab:lC}
\end{table}

\section{From $\sigma$-model to wave functions:  Total density observables}
\label{sec:wave_iw}
In Sec.~\ref{sec:wave_ops}, we developed two approaches to numerical determination of the exponents characterizing the generalized multifractality in class C. First, we have translated exact pure-scaling operators (obtained earlier by RG) to eigenstate expressions. The difficulty with direct application of this approach is that the corresponding subleading eigenstate combinations are sign-indefinite and fluctuate very strongly. We thus used it directly only for the (1,1) operator that has the form $H-2F$, see upper right panel of Fig.~\ref{fig:AC}. Even in this case of the simplest subleading observable, fluctuations are very strong and require very extensive numerical efforts. In view of this, we have used for determining other exponents a complementary approach based on studying strictly-positive Young-symmetrized combinations corresponding to pure-scaling operators of class A. However, this way does not allow us to access exponents that are on the more RG-irrelevant side of the spectrum, including (1,1).  In this section, we develop a third approach that, one one hand, deals with strictly positive observables and, on the other hand, allows us to determine numerically the (1,1) exponent and various other subleading exponents not accessible by the class-A-to-class-C approach of Sec.~\ref{sec:wave_ops}. The central idea of the  approach 
developed in this section is to exploit observables built out of $|\psi|=\sqrt{|\psi_\uparrow|^2+|\psi_\downarrow|^2}$ and thus involving both spin projections at each of the relevant spatial points $r_i$. Below we give a physical motivation of this approach, justify it by establishing a connection with sigma-model pure-scaling operators obtained by the Iwasawa decomposition, and demonstrate numerically its efficiency.
For those exponents that are determined numerically by two approaches (i.e., both here and in Sec.~\ref{sec:wave_ops}), we find a very good agreement. 

\subsection{Physical motivation}
\label{sec:app_iw}

In class C, the wave function  $\psi=(\psi_\uparrow, \psi_\downarrow)$ is an intrinsically two-component object: there is the combined spin and particle-hole space. In Sec. \ref{sec:wave_ops}, we used only one spin component (say, $\psi_\uparrow$) to construct pure-scaling observables. Here we will use the total density
\begin{align}
|\psi|\equiv\sqrt{|\psi_\uparrow|^2+|\psi_\downarrow|^2},
\label{eq:total-density}
\end{align} 
where both components at the same spatial point enter simultaneously. 
[Everywhere below in Sec.~\ref{sec:wave_iw} the notation $|\psi|$ has the meaning defined by Eq.~\eqref{eq:total-density}.]
To demonstrate why this is useful, let us first consider in detail observables that are of fourth order with respect to wave functions (which corresponds to $q=2$ in our classification).  There are two distinct representations $(2)$ and $(1,1)$, and we can construct operators in each of them as linear combinations of Hartree ($H$) and Fock ($F$) terms \eqref{O-H-F}.
In class A the pure-scaling observables are (here $\psi$ abbreviates the two wave functions involved)
\be
C_{(2)}^A[\psi] = H[\psi]+F[\psi] \,, \qquad \qquad C_{(1,1)}^A[\psi] = H[\psi]-F[\psi]\,,
\ee
 whereas in class C the pure-scaling combinations involving only $\psi_\uparrow$ are
 \be
 C_{(2)}^C[\psi_\uparrow] = H[\psi_\uparrow]+F[\psi_\uparrow]\,, \qquad \qquad C_{(1,1)}^C[\psi_\uparrow] = H[\psi_\uparrow]-2F[\psi_\uparrow]\,.
 \ee
 The only difference here is in the coefficient 2 in the subleading operator: $H-2F$ in class C instead of $H-F$ in class $A$. This seemingly small difference, is, however, connected in a profound way with properties of critical eigenstates. 
 
\begin{figure}
	\centering
	\includegraphics[width = .48\textwidth]{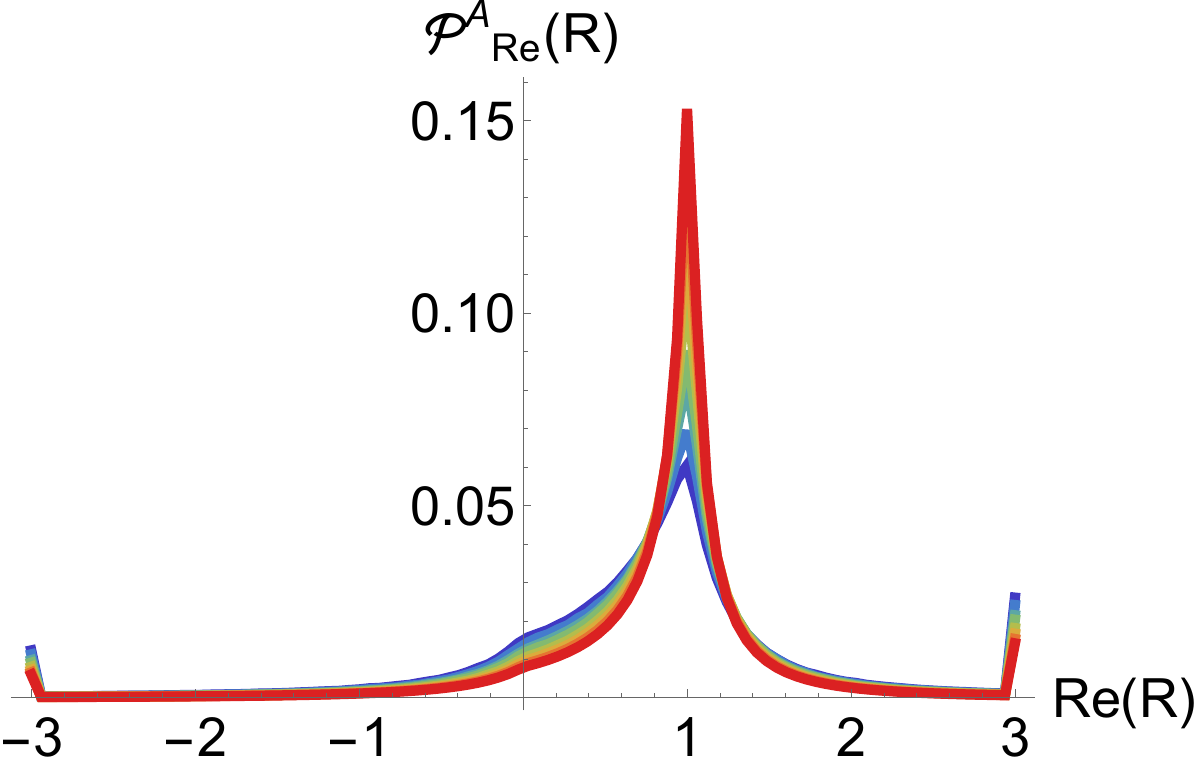}
	\includegraphics[width = .48\textwidth]{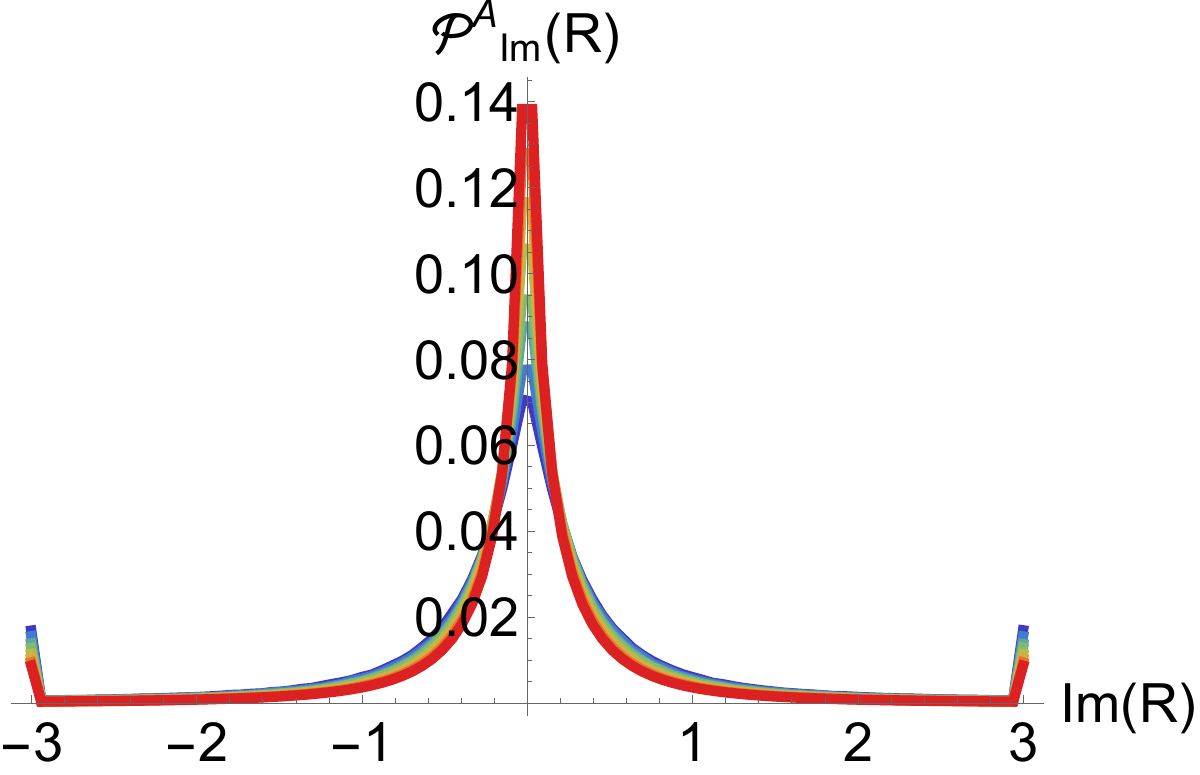}
	\caption{Distributions of the real part $\Re R$ (left) and the imaginary part $\Im R$ (right) of the ratio $R$, Eq.~\eqref{eq:ratio-R}, characterizing the ``local similarity'' of two eigenstates at the IQH transition (class A). Different colors correspond to different system sizes, varying from from $L=96$ (blue) to $L=1024$ (red).  The peak at $R=1$ becomes sharper with increasing $L$, implying that $R \to 1$  at $L\rightarrow \infty$ with probability unity. This visualizes the suppression of the pure-scaling subleading combination $H-F$ with respect to $H+F$ for any realization of disorder.
	}
	\label{fig:PA}
\end{figure}

Let us introduce the ratio $R$ characterizing the local similarity of the two eigenstates involved:
\be
R =  \dfrac{\psi_1(r_2) \psi_2(r_1)} { \psi_1(r_1) \psi_2(r_2)} \,.
\label{eq:ratio-R}
\ee
Note that the U(1) freedom in the definition of the eigenfunctions does not affect $R$. It is easy to see that the ratio of the Fock and Hartree terms is expressed through $R$ as follows:
\be
\dfrac{F[\psi]}{H[\psi]} =  \dfrac{R+R^*}{1+RR^*} \,.
\label{eq:ratio-HF}
\ee
Clearly, the ratio $F/H$ satisfies $-1 \le F/H \le 1$ and reaches its maximum value unity only for $R=1$. Thus, the class-A subleading observable $H-F$ satisfies $H-F \ge 0$ and is equal zero only for $R=1$.  Further, the average value $\langle H-F \rangle$ is suppressed by a power of $L$ in comparison with $\langle H+F \rangle$. It follows that the ratio $F/H$ should be nearly unity, and thus $R$ should be parametrically close to unity, in any disorder realization in class A. This is very well seen in Fig.~\ref{fig:PA}, which shows how the distribution of $R$ at the IQH transition point evolves with increasing $L$. The distribution tends, in the limit $L \to \infty$, to a delta-function at $R=1$. 

\begin{figure}
	\centering
	\includegraphics[width = .48\textwidth]{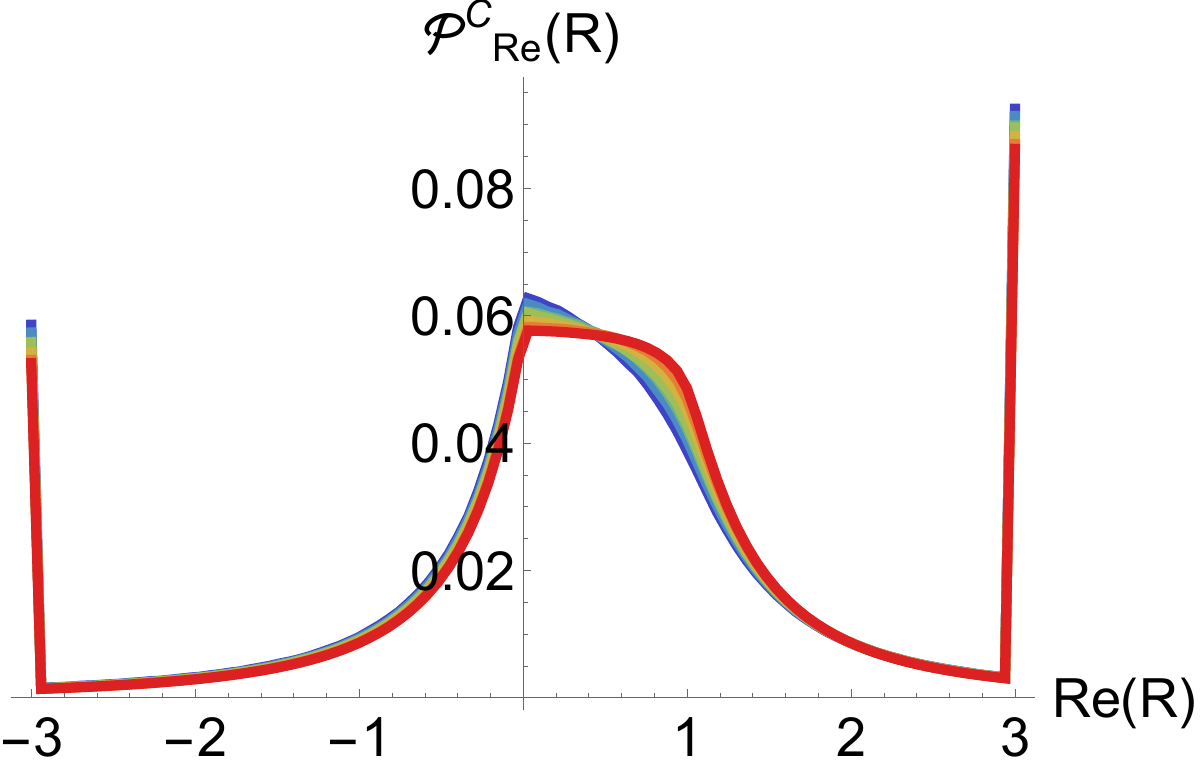}
	\includegraphics[width = .48\textwidth]{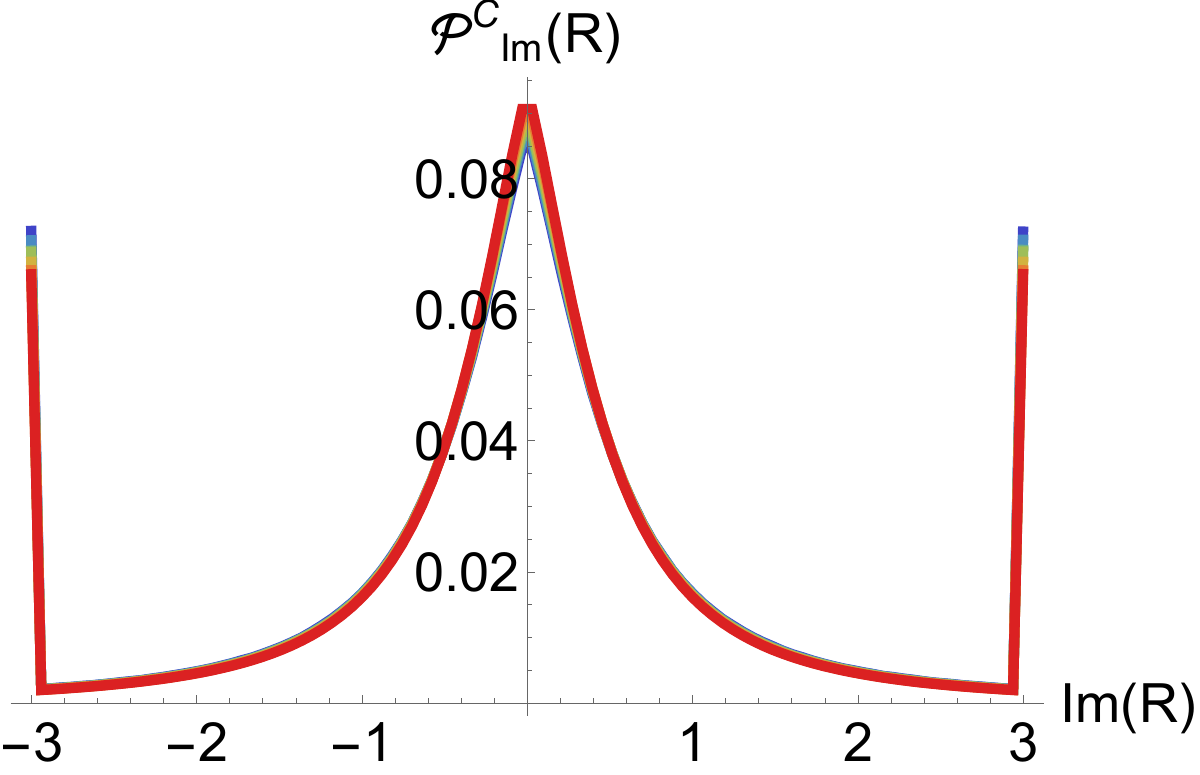}
	
	\includegraphics[width = .48\textwidth]{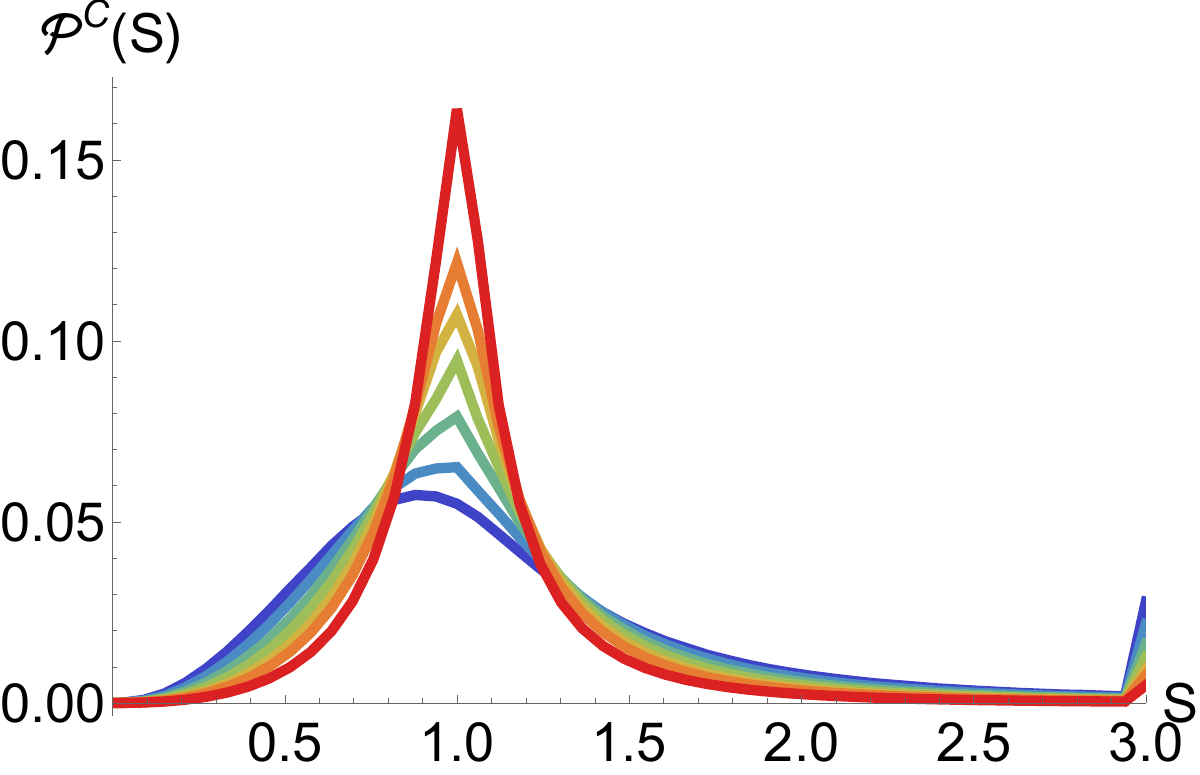}
	\caption{{\it Top:} Distributions of the real part $\Re R$ (left) and the imaginary part $\Im R$ (right) of the ratio $R$, Eq.~\eqref{eq:ratio-R}, at the SQH transition (class C).  The ratio $R$ is built in this case out of eigenfunctions components with fixed spin projection (all spins up). Different colors correspond to different system sizes, varying from from $L=96$ (blue) to $L=1024$ (red). It is seen that the distribution remains broad (with width of order unity) at $L \to \infty$. This illustrates that, while $H-2F$ is a subleading combination in average, it is typically of the same order as the leading combination $H+F$. 
	{\it Bottom:} Distribution of the ratio $S$, Eq.~\eqref{eq:ratio-S}, characterizing local similarity of total densities (spin up and down) of two eigenstates.    The peak at $S=1$ becomes sharper with increasing $L$, in close similarity to the behavior of the distribution of $R$ in class A, Fig.~\ref{fig:PA}.  This implies that the combination  $H[|\psi|] - F[|\psi|]$  does not contain the leading operator [Young diagram (2)] and scales according to the subleading one [Young diagram (1,1)], see  top left panel of Fig.~\ref{fig:CS}.  }
	\label{fig:PC}
\end{figure}

The situation in class C is very different. Here $\langle H[\psi_\uparrow]\rangle_V = 2\langle F[\psi_\uparrow] \rangle_V$ in the limit $L \to \infty$. Since the value $1/2$ is located in the middle of allowed values of $F/H$, the ratio $\langle F \rangle / \langle H \rangle = 1/2$ does not by itself tell anything about the values of $F/H$ in individual realizations of disorder. As shown in Fig. \ref{fig:PC}, the distribution of the ratio $R$  built out of $\psi_\uparrow$ remains broad (with width of order unity) in the large-$L$ limit. Consequently, the distribution of $F/H$ is broad as well. This is exactly the property that we have already emphasized several times: while 
$H[\psi_\uparrow] - 2 F[\psi_\uparrow] $ is parametrically small after disorder averaging, it is of the same order as 
$H[\psi_\uparrow] + F[\psi_\uparrow] $ in a typical disorder realization.

We thus face the following observation: while two adjacent-in-energy critical eigenstates are nearly identical locally in class A ($R \approx 1$), this does not hold for class C. At the same time, it is natural to expect that some form of a strong local similarity of eigenstates is a general property of Anderson transitions. A qualitative difference between classes A and C is the existence of the spin degree of freedom in class C. These considerations suggest to look at local correlations between total densities \eqref{eq:total-density} of two eigenstates. 
We define the corresponding ratio $S$ (which is real and positive):
\be
S = \dfrac{|\psi_1(r_2)| \, |\psi_2(r_1)|} { |\psi_1(r_1)| \, |\psi_2(r_2)|} \,,
\label{eq:ratio-S}
\ee
with $|\psi|$ as defined by Eq.~\eqref{eq:total-density}.
The distribution of $S$ at the SQH critical point is shown in the bottom panel of Fig.~\ref{fig:PC}. Remarkably, we see the same behavior as for the distribution of $R$ in class A: evolution, in the large-$L$ limit, towards the delta-distribution at $S=1$. Thus, total densities $|\psi_1|$ and $|\psi_2|$ are indeed strongly correlated locally. Broad fluctuations of the ratio $R$ (characterizing one spin component) are therefore related to relative rotations in the spin space between two eigenfunctions. 

Since $S$ in class C behaves in the same way as $R$ in class A, it is natural to build Hartree and Fock combinations from total densities.
In full analogy with Eq.~\eqref{eq:ratio-HF},
\begin{align}
\dfrac{F[|\psi|]}{H[|\psi|]} &=  \dfrac{2S}{1+S^2} \,.
\label{eq:ratio-SHF}
\end{align}
In the large-$L$ limit, we have $S \to 1$, and the difference $H[|\psi|] -F [|\psi|]$ yields the subleading combination (1,1).  This was exactly our goal: we have determined the combination that is expected to scale as (1,1) and is strictly positive in any realization of disorder.

Below we demonstrate numerically that this approach indeed works very well, extend it to other subleading combinations, and support it by an analytical derivation connecting $|\psi|$ observables with the sigma-model operators.

\subsection{Young-symmetrized combinations of $|\psi|$: Relation to the sigma model}
\label{sec:sigma_iw}

As motivated above, we consider squared Young-symmetrized combinations $\left|\Psi_\lambda\left[|\psi|\right]\right|^2$ analogous to  \eqref{eq:waves} but constructed on absolute values $|\psi|$ of eigenstates that involve both spin projections, Eq.~\eqref{eq:total-density}. 
For the lowest subleading operator (1,1), this combination is 
\be
\left|\Psi_{(1,1)}\left[|\psi|\right]\right|^2 = \left (    |\psi_a(r_1)| \, |\psi_b(r_2)| -|\psi_b(r_1)| \, |\psi_a(r_2)| \right) ^2 \,.
\label{Psi-mod-psi-11}
\ee
In order to rigorously prove analytically that this combination yields the (1,1) scaling, we should map it onto a sigma-model composite operator. While this is in principle possible, the mapping is technically cumbersome in view of the square involved in the definition of $|\psi|$. We thus restrict ourselves to showing how this derivation works in a closely related but technically simpler case.

Consider the Young-symmetrized expression built on $|\psi|$ and corresponding to the Young diagram (3,1):
\begin{align}
\left|\Psi_{(3,1)}\left[|\psi|\right]\right|^2 &= \left (  |\psi_a(r_1)|^2 |\psi_b(r_2)|^2 - |\psi_b(r_1)|^2 |\psi_a(r_2)|^2 \right ) ^2\nonumber\\
&=\left(|\psi_a(r_1)|^2\right)^2\left(|\psi_b(r_2)|^2\right)^2 +
\left(|\psi_b(r_1)|^2\right)^2\left(|\psi_a(r_2)|^2\right)^2 \nonumber \\
& -2|\psi_a(r_1)|^2 |\psi_b(r_1)|^2 |\psi_a(r_2)|^2 |\psi_b(r_2)|^2.
\label{Psi-mod-psi-31} 
\end{align}
Note that it can be written as 
\bea
\left|\Psi_{(3,1)}\left[|\psi|\right]\right|^2 &=& \left|\Psi_{(1,1)}\left[|\psi|\right]\right|^2  \left|\Psi_{(2)}\left[|\psi|\right]\right|^2 \nonumber \\
& = & 
\left (    |\psi_a(r_1)| \, |\psi_b(r_2)| -|\psi_b(r_1)| \, |\psi_a(r_2)| \right) ^2 \left (  |\psi_a(r_1)| \, |\psi_b(r_2)| + |\psi_b(r_1)| \, |\psi_a(r_2)| \right) ^2  \,,
\nonumber \\
\eea
i.e., it can be intuitively viewed as a result of fusion of \eqref{Psi-mod-psi-11} with $\left|\Psi_{(2)}\left[|\psi|\right]\right|^2$ that has the leading scaling corresponding to representation (2).  By construction,  the combination \eqref{Psi-mod-psi-31}  is manifestly positive and free of square roots. We can thus translate it to the sigma-model language by using the rules derived above.

First, we perform the translation to vector field variables $S$   by using Eq. \eqref{eq:psiS}: \\
$ \left\langle \left|\Psi_{(3,1)}\left[|\psi|\right]\right|^2 \right\rangle_V  \sim   \left\langle \left\langle C_{(3,1)}[S]\right\rangle_{S_0}\right\rangle_V $, where
\begin{align}
C_{(3,1)}[S] &= \left(|S_a|^2(r_1)\right)^2\left(|S_b|^2(r_2)\right)^2+\left(|S_b|^2(r_1)\right)^2\left(|S_a|^2(r_2)\right)^2-2|S_a|^2(r_1)|S_b|^2(r_1)|S_b|^2(r_2)|S_a|^2(r_2).
\end{align}
Here $|S|^2 \equiv |S_\uparrow|^2+|S_\downarrow|^2$.  We further translate this to the sigma-model language as outlined in Sec.~ \ref{sec:from-phi-to-Q}. Discarding contractions of $S$ variables at different points, we find:
\begin{align}
\langle C_{(3,1)}[S] \rangle_{S[\phi,\phi^*,Q]}  &= 
2\langle \left(|S_a|^2\right)^2 \rangle_{S[\phi,\phi^*,Q]} 
\langle \left(|S_b|^2\right)^2 \rangle_{S[\phi,\phi^*,Q]}-2\langle |S_a|^2 |S_b|^2 \rangle_{S[\phi,\phi^*,Q]} 
\langle |S_a|^2 |S_b|^2 \rangle_{S[\phi,\phi^*,Q]} \label{eq:oiw31}.
\end{align}
Using Eqs. \eqref{eq:trans2}--\eqref{eq:trans3}, we can translate the building block for generic combinations containing $|\psi|^2_i$ at distinct sites $r_i$:
\begin{align}
\langle |S_a|^2 |S_b|^2 \rangle_{S[\phi,\phi^*,Q]} &= \sum_{\sigma\sigma'} \left(\langle|S_{a\sigma}|^2\rangle\langle|S_{b\sigma'}|^2\rangle + \langle S_{a\sigma}^*S_{b\sigma'}\rangle\langle S_{b\sigma'}^*S_{a\sigma}\rangle + \langle S_{a\sigma}^*S_{b\sigma'}^*\rangle\langle S_{b\sigma'}S_{a\sigma}\rangle \right)\nonumber\\
&= 4\mathcal{Q}^{00}_{aa}\mathcal{Q}^{00}_{bb}+2\mathcal{Q}^{00}_{ab}\mathcal{Q}^{00}_{ba}+2\mathcal{Q}^{01}_{ab}\mathcal{Q}^{10}_{ba} \,.
\label{eq:S-total-dens-to-Q}
\end{align}
This is valid also for equal replica indices, $a=b$, in which case the last term vanishes.
Substituting Eq.~\eqref{eq:S-total-dens-to-Q} into Eq. \eqref{eq:oiw31}, we obtain
\begin{align}
\langle C_{(3,1)}[S] \rangle_{S[\phi,\phi^*,Q]} &= 72\left(\mathcal{Q}^{00}_{aa}\mathcal{Q}^{00}_{bb}\right)^2-8\left(2\mathcal{Q}^{00}_{aa}\mathcal{Q}^{00}_{bb}+\mathcal{Q}^{00}_{ab}\mathcal{Q}^{00}_{ba}+\mathcal{Q}^{01}_{ab}\mathcal{Q}^{10}_{ba}\right)^2\nonumber\\
&=8\left(5\mathcal{Q}^{00}_{aa}\mathcal{Q}^{00}_{bb}+\mathcal{Q}^{00}_{ab}\mathcal{Q}^{00}_{ba}+\mathcal{Q}^{01}_{ab}\mathcal{Q}^{10}_{ba}\right)\left(\mathcal{Q}^{00}_{aa}\mathcal{Q}^{00}_{bb}-\mathcal{Q}^{00}_{ab}\mathcal{Q}^{00}_{ba}-\mathcal{Q}^{01}_{ab}\mathcal{Q}^{10}_{ba}\right).
\label{eq:C31-expansion}
\end{align}
This contains the operator 
\be
\mathcal{P}^C_{(1,1)} = \mathcal{Q}^{00}_{aa}\mathcal{Q}^{00}_{bb}-\mathcal{Q}^{00}_{ab}\mathcal{Q}^{00}_{ba}-\mathcal{Q}^{01}_{ab}\mathcal{Q}^{10}_{ba}
\label{eq:PC-11-Pfaffian}
\ee
 as a factor. It follows from the construction based on the Iwasawa decomposition in Sec. \ref{sec:iwasawa} that $\mathcal{P}^C_{(1,1)}$ in this form is exactly the (1,1) composite operator that satisfies Abelian fusion. Indeed, according to  Eq. \eqref{eq:nradial}, a complete family of eigenoperators satisfying Abelian fusion is generated by Pfaffians of the matrices
\begin{align}
\left(\left(T(\tilde{Q} \Lambda)\right)^{AA} \Sigma_{2}\right)_{m}
=\begin{pmatrix}
\mathcal{Q}^{01}&\mathcal{Q}^{00}\\
-\left(\mathcal{Q}^{00}\right)^T&\mathcal{Q}^{10}
\end{pmatrix}_{m} \,.
\label{eq:matrix-for-Pfaffian}
\end{align}
Here the subscript $m$ indicates a projection to first $m$ replicas (out of their total number of $n$). On the left-hand side, $T$ is the conjugation defined by Eq. \eqref{T-rotation}, which is applied to $\tilde{Q}$ matrix satisfying Eq. \eqref{eq:symopQiw}. Since the matrix is restricted to the advanced-advanced block and the first $m$ replicas, its size is $2m\times 2m$. The right-hand side is obtained by computing the action of $T$ explicitly and choosing appropriate phases $\alpha_{a_i}$ in Eqs.\eqref{eq:trans2}--\eqref{eq:trans3}, where $\mathcal{Q}^{ij}$ is introduced. The Pfaffian of the $2m\times 2m$ matrix \eqref{eq:matrix-for-Pfaffian} is a scaling operator in the representation $(1,\ldots,1)_m$.  For $m=2$ we thus find that
\begin{align}
\mathrm{Pf}\left(\left(T(\tilde{Q} \Lambda)\right)^{AA} \Sigma_{2}\right)_{2} &= -\mathcal{Q}^{00}_{aa}\mathcal{Q}^{00}_{bb}+\mathcal{Q}^{00}_{ab}\mathcal{Q}^{00}_{ba}+\mathcal{Q}^{01}_{ab}\mathcal{Q}^{10}_{ba}
\end{align}
is a pure scaling operator in representation $(1,1)$ that satisfies Abelian fusion. This is exactly what we stated below Eq.~\eqref{eq:PC-11-Pfaffian}.

The other factor in Eq.~\eqref{eq:C31-expansion} is a linear combination of operators belonging to the  representations $(2)$ and $(1,1)$. In view of the Abelian rules, fusion of the (1,1) operator, Eq.~\eqref{eq:PC-11-Pfaffian}, and the dominant contribution (2)  from the other factor yields (3,1).  Therefore,  the leading behavior of $\left|\Psi_{(3,1)}\left[|\psi|\right]\right|^2$ is indeed governed by an operator from the representation $(3,1)$.

\subsection{Numerical results}
\label{sec:numerics_iw}

\begin{figure}
	\centering
	\includegraphics[width = .48\textwidth]{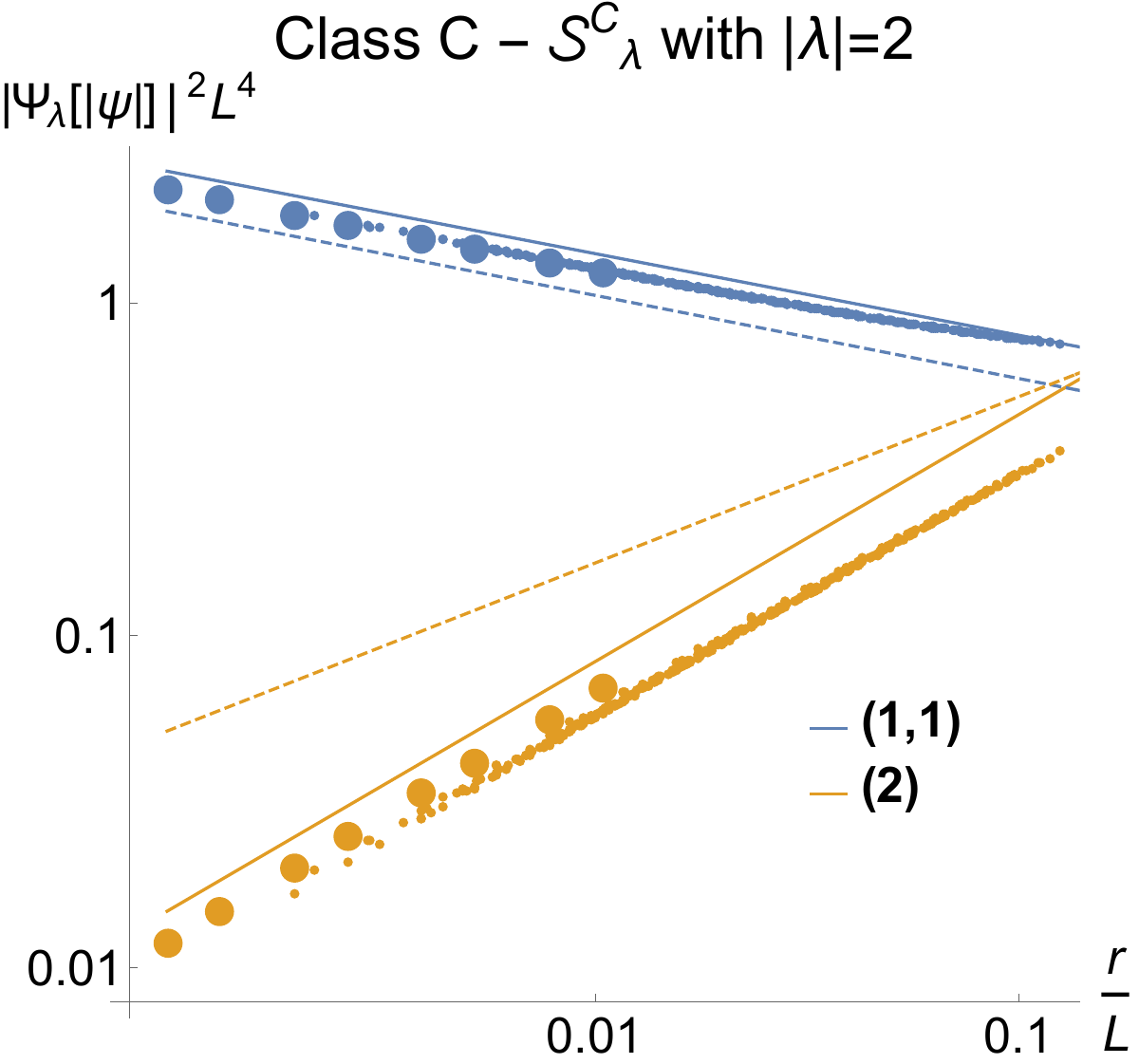}
	\includegraphics[width = .48\textwidth]{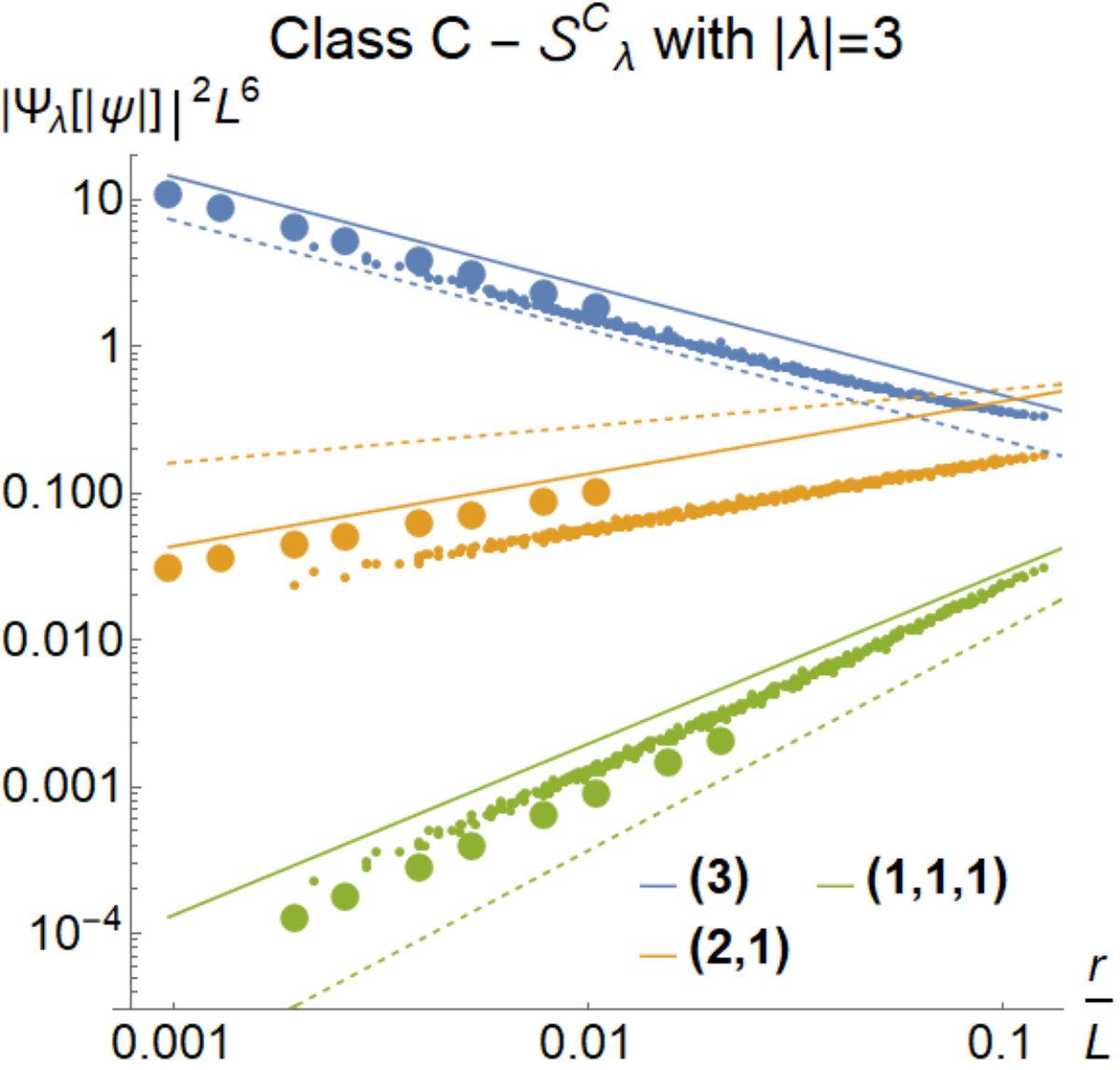}
	
	\includegraphics[width = .48\textwidth]{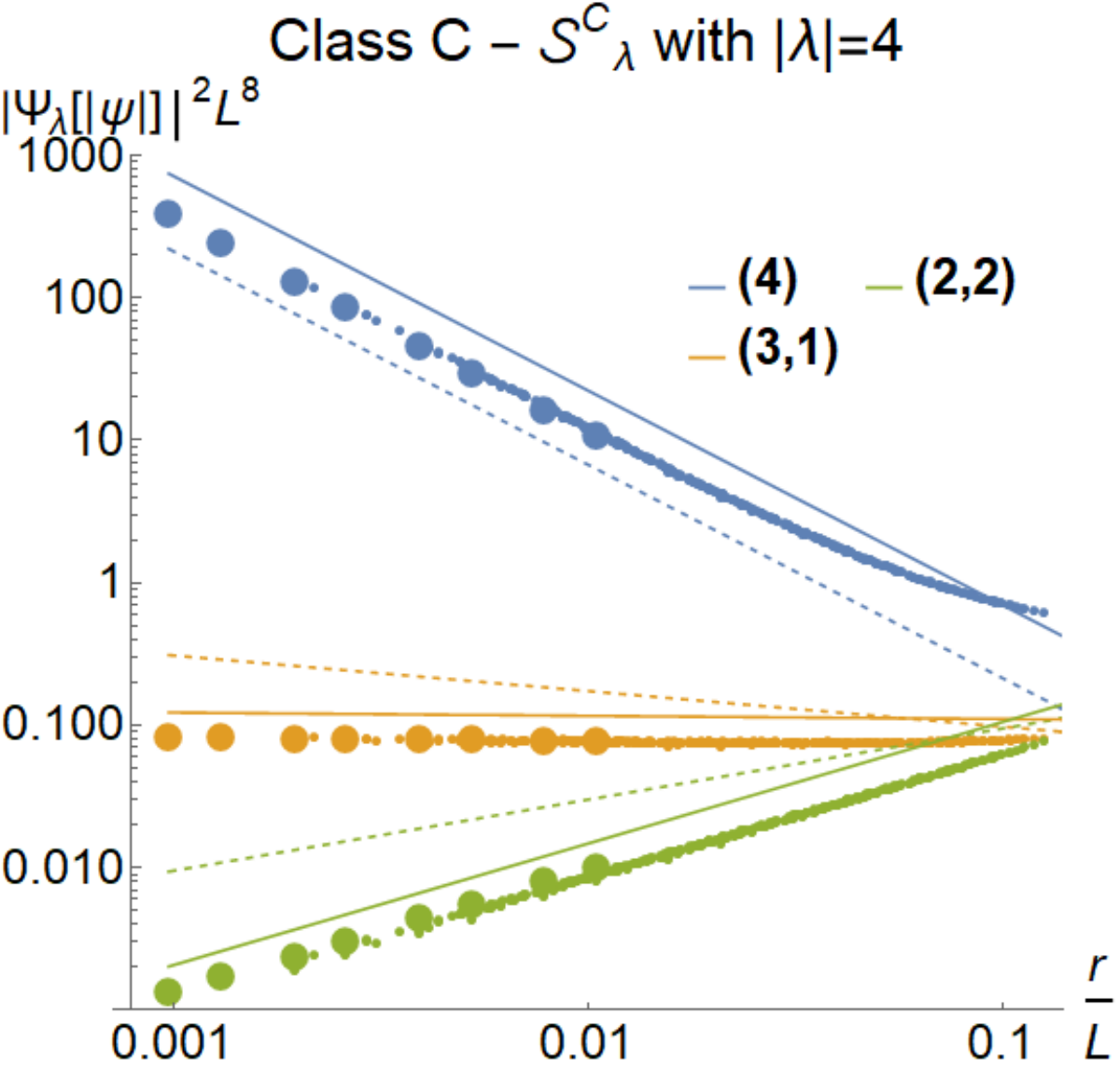}
	\includegraphics[width = .48\textwidth]{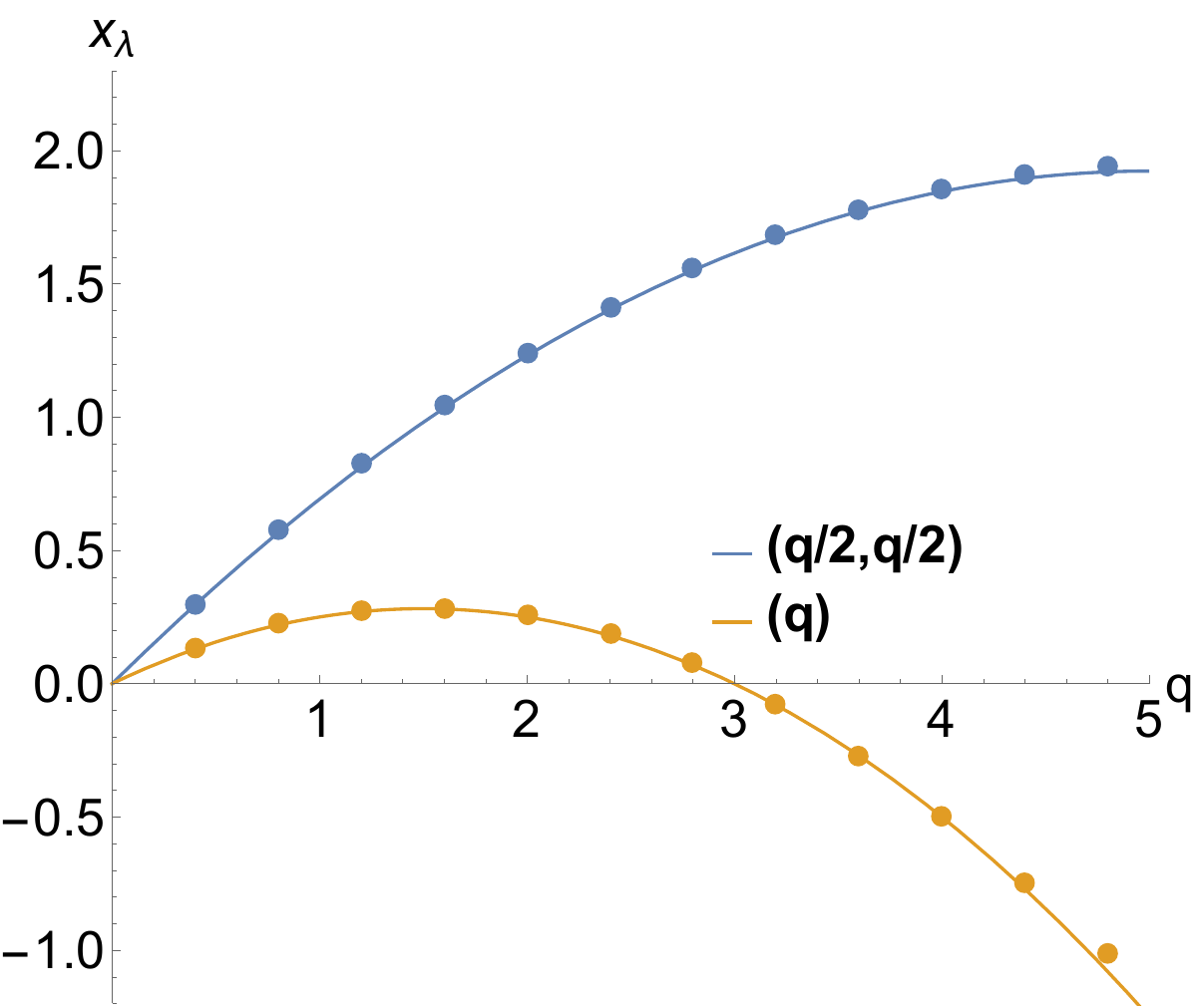}
	
	\caption{Generalized multifractality at the SQH transition (class C) studied via Young-symmetrized eigenstate combinations built out of the total density  $|\psi|(r)$,  Eq.~\eqref{eq:total-density}, for observables of order $q=2$ (top left), $q=3$ (top right), and $q=4$ (bottom left). Full lines are fits to the data; the corresponding exponents $\Delta^{{\rm num}, \, |\psi|}_\lambda$ are given in Table \ref{tab:lC} and are fully consistent with exponents $\Delta^{{\rm num}}_\lambda$ obtained from single-spin observables, Fig.~\ref{fig:AC}. The dashed lines correspond to generalized parabolicity \eqref{eq:xq}; the corresponding exponents $\Delta^{\rm para}_\lambda$ are also listed in Table \ref{tab:lC}. A strong violation of the generalized parabolicity at the SQH transition is evident. 		
	\textit{Bottom right:} Scaling dimensions $x_{(q)}$ and $x_{(q/2,q/2)}$ extracted from eigenstate combinations $|H+F|^{q/2}$ and $|H-F|^{q/2}$, respectively, formed with $|\psi|$. The Weyl symmetry dictates that $x_{(q)}$ is symmetric around $q=3/2$, and $x_{(q/2,q/2)}$ is symmetric around $q=5$. Solid lines are parabolas $x_{(q)}^{\rm para}=b q(3-q)$ and $\tilde{x}_{(q/2,q/2)}^{\rm para}=\tilde{b} q(10-q)/2$ with $b =1/8$ and $\tilde{b} = 2/13 = 0.154$. They serve as guides to the eye, illustrating that the numerical data fulfil the Weyl symmetry very well. Note that the parabolicity of $x_{(q)}$  [and, most likely, of $x_{(q/2,q/2)}$ as well] is only approximate.
Deviations from parabolicity in $x_{(q)}$ at relatively small $q$ were explored systematically in Refs.~\cite{mirlin2003wavefunction, puschmann2021quartic}; they are seen much better if one plots $x_{(q)} / q(3-q)$. The fact that $\tilde{b}$ [chosen to optimize the fit to
$x_{(q/2,q/2)}$] is substantially different from $b$ is a manifestation of strong violation of generalized parabolicity.}
	\label{fig:CS}
\end{figure}

In Fig. \ref{fig:CS} we present results of the numerical study of the $r/L$ dependence of $\left|\Psi_\lambda\left[|\psi|\right]\right|^2$, where $\Psi_\lambda\left[|\psi|\right]$ are Young-symmetrized combinations of $|\psi|$, on the class-C network model.  For $q=2$ and $q=3$, we obtain the scaling of all the corresponding class-C operators: (2) and (1,1)  for $q=2$ as well as (3), (2,1), and (1,1,1) for $q=3$. For $q=4$ we get in this way the scaling of (4), (3,1) and (2,2). At the same time, we do not get access to (2,1,1) and (1,1,1,1) scaling in this way, as 
the corresponding $\left|\Psi_\lambda\left[|\psi|\right]\right|^2$ numerically shows the admixture of the more relevant contribution (2,2). An interesting question is how to improve the construction to get access to all generalized-multifractality exponents via strictly positive observables; we do not have an answer to this question at the present stage. 

The exponents $\Delta^{{\rm num}, \, |\psi|}_\lambda$ obtained in this way are presented in Table \ref{tab:lC} and are in a very good agreement with exponents $\Delta^{{\rm num}}_\lambda$ obtained from single-spin observables, Fig.~\ref{fig:AC}.  For $\lambda= (1,1)$, the accuracy of $\Delta^{{\rm num}, \, |\psi|}_{(1,1)}$ is somewhat higher than that of $\Delta^{{\rm num}}_{(1,1)}$ since fluctuations are not so strong because we deal here with a strictly positive observable. The analytically known exponents $\Delta_{(2)}$ and $\Delta_{(3)}$ are perfectly reproduced. Further, the Weyl-symmetry relation $\Delta_{(2,1)} = \Delta_{(1,1)} - 1/4$ is also excellently fulfilled. 

In the bottom right panel of Fig. \ref{fig:CS}, we show the numerical data for $\Psi_{(q/2,q/2)}[|\psi|]$ obtained out of two eigenstates as $(\Psi_{(1,1)}[|\psi|])^{q/2}  = (H[|\psi|]-F[|\psi|] )^{q/2}$. 
Since $\Psi_{(1,1)}[|\psi|]$ is a strictly positive observable corresponding to the representation (1,1), we expect that $(\Psi_{(1,1)}[|\psi|])^{q/2}$ should exhibit the scaling of $(q/2,q/2)$. All numerical tests perfectly confirm this. First, the value of the exponent $\Delta_{(2,2)}$ obtained in this way is in excellent agreement with $\Delta^{{\rm num}, \, |\psi|}_{(2,2)} = 0.86 $ found by using four eigenstates. Second, the Weyl symmetry requires that $x_{(q/2,q/2)}$ is invariant with respect to the symmetry transformation $q \to 10-q$. While we were able to perform the numerics controllably up to $q=5$ only (for larger $q$ a still larger number of disorder realizations is needed), it is sufficient to see that the behavior of $x^{{\rm num}, \, |\psi|}_{(q/2,q/2)}$ is fully consistent with the required maximum at $q=5$. In fact, $x^{{\rm num}, \, |\psi|}_{(q/2,q/2)}$ turns out to be approximated rather well by a parabola $\tilde{x}_{(q/2,q/2)}^{\rm para} = \tilde{b} q(10-q)/2$  (which clearly satisfies the Weyl symmetry) with $\tilde{b} = 2/13 = 0.154$. For comparison, we show in the same panel also the conventional (leading) multifractal spectrum $x^{{\rm num}, \, |\psi|}_{(q)}$ obtained from the scaling of  $(H[|\psi|] + F[|\psi|] )^{q/2}$. The Weyl symmetry for this spectrum is $q \to 3-q$ and is again pefectly fulfilled. We show in the figure also the parabolic approximation $x_{(q)}^{\rm para}=b q(3-q)$. While it appears to describe the data  rather well, the parabolicity was found to be only approximate in Refs.~\cite{mirlin2003wavefunction, puschmann2021quartic}. Deviations from parabolicity in $x_{(q)}$ are seen more clearly if one plots the ratio $x_{(q)} / q(3-q)$; we do not study them here since they were explored recently in great detail and with very high accuracy in Ref.~\cite{puschmann2021quartic}.

As we have already emphasized above, the numerical results for generalized-miltifractality scaling exponents collected in Table \ref{tab:lC} demonstrate a strong vioaltion of generalized parabolicity. The exponents corresponding to representations $(1,1)$, $(2,1)$ , $(3,1)$ and $(2,2)$ strongly deviate from their ``parabolic values''.  Another manifestation of this fact is a substantial difference of the prefactors $b$ and $\tilde{b}$ in the parabolic fits in the bottom  right panel of Fig. \ref{fig:CS}. Indeed, the generalized parabolicity would imply not only that these parabolic fits are exact but also that $\tilde{b} = b$.  

The agreement between different numerical approaches and the fulfilment of all exact analytical results (values of exponents and Weyl-symmetry relations) make us confident that our numerical results for the exponents are accurate. Thus, the strong violation of generalized parabolicity at the SQH transition observed in our numerics is indeed the genuine property of the SQH critical point and not a finite-size effect.  In combinations with the results Sec. \ref{sec:CFT-2D}, this means that the local conformal invariance must be violated in the field theory describing the SQH critical point. Indeed, the Abelian fusion was implicitly demonstrated In Sec.~\ref{sec:iwasawa}. Thus, as was emphasized in Sec.~\ref{sec:gen-parab-hallmark-conf-inv}, violation of generalized parabolicity of the generalized-multifractality spectrum of a 2D system implies violation of local conformal invariance.

\section{Summary and outlook}
\label{sec:summary}
In this article, we have discussed the generalized multifractality that refers to scaling of a broad family of observables characterizing eigenstates at critical points of Anderson-localization transitions. While our main focus was on the SQH transition point in 2D superconducting systems, some parts of the article are more general. Our main results are as follows.

\begin{enumerate}

\item
We have reviewed the framework of generalized multifractality and its relations with conformal invariance. In particular, we have shown that, in the case of 2D systems, the corresponding spectrum of scaling exponents should exhibit generalized parabolicity (parametrized by a single constant) if the associated field theory satisfies two conditions: (i) existence of a family of composite operators realizing the generalized multifractality and satisfying the Abelian fusion rules, and (ii) local conformal invariance.

\item  Within the field-theoretic framework of the sigma-model for superconducting systems of class C, we have explicitly constructed pure-scaling composite operators that realize generalized multifractality.
By using two approaches---heighest-weight vectors and the Iwasawa decomposition---we have built families of these operators that obey the Abelian fusion rules.

\item  We have derived one-loop RG rules for composite operators of class C that have polynomial structure in the $Q$-field of the sigma-model. We have used the RG approach as one more way towards pure-scaling composite operators.

\item  We have developed a systematic ``translation'' of the polynomial pure-scaling operators obtained from sigma-model RG to observables built out of eigenstates. In addition, we have established a relation (although for particular cases only) between another class of eigenstate observables (involving total density in spin space) and pure-scaling operators from the Iwasawa construction.

\item Using network models of classes C and A, we have performed a numerical investigation of the generalized multifractality of eigenfunction observables at SQH and IQH transitions.
For class C, several complementary numerical approaches were used, which were all in agreement to each other whenever a comparison was possible.
The numerical results have confirmed all analytical predictions and have allowed us to find many exponents characterizing the generalized multifractality at the SQH and IQH transitions.

\item One of central results of the paper is a strong violation of generalized parabolicity by numerically determined exponents characterizing the SQH transition. In combination with the above analytical results, this
leads to a very fundamental implication: a violation of the local conformal invariance at the SQH critical point.  Our result is consistent with deviations from parabolicity found in the conventional SQH multifractal spectrum \cite{mirlin2003wavefunction, puschmann2021quartic}. The violation of parabolicity in subleading exponents found in this work is, however, many times stronger.

In view of the importance of the conclusions (absence of local conformal invariance), it is worth commenting on the status of the numerically determined exponents. Of course,
whatever the size of the numerically studied system is, it is always finite, and, if one wants to be absolutely rigorous, one should always keep in mind a possibility that the extracted exponents strongly deviate from their true values due to finite-size effects. Specifically, this would happen in a situation when the critical system, even for largest system sizes, is still far from the fixed point. There is, however, a strong evidence that this is not the case for the network model that we have used for numerical simulations. Indeed, all exact analytical results, including values of the exponents and Weyl-symmetry relations are perfectly fulfilled by numerical data. Thus, large finite-size errors in exponent values are highly improbable.

\item For the IQH transition, our numerical results yield relatively small but clear deviations from the WZNW theory conjectured in Ref.~\cite{zirnbauer2019integer}
(generalized parabolicity with prefactor $b=1/4$), in consistency with previous works \cite{Obuse-Boundary-2008, evers2008multifractality}.

\end{enumerate}

The following comment is in order here.  Our results on violation of the local conformal invariance generated by Virasoro algebra do not exclude a possibility that some aspects of the conformal invariance---i.e., symmetry with respect to some subclass of conformal transformation or for some particular observables--do hold. As an example, it is known that some 2D models possess global but not local conformal invariance \cite{Riva-Scale-2005, Nakayama-Hidden-2016}. In fact, a number of previous works have carried out numerical tests of some manifestations of conformal invariance at 2D Anderson transitions. A particular focus has been put on the relation between the parameter $\alpha_0$ characterizing the scaling of typical LDOS and the typical localization length in a quasi-1D cylinder geometry \cite{Janssen-Multifractal-1994, Dohmen-Disordered-1996, Janssen-Statistics-1998, Merkt-Network-1998, evers2001multifractality, Obuse-Multifractality-2007, Obuse-Conformal-2010}. It was found that this relation holds with a very good accuracy. It is worth noting that the corresponding conformal map has direct counterparts for arbitrary spatial dimensionality ($d>2$) \cite{cardy1985universal} in analogy with global conformal transformations.  In Refs.~\cite{Obuse-Multifractality-2007, Obuse-Corner-2008} a relation between the surface and corner multifractality spectra was tested. While the results are consistent with conformal invariance near $q=0$, strong deviations away from $q=0$ were found. It is not known at this stage whether these deviations are a manifestation of violation of conformal invariance or are related to numerical errors.

Our work paves the way for further analytical and numerical studies of generalized multifractality at Anderson transitions. We close the paper by briefly discussing some of prospective research directions:

\begin{enumerate}

\item
We expect that further work will extend the analysis to other symmetry classes.  Among them, class D describing disordered Majorana systems (that undergo a metal-insulator transition as well as a thermal quantum Hall transition in 2D)  attracts particular interest. The RG flow and the pure-scaling composite operators for class D can be obtained from those for class C (explored in this work) by using the duality of the corresponding sigma models. Further, preliminary results indicate interesting relations with other symmetry classes. In particular, the pure-scaling operators for the symplectic Wigner-Dyson class AII, which also hosts an Anderson metal-insulator transition in 2D geometry, appear to have the same form as for class C. The class AII is in turn dual to the conventional orthogonal Wigner-Dyson class AI that was studied in Ref.~\cite{burmistrov2016mesoscopic}.

\item
As discussed in Sections \ref{sec:wave_ops} and \ref{sec:wave_iw}, there is a qualitative difference between the pure-scaling observables in classes A and C that is related to the presence of spin in class C. This makes numerical studies of generalized multifractality at the SQH transition much more demanding than at the IQH transition. Specifically, simplest pure-scaling observables have indefinite sign and fluctuate wildly from one configuration of disorder to another.  A natural conjecture is that this property applies also to other symmetry classes for which the spin degree of freedom is essential. The challenge is to better understand the character of correlations between nearby-in-energy eigenstates in such spinful systems at criticality.

\item
Numerical studies of generalized multifractality at other 2D Anderson-localization transitions would be very interesting. In particular, it  is important to understand how generic are violations of generalized parabolicity and of the local conformal invariance.

\item
Finally, many interesting questions arise when one ``switches on'' the electron-electron interaction. They include mutual influence of interaction and generalized multifractality in systems of various symmetry classes.
In this context, one should distinguish between a long-range (Coulomb) and a short-range (screened) interaction. For Wigner-Dyson classes, 
it was shown in Refs. \cite{burmistrov2013multifractality} and \cite{burmistrov2016mesoscopic} within a RG analysis that the generalized multifractality holds even in the presence of Coulomb interaction.

\end{enumerate}

\section{Acknowledgments}
We thank S. Bera, F. Evers, and D. Hernang\'omez-P\'erez for collaboration on related projects and for useful discussions. We also acknowledge useful discussions with I. Burmistrov. Further, we thank F. Evers and M. Foster as well as S. Bera, I.  Burmistrov, H. Obuse, T. Ohtsuki, and K. Slevin for insightful comments on the manuscript.
JFK acknowledges funding by Graduate Funding from the German States awarded by KHYS.

\appendix

\section{Analytical expressions for $K$-invariant eigenoperators in class A}
\label{appendix:class-A-eigenop}

In this Appendix, we derive analytical expressions for the coefficients of the $K$-invariant scaling operators $\mathcal{P}^A_{\lambda}$ for arbitrary Young diagrams $\lambda = (q_1, \ldots, q_n)$. These operators for $ |\lambda| \equiv q \equiv q_1 + \ldots + q_n = 2$, 3, and 4 are determined in Sec.~\ref{sec:inv-op-class-A}, see  Eq. \eqref{eq:ops_a}. 

To find $\tilde{\mathcal{P}}^A_{\lambda}$ analytically, we use Young-symmetrized expressions for eigenoperators in Ref. \cite{gruzberg2013classification} and express them in terms of the characters $\chi$ of representations of the symmetric group $S_q$. These operators (that will be denoted as $\tilde{\mathcal{P}}^A_\lambda$) are plane waves in the Iwasawa construction and also highest-weight vectors, and are not $K$-invariant. We perform then the $\text{U}(n)\times \text{U}(n)$-averaging  of these combinations to derive the $K$-invariant operators.

We define $\mathcal{Q} = Q^{RR}- Q^{AA}+ Q^{RA}-Q^{AR}$. Further, let $\nu_{ij} =\mathrm{tr}\left(E_{ij}\mathcal{Q}\right)$ be matrix elements of $\mathcal{Q}$, and $S_q$ the symmetric group. 
In Ref. \cite{gruzberg2013classification}, expressions for eigenoperators in terms of Young-symmetrized products of matrix elements $\nu_{ij}$ of $\mathcal{Q}$  were derived. As an example, let us start with 
 the most antisymmetric operators $(1,1,\ldots,1)$. According to Ref. \cite{gruzberg2013classification}, we know that the determinant of the $q\times q$ subblock $\hat{\nu}_q$ of the matrix $\hat{\nu}$,
\begin{align}
\tilde{\mathcal{P}}^A_{(1,1,\ldots,1)} &\equiv \det\hat{\nu}_q = \sum_{\sigma\in S_q} \mathrm{sign}(\sigma)\prod_{i=1}^q \nu_{i\,\sigma(i)} \,,
\end{align} 
belongs to the  representation $(1,1,\ldots,1)$. The opposite extreme is the most symmetrized operator, which is given by the permanent of the same matrix,
\begin{align}
\tilde{\mathcal{P}}^A_{(q)} & = \mathrm{perm} \, \hat{\nu}_q  \equiv \sum_{\sigma\in S_q} \prod_{i=1}^q \nu_{i\,\sigma(i)} \,,
\end{align} 
and belongs to the representation $(q)$. For a generic representation (Young diagram) $\lambda$, the operator $\tilde{\mathcal{P}}^A_{\lambda}$ is obtained by applying the corresponding combination of symmetrization and antisymmetrization operations \cite{gruzberg2013classification}. We can express the scaling operators $\tilde{\mathcal{P}}^A_{\lambda}$ in terms of the 
immanants of the matrix $\hat{\nu}_q$ (multilinear forms invariant under permutations generalizing the determinant):
\begin{align}
\tilde{\mathcal{P}}^A_{\lambda} &= \mathrm{Imm}_{\lambda} \, \hat{\nu} \equiv \sum_{\sigma\in S_q} \chi_\lambda(\sigma)\prod_i \nu_{i\,\sigma(i)} \,,
\label{eq:opsa}
\end{align}
where $\chi_\lambda$ is the character of the $\lambda$ representation of $S_q$. In the special case of $\lambda=(1,1,\ldots,1)$, this character reduces to the sign function on $S_q$. For $\lambda = (q)$ we obtain $\chi_{(q)}=1$. This formula can be found from mapping the generic Young-symmetrized wave-function combinations in Ref. \cite{gruzberg2013classification} to sigma-model operators. The Young symmetrization makes the characters $\chi_\lambda(\sigma)$ to appear naturally in the above expression \cite{fulton_harris}. 

Now we perform the $\text{U}(n)\times \text{U}(n)$ averaging of Eq.~\eqref{eq:opsa} to obtain the $K$-invariant operators.  In the cycle decomposition of $S_q$, each equivalence class can be labeled by a Young diagram $\mu$ with $|\mu|=q$. Let $[\mu]$ be the equivalence class containing $\sigma$. Under $\text{U}(n)\times \text{U}(n)$ averaging, and to leading order in the replica limit $n \to 0$, we have
\begin{align}
\left\langle \prod_{i=1}^q \nu_{i\,\sigma(i)}\right\rangle_{\text{U}(n)\times \text{U}(n)} &\simeq n^{-q}O_\mu \,,
\end{align}
where $O_\mu$ are basis $K$-invariant operators defined  in Eq. \eqref{eq:rg-kinv}.  Thus, we obtain for the expansion of $K$-invariant operators $\mathcal{P}^A_{\lambda}$ in basis operators $O_\mu$
\begin{align}
\mathcal{P}^A_{\lambda} &= \sum_{\mu} \#[\mu]\chi_\lambda(\mu) O_\mu \,,
\end{align}
where $\#[\mu]$ denotes the number of elements in the equivalence class $[\mu]$. We can read off the matrices of coefficients $(P^A_q)_{\lambda\mu}\equiv \#[\mu]\chi_\lambda(\mu)$ of the $K$-invariant operators. In Eq.~\eqref{eq:ops_a}, these matrices of coefficients, as obtained from RG, were shown for $q=2$, 3, and 4. 

\section{Transformation from the sigma-model field $Q$ to the field $\tilde{Q}$ in Sec.~\ref{sec:from-phi-to-Q} }
\label{appendix:Q-to-tilde-Q}

In this Appendix, we provide technical details to the transformation from the sigma-model field $Q$ to the field $\tilde{Q}$ that is most convenient for the derivation of RG equations. This transformation is used in Sec.~\ref{sec:from-phi-to-Q} in course of establishing the mapping between sigma-model composite operators and eigenstate observables. The fields $Q$ and $\tilde{Q}$ represent two different parametrization of $G/K$
[see Eqs.~\eqref{eq:symopQ} and  \eqref{eq:symopQiw}]; they are related by the rotation $\tilde{Q}=U_\Sigma Q U_\Sigma $,  Eq.\eqref{Q-tilde}, with the matrix $U_\Sigma$ given by $U_\Sigma =\mathrm{diag}(1, \Sigma_1)_\tau$, Eq.~\eqref{eq:U-Sigma}. The operators in the $Q$-representation in Sec.~\ref{sec:from-phi-to-Q}  contain products of traces of the type $O=\mathrm{tr}(\cdots E_{a_ia_i} \Lambda Q\cdots )$, see Eq.~\eqref{eq:sigma_mod_basis_comp_ops} and its $S_q$-symmetrized version, Eq.~\eqref{eq:ops_rep}. When transformed to the $\tilde{Q}$-representation, this becomes
$O=\mathrm{tr}(\cdots U_\Sigma E_{a_ia_i} \Lambda U_\Sigma \tilde{Q}\cdots )$. When deriving the RG rules above, we have assumed that the matrices multiplying the sigma-model field are odd with respect to the operation \eqref{eq:bar-operation}; see Eq.~\eqref{eq:RG-classC-A-B-odd}. 
A subtlety arises since the matrix $U_\Sigma E_{a_ia_i} \Lambda U_\Sigma$ does not satisfy this requirement. We show, however, in this Appendix that this is not essential, and the operators \eqref{eq:ops_rep} undergo the same RG equations as derived in Sec.~\ref{sec:RG-class_C-arbitrary_q}, so that the pure-scaling observables are constructed from them according to Eq.~\eqref{rg:higher_ord}.  To prove this, we show that the operators $O$ are equivalent to $\tilde{O}=\mathrm{tr}(\cdots \tilde{E}_{a_ia_i} \Lambda\tilde{Q}\cdots )$ by using gauge transformations $\langle O\rangle_{\text{U}_b} =\tilde{O}$ with a certain subgroup $\text{U}_b\subset K \equiv \rm{U}(2n)$. The newly introduced matrices 
\be
\tilde{E}_{a_ia_i}\Lambda =\frac12 \mathrm{diag}(E_{a_i,a_i}+E_{-a_i,-a_i}, -E_{a_i,a_i}-E_{-a_i,-a_i})_\tau
\label{eq:tilde-E}
\ee
 are odd under the symmetry operation \eqref{eq:bar-operation} as required.

We compute the average of $O$ over the subgroup $\text{U}_b$ of the full gauge group $K = \rm{U}(2n)$: 
\begin{align}
\langle \mathrm{tr}(\cdots E_{a_ia_i} \Lambda Q\cdots )\rangle_{\text{U}_b}  &= \int_{\text{U}_b} d\mu(U)\mathrm{tr}(\cdots U_\Sigma U E_{a_ia_i}\Lambda  U^\dagger U_\Sigma\tilde{Q}\cdots ) \,,
\end{align}
where $\mu$ is the Haar measure on $\text{U}_b$.  We choose $\text{U}_b$ to be the block-diagonal subgroup $\text{U}_b = \mathrm{U}(2)_\Sigma^n\subset \mathrm{U}(2n)$ with $n$ two-by-two blocks in particle-hole space $\Sigma$.  These $n$ blocks correspond to $n$ replica indices $1,\ldots n$. Here we need to  distinguish between the matrix $E_{a_ia_i}$ living in $2n\times 2n$ combined replica and particle-hole ($\Sigma$) space 
(that we represent by associating with each replica $a_i$ a replica $-a_i$, see Sec.~\ref{sec:sigma})
and the matrix $E_{a_i,a_i}^{n\times n}$ living in the $n\times n $ replica space. They are related by $E_{a_ia_i} = E_{a_i,a_i}^{n\times n}\Sigma_+$, where $\Sigma_\pm = (1_\Sigma\pm\Sigma_3)/2$  and $1_\Sigma$ is the identity matrix in $\Sigma$ space. We prepare for computing the average over the matrix $U \in \text{U}_b$ by simplifying:
\begin{align}
U_\Sigma U E_{a_ia_i}\Lambda  U^\dagger U_\Sigma&= \begin{pmatrix}
1 & \\
&\Sigma_1
\end{pmatrix}_\tau\begin{pmatrix}
 U_{a_i} & \\
 &U_{a_i}^*
 \end{pmatrix}_\tau\begin{pmatrix}
 E_{a_i,a_i}^{n\times n}\Sigma_+ & \\
 &E_{a_i,a_i}^{n\times n}\Sigma_+
 \end{pmatrix}_\tau\begin{pmatrix}
 1 & \\
  &-1
 \end{pmatrix}_\tau \begin{pmatrix}
 U_{a_i}^\dagger  & \\
 &U_{a_i}^T
 \end{pmatrix}_\tau\begin{pmatrix}
 1 & \\
 &\Sigma_1
 \end{pmatrix}_\tau
\nonumber\\
 &= \begin{pmatrix}
 E_{a_i,a_i}^{n\times n}U_{a_i}\Sigma_+U_{a_i}^\dagger & \\
 &-E_{a_i,a_i}^{n\times n}\Sigma_1U_{a_i}^*\Sigma_+U_{a_i}^T\Sigma_1
 \end{pmatrix}_\tau.
\end{align}
Here $U_{a_i}$ is the two-by-two $\Sigma$-space block of the matrix $U$ corresponding to the replicas $a_i$ and $-a_i$. 

Now we can integrate over the $n$ independent $\mathrm{U}(2)$ subgroups, out of which $\text{U}_b$ is composed.  Since all $a_i$ are assumed to be distinct in the considered composite operators $O$, the integrals decouple and we find $\langle U_{a_i}\Sigma_+U_{a_i}^\dagger\rangle_{\mathrm{U}(2)} = \langle U_{a_i}^*\Sigma_+U_{a_i}^T\rangle_{\mathrm{U}(2)} = \frac12 1_\Sigma$. Since $1_\Sigma=\Sigma_++\Sigma_-$ and  $E_{a_i,a_i}^{n\times n}\Sigma_\pm = E_{\pm a_i,\pm a_i}$, we obtain
\begin{align}
\langle \mathrm{tr}(\cdots E_{a_ia_i} \Lambda Q\cdots )\rangle_{\text{U}_b}  &= \mathrm{tr}(\cdots \tilde{E}_{a_ia_i} \Lambda \tilde{Q}\cdots ) \,,
\end{align}
with the matrix  $\tilde{E}_{a_ia_i} \Lambda$ defined in Eq.~\eqref{eq:tilde-E}.  
This proves that the operators $O=\mathrm{tr}(\cdots E_{a_ia_i} \Lambda Q\cdots )$ undergo the same RG flow as $\tilde{O}=\mathrm{tr}(\cdots \tilde{E}_{a_ia_i} \Lambda\tilde{Q}\cdots )$ since they coincide up to terms that vanish upon the averaging over a subgroup of the gauge group.
Since $\tilde{E}_{a_ia_i} \Lambda$ is odd under the operation \eqref{eq:bar-operation}, the RG flow equations derived in Sec.~\ref{sec:RG-class_C-arbitrary_q} apply.   This completes the justification of applying Eq.~\eqref{rg:higher_ord}  for pure-scaling operators of class C (derived in $\tilde{Q}$ parametrization) to eigenstate basis combinations corresponding to operators in $Q$-representation, Eq.~\eqref{eq:O-psi-correspondence}. The corresponding analytical predictions are verified numerically in Sec.~\ref{sec:numerical_results}.

\bibliography{class-C,parabolicity}

\begin{thebibliography}{100}
\expandafter\ifx\csname url\endcsname\relax
  \def\url#1{\texttt{#1}}\fi
\expandafter\ifx\csname urlprefix\endcsname\relax\def\urlprefix{URL }\fi
\expandafter\ifx\csname href\endcsname\relax
  \def\href#1#2{#2} \def\path#1{#1}\fi

\bibitem{anderson58}
P.~W. Anderson, {Absence of diffusion in certain random lattices}, Phys. Rev.
  109~(5) (1958) 1492.

\bibitem{50_years_of_localization}
E.~Abrahams (Ed.), {50 Years of Anderson Localization}, World Scientific, 2010.

\bibitem{evers08}
F.~Evers, A.~D. Mirlin, {Anderson transitions}, Reviews of Modern Physics
  80~(4) (2008) 1355.

\bibitem{altland1997nonstandard}
A.~Altland, M.~R. Zirnbauer, {Nonstandard symmetry classes in mesoscopic
  normal-superconducting hybrid structures}, Phys. Rev. B 55 (1997) 1142--1161.

\bibitem{zirnbauer1996riemannian}
M.~R. Zirnbauer, {Riemannian symmetric superspaces and their origin in
  random‐matrix theory}, Journal of Mathematical Physics 37~(10) (1996)
  4986--5018.

\bibitem{heinzner2005symmetry}
P.~Heinzner, A.~Huckleberry, M.~Zirnbauer, {Symmetry Classes of Disordered
  Fermions}, Commun. Math. Phys. 257 (2005) 725--771.

\bibitem{rodriguez2010critical}
A.~Rodriguez, L.~J. Vasquez, K.~Slevin, R.~A. R\"omer, {Critical Parameters
  from a Generalized Multifractal Analysis at the Anderson Transition}, Phys.
  Rev. Lett. 105 (2010) 046403.

\bibitem{rodriguez2011multifractal}
A.~Rodriguez, L.~J. Vasquez, K.~Slevin, R.~A. R{\"o}mer, {Multifractal
  finite-size scaling and universality at the Anderson transition}, Physical
  Review B 84~(13) (2011) 134209.

\bibitem{burmistrov2013multifractality}
I.~S. Burmistrov, I.~V. Gornyi, A.~D. Mirlin, {Multifractality at Anderson
  Transitions with Coulomb Interaction}, Phys. Rev. Lett. 111 (2013) 066601.

\bibitem{ghorashi2018critical}
S.~A.~A. Ghorashi, Y.~Liao, M.~S. Foster, {Critical Percolation without
  Fine-Tuning on the Surface of a Topological Superconductor}, Phys. Rev. Lett.
  121 (2018) 016802.

\bibitem{ghorashi2020criticality}
S.~A.~A. Ghorashi, J.~F. Karcher, S.~M. Davis, M.~S. Foster, {Criticality
  across the energy spectrum from random artificial gravitational lensing in
  two-dimensional Dirac superconductors}, Phys. Rev. B 101 (2020) 214521.

\bibitem{sbierski2020spectrum-wide}
B.~Sbierski, J.~F. Karcher, M.~S. Foster, {Spectrum-Wide Quantum Criticality at
  the Surface of Class AIII Topological Phases: An ``Energy Stack'' of Integer
  Quantum Hall Plateau Transitions}, Phys. Rev. X 10 (2020) 021025.

\bibitem{karcher2021how}
J.~F. Karcher, M.~S. Foster, {How spectrum-wide quantum criticality protects
  surface states of topological superconductors from Anderson localization:
  Quantum Hall plateau transitions (almost) all the way down}, Annals of
  Physics (2021) 168439.

\bibitem{faez2009observation}
S.~Faez, A.~Strybulevych, J.~H. Page, A.~Lagendijk, B.~A. van Tiggelen,
  {Observation of Multifractality in Anderson Localization of Ultrasound},
  Phys. Rev. Lett. 103 (2009) 155703.

\bibitem{richardella2010visualizing}
A.~Richardella, P.~Roushan, S.~Mack, B.~Zhou, D.~A. Huse, D.~D. Awschalom,
  A.~Yazdani, {Visualizing Critical Correlations Near the Metal-Insulator
  Transition in Ga1-xMnxAs}, Science 327~(5966) (2010) 665--669.

\bibitem{sacepe2008disorder-induced}
B.~Sac\'ep\'e, C.~Chapelier, T.~I. Baturina, V.~M. Vinokur, M.~R. Baklanov,
  M.~Sanquer, {Disorder-Induced Inhomogeneities of the Superconducting State
  Close to the Superconductor-Insulator Transition}, Phys. Rev. Lett. 101
  (2008) 157006.

\bibitem{noat2013unconventional}
Y.~Noat, V.~Cherkez, C.~Brun, T.~Cren, C.~Carbillet, F.~Debontridder, K.~Ilin,
  M.~Siegel, A.~Semenov, H.-W. H\"ubers, D.~Roditchev, {Unconventional
  superconductivity in ultrathin superconducting NbN films studied by scanning
  tunneling spectroscopy}, Phys. Rev. B 88 (2013) 014503.

\bibitem{feigel2007eigenfunction}
M.~V. Feigel'man, L.~B. Ioffe, V.~E. Kravtsov, E.~A. Yuzbashyan, {Eigenfunction
  Fractality and Pseudogap State near the Superconductor-Insulator Transition},
  Phys. Rev. Lett. 98 (2007) 027001.

\bibitem{feigel2010fractal}
M.~Feigel'man, L.~Ioffe, V.~Kravtsov, E.~Cuevas, {Fractal superconductivity
  near localization threshold}, Annals of Physics 325~(7) (2010) 1390--1478.

\bibitem{garcia-garcia2020superconductivity}
B.~Fan, A.~M. Garc\'{\i}a-Garc\'{\i}a, {Superconductivity at the
  three-dimensional Anderson metal-insulator transition}, Phys. Rev. B 102
  (2020) 184507.

\bibitem{burmistrov2012enhancement}
I.~Burmistrov, I.~Gornyi, A.~Mirlin, {Enhancement of the critical temperature
  of superconductors by Anderson localization}, Physical Review Letters 108~(1)
  (2012) 017002.

\bibitem{burmistrov2015superconductor}
I.~S. Burmistrov, I.~V. Gornyi, A.~D. Mirlin, {Superconductor-insulator
  transitions: Phase diagram and magnetoresistance}, Phys. Rev. B 92 (2015)
  014506.

\bibitem{mayoh2015global}
J.~Mayoh, A.~M. Garc{\'\i}a-Garc{\'\i}a, {Global critical temperature in
  disordered superconductors with weak multifractality}, Physical Review B
  92~(17) (2015) 174526.

\bibitem{fan2020enhanced}
B.~Fan, A.~M. Garc\'{\i}a-Garc\'{\i}a, {Enhanced phase-coherent multifractal
  two-dimensional superconductivity}, Phys. Rev. B 101 (2020) 104509.

\bibitem{burmistrov2021multifractallyenhanced}
I.~Burmistrov, I.~Gornyi, A.~Mirlin, Multifractally-enhanced superconductivity
  in thin films, Annals of Physics (2021) 168499.

\bibitem{zhao2019disorder-induced}
K.~Zhao, H.~Lin, X.~Xiao, W.~Huang, W.~Yao, M.~Yan, Y.~Xing, Q.~Zhang, Z.-X.
  Li, S.~Hoshino, J.~Wang, S.~Zhou, L.~Gu, M.~S. Bahramy, H.~Yao, N.~Nagaosa,
  Q.-K. Xue, K.~T. Law, X.~Chen, S.-H. Ji, {Disorder-induced multifractal
  superconductivity in monolayer niobium dichalcogenides}, Nature Physics
  15~(9) (2019) 904--910.

\bibitem{rubio-verdu2020visualization}
C.~Rubio-Verdu, A.~M. Garcia-Garcia, H.~Ryu, D.-J. Choi, J.~Zaldivar, S.~Tang,
  B.~Fan, Z.-X. Shen, S.-K. Mo, J.~I. Pascual, M.~M. Ugeda, {Visualization of
  Multifractal Superconductivity in a Two-Dimensional Transition Metal
  Dichalcogenide in the Weak-Disorder Regime}, Nano Letters 20~(7) (2020)
  5111--5118.

\bibitem{foster2012interaction}
M.~S. Foster, E.~A. Yuzbashyan, {Interaction-Mediated Surface-State Instability
  in Disordered Three-Dimensional Topological Superconductors with Spin SU(2)
  Symmetry}, Phys. Rev. Lett. 109 (2012) 246801.

\bibitem{foster2014topological}
M.~S. Foster, H.-Y. Xie, Y.-Z. Chou, {Topological protection, disorder, and
  interactions: Survival at the surface of three-dimensional topological
  superconductors}, Phys. Rev. B 89 (2014) 155140.

\bibitem{kettemann2012kondo}
S.~Kettemann, E.~R. Mucciolo, I.~Varga, K.~Slevin, {Kondo-Anderson
  transitions}, Phys. Rev. B 85 (2012) 115112.

\bibitem{kettemann2009critical}
S.~Kettemann, E.~Mucciolo, I.~Varga, {Critical metal phase at the anderson
  metal-insulator transition with Kondo impurities}, Physical Review Letters
  103~(12) (2009) 126401.

\bibitem{slevin2019multifractality}
K.~Slevin, S.~Kettemann, T.~Ohtsuki, {Multifractality and the distribution of
  the Kondo temperature at the Anderson transition}, The European Physical
  Journal B 92~(12) (2019) 281.

\bibitem{mirlin94a}
A.~D. Mirlin, Y.~V. Fyodorov, {Distribution of local densities of states, order
  parameter function, and critical behavior near the {A}nderson transition},
  Physical Review Letters 72~(4) (1994) 526.

\bibitem{mirlin94b}
A.~D. Mirlin, Y.~V. Fyodorov, {Statistical properties of one-point {G}reen
  functions in disordered systems and critical behavior near the {A}nderson
  transition}, Journal de Physique I 4~(5) (1994) 655--673.

\bibitem{fyodorov2004statistics}
Y.~V. Fyodorov, D.~V. Savin, {Statistics of impedance, local density of states,
  and reflection in quantum chaotic systems with absorption}, Journal of
  Experimental and Theoretical Physics Letters 80~(12) (2004) 725--729.

\bibitem{savin2005universal}
D.~V. Savin, H.-J. Sommers, Y.~V. Fyodorov, {Universal statistics of the local
  Green's function in wave chaotic systems with absorption}, Journal of
  Experimental and Theoretical Physics Letters 82~(8) (2005) 544--548.

\bibitem{fyodorov2005scattering}
Y.~V. Fyodorov, D.~V. Savin, H.-J. Sommers, {Scattering, reflection and
  impedance of waves in chaotic and disordered systems with absorption},
  Journal of Physics A: Mathematical and General 38~(49) (2005) 10731--10760.

\bibitem{mirlin2006exact}
A.~D. Mirlin, Y.~V. Fyodorov, A.~Mildenberger, F.~Evers, {Exact Relations
  between Multifractal Exponents at the Anderson Transition}, Phys. Rev. Lett.
  97 (2006) 046803.

\bibitem{gruzberg2011symmetries}
I.~A. Gruzberg, A.~W.~W. Ludwig, A.~D. Mirlin, M.~R. Zirnbauer, {Symmetries of
  Multifractal Spectra and Field Theories of Anderson Localization}, Phys. Rev.
  Lett. 107 (2011) 086403.

\bibitem{gruzberg2013classification}
I.~A. Gruzberg, A.~D. Mirlin, M.~R. Zirnbauer, {Classification and symmetry
  properties of scaling dimensions at Anderson transitions}, Phys. Rev. B 87
  (2013) 125144.

\bibitem{hoef1986calculation}
D.~Höf, F.~Wegner, {Calculation of anomalous dimensions for the nonlinear
  sigma model}, Nuclear Physics B 275~(4) (1986) 561--579.

\bibitem{wegner1987anomalous1}
F.~Wegner, {Anomalous dimesions for the nonlinear sigma-model in 2 + $\epsilon$
  dimensions (I)}, Nuclear Physics B 280 (1987) 193--209.

\bibitem{wegner1987anomalous2}
F.~Wegner, {Anomalous dimensions for the nonlinear sigma-model, in 2 +
  $\epsilon$ dimensions (II)}, Nuclear Physics B 280 (1987) 210--224.

\bibitem{lee1996effects}
D.-H. Lee, Z.~Wang, {Effects of Electron-Electron Interactions on the Integer
  Quantum Hall Transitions}, Phys. Rev. Lett. 76 (1996) 4014--4017.

\bibitem{wang2000short-range}
Z.~Wang, M.~P.~A. Fisher, S.~M. Girvin, J.~T. Chalker, {Short-range
  interactions and scaling near integer quantum Hall transitions}, Phys. Rev. B
  61 (2000) 8326--8333.

\bibitem{burmistrov2011wave}
I.~S. {Burmistrov}, S.~{Bera}, F.~{Evers}, I.~V. {Gornyi}, A.~D. {Mirlin},
  {Wave function multifractality and dephasing at metal–insulator and quantum
  Hall transitions}, Annals of Physics 326~(6) (2011) 1457--1478.

\bibitem{evers2001multifractality}
F.~Evers, A.~Mildenberger, A.~D. Mirlin, {Multifractality of wave functions at
  the quantum Hall transition revisited}, Phys. Rev. B 64 (2001) 241303.

\bibitem{Obuse-Boundary-2008}
H.~{Obuse}, A.~R. {Subramaniam}, A.~{Furusaki}, I.~A. {Gruzberg}, A.~W.~W.
  {Ludwig}, {Boundary Multifractality at the Integer Quantum Hall Plateau
  Transition: Implications for the Critical Theory}, Phys. Rev. Lett. 101~(11)
  (2008) 116802.

\bibitem{evers2008multifractality}
F.~Evers, A.~Mildenberger, A.~D. Mirlin, {Multifractality at the Quantum Hall
  Transition: Beyond the Parabolic Paradigm}, Phys. Rev. Lett. 101 (2008)
  116803.

\bibitem{zirnbauer1994towards}
M.~R. Zirnbauer, {Towards a theory of the integer quantum Hall transition: From
  the nonlinear sigma model to superspin chains}, Annalen der Physik
  506~(7‐8) (1994) 513--577.

\bibitem{zirnbauer1997toward}
M.~R. Zirnbauer, {Toward a theory of the integer quantum Hall transition:
  Continuum limit of the Chalker–Coddington model}, Journal of Mathematical
  Physics 38~(4) (1997) 2007--2036.

\bibitem{janssen1999point-contact}
M.~Janssen, M.~Metzler, M.~R. Zirnbauer, {Point-contact conductances at the
  quantum Hall transition}, Phys. Rev. B 59 (1999) 15836--15853.

\bibitem{zirnbauer1999conformal}
M.~R. Zirnbauer, {Conformal field theory of the integer quantum Hall plateau
  transition} (1999).
\newblock \href {http://arxiv.org/abs/hep-th/9905054}
  {\path{arXiv:hep-th/9905054}}.

\bibitem{kettemann1999information}
S.~Kettemann, A.~Tsvelik, {Information about the Integer Quantum Hall
  Transition Extracted from the Autocorrelation Function of Spectral
  Determinants}, Phys. Rev. Lett. 82 (1999) 3689--3692.

\bibitem{bhaseen2000towards}
M.~J. Bhaseen, I.~I. Kogan, O.~A. Soloviev, N.~Taniguchi, A.~M. Tsvelik,
  {Towards a field theory of the plateau transitions in the integer quantum
  Hall effect}, Nuclear Physics B 580~(3) (2000) 688--720.

\bibitem{tsvelik2001wave}
A.~M. Tsvelik, {Wave Functions Statistics at Quantum Hall Critical Point}
  (2001).
\newblock \href {http://arxiv.org/abs/cond-mat/0112008}
  {\path{arXiv:cond-mat/0112008}}.

\bibitem{tsvelik2007evidence}
A.~M. Tsvelik, Evidence for the psl(2\ensuremath{\mid}2)
  wess-zumino-novikov-witten model as a model for the plateau transition in the
  quantum hall effect: Evaluation of numerical simulations, Phys. Rev. B 75
  (2007) 184201.

\bibitem{zirnbauer2019integer}
M.~R. Zirnbauer, {The integer quantum Hall plateau transition is a current
  algebra after all}, Nuclear Physics B 941 (2019) 458--506.

\bibitem{bondesan2017gaussian}
R.~Bondesan, D.~Wieczorek, M.~Zirnbauer, {Gaussian free fields at the integer
  quantum Hall plateau transition}, Nuclear Physics B 918 (2017) 52--90.

\bibitem{kagalovsky1999quantum}
V.~Kagalovsky, B.~Horovitz, Y.~Avishai, J.~T. Chalker, {Quantum Hall Plateau
  Transitions in Disordered Superconductors}, Phys. Rev. Lett. 82 (1999)
  3516--3519.

\bibitem{senthil1999spin}
T.~Senthil, J.~B. Marston, M.~P.~A. Fisher, {Spin quantum Hall effect in
  unconventional superconductors}, Phys. Rev. B 60 (1999) 4245--4254.

\bibitem{gruzberg1999exact}
I.~A. Gruzberg, A.~W.~W. Ludwig, N.~Read, {Exact Exponents for the Spin Quantum
  Hall Transition}, Phys. Rev. Lett. 82 (1999) 4524--4527.

\bibitem{beamond2002quantum}
E.~J. Beamond, J.~Cardy, J.~T. Chalker, {Quantum and classical localization,
  the spin quantum Hall effect, and generalizations}, Phys. Rev. B 65 (2002)
  214301.

\bibitem{mirlin2003wavefunction}
A.~D. Mirlin, F.~Evers, A.~Mildenberger, {Wavefunction statistics and
  multifractality at the spin quantum Hall transition}, Journal of Physics A:
  Mathematical and General 36~(12) (2003) 3255--3279.

\bibitem{evers2003multifractality}
F.~Evers, A.~Mildenberger, A.~D. Mirlin, {Multifractality at the spin quantum
  Hall transition}, Phys. Rev. B 67 (2003) 041303.

\bibitem{puschmann2021quartic}
M.~Puschmann, D.~Hernang\'omez-P\'erez, B.~Lang, S.~Bera, F.~Evers, Quartic
  multifractality and finite-size corrections at the spin quantum hall
  transition, Phys. Rev. B 103 (2021) 235167.

\bibitem{Duplantier-Multifractals-1991}
B.~{Duplantier}, A.~W.~W. {Ludwig}, {Multifractals, operator-product expansion,
  and field theory}, Physical Review Letters 66~(3) (1991) 247--251.

\bibitem{mirlin00}
A.~D. Mirlin, {Statistics of energy levels and eigenfunctions in disordered
  systems}, Physics Reports 326~(5) (2000) 259--382.

\bibitem{Yellow-book}
P.~Di~Francesco, P.~Mathieu, D.~S\'en\'echal, Conformal Field Theory,
  Springer-Verlag, 1997.

\bibitem{Nakayama-Scale-2015}
Y.~{Nakayama}, {Scale invariance vs conformal invariance}, Physics Reports 569
  (2015) 1--93.

\bibitem{Bondesan-Pure-2014}
R.~{Bondesan}, D.~{Wieczorek}, M.~R. {Zirnbauer}, {Pure Scaling Operators at
  the Integer Quantum Hall Plateau Transition}, Physical Review Letters
  112~(18) (2014) 186803.

\bibitem{Lewellen-Constraints-1989}
D.~C. {Lewellen}, {Constraints for conformal field theories on the plane:
  Reviving the conformal bootstrap}, Nuclear Physics B 320~(2) (1989) 345--376.

\bibitem{mildenberger2002dimensionality}
A.~Mildenberger, F.~Evers, A.~Mirlin, {Dimensionality dependence of the
  wave-function statistics at the Anderson transition}, Physical Review B
  66~(3) (2002) 033109.

\bibitem{rodriguez2008multifractal}
A.~Rodriguez, L.~J. Vasquez, R.~A. R\"omer, {Multifractal analysis of the
  metal-insulator transition in the three-dimensional Anderson model. II.
  Symmetry relation under ensemble averaging}, Phys. Rev. B 78 (2008) 195107.

\bibitem{tarquini2017critical}
E.~Tarquini, G.~Biroli, M.~Tarzia, {Critical properties of the Anderson
  localization transition and the high-dimensional limit}, Physical Review B
  95~(9) (2017) 094204.

\bibitem{suslov2016strict}
I.~M. Suslov, Strict parabolicity of the multifractal spectrum at the anderson
  transition, Journal of Experimental and Theoretical Physics 123~(5) (2016)
  845--850.

\bibitem{wegner1979the}
F.~Wegner, The mobility edge problem: continuous symmetry and a conjecture,
  Zeitschrift für Physik B 35 (1979) 207--210.

\bibitem{jungling1979effects}
K.~Jüngling, R.~Oppermann, Effects of spin interactions in disordered
  electronic systems: Loop expansions and exact relations among local gauge
  invariant models, Zeitschrift für Physik B 35 (1979) 207--210.

\bibitem{efetov1983supersymmetry}
K.~Efetov, Supersymmetry and theory of disordered metals, Advances in Physics
  32~(1) (1983) 53--127.

\bibitem{efetov1997supersymmetry}
K.~Efetov, Supersymmetry in Disorder and Chaos, Cambridge University Press,
  Cambridge,England, 1997.

\bibitem{verbaarschot1985grassmann}
J.~Verbaarschot, H.~Weidenmüller, M.~Zirnbauer, Grassmann integration in
  stochastic quantum physics: The case of compound-nucleus scattering, Physics
  Reports 129~(6) (1985) 367--438.

\bibitem{altland2000field}
A.~Altland, B.~D. Simons, D.~Taras~Semchuk, Field theory of mesoscopic
  fluctuations in superconductor-normal-metal systems, Advances in Physics
  49~(3) (2000) 321--394.

\bibitem{taras_semchuk2001quantum}
D.~Taras-Semchuk, A.~Altland, Quantum interference and the formation of the
  proximity effect in chaotic normal-metal/superconducting structures, Phys.
  Rev. B 64 (2001) 014512.

\bibitem{liao2017response}
Y.~Liao, A.~Levchenko, M.~S. Foster, Response theory of the ergodic many-body
  delocalized phase: Keldysh finkel’stein sigma models and the 10-fold way,
  Annals of Physics 386 (2017) 97--157.

\bibitem{Knapp-Lie-2002}
A.~W. Knapp, Lie Groups Beyond an Introduction, {Birkh\"auser}, Boston, Basel,
  Berlin, 2002.

\bibitem{burmistrov2016mesoscopic}
E.~V. Repin, I.~S. Burmistrov, Mesoscopic fluctuations of the single-particle
  green's function at anderson transitions with coulomb interaction, Phys. Rev.
  B 94 (2016) 245442.

\bibitem{Burmistrov-Magnetic-2018}
I.~S. Burmistrov, M.~A. Skvortsov, Magnetic disorder in superconductors:
  Enhancement by mesoscopic fluctuations, Phys. Rev. B 97 (2018) 014515.

\bibitem{mello1990averages}
P.~A. Mello, Averages on the unitary group and applications to the problem of
  disordered conductors, Journal of Physics A: Mathematical and General 23~(18)
  (1990) 4061--4080.

\bibitem{weingarten1978asymptotic}
D.~Weingarten, Asymptotic behavior of group integrals in the limit of infinite
  rank, Journal of Mathematical Physics 19~(5) (1978) 999--1001.

\bibitem{friedan1980nonlinear}
D.~Friedan, Nonlinear models in $2+\ensuremath{\epsilon}$ dimensions, Phys.
  Rev. Lett. 45 (1980) 1057--1060.

\bibitem{chalker1988percolation}
J.~T. Chalker, P.~D. Coddington, Percolation, quantum tunnelling and the
  integer hall effect, Journal of Physics C: Solid State Physics 21~(14) (1988)
  2665--2679.

\bibitem{kramer2005random}
B.~Kramer, T.~Ohtsuki, S.~Kettemann, Random network models and quantum phase
  transitions in two dimensions, Physics Reports 417~(5) (2005) 211--342.

\bibitem{klesse1995universal}
R.~Klesse, M.~Metzler, Universal multifractality in quantum hall systems with
  long-range disorder potential, Europhysics Letters ({EPL}) 32~(3) (1995)
  229--234.

\bibitem{Riva-Scale-2005}
V.~{Riva}, J.~{Cardy}, {Scale and conformal invariance in field theory: a
  physical counterexample [rapid communication]}, Physics Letters B 622~(3-4)
  (2005) 339--342.

\bibitem{Nakayama-Hidden-2016}
Y.~{Nakayama}, {Hidden global conformal symmetry without Virasoro extension in
  theory of elasticity}, Annals of Physics 372 (2016) 392--396.

\bibitem{Janssen-Multifractal-1994}
M.~{Janssen}, {Multifractal analysis of broadly-distributed observables at
  criticality}, International Journal of Modern Physics B 8~(8) (1994)
  943--984.

\bibitem{Dohmen-Disordered-1996}
A.~{Dohmen}, P.~{Freche}, M.~{Janssen}, {Disordered electrons in a strong
  magnetic field: transfer matrix approaches to the statistics of the local
  density of states}, Phys. Rev. Lett. 76~(22) (1996) 4207--4210.

\bibitem{Janssen-Statistics-1998}
M.~{Janssen}, {Statistics and scaling in disordered mesoscopic electron
  systems}, Physics Reports 295~(1) (1998) 1--91.

\bibitem{Merkt-Network-1998}
R.~{Merkt}, M.~{Janssen}, B.~{Huckestein}, {Network model for a two-dimensional
  disordered electron system with spin-orbit scattering}, Phys. Rev. B 58~(8)
  (1998) 4394--4405.

\bibitem{Obuse-Multifractality-2007}
H.~{Obuse}, A.~R. {Subramaniam}, A.~{Furusaki}, I.~A. {Gruzberg}, A.~W.~W.
  {Ludwig}, {Multifractality and conformal invariance at 2D metal-insulator
  transition in the spin-orbit symmetry class}, Phys. Rev. Lett. 98~(15) (2007)
  156802.

\bibitem{Obuse-Conformal-2010}
H.~{Obuse}, A.~R. {Subramaniam}, A.~{Furusaki}, I.~A. {Gruzberg}, A.~W.~W.
  {Ludwig}, {Conformal invariance, multifractality, and finite-size scaling at
  Anderson localization transitions in two dimensions}, Phys. Rev. B 82~(3)
  (2010) 035309.

\bibitem{cardy1985universal}
J.~L. Cardy, Universal amplitudes in finite-size scaling: generalisation to
  arbitrary dimensionality, Journal of Physics A: Mathematical and General
  18~(13) (1985) L757--L760.

\bibitem{Obuse-Corner-2008}
H.~{Obuse}, A.~R. {Subramaniam}, A.~{Furusaki}, I.~A. {Gruzberg}, A.~W.~W.
  {Ludwig}, {Corner multifractality for reflex angles and conformal invariance
  at 2D Anderson metal insulator transition with spin orbit scattering},
  Physica E Low-Dimensional Systems and Nanostructures 40~(5) (2008)
  1404--1406.

\bibitem{fulton_harris}
W.~Fulton, J.~Harris, Representation Theory: A First Course, Springer, New
  York, NY, 2004.

\end{thebibliography}

\end{document}